\RequirePackage[l2tabu, orthodox]{nag}

\documentclass[12pt,phd,a4paper,oneside]{ucl_thesis} 

\usepackage[utf8]{inputenc}

\usepackage[english]{babel} 
\usepackage{xspace} 
\usepackage{upgreek} 
\usepackage{emptypage}

\usepackage{tocloft}
\usepackage{setspace} 
\usepackage{draftwatermark}
\SetWatermarkText{\shortstack{\textsc{}\\ \strut \\ \strut \\ \strut}}
\SetWatermarkScale{0}
\SetWatermarkAngle{90}

\usepackage{ifdraft}
\ifdraft{
 \usepackage{draftwatermark}
 \SetWatermarkText{\shortstack{\textsc{Draft Mode}\\ \strut \\ \strut \\ \strut}}
 \SetWatermarkScale{0.5}
 \SetWatermarkAngle{90}
}

\usepackage{bibentry}
\makeatletter\let\saved@bibitem\@bibitem\makeatother
\let\OLDthebibliography\thebibliography
\renewcommand\thebibliography[1]{
  \OLDthebibliography{#1}
  \setlength{\parskip}{0pt}
  \setlength{\itemsep}{3pt plus 0.3ex}
}

\usepackage[pdftex,hidelinks]{hyperref} 
\usepackage{bookmark}  
\usepackage[printonlyused, withpage]{acronym} 

\usepackage{amsmath} 
\usepackage{subdepth} 
\usepackage{siunitx} 
\usepackage{mathtools} 
\usepackage{amsfonts} 
\usepackage{amssymb}  
\usepackage{gensymb} 
\usepackage{textcomp} 

\usepackage[capitalise]{cleveref} 
\Crefname{section}{Section}{Sections}
\Crefname{chapter}{Chapter}{Chapters}
\Crefname{equation}{Equation}{Equations}
\Crefname{figure}{Figure}{Figures}
\Crefname{tabular}{Table}{Tables}
\usepackage{apptools}
\crefname{subappendix}{\IfAppendix{Section}{Appendix}}{\IfAppendix{Sections}{Appendices}s} 
\usepackage{notoccite} 

\usepackage{graphicx} 
\usepackage{float} 
\usepackage[format=hang,font=small,labelfont=bf]{caption} 
\usepackage{sidecap} 
\usepackage{subfig} 
\usepackage{array} 
\usepackage{multirow} 
\usepackage{booktabs}
\usepackage{tabularx} 
\newcolumntype{L}{>{\raggedright\arraybackslash}X}

\usepackage{fancyhdr} 
\makeatletter
\g@addto@macro\bfseries\boldmath 
\makeatother

\usepackage{dirtytalk} 
\usepackage{nicefrac} 

\usepackage{listings}
\usepackage{color}
\definecolor{codegreen}{rgb}{0,0.6,0}
\definecolor{codegray}{rgb}{0.5,0.5,0.5}
\definecolor{codepurple}{rgb}{0.58,0,0.82}
\definecolor{backcolour}{rgb}{0.95,0.95,0.92}
\lstdefinestyle{mystyle}{
    backgroundcolor=\color{backcolour},   
    commentstyle=\color{codegreen},
    keywordstyle=\color{magenta},
    numberstyle=\tiny\color{codegray},
    stringstyle=\color{codepurple},
    basicstyle=\footnotesize,
    breakatwhitespace=false,         
    breaklines=true,                 
    captionpos=b,                    
    keepspaces=true,                 
    numbers=left,                    
    numbersep=5pt,                  
    showspaces=false,                
    showstringspaces=false,
    showtabs=false,                  
    tabsize=2
}
\newcommand{\gm}[1]{$g-{#1}$\xspace}

\newcommand{\R}[1]{Run-{#1}\xspace}
\newcommand{\tms}{{\SI{30}{\micro\second}}\xspace}
\newcommand{\tdr}{{technical design report}\xspace}
\newcommand{\fom}{{figure-of-merit}\xspace}

\newcommand{\eV}{\text{e\kern-0.15ex V}\xspace}
\newcommand{\MeV}{\text{M\eV}\xspace}
\newcommand{\GeV}{\text{G\eV}\xspace}
\newcommand{\TeV}{\text{T\kern-0.1ex \eV}\xspace}
\newcommand{\pval}{{$p\textnormal{-value}$}\xspace}
\newcommand{\pvals}{{$p\textnormal{-values}$}\xspace}

\newcommand{\cpp}{{\texttt{C++}}\xspace}

\newcommand{\mpt}{{\texttt{Millepede-II}}\xspace}
\newcommand{\pede}{\texttt{PEDE}\xspace}
\newcommand{\python}{{\texttt{Python}}\xspace}
\newcommand{\mille}{{\texttt{Mille}}\xspace}

\newcommand{\art}{{\textit{art}}\xspace}

\newcommand{\urlNewWindow}[1]{\href[pdfnewwindow=true]{#1}{\nolinkurl{#1}}}

\begin{document}

\graphicspath{{fig/}}

\title{Alignment of the straw tracking detectors for the Fermilab Muon \texorpdfstring{\gm2}~experiment and systematic studies for a muon electric dipole moment measurement}

\author{\textbf{\LARGE{Gleb Lukicov}}}
\department{Department of Physics and Astronomy}
\pdfbookmark[section]{Titlepage}{title} 
\maketitle
\makedeclaration 

\pdfbookmark[section]{Abstract}{abstract}
\begin{abstract} 
The Fermilab Muon \gm2 experiment is currently preparing for its fourth data-taking period (\R4). The experiment-wide effort on the analysis of \R1 data is nearing completion, with the announcement of the first result expected in the coming months. The final goal of the experiment is to determine the muon magnetic anomaly, $a_{\mu}=\frac{g-2}{2}$, to a precision of 140 ppb. This level of precision will provide indirect evidence of new physics, if the central value agrees with the previously-measured value of $a_{\mu}$. Essential in reducing the systematic uncertainty on $a_{\mu}$, through measurements of the muon beam profile, are the in-vacuum straw tracking detectors. A crucial prerequisite in obtaining accurate distributions of the beam profile is the internal alignment of the tracking detectors, which is described in this thesis. As a result of this position calibration, the tracking efficiency has increased by $3\%$, while the track quality increased by $4\%$.

This thesis also discusses an additional measurement that will be made using the tracking detectors: a search for an electric dipole moment (EDM) of the muon, through the direct detection of an oscillation in the average vertical angle of the $e^+$ from the $\mu^+$ decay. An observation of a muon EDM would be evidence of new physics and would provide a new source of CP violation in the charged lepton sector. Essential in measuring the EDM, as well as $a_{\mu}$, are accurate and precise estimations of potential non-zero radial and longitudinal magnetic fields, which were estimated using the \R1 data. In addition, a preliminary analysis using the \R1 data was undertaken to estimate the available precision for the $a_{\mu}$ measurement using the tracking detectors.
\end{abstract}

\pdfbookmark[section]{Impact statement}{impact}
\begin{impact}
The results of the fundamental research increase humanity's understanding of the Universe and are of great value for society. The Fermilab \gm2 experiment will make two measurements to establish the presence of new physics -- new particles or forces. The work described in this thesis has had a direct contribution in making these measurements possible. Should signs of new physics be observed, it would result in the first fundamental discovery in the field of experimental particle physics since 2012, and would set a clear pathway for future experiments to explore this new potential phenomenon more rigorously.

As has been the case in the past, fundamental discoveries in particle physics were later able to significantly benefit humanity technologically, beyond their original scope of research. For example, proton therapy is now an established procedure to cure cancer, with its origin rooted in particle accelerators. Moreover, an emerging challenge for a truly globalised world of the 21st century is global security. Notably, research into muons allowed for the development of muon tomography, enabling to improve the detection of nuclear materials. It remains to be seen what technological advancements will be enabled by new physics.

The technological expertise developed during this thesis is directly applicable to tackle one of the four Grand Challenges\footnote{HM Government, \textit{Industrial Strategy White Paper} (2017).} to put the UK at the forefront of the industries of the future -- AI and Data Economy. By 2030, the AI industry is projected to expand the UK economy by 22\%\footnote{McKinsey Global Institute, \textit{Artificial intelligence in the United Kingdom} (2019).}.


\end{impact}

\pdfbookmark[section]{Acknowledgements}{acknowledgements}
\begin{acknowledgements}
First and foremost, I would like to express my deepest gratitude to my supervisor, \textbf{Mark Lancaster}, who, through his outstanding leadership, made my PhD journey a truly remarkable experience. Mark, I greatly value all the advice, support, and encouragement that you have given me over the years -- thank you!
\vspace{0.25cm}

A special recognition should be given to \textbf{Rebecca Chislett}, \textbf{Joe Price}, and \textbf{James Mott}. I am truly grateful for the help that you have provided in tackling many of the arising challenges in this thesis, and for investing your time in me. Your ability to solve difficult tasks with ease and a friendly approach serves as an inspiration to me.
\vspace{0.25cm}

\textbf{Nicholas Kinnaird} and \textbf{Saskia Charity}, I appreciate your time in answering many of my questions on tracking and data analysis, especially when you were busy finishing your own theses. You both have set a tremendous example for me to follow. 
\vspace{0.25cm}

I would like to thank the members of the \gm2 tracker group: \textbf{Brendan Casey}, \textbf{Tammy Walton}, \textbf{Alessandra Luca}, and others, for creating a fantastic environment to collaborate and share ideas. Past and present members of the UCL \gm2 team have also provided ample support and inspiration to me: \textbf{Tom Stuttard}, \textbf{Gavin Hesketh}, \textbf{Samer Al-Kilani}, \textbf{Matt Warren}, \textbf{Erdem Motuk}, and \textbf{Dominika Vasilkova}.

\clearpage

Moreover, I would like to thank the entire \gm2 collaboration: I could never have imagined to have a group of such talented, hard-working, attentive and friendly colleagues is possible -- you are all absolutely brilliant! In particular, I would like to thank -- for many exciting discussions, interactions and shenanigans -- \textbf{Sam Grant}, \textbf{Sophie Middleton}, \textbf{Sean Foster}, \textbf{Will Turner}, \textbf{Tabitha Halewood-Leagas}, \textbf{Talal Albahri}, \textbf{Meghna Bhattacharya}, \textbf{Sudeshna Ganguly}, \textbf{Jarek Kaspar}, \textbf{Anna Driutti}, \textbf{Chris Polly}, \textbf{Karie Badgley}, \textbf{Mandy Rominsky}, and \textbf{Brendan Kiburg}.
\vspace{0.25cm}

A big thanks to \textbf{Leah Welty-Rieger} and the \say{\gm2 Tevatron ring-run crew}, as well as coach \textbf{Sonny Ross}, at the Prime Muay Thai club. The intense exercise provided a welcome distraction from the never-ending data analysis. I would also like to thank \textbf{Tania Claudette}, at the Fermilab's Frontier Pub, for the endless supply of IPAs and for playing \textit{the Black Keys} on demand!
\vspace{0.25cm}

Finally, I would like to thank my parents for their endless support, love, and understanding. It would not have been possible to get here without you, and I am truly fortunate to have you in my life.
\end{acknowledgements}
\clearpage

\pdfbookmark[section]{\contentsname}{toc}
\tableofcontents

\chapter*{List of acronyms}
\begin{acronym}\itemsep-30pt
\small
\acro{ASDQ}[ASDQ]{amplified shaper discriminator charge}
\acro{BE}[BE]{back-end}
\acro{BNL}[BNL]{Brookhaven National Laboratory}
\acro{CBO}[CBO]{coherent betatron oscillation}
\acro{CP}[CP]{charge parity}
\acro{DAQ}[DAQ]{data acquisition}
\acro{DCA}[DCA]{distance of closest approach}
\acro{DoF}[DoF]{degrees of freedom}
\acro{EDM}[EDM]{electric dipole moment}  
\acro{ESQ}[ESQs]{electrostatic quadrupoles}
\acro{FE}[FE]{front-end}
\acro{FFT}[FFT]{fast Fourier transform}
\acro{FLOBBER}[FLOBBER]{front-end low voltage optical box to back-end readout}
\acro{HV}[HV]{high voltage}
\acro{IoV}[IoV]{interval of validity}
\acro{LV}[LV]{low voltage}
\acro{LR}[LR]{left-right}
\acro{LSR}[LSR]{least squares regression}
\acro{NMR}[NMR]{nuclear magnetic resonance}
\acro{OSG}[OSG]{Open Science Grid}
\acro{POMS}[POMS]{Production Operations Management Service}
\acro{ppb}[ppb]{parts-per-billion}
\acro{ppm}[ppm]{parts-per-million}
\acro{SAM}[SAM]{sequential access via metadata}
\acro{SC}[SC]{slow control}
\acro{SD}[SD]{standard deviation}
\acro{SiPM}[SiPM]{silicon photomultiplier}
\acro{SM}[SM]{Standard Model}
\acro{TDC}[TDC]{time-to-digital converter}
\acro{VW}[VW]{vertical waist}
\normalsize
\addcontentsline{toc}{chapter}{List of acronyms}
\end{acronym}
\clearpage

\addcontentsline{toc}{chapter}{List of figures}
\listoffigures
\clearpage

\addcontentsline{toc}{chapter}{List of tables}
\listoftables

\graphicspath{{fig/}}

\chapter{Introduction}\label{ch:intro}

Despite many successes of the current theoretical framework, the \ac{SM}, that describes known fundamental particles and their interactions, there exist several unexplained phenomena in physics that motivate searches for new particles or forces. The last discovered fundamental particle, the Higgs boson, in 2012 further verified the predictive power of the \ac{SM}. However, the search for new physics remains highly motivated, as fundamental questions such as the origin of the universe's matter-antimatter asymmetry, the source of neutrino mass, and the origin of the dark matter remain unanswered. 

Many experiments are trying to discover signs of new physics. One such project is the Fermilab Muon \gm2 experiment, which will provide a stringent test of the \ac{SM} through a precise comparison of the theoretical prediction with an experimentally measured value. If the two values disagree, this would be indicative that the theory has not accounted for an effect seen by the experiment, providing indirect evidence of new physics. The measurement of interest is the \mbox{so-called} muon magnetic anomaly, $a_{\mu}$. The current world's best measurement of $a_{\mu}$ at \ac{BNL}~\cite{BNL_AMM} yielded a discrepancy between the theoretically predicted and experimentally measured values of more than $3\sigma$: a possible indication of new physics, but below the $5 \sigma$ \say{discovery threshold}. The Fermilab Muon \gm2 experiment will determine $a_{\mu}$ to a precision of 140 \ac{ppb}, sufficient to establish the presence of new physics at a significance of $6\sigma$ should the same central value be measured. 

Additionally, the experiment will search for a muon \ac{EDM}, with at least a factor of 10 improvement in sensitivity compared to the \ac{BNL} \gm2 experiment. An observation of a muon \ac{EDM} would provide a new source of \ac{CP} violation in the charged lepton sector -- a potential candidate to explain the matter-antimatter asymmetry of the universe.

This thesis describes the author's contributions to the Fermilab Muon \gm2 experiment in making these measurements possible. \cref{ch:theory} gives a brief overview of the theory of the two dipole moments of the muon. The Fermilab Muon \gm2 experiment and the methodology of the measurements are presented in \cref{ch:experiment}. \cref{ch:daq,ch:tracker} give a more detailed description of the data acquisition system and the tracking detector. The estimation of the systematic uncertainty on the beam measurements due to misalignment of the tracking detector is presented in \cref{ch:align_error}. The internal alignment strategy and results are given in \cref{ch:align}. \cref{ch:align_curvature} gives the results of constraining the radial curvature of the tracking detector. A preliminary $a_{\mu}$ analysis with data from the tracking detector is presented in \cref{ch:wiggle}. \cref{ch:edm} contains analysis work on the search for a muon \ac{EDM}. Finally, in \cref{ch:end} the results achieved in this thesis are summarised. 

\small
\section{Personal contributions}
The work of hundreds of scientists, engineers, graduate students and interns over more than a decade went into developing, building, and running the experiment, as well as the processing and analysing of the experimental data. Any work presented in this thesis that was not made by the author is indicated and cited. The author's direct contribution to the success of the experiment is listed below, for clarity:
\begin{enumerate} 
    \item \textbf{Systematic studies for the EDM and $a_{\mu}$ measurements.} The residual radial and longitudinal magnetic fields were estimated using data from the tracking detector. This work is presented in \cref{ch:edm}, with the produced analysis code available~\cite{Gleb_git_2}. Moreover, the first effort to begin the $a_{\mu}$ analysis with data from the tracking detector was made in \cref{ch:wiggle}.
    \vspace{-0.35cm}
    \clearpage
    
    \item \textbf{Internal alignment of the tracking detector.} The internal alignment, of the two tracker stations in Run-1 and Run-2, has been successfully determined. An alignment manual~\cite{Gleb_manual} allowing future alignment determinations, has been produced. The derived alignment constants were written into a \verb!PostgreSQL! database, where each set of constants is associated with a given range of runs. This work was presented at the American Physical Society division conference, with an associated conference proceeding paper~\cite{Gleb_paper}. The produced analysis code is also available~\cite{Gleb_git_1}. This work is presented in \cref{ch:align}.

    \item \textbf{Alignment contribution to the measurement of the spatial and temporal distribution of the stored beam.}  If the internal misalignment of the tracking detector is not determined and corrected for, there is a systematic effect on the measurement of the beam profile. The uncertainty on the mean radial and vertical extrapolated beam positions were estimated in \cref{ch:align_error}.

    \item \textbf{Estimation of the radial curvature of the tracking detector and its contribution to the measurements of the stored beam.} The accuracy of beam position determination is affected by detector effects, such as the radial detector curvature. This effect was estimated in \cref{ch:align_curvature}.
    
    \item \textbf{Data reconstruction on the UK grid.} To speed-up the experiment-wide data reconstruction effort, the track reconstruction code was configured to run on grid resources in the UK, as detailed in \cref{sub:track_production_in_the_uk}.

    \item \textbf{Data acquisition system (DAQ) expert on-call.} To support the smooth operation of the experiment and ensure continuous data taking, a team of DAQ experts are available for 24/7 support. The author actively participated as the DAQ on-call expert during \R1 and \R2. The DAQ of the \gm2 experiment is described in \cref{ch:daq}.

    \item \textbf{Tracking detector testing.} The tracking detector was developed by the University of Liverpool, with the UCL \gm2 team developing the tracker DAQ; both were subsequently installed and tested at Fermilab. Some of this work is described in \cref{ch:tracker}.
\normalsize

\end{enumerate}

\graphicspath{{fig/}}

\chapter{Theory}\label{ch:theory}

The search for new physics in the charged lepton sector provides an exciting and promising avenue for discovery. In the \gm2 experiment, two such searches will be performed: a measurement of the muon magnetic anomaly and the search for a non-zero muon \ac{EDM}.

\section{Muon magnetic anomaly}\label{sc:mma}
The muon has an intrinsic magnetic dipole moment, $\boldsymbol{\mu}$, that is coupled to its spin~\cite{LDM}, $\boldsymbol{s}$, by the $g$-factor, $g_{\mu}$,   
\begin{equation}
    \boldsymbol{\mu}=g_\mu\left(\frac{e}{2m_{\mu}}\right)\boldsymbol{s},
    \label{eq:mdm}
\end{equation}
where $m_{\mu}$ is the mass of the muon and $e$ is the elementary charge. 
The Dirac equation predicts the value of $g_{\mu}$ to be exactly equal to 2. However, additional radiative corrections, due to the contribution of virtual particles, cause it to be slightly larger than 2. This difference, $a_{\mu}$, is defined as
\begin{equation}
a_{\mu}=\frac{g_{\mu}-2}{2},
\label{eq:AMM}
\end{equation}
and is known as the muon magnetic anomaly. Defining $a_{\mu}$ this way allows for \cref{eq:mdm} to be written in the form
\begin{equation}
\boldsymbol{\mu}=(1+a_{\mu}) \left(\frac{e}{m_{\mu}}\right)\boldsymbol{s},
\end{equation}
which clearly demonstrates the influence of the muon magnetic anomaly on the magnetic dipole moment.
\clearpage

\begin{figure}[htpb]
\centering
\subfloat[]{\includegraphics[height=1.2in]{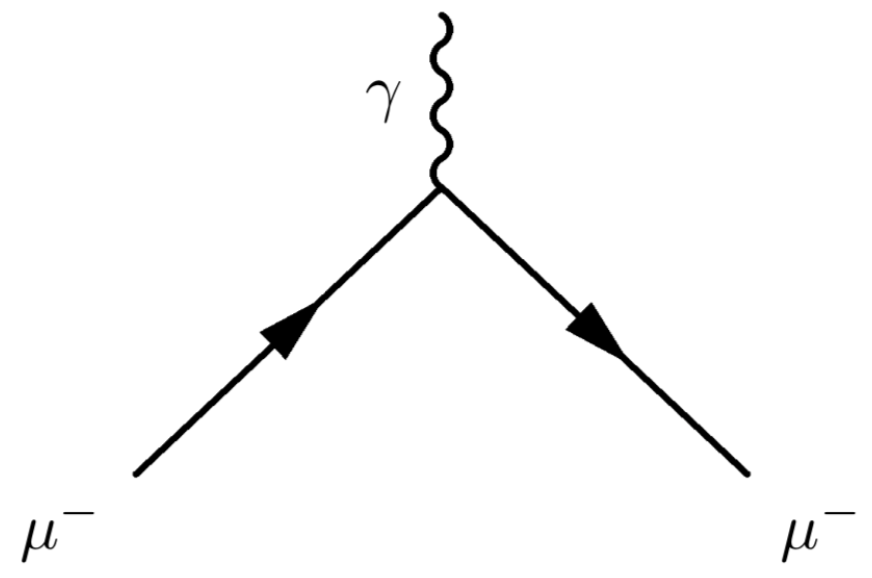}}
\subfloat[]{\includegraphics[height=1.2in]{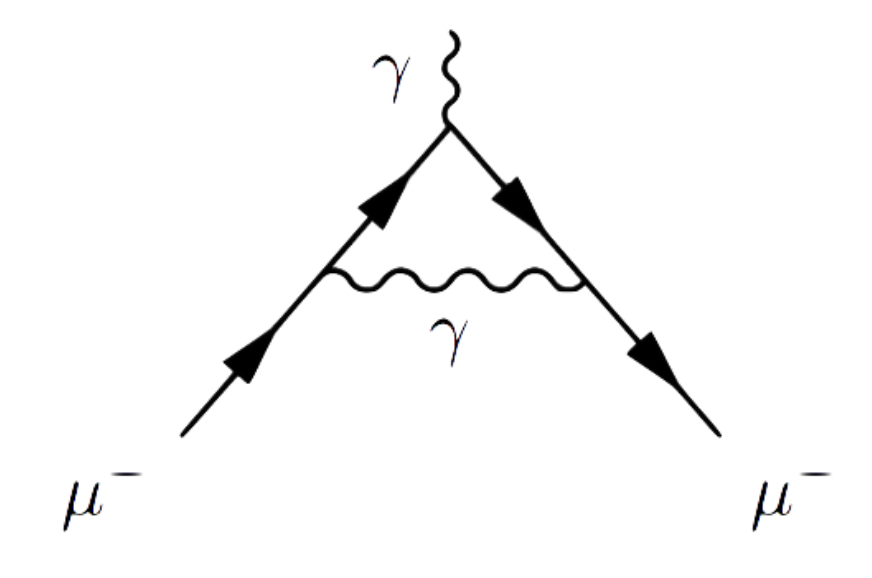}}
\subfloat[]{\includegraphics[height=1.2in]{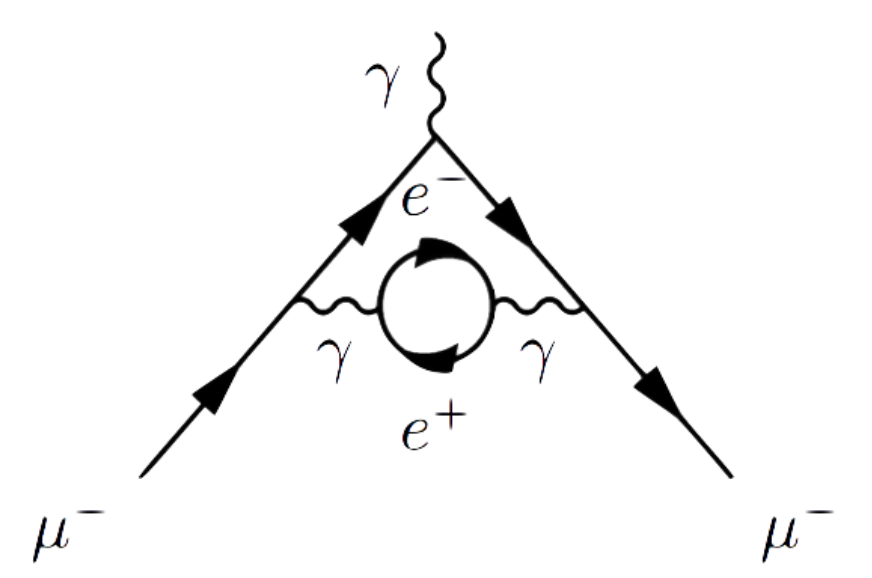}} \\
\subfloat[]{\includegraphics[height=1.2in]{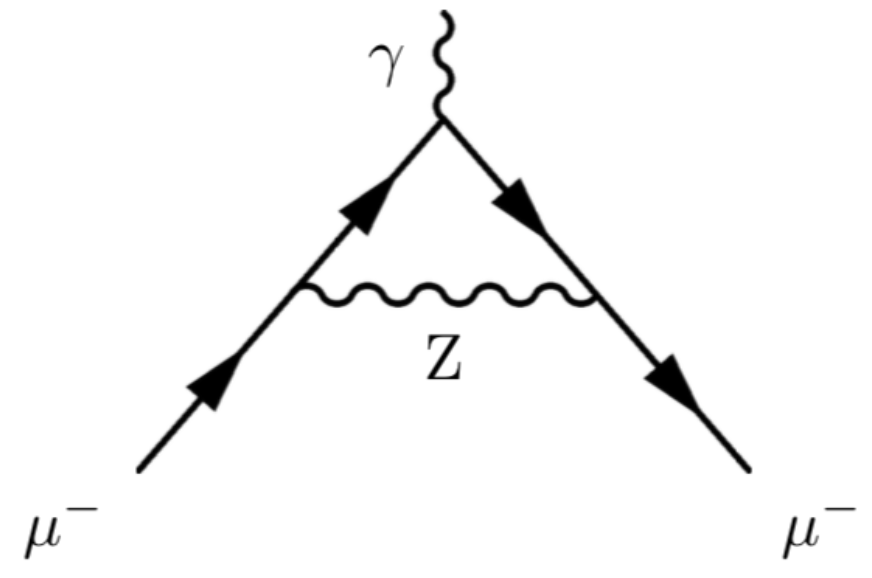}}
\subfloat[]{\includegraphics[height=1.2in]{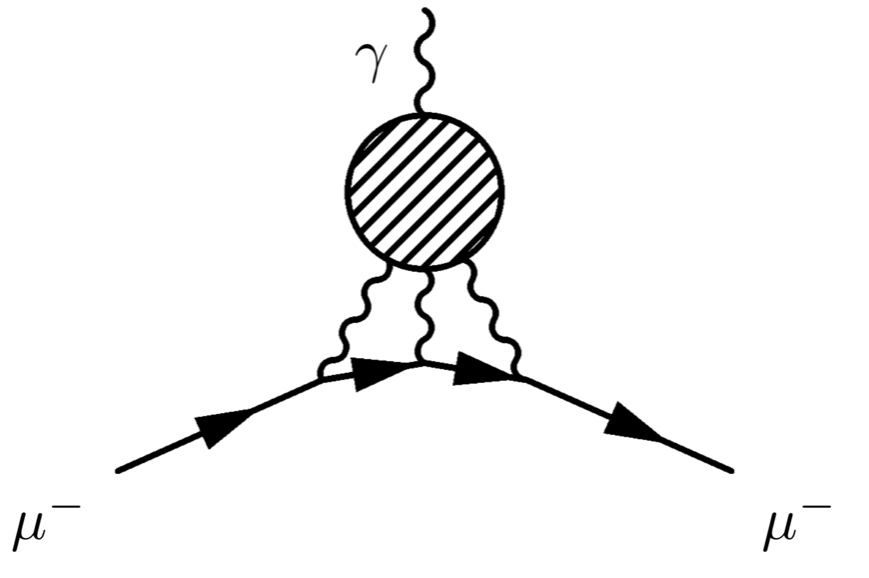}}
\subfloat[]{\includegraphics[height=1.2in]{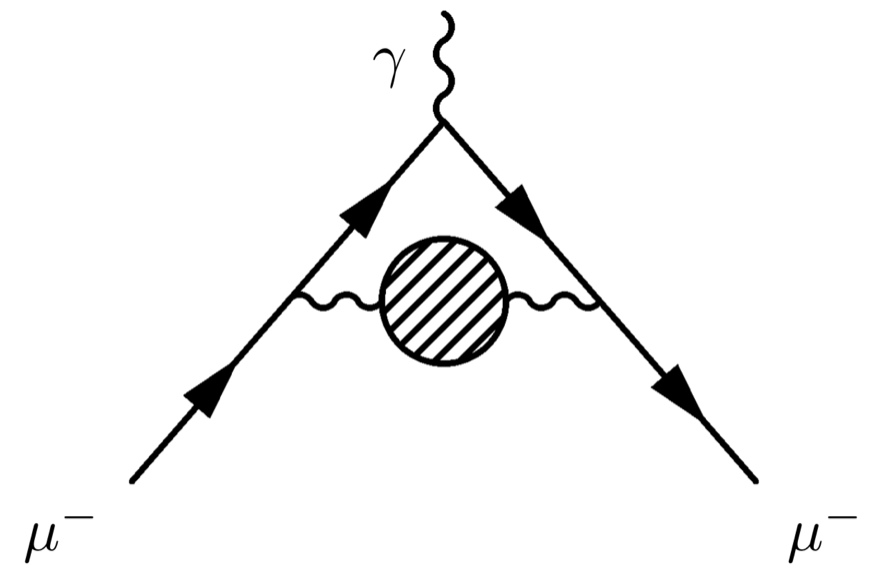}}
\caption[Examples of SM contributions to $a_{\mu}$]{Examples of SM contributions to $a_{\mu}$: (a) Dirac (b) Schwinger (c) higher order QED (d) electroweak (e) hadronic light-by-light (f) hadronic vacuum polarisation.}
\label{fig:feynam}
\end{figure}

The current theoretical calculations of $a_{\mu}$ is 0.00116591810(43)~\cite{TheoryInitiative}, with a precision of 369 \ac{ppb}. This calculation comprises several contributions:
\begin{equation}
a^{SM}_{\mu}=a^{\mathrm{QED}}_{\mu}+a^{\mathrm{EW}}_{\mu}+a^{\mathrm{Hadron}}_{\mu},
\label{eq:a}
\end{equation}
where $a^{\mathrm{QED}}_{\mu}$, $a^{\mathrm{EW}}_{\mu}$, and $a^{\mathrm{Hadron}}_{\mu}$ are contributions from the electromagnetic, electroweak, and hadron sectors respectively, as shown in \cref{fig:feynam}. 

\begin{figure}[htpb]
\centering
\includegraphics[height=5cm]{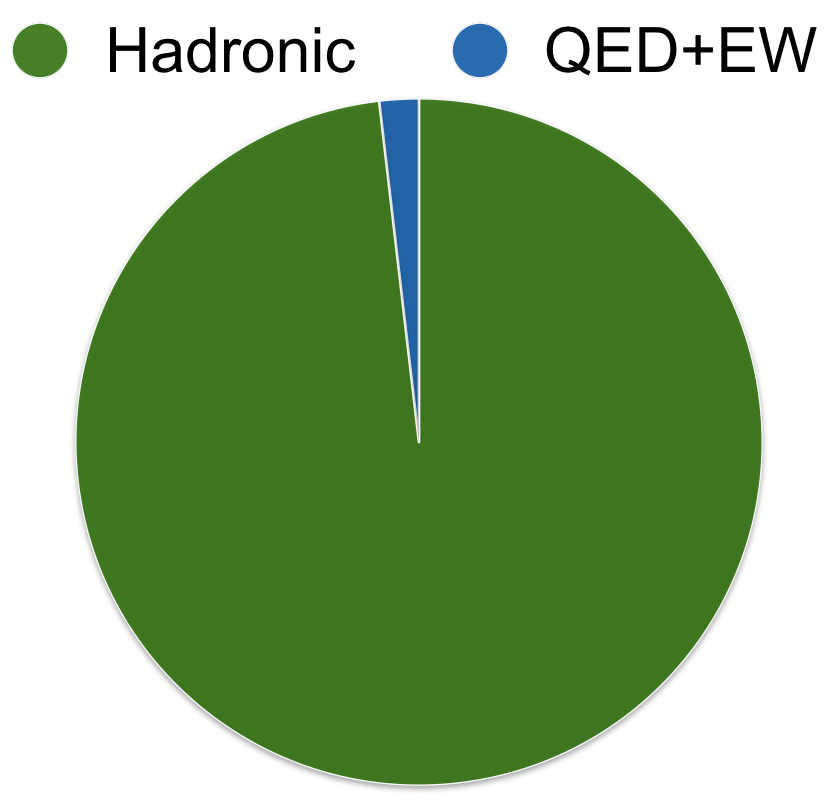}
\caption[The uncertainty on the theoretical prediction of $a_{\mu}$]{The uncertainty on the theoretical prediction of $a_{\mu}$. It is dominated by the hadronic interactions.}
\label{fig:error}
\end{figure}

$a^{\mathrm{QED}}_{\mu}$ is the dominant contribution, while the uncertainty is dominated by the hadronic part (lowest order hadronic vacuum polarisation and light-by-light interactions), as shown in \cref{fig:error}. A more detailed account of \ac{SM} contributions to $a_{\mu}$, and their corresponding theoretical uncertainties, is given by T. Aoyama et al.~\cite{TheoryInitiative} and A. Keshavarzi~\cite{Alex}. 
\clearpage

The theoretical accuracy of $a_{\mu}$ is comparable to the experimental accuracy of the previous Muon \gm2 experiment at \ac{BNL}, with the final result of 0.00116592091(63)~\cite{BNL_AMM}, having a precision of 540 ppb. The difference between the \ac{BNL} \gm2 experiment and theory is given by
\begin{equation}
\delta a_{\mu} = a_{\mu}^{\mathrm{experiment}} - a_{\mu}^{\mathrm{SM}} = 281(76) \times 10^{-11}
\end{equation}
which corresponds to a deviation between the experiment and the theory of 3.7$\sigma$.
To account for this discrepancy, a new physics correction ($a^{\mathrm{NP}}_{\mu}$) to $a^{\mathrm{SM}}_{\mu}$ may be required. One of such possible corrections, using an extension to the \ac{SM} known as supersymmetry, is discussed in~\cite{MSSM}.

\subsection{Muon's \texorpdfstring{$a_{\mu} \:$} ~sensitivity to new physics}
Other fundamental particles, such as an electron, also posses a magnetic dipole moment, which can be used in searches for new physics. However, the contribution of the virtual particles to the magnetic anomaly, $a^{\mathrm{NP}}_l$, of leptons scale with mass~\cite{NP_scale} according to
\begin{equation}
  a^{\mathrm{NP}}_l \sim \frac{m_l^2}{\Lambda^2},
\end{equation}
where $m_l$ is the mass of a lepton, and $\Lambda$ is the mass scale of new physics. Given the muon-to-electron mass ratio ($\nicefrac{m_{\mu}^2}{m_{e}^2}$) of $\mathcal{O}(10^4)$, the search for new physics with muons is much more effective than with electrons, given the relative experimental precision of these measurements. 
\clearpage

\begin{figure}[htpb]
  \centering
  \subfloat[]{\includegraphics[width=0.45\linewidth]{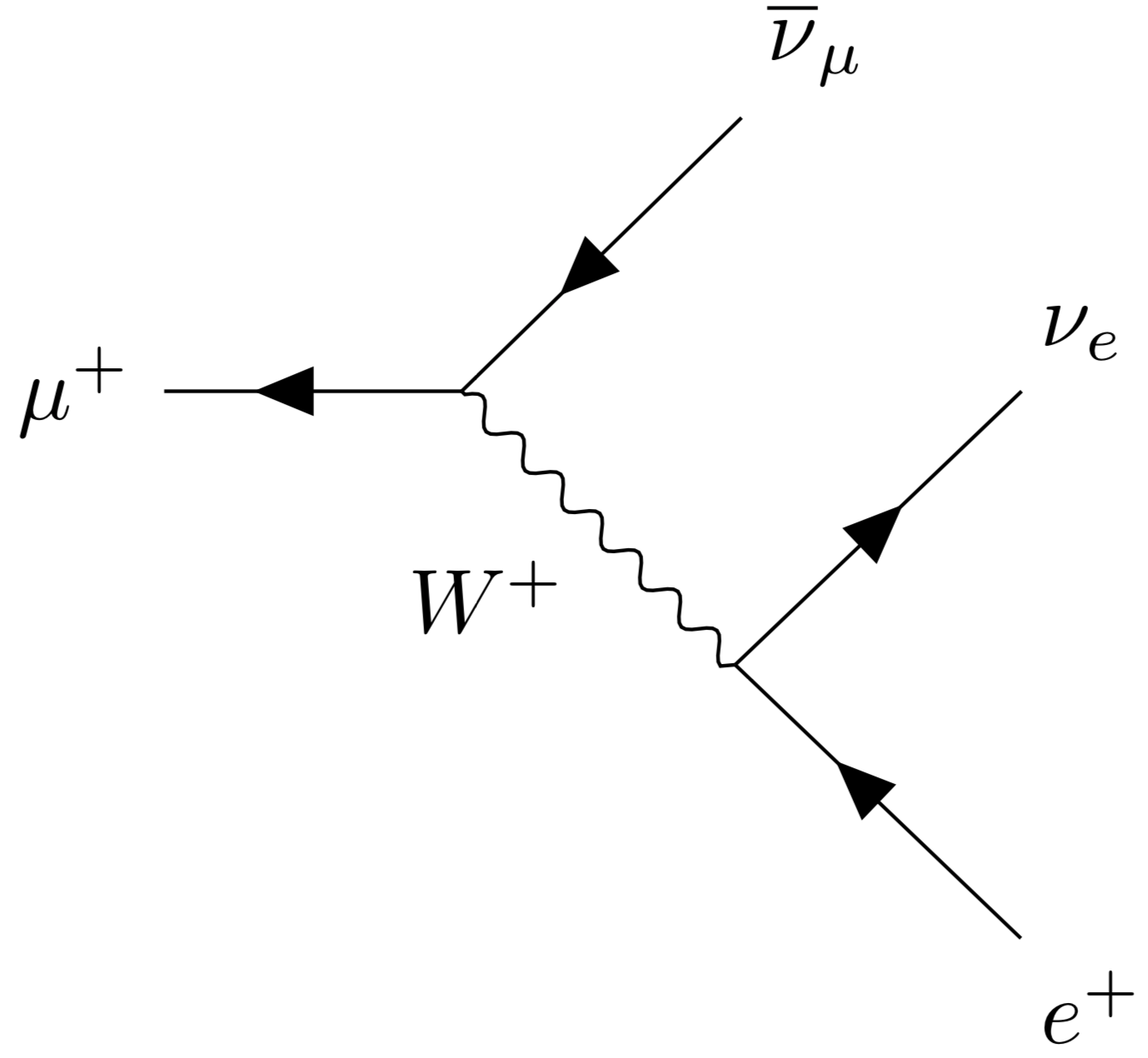}}
  \subfloat[]{\hspace*{0.05\linewidth}\raisebox{12mm}{\includegraphics[width=0.45\linewidth]{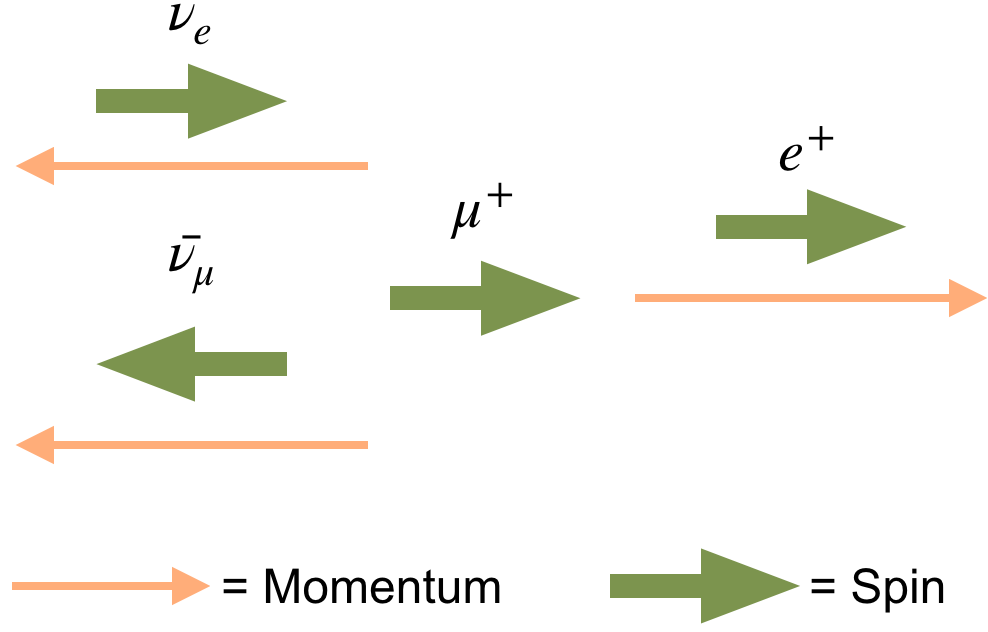}}}
  \caption[Decay of a positive muon]{(a) Feynman diagram for the decay of a positive muon. (b) Helicity diagram of positive muon decay in its rest frame. Bold arrows indicate spin, and long arrows indicate momentum.}
 \label{fig:mu_decay}
\end{figure}

What makes this search different, compared to the one with the electron, is the fact that the muon is an unstable particle. The muon decay channel  of interest to the experiment is $\mu^+ \rightarrow e^+\nu_e\bar{\nu_{\mu}}$. This process is illustrated in \cref{fig:mu_decay}.

The three-body decay of the muon in \cref{fig:mu_decay} provides information on the muon spin orientation at the time of the decay. The $V-A$ structure of the weak interaction means that the highest energy positrons from the decay of the $\mu^+$ are preferentially emitted along the direction of the muon spin.

\section{Electric dipole moment}\label{sec:edm} 
Analogous to $\boldsymbol{\mu}$ in \cref{eq:mdm}, a potential muon \ac{EDM}, $\boldsymbol{d_{\mu}}$, is defined by 
\begin{equation}
\boldsymbol{d_{\mu}}=\eta \left(\frac{e\hbar}{4m_{\mu}c}\right)\boldsymbol{s},
\label{eq:edm_1}
\end{equation}
where $\eta$ is a dimensionless constant~\cite{FNAL_TDR}, analogous to $g_\mu$ in \cref{eq:mdm}. 

The \ac{SM} prediction of the electron EDM, $\boldsymbol{d_{e}}$, is $10^{-38}~e\cdot$cm~\cite{EDM}. The scaling law~\cite{EDM_2} suggests that the corresponding value for the muon \ac{EDM} would be 
\begin{equation}
  \boldsymbol{d_{\mu}} \simeq \boldsymbol{d_{e}}\frac{m_{\mu}}{m_{e}}.
\end{equation}
This yields a value of $\boldsymbol{d_{\mu}}$ of $10^{-36}~e\cdot$cm, well below the current experimental reach. Hence, any observation of $\boldsymbol{d_{\mu}}$ would be evidence of NP. 

\clearpage
The goal of the Fermilab \gm2 experiment is to measure $\boldsymbol{d_{\mu}}$ with a sensitivity greater than $10^{-21}~e\cdot$cm~\cite{Becky}, more than a factor of 10 improvement on the current best experimental limit from the BNL experiment~\cite{BNL_EDM}.

The transformation properties of the dipole moments reveal an important conclusion. The Hamiltonian~\cite{LDM} for the muon in an applied magnetic, $\boldsymbol{B}$, and electric, $\boldsymbol{E}$, fields is given by
\begin{equation}
\mathcal{H} = -\boldsymbol{\mu}\cdot\boldsymbol{B}-\boldsymbol{d_{\mu}}\cdot\boldsymbol{E}.
\label{eq:H}
\end{equation}
Transformations under time reversal (T) are odd for axial vectors (i.e. $\boldsymbol{d}, \boldsymbol{\mu},$ and $\boldsymbol{B}$) and even for polar vectors (i.e. $\boldsymbol{E}$), while the reverse is true for charge (C) and parity (P) transformations, as shown in \cref{table:trans}.
\begin{table}[!ht]  
  \centering
  \begin{tabular}{c|ccc}
    \toprule
           & $\boldsymbol{E}$ & $\boldsymbol{B}$ & $\boldsymbol{\mu}$ or $\boldsymbol{d}$ \\ \midrule
    P  & $-$  & $+$ & $+$  \\ 
   C  & $-$  & $-$ & $-$  \\ 
   T  & $+$  & $-$ & $-$ \\ \bottomrule
  \end{tabular}
    \caption[Transformations of $\boldsymbol{E}$, $\boldsymbol{B}$, $\boldsymbol{\mu}$ and $\boldsymbol{d}$ under the P, C and T operators]{Transformations of $\boldsymbol{E}$, $\boldsymbol{B}$, $\boldsymbol{\mu}$ and $\boldsymbol{d}$ under the P, C and T operators: (+) denotes an even transformation and (-) an odd.}
\label{table:trans}
\end{table}

The first term in \cref{eq:H} is invariant under a T transformation
\begin{equation}
\mathcal{H_B} = -\boldsymbol{\mu}\cdot\boldsymbol{B} \rightarrow^\mathrm{T} -(-)\boldsymbol{\mu}\cdot-\boldsymbol{B} = -\boldsymbol{\mu}\cdot\boldsymbol{B}.
\end{equation}
 However, the second term is not 
 \begin{equation}
\mathcal{H_E} = -\boldsymbol{d}\cdot\boldsymbol{E} \rightarrow^\mathrm{T} -(-)\boldsymbol{d}\cdot\boldsymbol{E} = \boldsymbol{d}\cdot\boldsymbol{E}.
\end{equation}
Therefore, assuming an overall CPT invariance, T violation implies CP violation. Hence, an EDM measurement by the \gm2 experiment (above $10^{-36}~e\cdot$cm) would provide clear evidence of CP violation in the charged lepton sector. A more detailed discussion of the muon EDM is available from B. Roberts and Y. Semertzidis~\cite{LDM}. The methodology of the \ac{EDM} measurement with the tracking detector is given in \cref{sec:edm_track}.
\clearpage

\graphicspath{{fig/}}  

\chapter{The Muon \texorpdfstring{\gm2}~experiment at Fermilab}\label{ch:experiment}
This chapter describes the methodology of the measurement of the muon magnetic anomaly by the Fermilab \gm2 experiment. \cref{sc:g-2} provides an overview of the entire experiment, with a more detailed description of the beamline, storage ring components, spatial and temporal distribution of the stored beam, magnetic field, and detectors given in \cref{sc:beamline,sc:sr,sc:bd,sc:field,sc:det}. A more in-depth description of the \ac{DAQ} and the tracking detectors is given separately in \cref{ch:daq,ch:tracker}, respectively, while the methodology of the \ac{EDM} measurement with the tracking detector is given in \cref{sec:edm_track}.

The performance of the experiment in the data taking periods from 2017 to 2020 is discussed \cref{sc:runs}. The expectation for future data taking periods is also given. An overview of the experimental data reconstruction is given in \cref{sc:prod}, where data quality cuts are also discussed, along with an alternative implementation of tracker data reconstruction in the UK. Finally, the experiment's simulation framework is described in \cref{sec:simulation_framework}.

\section{Overview}\label{sc:g-2}
The principal goal of the Fermilab Muon \gm2 experiment is the reduction of the experimental uncertainty on the measurement of $a_{\mu}$, as compared with the \ac{BNL} experiment, using a measurement strategy similar to the one at the \ac{BNL}~\cite{BNL_AMM}. This reduction in the uncertainty on $a_{\mu}$ will be achieved via reductions in the statistical uncertainty due to the use of the Fermilab muon beam~\cite{Diktys}, and the systematic uncertainty due to improvements in the detectors, such as the in-vacuum tracking detectors~\cite{Tom}, and more finely segmented calorimeters~\cite{Jarek}, a better field uniformity~\cite{Mark}, and a laser calibration system~\cite{Anastasi}, as well as improved analysis techniques.

Achieving the experimental goal requires a dataset containing $1.6 \times 10^{11}$ positrons with energies above 1.8~\GeV. The momentum of the injected muons is 3.09~\GeV: the so-called \say{magic momentum} at which the effect of the motional magnetic field is zero (see \cref{sc:scraping}). This momentum corresponds to a boosted lifetime of the muon of \SI{64}{\micro\second}. To perform the measurement of $a_{\mu}$, two measurable quantities are of interest: the anomalous precession frequency, $\boldsymbol{\omega_a}$, and the applied magnetic field, $\boldsymbol{B}$. $\boldsymbol{\omega_a}$ is defined as the difference between the spin and cyclotron frequencies (see \cref{sc:scraping}). In the case of a uniform magnetic field and negligible betatron oscillations, the relationship between $a_{\mu}$, $\boldsymbol{\omega_a}$, and $\boldsymbol{B}$ can be written as follows
\begin{equation}
\boldsymbol{\omega_a}=a_{\mu}\frac{e}{m_{\mu}}\boldsymbol{B}=\left(\frac{g-2}{2}\right)\frac{e}{m_{\mu}}\boldsymbol{B}.
\label{eq:mub}
\end{equation}

\begin{figure}[htpb]
\centering
\subfloat[]{\hspace*{1.5cm}\includegraphics[width=0.748\linewidth]{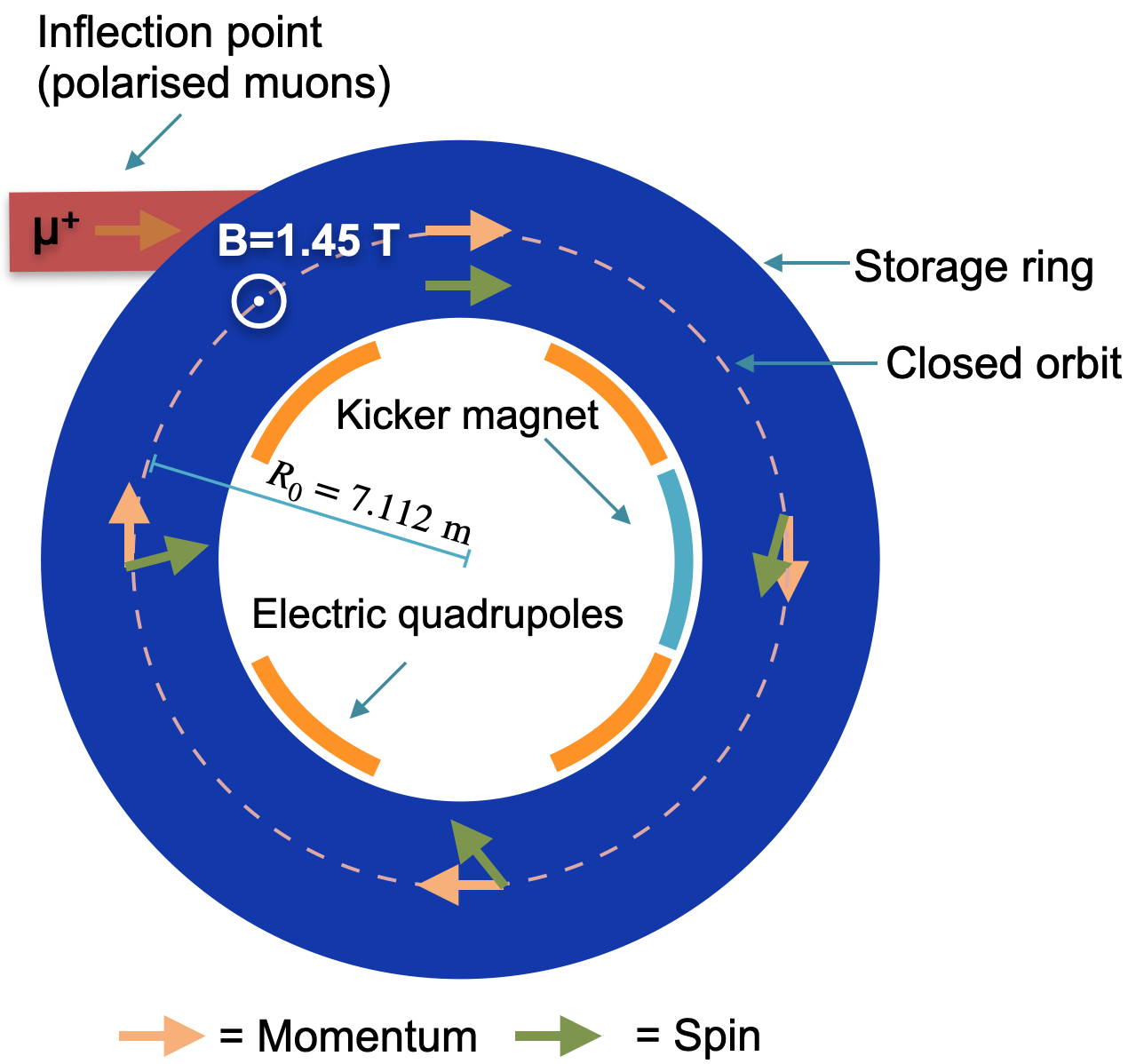}\label{fig:ring}} \\
\subfloat[]{\includegraphics[width=0.75\linewidth]{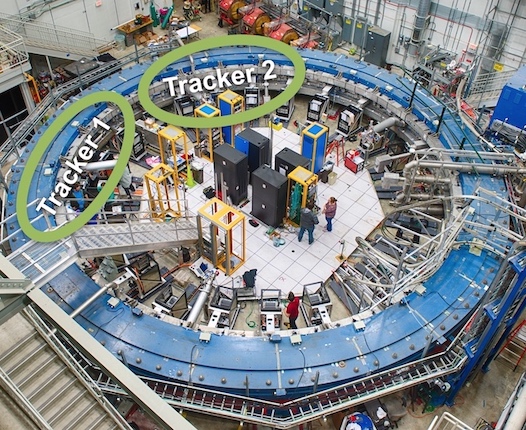}\label{fig:ring_pic}}
\caption[The \gm2 storage ring]{The \gm2 storage ring. (a) Superconducting magnets provide a uniform vertical magnetic dipole field of 1.45 T. The muon beam enters the storage ring via the inflector. The direction of the beam is clockwise in this orientation. (b) A photograph of the ring with the location of the two tracker stations indicated. The 24 calorimeter readout crates (shown in black) are distributed uniformly on the inside of the ring.}
\end{figure}

Longitudinally polarised muons are injected into the storage ring, as shown in \cref{fig:ring}, where they ultimately follow a circular orbit of mean radius, $R_0$, of 7.112~m in the magnetic dipole field and are vertically focused using \ac{ESQ}. The muon beam enters the ring through the inflector magnet, designed to cancel out the dipole magnetic field, at a slightly larger radius than the ideal orbit. The kicker magnets are used to deflect the beam onto the correct orbit.

\begin{figure}[htpb]
\centering
\subfloat[]{\raisebox{10mm}{\includegraphics[width=0.49\linewidth]{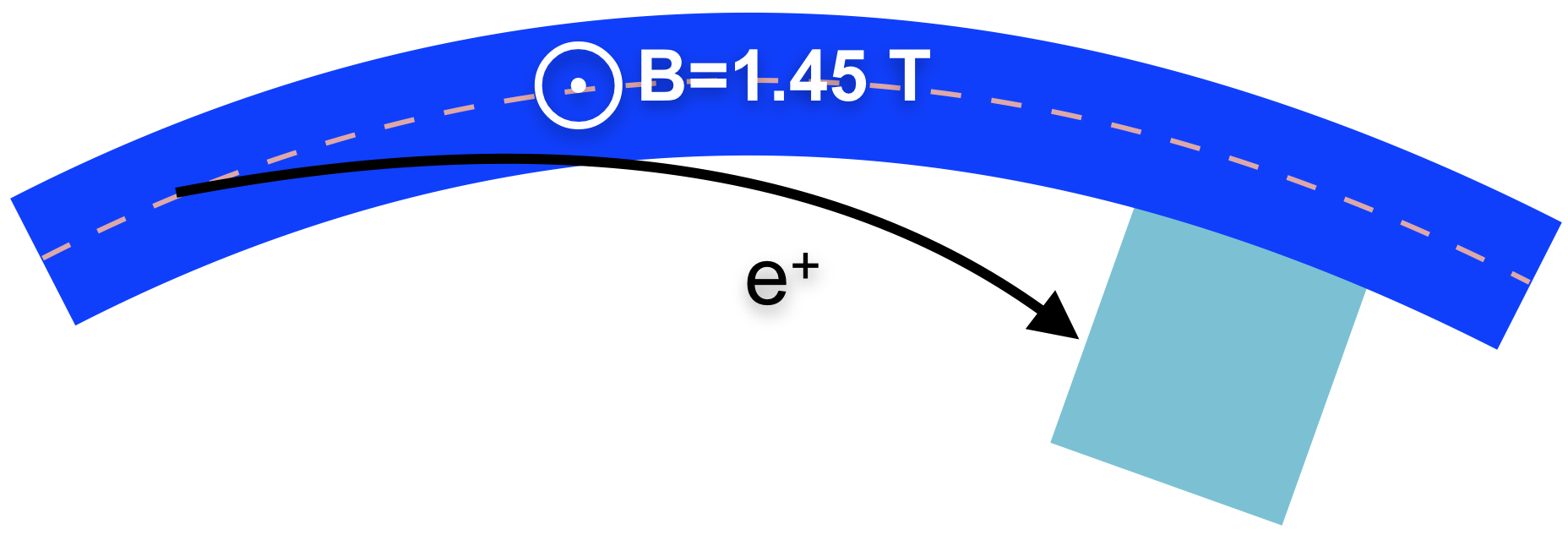}\label{fig:calo}}}
\subfloat[]{\includegraphics[width=0.49\linewidth]{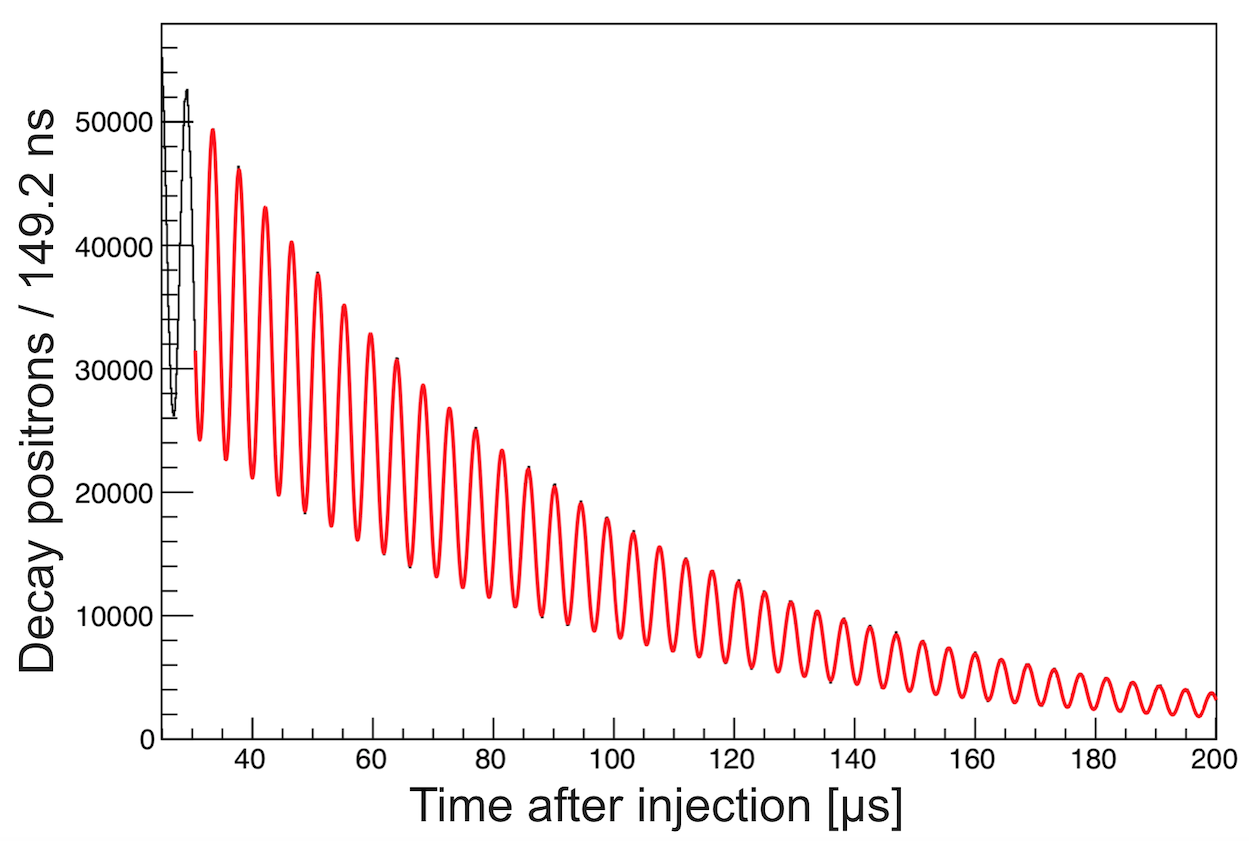}\label{fig:wiggle_example}}
\caption[Measurement of $\boldsymbol{\omega_a}$]{Measurement of $\boldsymbol{\omega_a}$: (a) a decay positron entering one of the 24 calorimeters, (b) $\boldsymbol{\omega_a}$ is extracted from a fit to calorimeter events after $\SI{30}{\micro\second}$, when the beam is stable. Plot (b) courtesy of J. Hempstead~\cite{Jason_wiggle}.}
\end{figure}

Since $a_{\mu} > 0 $, the muon spin precesses faster than the momentum vector in the storage ring. Due to the $V-A$ structure of the weak interaction, the highest-energy positrons, from the decay of the $\mu^+$, are preferentially emitted along the direction of the muon spin, as shown in \cref{fig:mu_decay}. The positrons curl inwards, where they can interact with one of the 24 calorimeters placed around the interior of the storage ring. This is shown in \cref{fig:calo}. High energy positron events are then histogrammed as a function of time to extract $\boldsymbol{\omega_a}$, as shown in \cref{fig:wiggle_example}.
\clearpage

\clearpage

The magnetic field is determined through a frequency measurement: the Larmor frequency of a free proton, $\omega_p$. It is measured using \ac{NMR} and the relationship 
\begin{equation}
\omega_p = \gamma_p\langle\boldsymbol{B}\rangle,
\label{eq:omega_p}
\end{equation}
where $\gamma_p$ is the proton gyromagnetic moment ratio~\cite{CODATA}. Using \cref{eq:omega_p}, $a_{\mu}$ can be expressed as a function of the two experimentally measured frequencies and well-determined ratios
\begin{equation}
a_{\mu} = \frac{\nicefrac{\omega_a}{\omega_p}}{\nicefrac{\mu_{\mu}}{\mu_p}-\nicefrac{\omega_a}{\omega_p}}=\frac{\omega_a}{\omega_p}\left(\frac{g_e}{2}\right)\left(\frac{m_{\mu}}{m_e}\right)\left(\frac{\mu_p}{\mu_e}\right),
\end{equation}
where $g_e$, $\nicefrac{\mu_p}{\mu_e}$, and $\nicefrac{m_{\mu}}{m_e}$ are known~\cite{CODATA} with uncertainties of 0.00026 ppb, 3.0 ppb, and 22 ppb, respectively.

The simple harmonic motion of the stored muon beam in the horizontal plane, known as the \ac{CBO}~\cite{CBO}, contribute significantly to the systematic uncertainty on $\boldsymbol{\omega_a}$. The fraction of events where a positron is detected by the calorimeters depends on the radius of the muon beam at the point of decay. The \ac{CBO} of the stored beam can thus produce an amplitude modulation in the observed positron time spectrum, as described in \cref{sec:CBO}. One of the aims of the tracking detectors is to measure the \ac{CBO}.
\clearpage

\section{Beamline}\label{sc:beamline}
\begin{figure}[htpb]
    \centering
    \includegraphics[width=0.72\linewidth]{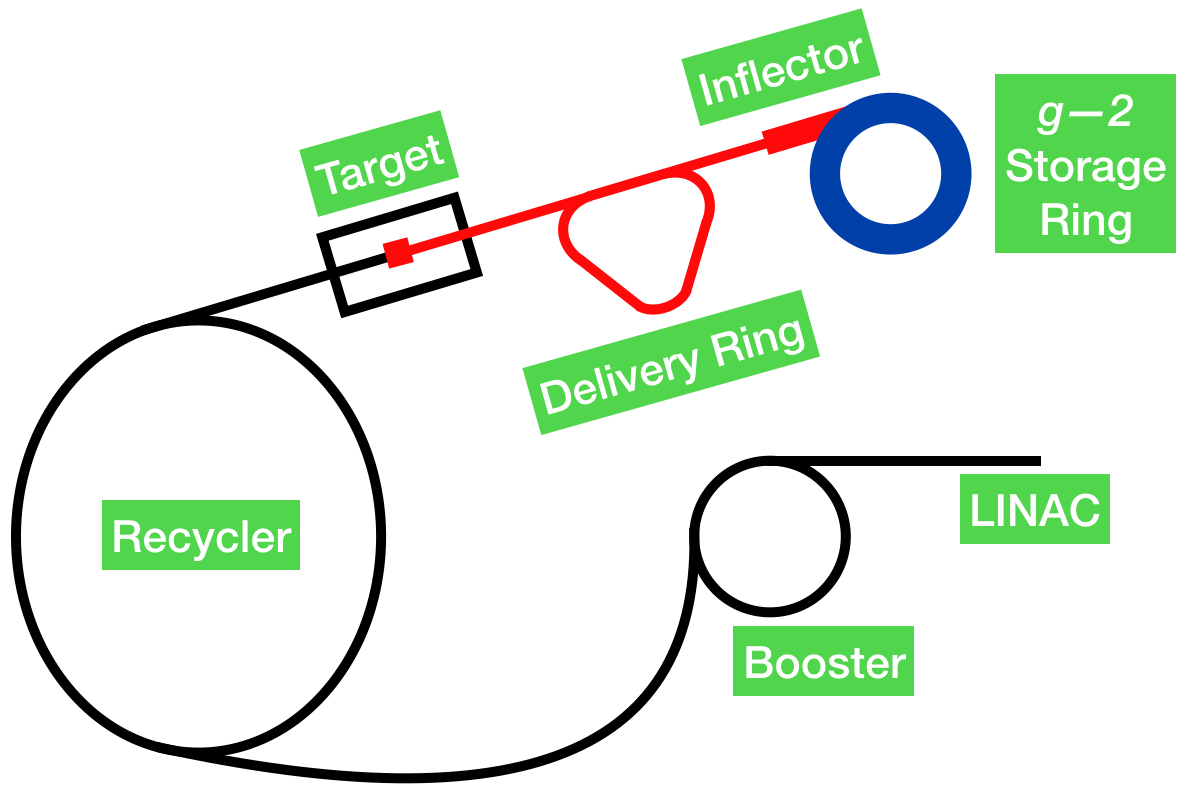}
    \caption[The accelerator complex at Fermilab]{The schematic above (not to scale) shows the accelerator complex at the Fermilab Muon Campus. The proton beamlines are shown in black, and secondary beamlines in red. The protons are accelerated through a linear accelerator (LINAC) and the Booster to reach 8.89~\GeV.}
    \label{fig:beamline}
\end{figure}

The aim of the accelerator complex, shown in \cref{fig:beamline}, is to deliver a beam of muons, which is largely free of contaminants (e.g. $p$, $e$, $d$, $\pi$), into the \gm2 storage ring. The muons in the \gm2 experiment come from pion decays
\begin{equation}
   \pi^{+}\rightarrow\mu^{+}+\nu_{\mu}, 
\end{equation}
which are produced by sending a proton beam with a momentum of 8.89~\GeV -- with $\mathcal{O}(10^{12})$ protons in a single \textit{bunch} of $\mathcal{O}(100)$~ns -- on an \textit{Inconel 600}~\cite{target} production target 
\begin{equation}
    p + p_{(target)} \rightarrow p + n + \pi^+.
\end{equation}
$\mu^+$ are used in the experiment, as the cross-section for $\pi^+$ production is greater than that for $\pi^-$. Each proton bunch delivered by the accelerator complex to the pion production target corresponds to a muon \textit{fill} in the \gm2 storage ring. The proton beam is bunched in the Recycler to fill the \gm2 storage ring optimally. The produced 3.11~\GeV pions, as well as any remaining protons, undergo five revolutions in the Delivery Ring. This is done to allow all pions to decay, and to separate muons from electrons and protons, as much as possible. Further details on the beamline are given by D. Stratakis et al.~\cite{Diktys}.
\clearpage

\section{Storage ring components}\label{sc:sr}
One muon fill ($\SI{700}{\micro\second}$) corresponds to a single accumulation of the muons and defines the measurement window for the experiment. The goal of the storage ring components described here is to keep the beam stable during this time. For a cyclotron period, $T_\mathrm{c}$, of 149~ns this corresponds to $\sim4,500$ revolutions of the beam around the ring.

\subsection{Inflector} \label{sc:inflector}
\begin{figure}[htpb]
\centering
\subfloat[]{\includegraphics[height=5.7cm]{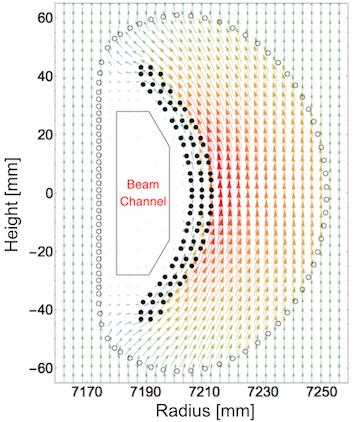}}
\subfloat[]{\hspace*{0.1cm}\includegraphics[height=5.7cm]{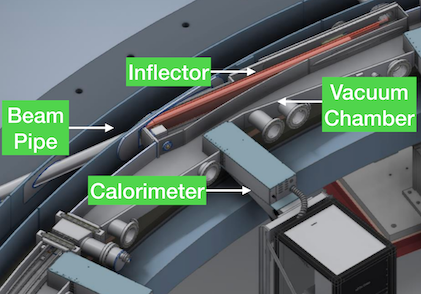}}
\caption[Inflector]{(a) Vector sum of inflector and storage-ring magnetic fields, through a cross-section of the inflector. The superposition creates a field-free region through which the injected muon beam enters the ring. Plot courtesy of N. Froemming et al.~\cite{inf1}. (b) A rendering of the inflector on the outside of the storage ring. The inflector is placed inside the vacuum chamber just before the beam pipe. Image courtesy of K. Badgley~\cite{inf2}.}
\label{fig:inf}
\end{figure}

The delivered muon beam to the \gm2 experiment passes through an inflector magnet which facilitates injection of the beam into the storage ring by cancelling the main magnetic dipole field (1.45~T) of the ring. The inflector provides a field-free region (beam channel) through which the injected muon beam passes without deflection, as shown in \cref{fig:inf}. This requirement on the inflector means that the beam is injected at a $1.25^{\circ}$ angle, meaning that the actual orbit is displaced by 77~mm radially outward from the closed orbit of the storage ring. The inflector is made of niobium-titanium-copper-aluminium superalloy, with a nominal design current of 2850~A. Further details of the inflector can be found in the \tdr~\cite{FNAL_TDR}.
\vspace{-0.2cm}
\clearpage

\subsection{Kicker} \label{sc:kicker}
\begin{figure}[htpb]
\centering
\subfloat[]{\includegraphics[height=4.8cm]{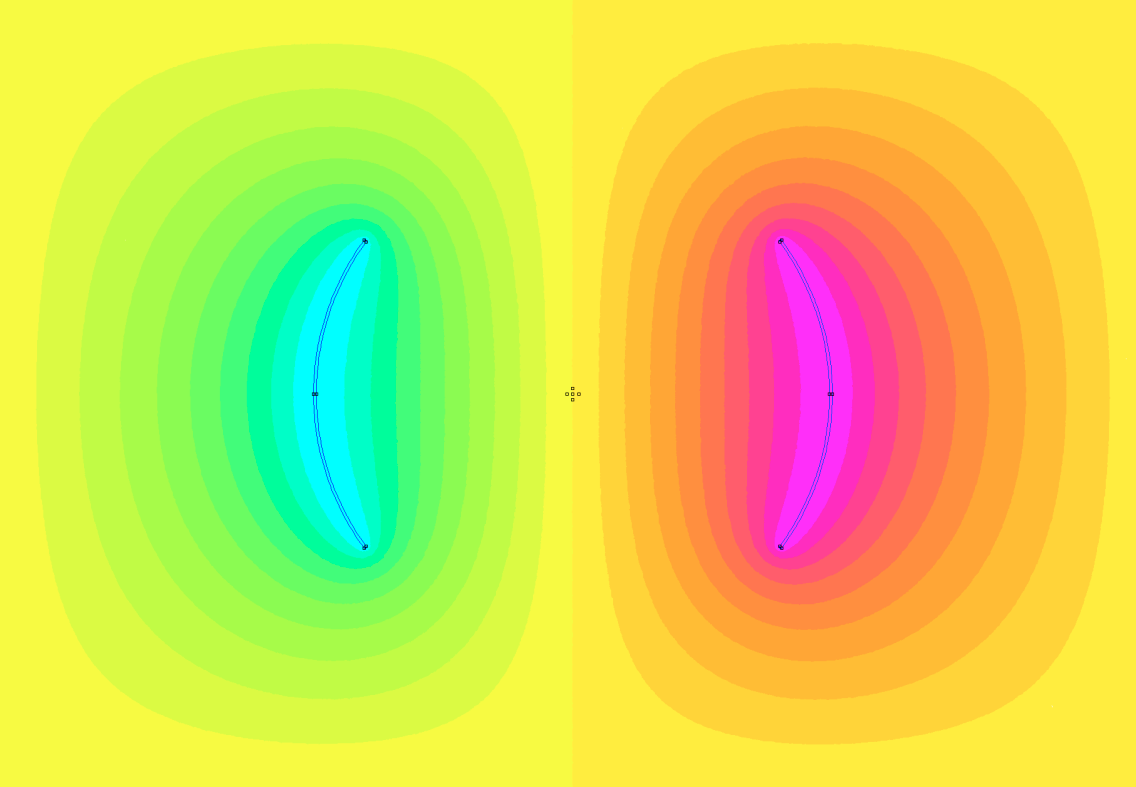}}
\subfloat[]{\hspace*{0.1cm}\raisebox{3mm}{\includegraphics[width=0.48\linewidth]{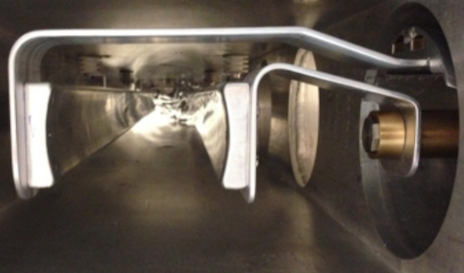}}}
\vspace{-0.05cm}
\caption[Kicker]{(a) Rendering of lines of the produced magnetic field around the two kicker plates, with colours representing relative field intensities. (b) The kicker plates mounted in the vacuum chamber of the storage ring. Images courtesy of D. Rubin~\cite{kick1}.}
\label{fig:kicker}
\end{figure}

The displaced orbit instantiated by the inflector is adjusted by the fast muon kicker, as shown in \cref{fig:kicker}. The kicker is a pulsed magnet with a vertical magnetic field that directs the muons onto the ideal orbit. The kick requires an integrated vertical magnetic field of 1.2~kG-m, for $\mathcal{O}(120)$~ns. After which the field must return to zero before the lead muons complete a single revolution and re-enter the kicker aperture 149~ns later. The kicker is further described by A. Schreckenberger et. al~\cite{kicker}.
\vspace{-0.2cm}
\vspace{-0.2cm}

The kicker and the inflector are crucial components in maximising the number of stored muons, as well as the position of the beam in the ring. Their optimisation is, therefore, critical to the operation of the experiment.

\subsection{Electrostatic quadrupoles} \label{sc:scraping}
\begin{figure}[htpb]
\centering
\subfloat[]{\includegraphics[height=5.7cm]{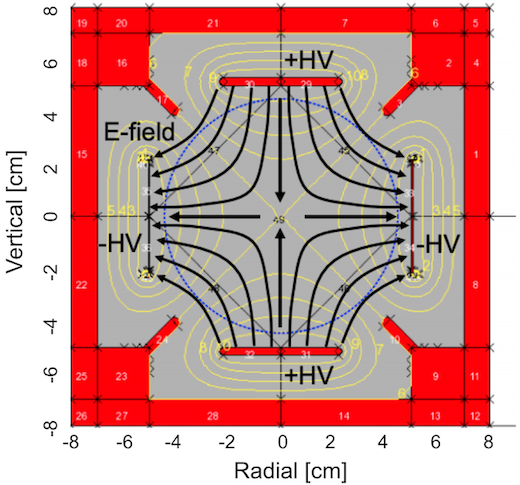}}
\subfloat[]{\hspace*{0.1cm}\raisebox{6.5mm}{\includegraphics[height=5cm]{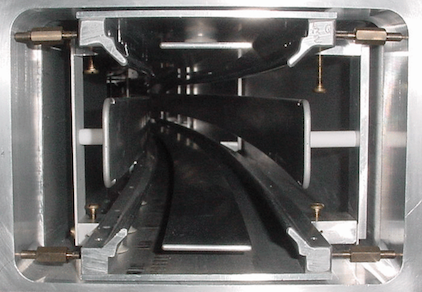}}}
\vspace{-0.1cm}
\caption[Electrostatic quadrupoles]{(a) The electric field lines produced by the \ac{ESQ}. (b) The \ac{ESQ} plates in the vacuum chamber of the storage ring. Images courtesy of the \gm2 collaboration~\cite{FNAL_TDR}.}
\label{fig:esq}
\end{figure}

\begin{figure}[htpb]
\centering
\subfloat[]{\raisebox{7mm}{\includegraphics[height=5cm]{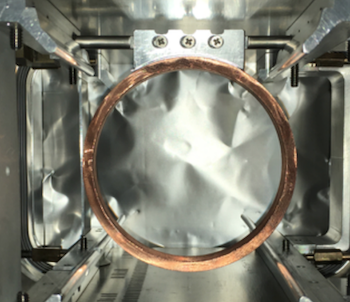}}}
\subfloat[]{\hspace*{0.1cm}\includegraphics[height=6cm]{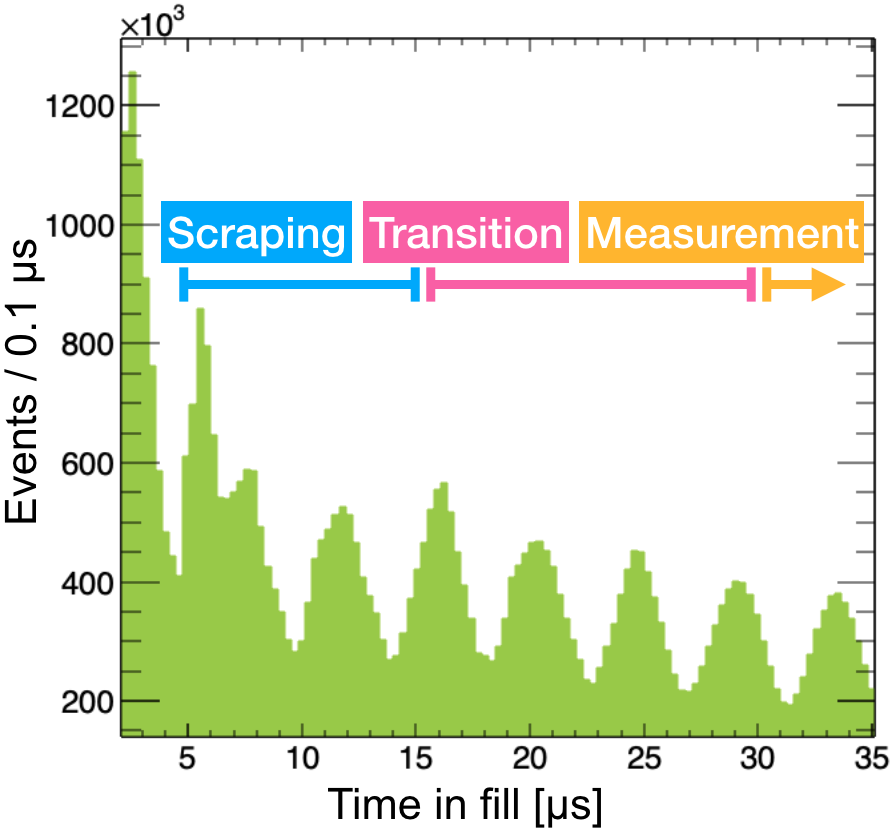}}
\vspace{-0.1cm}
\caption[Collimator]{(a) Photograph of a half-aperture (radius=45~mm) beam collimator, which is used for scraping, during installation. Image courtesy of J. George~\cite{Jimin}. (b) After scraping, the beam stabilises after \SI{30}{\micro\second} and can be used for performing the measurement of $\omega_a$.}
\label{fig:esq2}
\end{figure}

To provide a stable beam storage \ac{ESQ}, as shown in \cref{fig:esq}, are used to focus the beam vertically using an electric field of $\mathcal{O}(20)$~kV. Without the vertical focusing, the beam would diverge and would ultimately be lost after only a few turns around the ring. Moreover, the \ac{ESQ} are used to remove stored muons that fall outside of the storage region -- a process knows as \textit{scraping}. This involves adjusting the voltage on the quadrupoles to scrape the beam against \textit{collimators}, one of which is shown in \cref{fig:esq2}. The collimators have an inner radius of 45 mm, which defines the storage region aperture. After scraping, the beam stabilises after \SI{30}{\micro\second} and can be used for measuring $\omega_a$.
\clearpage

\vspace{-0.5cm}
\vspace{-0.1cm}

An important property of the \ac{ESQ} is the so-called \textit{field-index}, $n$, which characterises the strength of the electrostatic focusing in relation to the magnetic field strength (1.45~T), $B_0$ 
\begin{equation}
    n = \frac{\kappa R_0}{\beta B_0},
    \label{eq:field_index}
\end{equation}
where $\kappa$ is the electric quadrupole gradient, and $\beta \cdot c$ is the velocity of the muon beam. In the \gm2 experiment, a typical value of $n$ is $\sim 0.1$, which corresponds to the so-called weak focusing mode. 
\clearpage

The importance of the \ac{ESQ} is further demonstrated by considering the measurement of $\boldsymbol{\omega_a}$. The cyclotron $\boldsymbol{\omega_c}$ (149 ns) and spin precession $\boldsymbol{\omega_s}$ frequencies of the muon beam are given by~\cite{LDM}:
\begin{equation}
\boldsymbol{\omega_c}=-\frac{e}{m_{\mu}\gamma}\boldsymbol{B},
\end{equation}
\vspace{-0.4cm}
and
\begin{equation}
\boldsymbol{\omega_{s}}=\frac{-e}{\gamma m_{\mu}} \left[(1+\gamma a_{\mu})\boldsymbol{B}_{\bot} + (1+a_{\mu})\boldsymbol{B}_{\parallel} +  \left( \frac{1}{\gamma+1} + a_{\mu} \right)\frac{\gamma\boldsymbol{E}}{c} \times \boldsymbol{\beta} \right],
\end{equation}
for a muon moving in a horizontal plane of a magnetic storage ring, where $\gamma$ is the Lorentz factor, $\boldsymbol{\beta}=\nicefrac{\boldsymbol{v}}{c}$, with the magnetic field components resolved relative to the muon trajectory.
The anomalous precession frequency, $\boldsymbol{\omega_a}$ (i.e.~\say{the \gm2 frequency}), is defined as the difference of the spin
precession and cyclotron frequencies, which corresponds to the rate of muon spin precession relative to the momentum vector
\vspace{-0.2cm}
\begin{equation}
\boldsymbol{\omega_a}= \boldsymbol{\omega_s}- \boldsymbol{\omega_c}= -\frac{e}{\gamma m_{\mu}} \bigg[\gamma a_{\mu}\boldsymbol{B} - \frac{\gamma^2 a_{\mu}}{\gamma +1}(\boldsymbol{B}\cdot \boldsymbol{\beta})\boldsymbol{\beta} + \left(a_{\mu} - \frac{1}{\gamma^2-1}\right)\frac{\gamma \boldsymbol{E}}{c} \times \boldsymbol{\beta}   \bigg].
\label{eq:wa_Full}
\end{equation}
Assuming the muon velocities to be near perpendicular to $\boldsymbol{B}$, the inner product of $\boldsymbol{B}\cdot\boldsymbol{\beta} \rightarrow 0$. As the momentum of the injected muons has been deliberately set to the \say{magic momentum} of 3.09~\GeV ($\gamma=29.3$) the last term in the \cref{eq:wa_Full} vanishes. However, in reality these conditions are not realised exactly, and small corrections to \cref{eq:wa_Full} must be applied. These corrections are the pitch and electric field corrections, and are discussed next. Further details on the \ac{ESQ} can be found in the \tdr~\cite{FNAL_TDR}.

\section{Spatial and temporal distribution of the stored muon beam}\label{sc:bd}
Several important effects arise in the storage process of the muon beam in the experiment. In this section, effects such as vertical pitch, \ac{CBO}, fast rotation, closed-orbit distortions, and lost-muons are considered, as they are directly referenced in the further chapters of this thesis. The full description of the beam effects is given in~\cite{bd}.

\clearpage
\subsection{Vertical pitch}\label{sec:pitch}
\begin{figure}[htpb]
    \centering
    \subfloat[]{\raisebox{6mm}{\includegraphics[width=.48\linewidth]{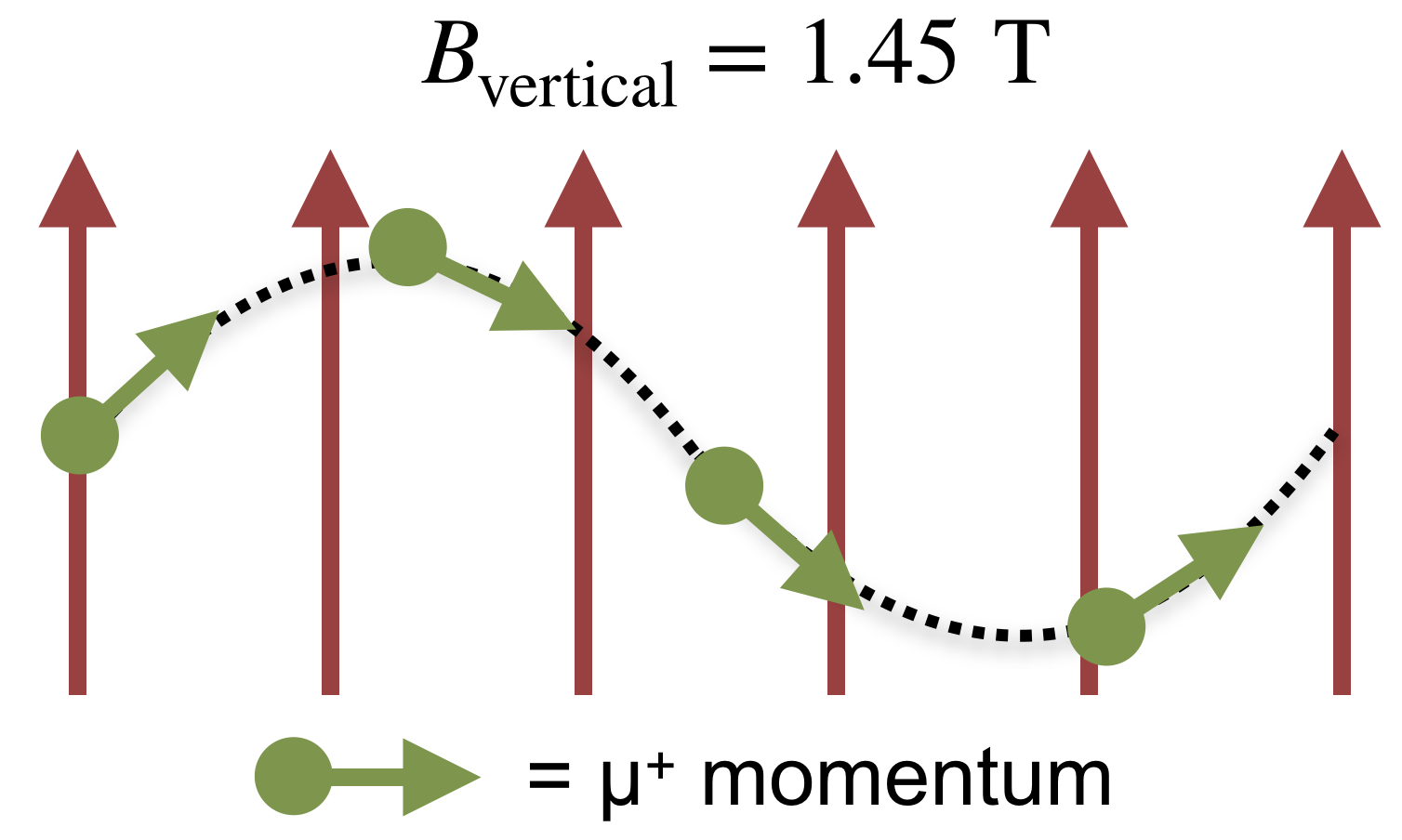} \label{fig:pitch}}}
    \subfloat[]{\includegraphics[width=.48\linewidth]{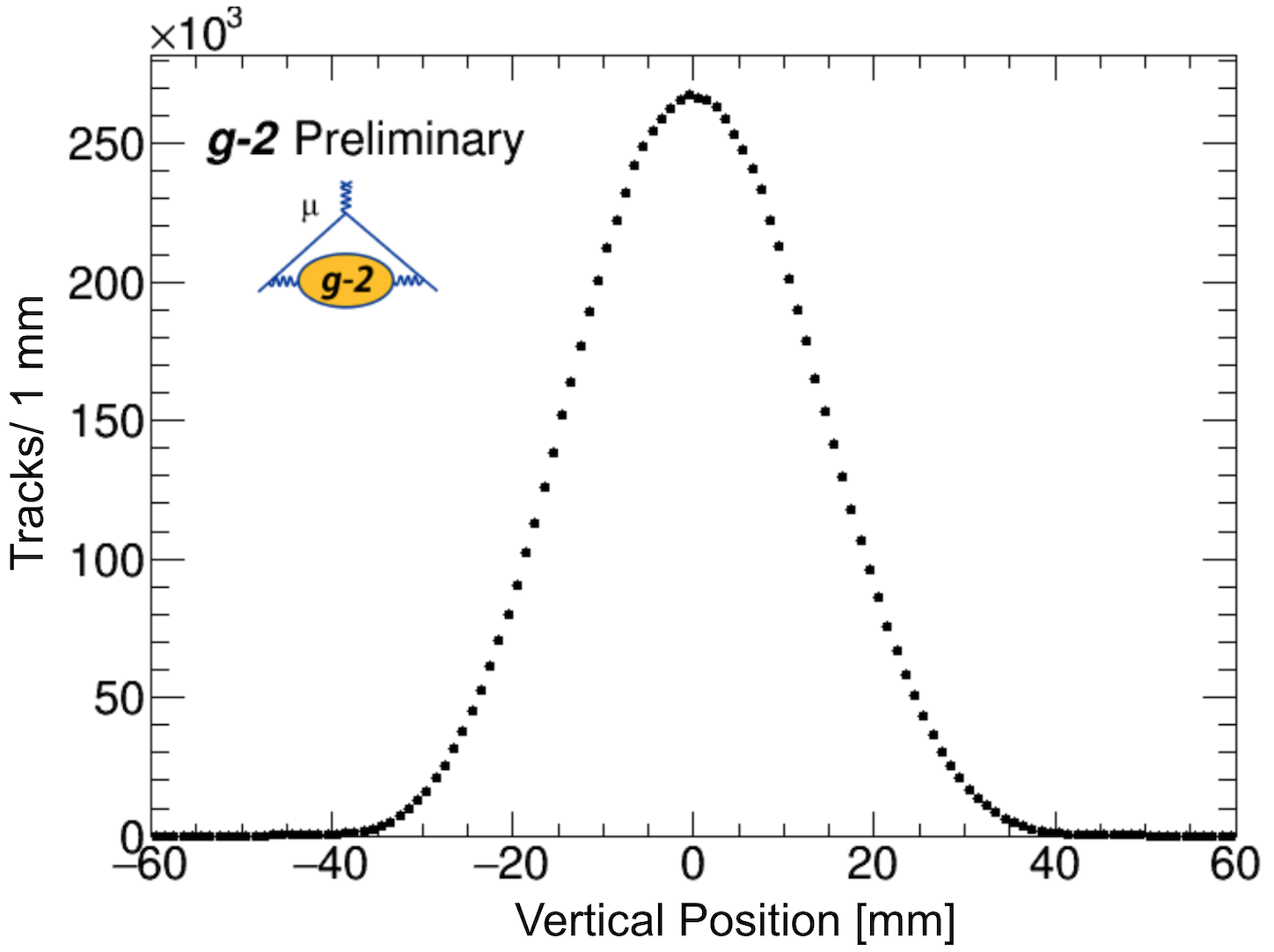} \label{fig:ver}}
    \vspace{-0.25cm}
    \caption[Vertical pitch]{Vertical pitch: a) the up-and-down motion experienced by the muons due to the quadrupole electric field, (b) Vertical position of the beam from the tracking detectors. Plot (b) courtesy of J. Mott~\cite{James_pitch}.}
\end{figure}

The muons in the storage ring undergo a vertical motion due to the focusing from the electrostatic quadruples, as shown in \cref{fig:pitch}. This introduces an additional term in the expression of $\boldsymbol{\omega_a}$ in \cref{eq:mub} that needs to be incorporated into the analysis
\begin{equation}
    \boldsymbol{\omega_a}= \frac{e}{m_{\mu}} \bigg[ a_{\mu}\boldsymbol{B} - \frac{\gamma a_{\mu}}{\gamma +1}(\boldsymbol{B}\cdot \boldsymbol{\beta})\boldsymbol{\beta}  \bigg].
\end{equation}

This changes the measured spin precession frequency, and so needs to be corrected for by the so-called \textit{pitch correction}, $C_{\mathrm{pitch}}$, which is proportional to the vertical width of the beam, $\sigma_{\mathrm{vertical}}$
\begin{equation}
C_{\mathrm{pitch}}= \frac{\Delta\omega_a}{\omega_a}  = -\frac{n}{2R_0^2} \langle y^2 \rangle =-\frac{n}{2R_0^2} \sigma_{\mathrm{vertical}}^{2}.
\label{eq:pitch}
\end{equation}
$\sigma_{\mathrm{vertical}}$ is measured by the tracking detectors, as shown in \cref{fig:ver}. 

A misalignment of the tracking detectors will bias the beam measurement, and hence the pitch correction. The alignment of the tracking detectors and its impact on the pitch correction is described in \cref{ch:align_error,ch:align,ch:align_curvature}.
\clearpage

\subsection{Coherent betatron oscillations (CBO)}\label{sec:CBO}
\begin{figure}[htpb]
    \centering
    \subfloat[]{\raisebox{6mm}{\includegraphics[width=.49\linewidth]{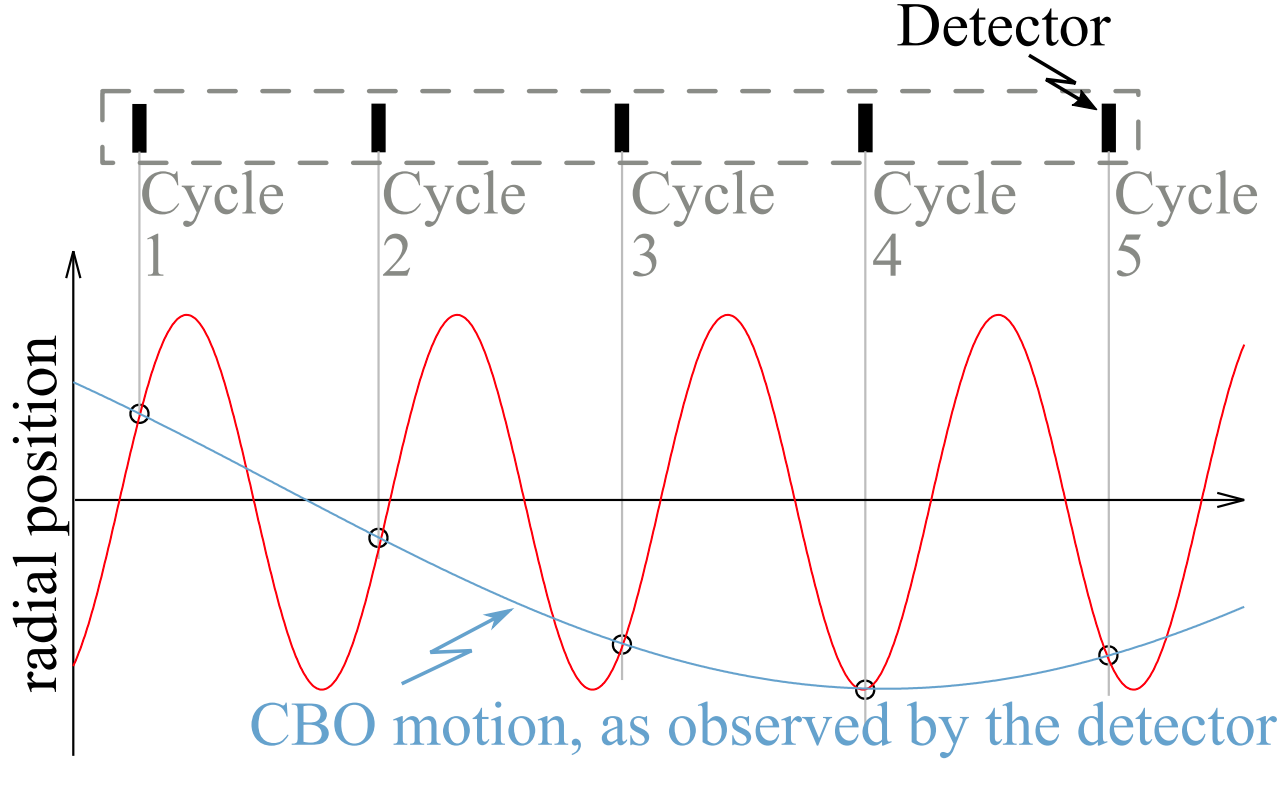}\label{fig:cbo_1}}}
    \subfloat[]{\includegraphics[width=.49\linewidth]{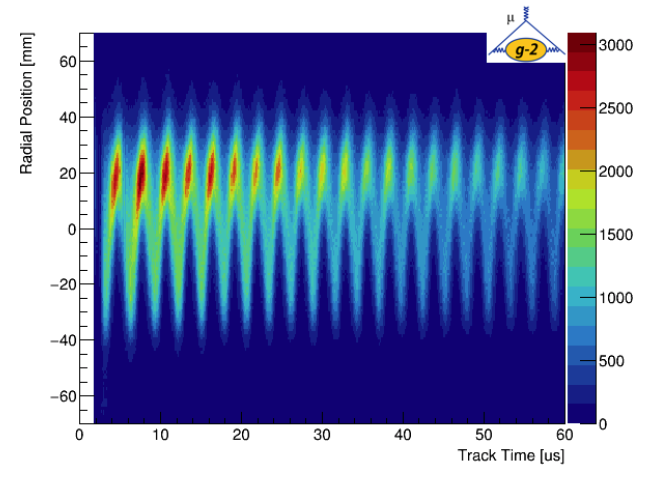}\label{fig:cbo_2}}
    \vspace{-0.2cm}
    \caption[Reconstructed radial position of the muon beam versus time]{a) The
betatron oscillation is shown in red. A detector sees the
motion at the \ac{CBO} frequency. The location of a single
detector is shown. Diagram  courtesy of O. Kim et al.~\cite{CBO}. b) Reconstructed radial position of the muon beam plotted against time, as measured by the tracking detectors. Plot courtesy of J. Mott~\cite{James}.}
\end{figure}

The fraction of events where a positron is detected by the calorimeters depends on the radial position of the muon beam in the storage ring. The beam has a finite momentum and position distribution, and so, any particles not having exactly the \say{magic momentum} or not positioned at the ideal orbit will be subject to restoring forces from the \ac{ESQ} and the magnetic field. These restoring forces cause simple harmonic motion of the beam, with the muons oscillating in-and-out radially. This results in slightly different cyclotron periods for different momenta. As each detector is effectively sampling the beam at the cyclotron frequency ($\sim6.71$ MHz),
\begin{equation}
    f_{\mathrm{c}} = \frac{vB_0e}{2\pi m_{\mathrm{\mu}}c},
\end{equation}
while the radial betatron frequency ($\sim6.34$ MHz)
\begin{equation}
    f_{x\mathrm{BO}} =  \sqrt{1-n} f_{\mathrm{c}},
\end{equation}
is larger than $f_{\mathrm{c}}/2$, there is an aliasing effect such that the radial betatron motion of the beam is instead observed as an apparent slow-moving oscillation (c.f. Nyquist frequency~\cite{Nyquist}). The actual frequency measured by a detector, as illustrated in \cref{fig:cbo_1}, is 
\begin{equation}
f_{\mathrm{CBO}} = f_{\mathrm{c}} - f_{x\mathrm{BO}} = f_{\mathrm{c}}(1-\sqrt{1-n}).
\end{equation}
\vspace{-1.5cm}
\clearpage

The radial position of the muon beam can be monitored as a function of time using the tracking detectors, as shown in \cref{fig:cbo_2}, in order to estimate the amplitude and frequency of the CBO. 

Analogous to the \ac{CBO} is the so-called \ac{VW}, with the corresponding vertical betatron frequency ($\sim2.21$ MHz)
\begin{equation}
    f_{y\mathrm{BO}} =  \sqrt{n} f_{\mathrm{c}}.
\end{equation}
Since $f_{y\mathrm{BO}}$ is smaller than $f_{\mathrm{c}}/2$, the \ac{VW} oscillation  
\begin{equation}
    f_{\mathrm{VW}} = f_{\mathrm{c}}- 2f_{y\mathrm{BO}} = f_{\mathrm{c}}(1-\sqrt{n}),
\end{equation}
is measured directly by the detectors (i.e. there is no aliasing effect), as described in more detail by T. Halewood-Leagas~\cite{Tabitha}.

Both the \ac{CBO} and \ac{VW} oscillations must be accounted for in the analysis: if these effects are not included, then significant residuals between the data and the fit to data, as discussed in \cref{ch:edm}, are obtained.  

\subsection{Fast rotation and electric field corrections}\label{sec:fast_rot}
\begin{figure}[htpb]    
    \centering  
    \subfloat[]{\includegraphics[width=0.49\linewidth]{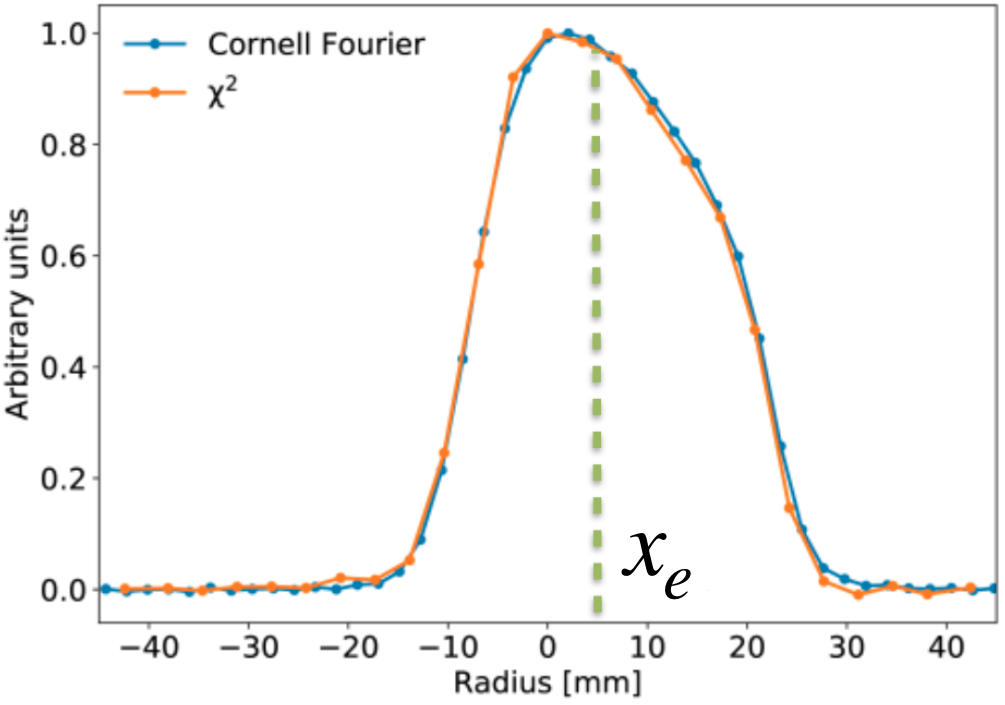}\label{fig:AxCBO1}}
    \subfloat[]{\raisebox{1mm}{\includegraphics[width=0.49\linewidth]{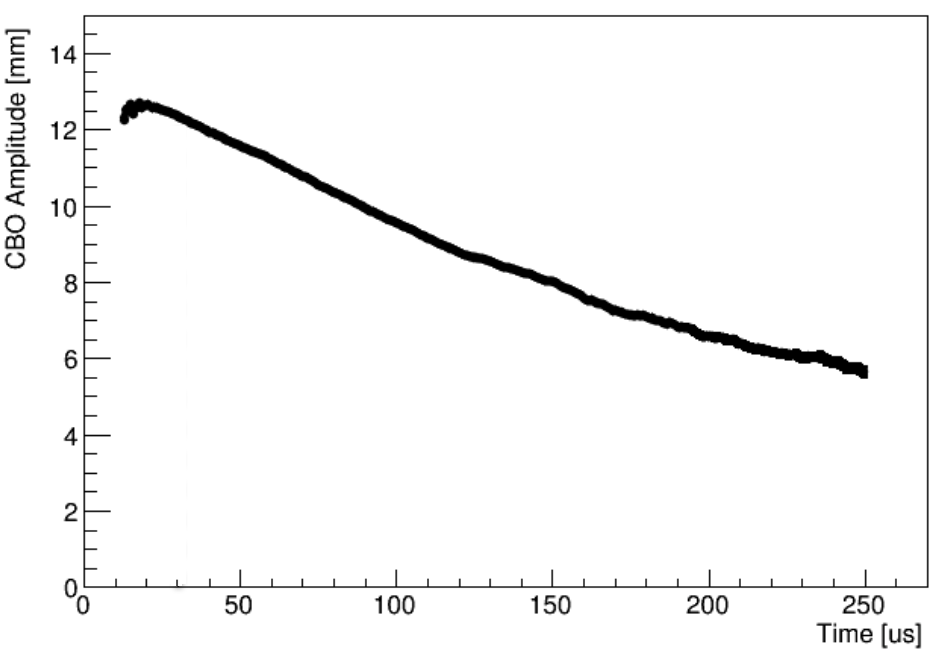}\label{fig:AxCBO2}}}
    \vspace{-0.2cm}
    \caption[The equilibrium beam position]{(a) The equilibrium beam position from the fast rotation analysis. Both analysis methods, Fourier and $\chi^2$ produced similar results ($x_e\sim\SI{6}{\milli\metre}$). Plot courtesy of A. Chapelain~\cite{Antoine}. (b) The measured CBO amplitude as a function of time in the fill. Plot courtesy of J. Mott~\cite{James}.}  
\end{figure}

To account for the fact that not all stored muons have the \say{magic momentum}, an electric field correction \cite{FNAL_TDR}, $C_{\mathrm{electric}}$, to \cref{eq:wa_Full} is implemented
\begin{equation}
    C_{\mathrm{electric}} = \frac{\Delta\omega_a}{\omega_a} = -2n(1-n)\beta^2\frac{\langle x_e^2\rangle}{R_0},
\end{equation}
where 
\begin{equation}
    \langle x_e^2\rangle = x_e^2 + \sigma^{2}_{\mathrm{radial}},
\end{equation}
and $x_e$ is the equilibrium radius of the beam (i.e. the average radial position of the beam), and $\sigma_{\mathrm{radial}}$ is the radial width of the beam. $\sigma_{\mathrm{radial}}$ and $x_e$ are extracted directly from the so-called fast rotation analysis, as shown in \cref{fig:AxCBO1}, which considers how stored muons at various momenta cause \say{de-bunching} of the beam. For example, for a group of muons with momentum larger than the \say{magic momentum}, the stored orbit will have a larger radius leading to a larger $f_{\mathrm{c}}$. The methodology of the fast rotation analysis is described in~\cite{Antoine}.
\clearpage

\vspace{-0.2cm}
\subsection{The effect of the \ac{CBO} and \texorpdfstring{$x_e$} ~ on the beam}\label{sub:improved_simulation_for_r1}
\begin{figure}[htpb]    
    \centering  
    \includegraphics[width = 0.56\linewidth]{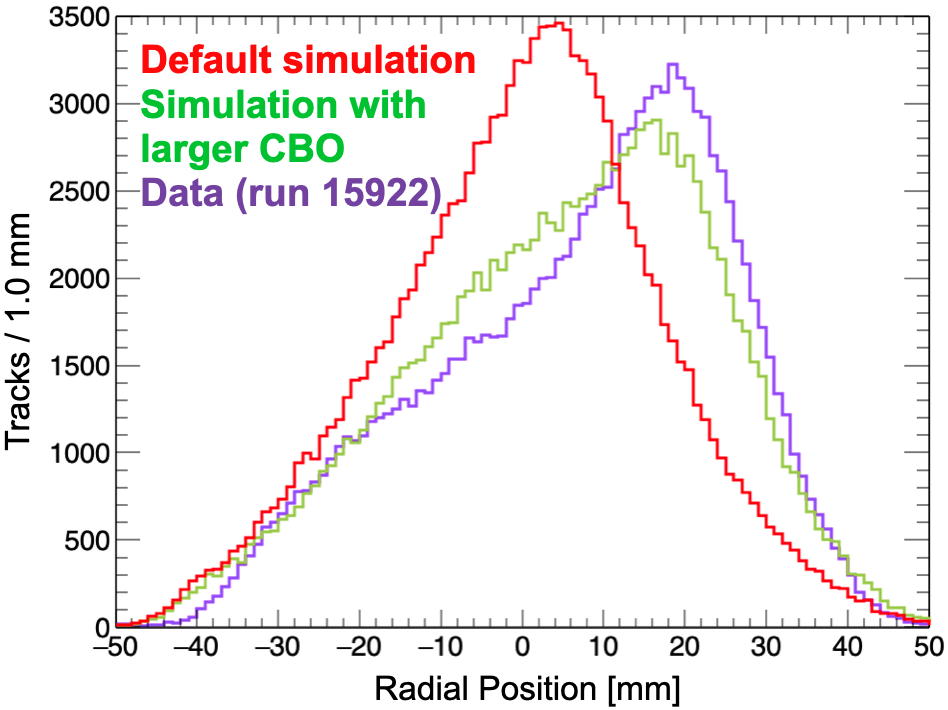}    
    \vspace{-0.2cm}
    \caption[Radial beam position in the modified simulation]{The modified simulation has an extrapolated radial beam position that more closely resembles the one seen in data ($x_e\sim\SI{7}{\milli\metre}$, in one of the two tracker stations), in comparison to the default simulation.} 
    \label{fig:rad_cbo} 
\end{figure}

The mean stored radial beam position is slightly larger than expected, due to the kicker providing a field below the design value. This is implemented in the simulation using data-derived parameters for $x_e$, using the fast rotation analysis, and the radial \ac{CBO} amplitude, as measured by the tracking detectors, as shown in \cref{fig:AxCBO2}. Using the measured CBO and $x_e$ produces a more realistic radial beam profile in the simulation, as shown in \cref{fig:rad_cbo}, where it is compared with the default simulation and data.   
\vspace{-0.2cm}
\clearpage

\subsection{Closed-orbit distortion}\label{sc:closed_orbit}
\begin{figure}[htpb]    
    \centering  
    \includegraphics[width = 0.7\linewidth]{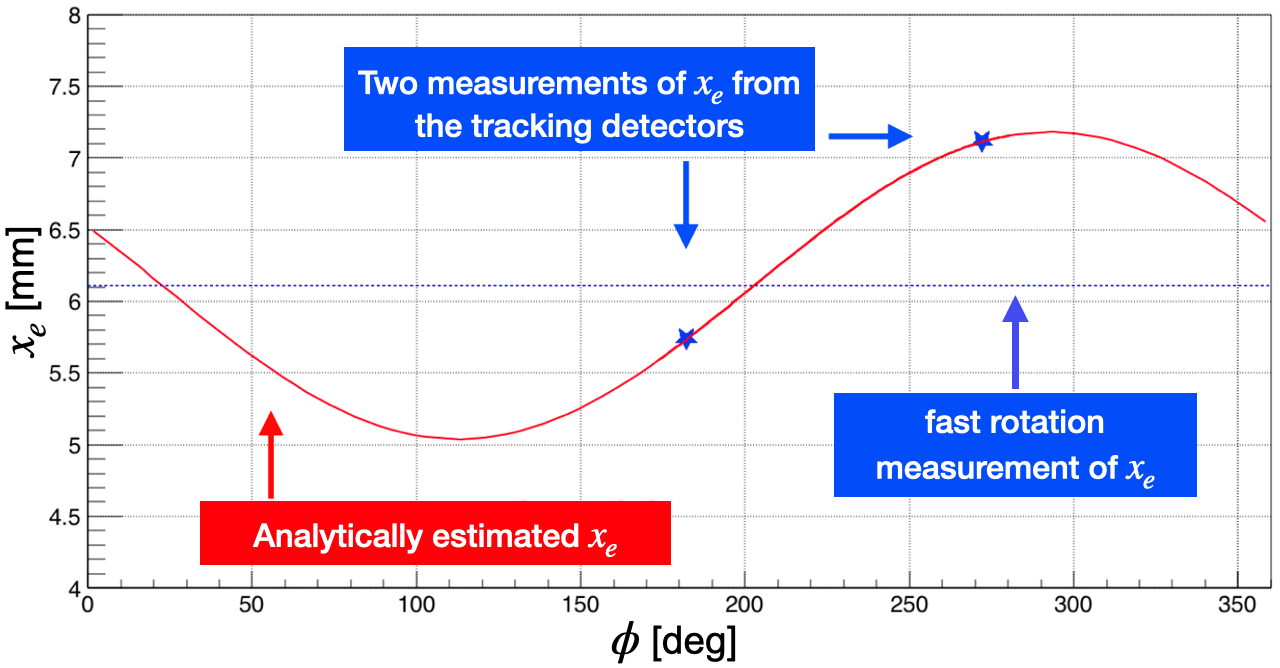}    
    \vspace{-0.2cm}
    \caption[Measurement of $x_e$]{$x_e$ as the function of azimuth ($\phi$) around the \gm2 storage ring. The three measurements are $x_e$ from the fast rotation analysis, and two radial beam position measurements from the two tracking detectors at $180^{\circ}$ and $270^{\circ}$. Plot courtesy of J. Grange~\cite{Grange}.} 
    \label{fig:closed} 
\end{figure}

Another assumption in \cref{eq:wa_Full} is the uniformity of the magnetic field throughout the ring. In practice, the field has small inhomogeneities, which contribute to an effect known as closed-orbit distortion, which results in different equilibrium positions of the beam around the ring. That is, $x_e$ varies with azimuth. The analytically estimated \cite{Grange} model and three measurements of $x_e$ are shown in \cref{fig:closed}. Two of these measurements come from the tracking detectors: each detector observes different segments of the beam's orbit. These measurements are sensitive to the global alignment, as described in \cref{sec:align_global}. Further details on the closed-orbit effect are presented in~\cite{Grange}.

\vspace{-0.2cm}
\subsection{Lost-muons}\label{sc:lost_muons}
Some of the stored muons in the ring will interact with material (e.g. collimators) emitting radiation and losing energy. These so-called \say{lost-muons} will curl inwards towards the centre of the ring and exit the storage ring. However, some of them will still deposit a small amount of energy in the calorimeters and fail the energy cut. Matching calorimeter and tracker measurements via $E/p$ can be used to distinguish between $e^+$ and $\mu^+$, as discussed in \cref{sc:track_soft}. A corresponding systematic uncertainty on the measurement of $\omega_a$ arises from time-dependence and momentum-dependence of the muon loss. Both of these effects introduce a time-dependent phase that biases $\omega_a$. Further details on the study of lost-muons are given in~\cite{lost}.
\clearpage

\section{Magnetic field system} \label{sc:field}
\begin{figure}[htpb]
    \centering
    \subfloat[]{\includegraphics[height=6.5cm]{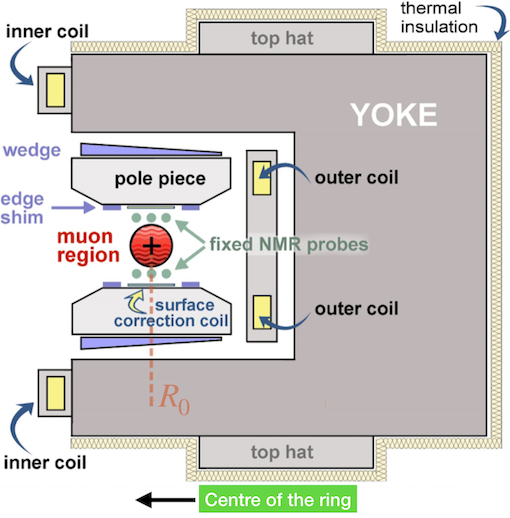}\label{fig:yoke}}
    \subfloat[]{\hspace*{0.2cm}\raisebox{8mm}{\includegraphics[height=5.5cm]{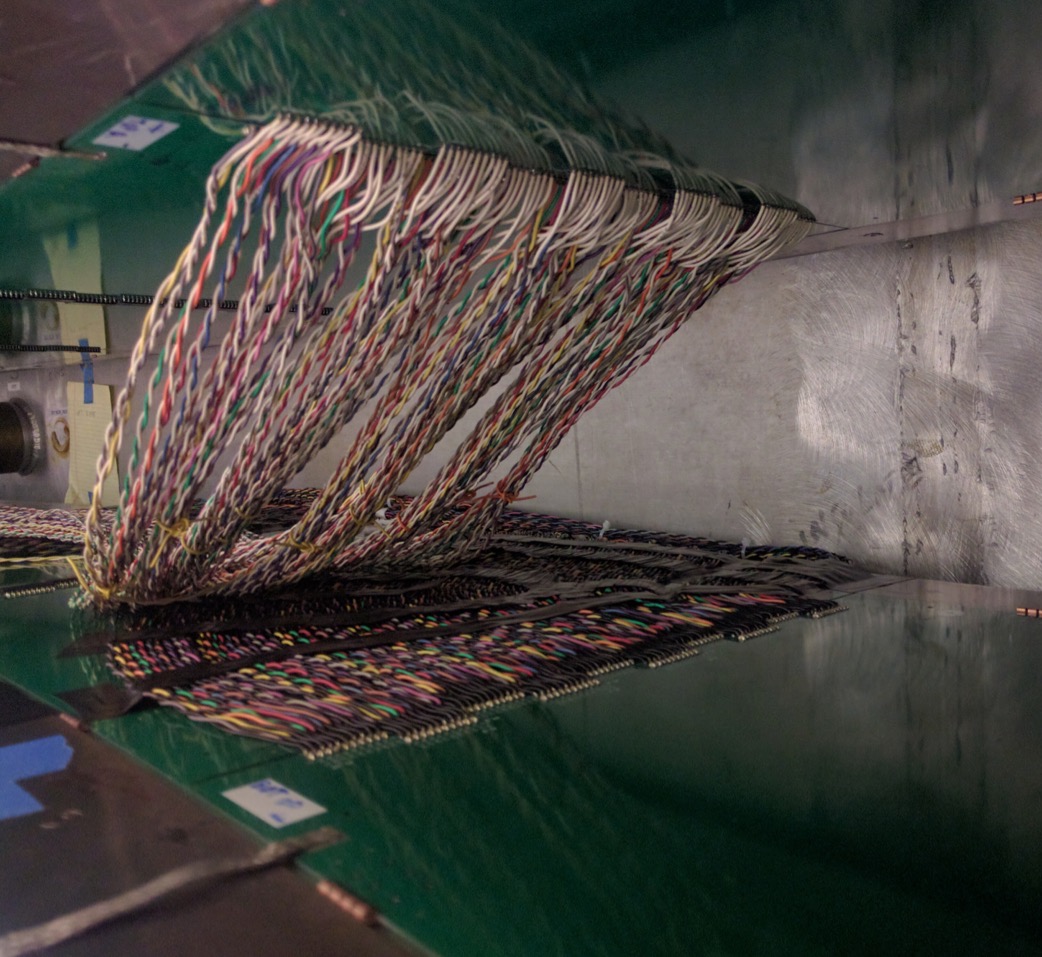}\label{fig:coil}}}
    \vspace{-0.2cm}
    \caption[Magnet cross-section and correction coils]{(a) Cross-section of the \gm2 magnet. The muon storage region is shown inside the magnet's yoke. Illustration courtesy of J. George~\cite{Jimin}. (b) The correction coils. The system consists of 100 concentric current carrying coils, above and below the
vacuum chambers.  Image courtesy of R. Osofsky~\cite{Rachel_sc}.}
\end{figure}

The magnetic field system aims to produce a uniform vertical magnetic field of 1.45~T and to measure it with an uncertainty of $70$~\ac{ppb}. This measurement is performed in the cross-section of the storage region, as shown in \cref{fig:yoke}, in the volume of 1.1~m$^3$. 

\begin{figure}[htpb]
    \centering
    \includegraphics[width=.75\linewidth]{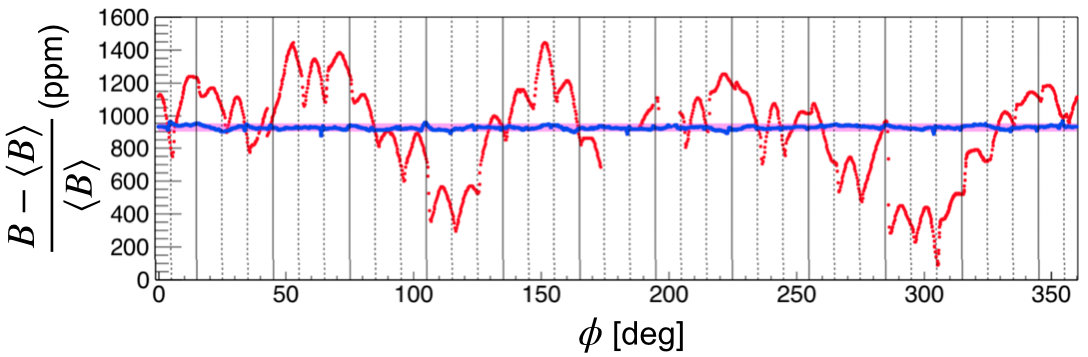} 
    \vspace{-0.2cm}
    \caption[Magnetic field shimming]{The normalised difference between the measured field ($B$) at a location and the average field ($\langle B \rangle$) as the function of azimuth ($\phi$) around the \gm2 storage ring. The field before the shimming is shown in red, the one after shimming in blue, and the goal of $25$~ppm is shown in purple. Plot courtesy of J. George~\cite{Jimin}.}
    \label{fig:shim}
\end{figure}

\vspace{-0.2cm}
A formidable technical challenge was the so-called \say{shimming} of the superconducting magnet, which involved physical modifications of the magnet to set a uniform field around the ring. The result of this procedure is shown in \cref{fig:shim}.
\vspace{-0.2cm}
\clearpage

To further refine the uniformity of the magnetic field, as well as to ensure stability over time, an active shimming was implemented. This was done using a power supply feedback to keep the dipole field constant using correction coils to smooth the field. This process achieved the field uniformity of 1~\ac{ppm} \cite{Jimin}, with the correction coils shown in \cref{fig:coil}.

\begin{figure}[htpb]
    \centering
    \includegraphics[width=.7\linewidth]{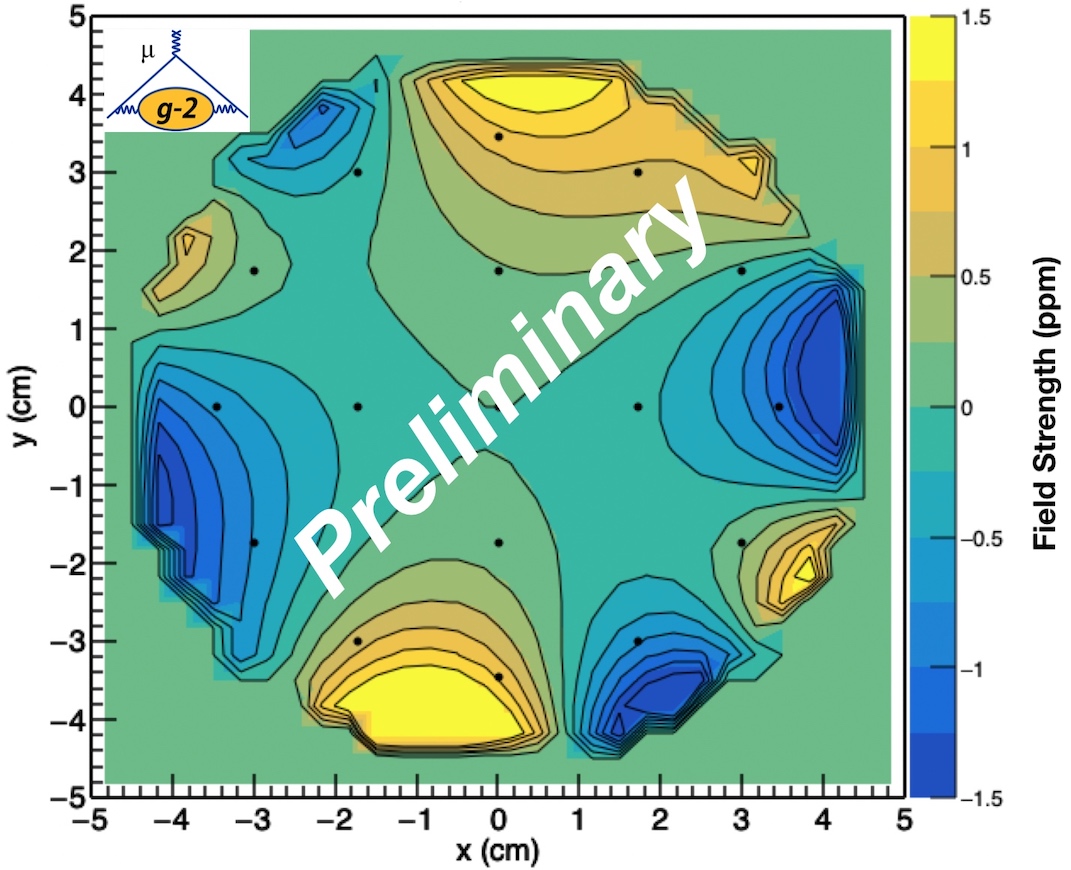} 
    \caption[A cross-section of the field]{A cross-section of the field in the storage region at one azimuthal location. Image courtesy of R. Osofsky~\cite{Rachel_map}.}
    \label{fig:field}
\end{figure}

When the muons are not stored in the ring, the magnetic field is measured by a trolley system, which contains 17 \ac{NMR} probes. The trolley completes a full revolution around the ring in approximately three hours every three days to map the field profile in the storage region, as shown in \cref{fig:field} for one azimuthal location.

When the muons are injected, the field is continuously measured by fixed \ac{NMR} probes located outside of the storage region. This is done in order to track the field between the trolley runs. This field map is then convoluted with the beam profile (see \cref{sec:field_convolution}) measured by the tracking detectors to find the average field experienced by the muons before decay. 

Further details of the field measurements, including the absolute field calibration, and the relative trolley calibration using a plunging probe, are given in~\cite{Matthias}.

\clearpage

\section{Detectors}\label{sc:det}
An in-depth description of the \ac{DAQ} and the tracking detectors is given in \cref{ch:daq,ch:tracker}, respectively, while the calorimeters and the auxiliary detectors are described in more detail in~\cite{Aaron}, as well as in the \tdr~\cite{FNAL_TDR}.

\subsection{Calorimeters}
\begin{figure}[htpb]
    \centering
    \subfloat[]{\includegraphics[width=0.5\linewidth]{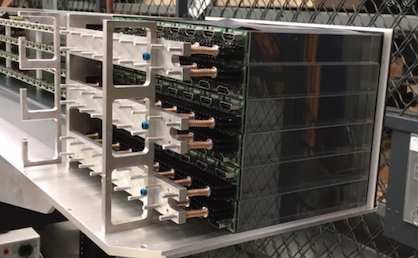}\label{fig:calo_det_a}}
    \subfloat[]{\hspace*{0.05cm}\includegraphics[width=0.465\linewidth]{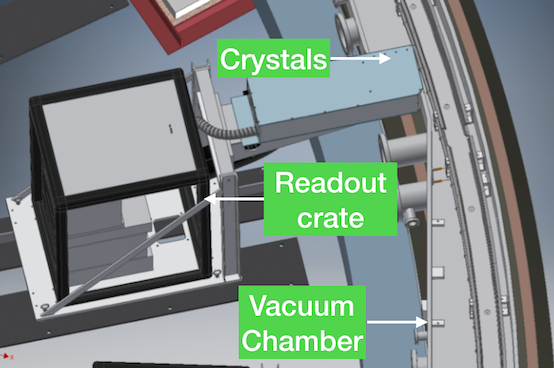}}
    \vspace{-0.2cm}
    \caption[Calorimeter crystals]{(a) The calorimeter crystals during assembly. (b) A rendering of the calorimeter outside of the storage ring. Images courtesy of J. Hempstead~\cite{Jason_wiggle}.}
    \label{fig:calo_det}
\end{figure}

The $\omega_a$ measurement will be realised by 24 electromagnetic calorimeter detectors, which are distributed uniformly on the inside of the storage ring. Each calorimeter comprises 54 segmented lead-fluoride Cherenkov detectors, as shown in \cref{fig:calo_det}. Each crystal is read out by a \textit{Hamamatsu} \ac{SiPM}, which are constantly gain-calibrated between the muon measurements by a laser calibration system \cite{Anastasi}. The calorimeters measure the energy and time of arrival of the positrons, from the muon decay.
\vspace{-0.2cm}

\vspace{-0.2cm}
The largest single systematic uncertainty associated with the calorimeters is pileup, which occurs when, for example, two low energy positrons deposit energy in the same crystal close together in time (within $\sim5$ ns). The tracking detectors can be used to investigate pileup in the adjacent calorimeters, as they can reconstruct independent trajectories of the positrons. The final goal for a systematic uncertainty on the $\omega_a$ measurement is 70~\ac{ppb}, while the experimental goal of collecting $1.6 \times 10^{11}$ positrons corresponds to a statistical uncertainty of 100~\ac{ppb}. Combining the statistical and the systematic uncertainty from the determination of $\omega_{a}$ in quadrature with the field measurement uncertainty gives the total experimental design uncertainty of 140~\ac{ppb}.
\clearpage

\section{Data-taking periods} \label{sc:runs}

\subsection{Commissioning run} \label{sub:cr}
To ensure that all components of the experiment were performing adequately, a commissioning run, which lasted from 31 May 2017 until 7 July 2017, was undertaken to collect preliminary data. During the commissioning run, protons and pions were also delivered to the storage ring, in addition to muons. This was due to the Delivery Ring not being used. 

\subsection{\R1, \R2 and \R3}
\begin{figure}[htpb]
    \centering
    \subfloat[]{\includegraphics[width=.5\linewidth]{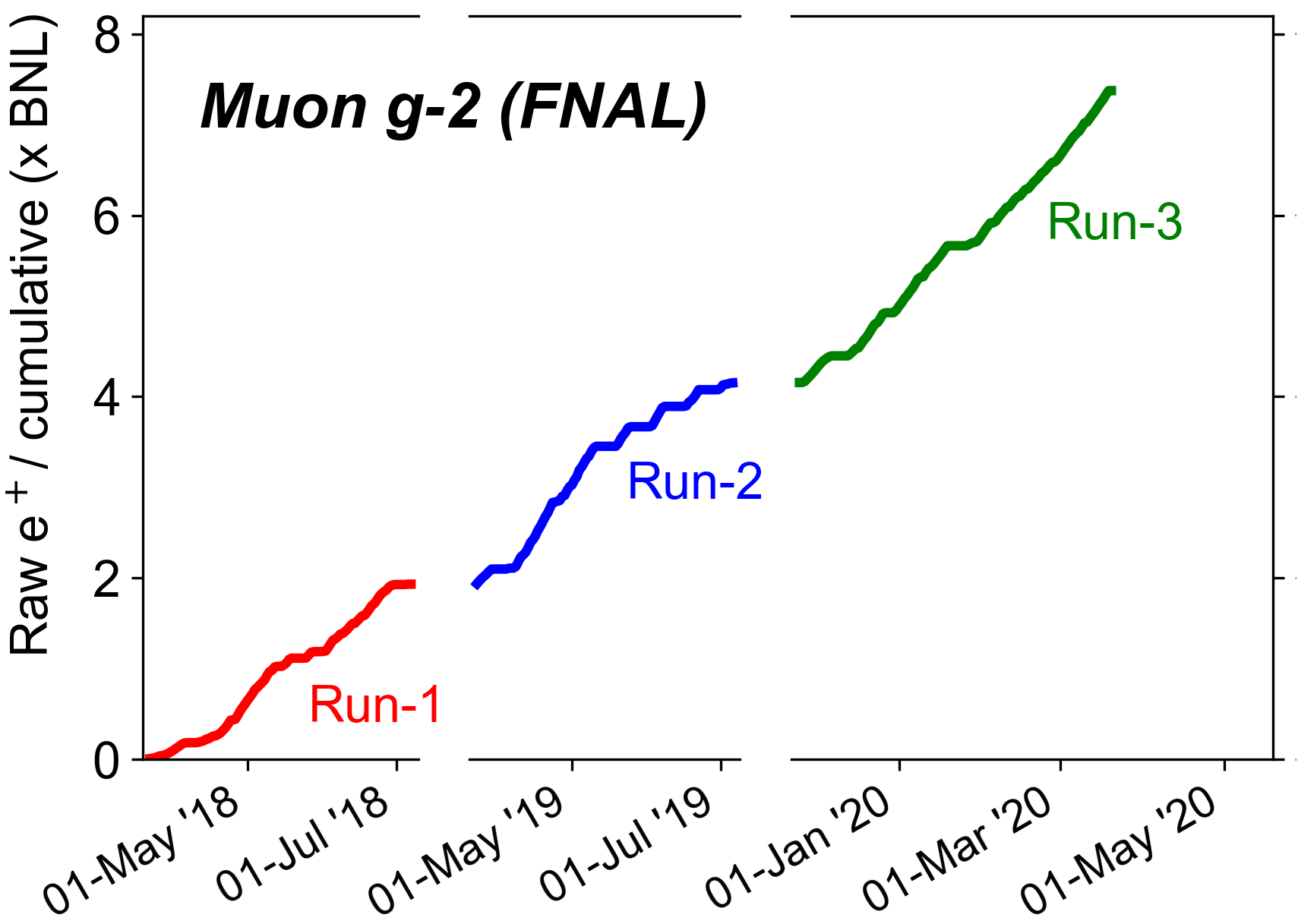}\label{fig:pot}} 
    \subfloat[]{\raisebox{4mm}{\includegraphics[width=.5\linewidth]{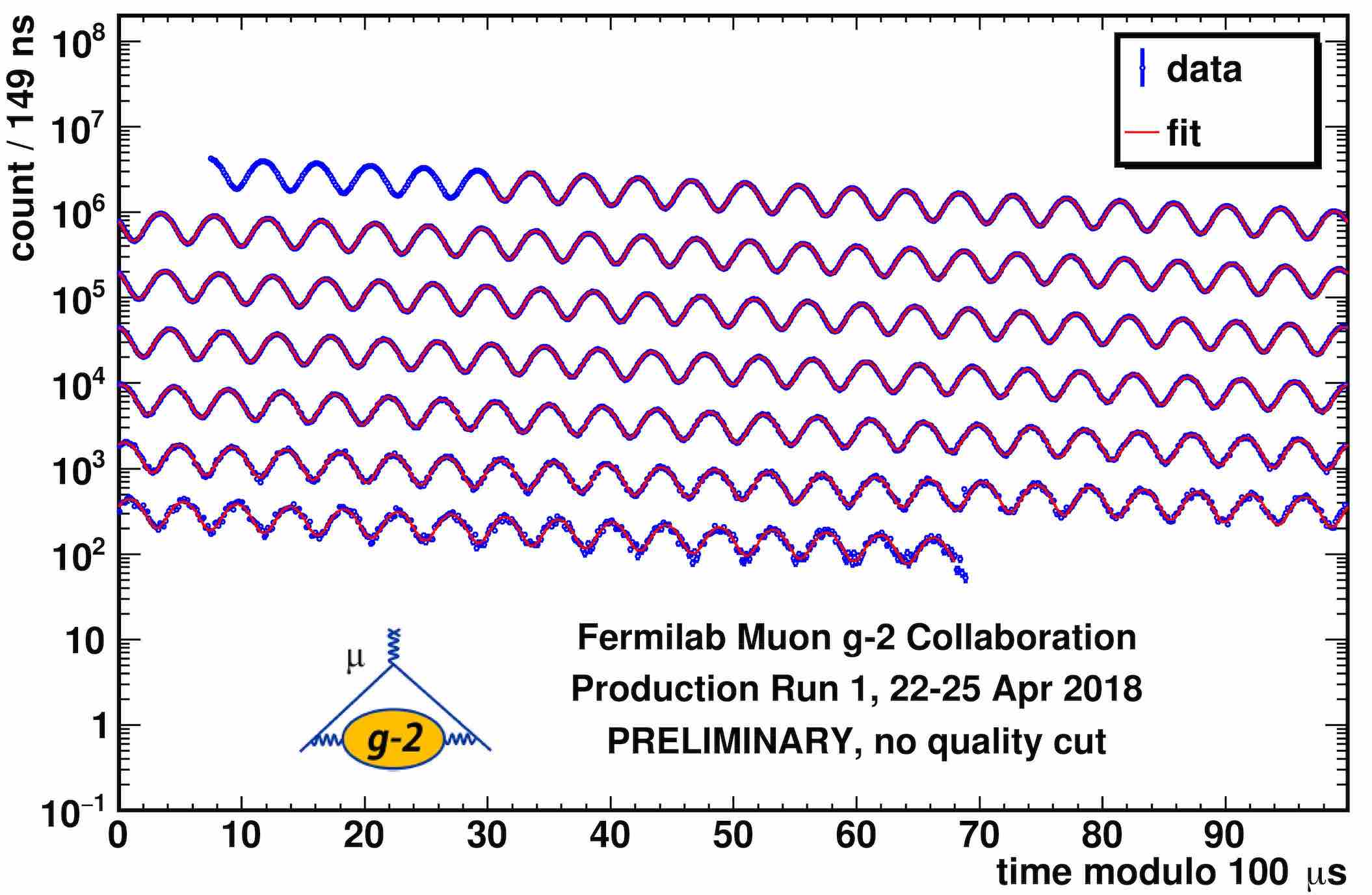}\label{fig:1a_wiggle}}}
    \caption[\R1, \R2 and \R3 results]{\R1, \R2 and \R3 results. (a) The number of recorded positrons in \R1, \R2 and \R3 as a fraction of total data collected by the \ac{BNL} experiment. Plot courtesy of M. Lancaster~\cite{Mark_pot}. (b) This plot was accumulated from 60 hours of data (dataset 1a), and has one billion positrons. The number of muons collected in this period is similar to the one achieved by the \ac{BNL} experiment in the entirety of 1999. Plot courtesy of the \gm2 collaboration~\cite{FNAL_TDR}.}   
\end{figure}

The first data-collection period (i.e. \R1) happened from 25 March 2018 until 8 July 2018, \R2 took place from 23 March 2019 until 5 July 2019, while \R3 happened between 25 November 2019 and 19 March 2020. The total number of collected $e^+$ in these periods, as a fraction of the total number collected at the \ac{BNL}, is shown in \cref{fig:pot}.

The work in this thesis will primarily focus on analysing the \R1 data, as described in \cref{ch:edm}. \R1 and \R2 data has also been used to check the calibration of the tracking detectors as described in \cref{ch:align}.

\clearpage

The datasets collected in \R1 are summarised in \cref{tab:DS}, with a preliminary $\omega_a$ analysis-level plot of the 1a dataset shown in \cref{fig:1a_wiggle}. The total amount of raw data collected, that will be used for the analysis, is roughly 0.5 PB in \R1.

\begin{table}[htpb]  
  \centering
  \begin{tabular}{ccccccc}
    \toprule
     Dataset & $e^+$ & Number of & Raw data & field \\
     name & ($E>1.7$~GeV) & tracks & size [TB] & index, $n$ \\ \toprule
    1a & $9.34\times10^8$ & $3.51\times10^7$ &  55 & 0.108 \\ \midrule
    1b & $8.70\times10^8$ & $4.83\times10^7$  & 75 & 0.120  \\ \midrule
    1c & $2.13\times10^9$ & $7.23\times10^7$ &  113 & 0.120 \\ \midrule
    1d & $4.10\times10^9$ & $1.40\times10^8$ & 273 & 0.108  \\ \toprule \toprule
    Total & $8.03\times10^9$ & $2.96\times10^8$ & 516 & -  \\ \bottomrule 
  \end{tabular} 
  \caption[\R1 datasets]{The collected data in \R1 is split into four datasets \cite{Kim_ds}. For each dataset, the total number of collected $e^+$ above the energy threshold, the number of tracks, the raw data size in TB, as well as the corresponding field-index (see \cref{eq:field_index}) are shown. The number of tracks is the number of the good-quality tracks, which are discussed in \cref{sc:track_quality}.}
  \label{tab:DS}
\end{table}

\subsection{\R4 and beyond}
\begin{figure}[htpb]
    \centering
    \subfloat[]{\includegraphics[width=.48\linewidth]{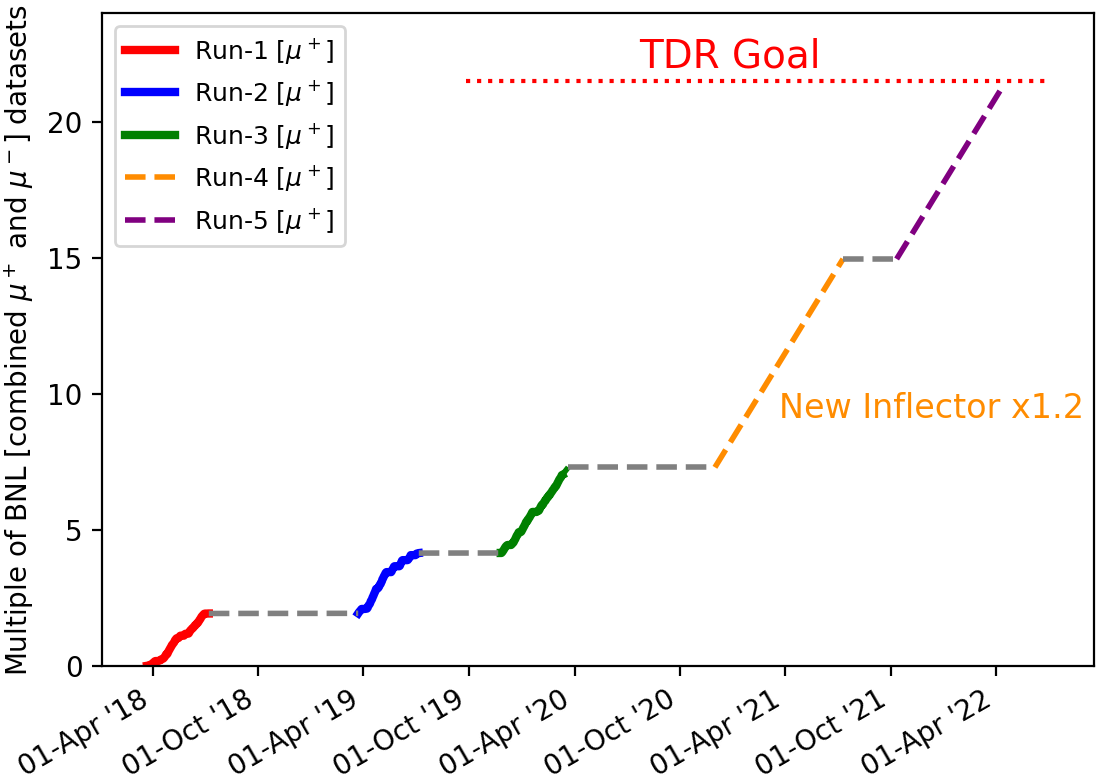}\label{fig:pot_R3}}
    \subfloat[]{\hspace*{0.05cm}\raisebox{6.5mm}{\includegraphics[width=0.478\linewidth]{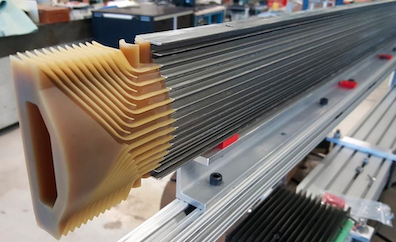}\label{fig:new_inf}}}
    \caption[\R4 and \R5 projections and new inflector]{(a) \R4 and \R5 projections. Plot courtesy of M. Lancaster~\cite{Mark_pot}. (b) The new inflector during construction in 2018. Image courtesy of K. Badgley~\cite{inf2}.}
    \label{fig:r3_new_inf}
\end{figure}

The next run will begin in December 2020 and will accumulate positrons at a higher rate due to the use of a new upgraded inflector \cite{inf2}, which is shown in \cref{fig:r3_new_inf}. The new inflector will increase the number of stored muons by $\sim20\%$ per fill.

\clearpage

\section{Data reconstruction}\label{sc:prod}
\begin{figure}[htpb]
    \centering
    \includegraphics[height=7.5cm]{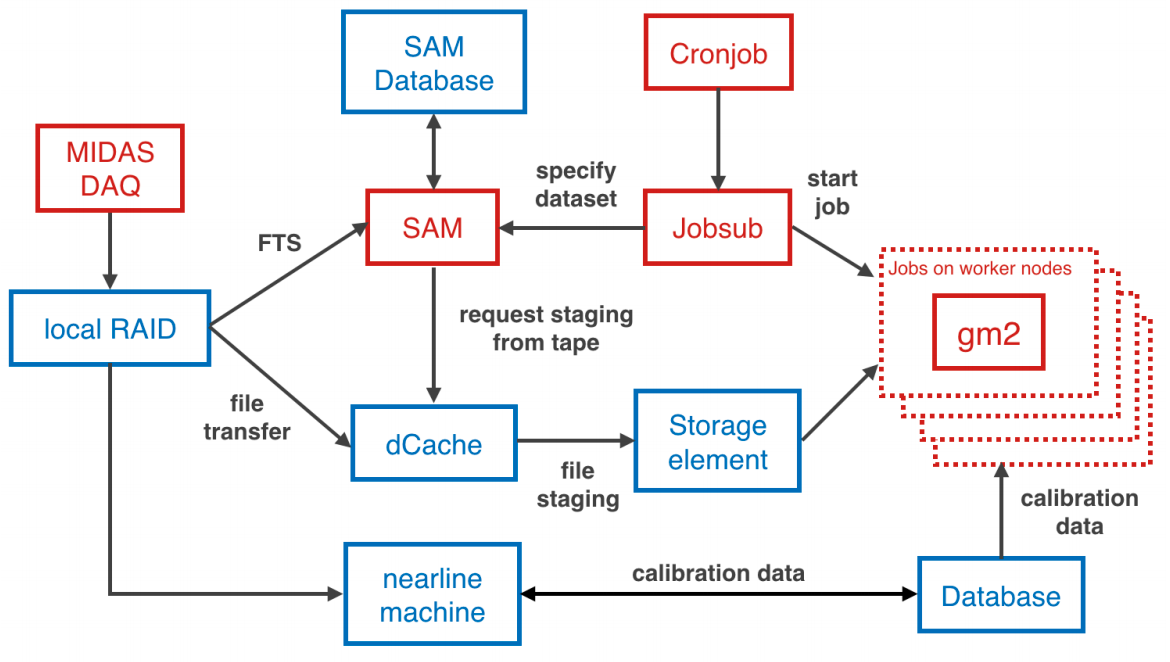}
    \caption[Data reconstruction]{The workflow for the data reconstruction in the experiment. The raw output data from the DAQ is stored on a local RAID. It is then catalogued (via SAM) and stored on tape (dCache). During the reconstruction on distributed worker grid nodes, the calibration constants are loaded from a database. Diagram courtesy of R. Fatemi et al.~\cite{prod}.}
    \label{fig:prod}
\end{figure}

Given the large amount of raw data collected by the experiment, an automated process, known as \say{data production}\cite{prod}, which involves data transfer, reconstruction and calibration, is used, as demonstrated in \cref{fig:prod}. Technologies such as dCache and \ac{SAM}~\cite{POMS_SAM} allow for robust storage and cataloguing of data.

The CPU-intensive process of data reconstruction involves the processing of raw data to form physics-level data objects (e.g. tracks). This process is facilitated using distributed computing resources on the FermiGrid~\cite{FermiGrid} and the \ac{OSG}~\cite{OSG}. To simplify this workflow, a \ac{POMS}~\cite{POMS_SAM} is used, that allows automated scheduling and monitoring of data production. 

\clearpage

\subsection{Data quality control} \label{sec:dqc}
\begin{figure}[htpb]
    \centering
    \includegraphics[height=6.5cm]{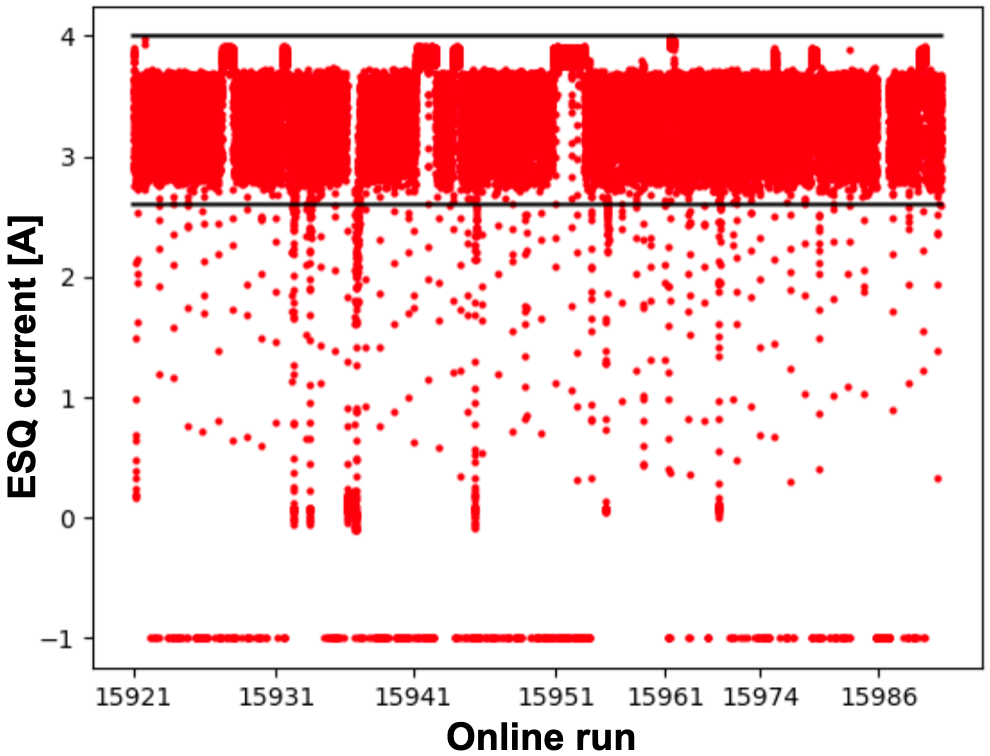}
    \caption[Data quality control]{The current fluctuations of the ESQs during the 1a dataset as a function of the online run number. The acceptable variation is indicated by two horizontal lines, which contain $94.8\%$ of the data points. Plot courtesy of F. Gray~\cite{Fred}.}
    \label{fig:dqc}
\end{figure}

It is essential that all data used in the analysis (c.f. \cref{tab:DS}) is of good quality. This good quality condition is defined by all subsystems of the experiment performing at their nominal operational settings. If an apparatus, for example, the \ac{ESQ}, recorded an abnormal drop in its current, the recorded data during this time is marked as being of bad quality and is not used during the analysis. This is shown in \cref{fig:dqc}.

\subsection{Track reconstruction in the UK}\label{sub:track_production_in_the_uk}
The processing of large datasets is an intensive and costly task. In the context of the \gm2 data reconstruction, the tracking algorithms use up to $50\%$ of the CPU, while only accounting for $10\%$ of the raw data volume. Therefore, the addition of extra grid computing resources dedicated to tracking is well-motivated.

\begin{figure}[htpb]
    \centering
    \includegraphics[width=\linewidth]{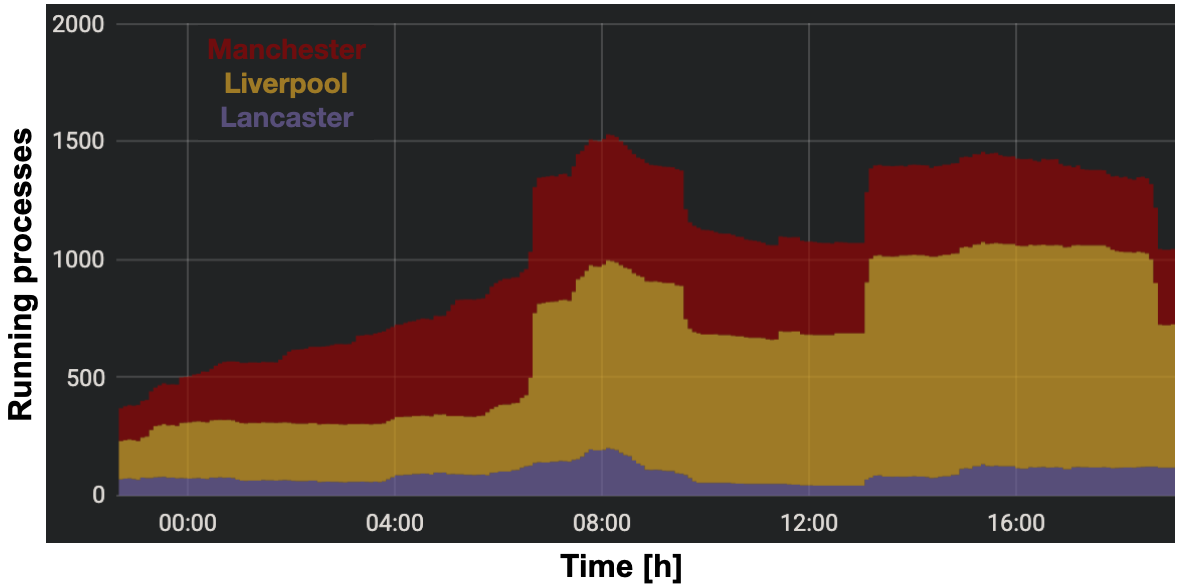}
    \caption[The UK grid nodes]{The UK grid nodes running \gm2 reconstruction on 13 March 2020.}
    \label{fig:pool}
\end{figure}

In \R2, 3 TB of raw track data was recorded, while the reconstruction step will produce 50 TB of track data. To facilitate this, grid nodes at the UK universities in Manchester, Liverpool and Lancaster were added to the \say{\gm2 pool}, as shown in \cref{fig:pool}. \verb!Singularity!~\cite{Singularity} was used to ensure that the \gm2 software could execute on the UK grid nodes.

\clearpage

\section{The simulation framework} \label{sec:simulation_framework}
\begin{figure}[htpb]
    \centering
    \includegraphics[width=0.85\linewidth]{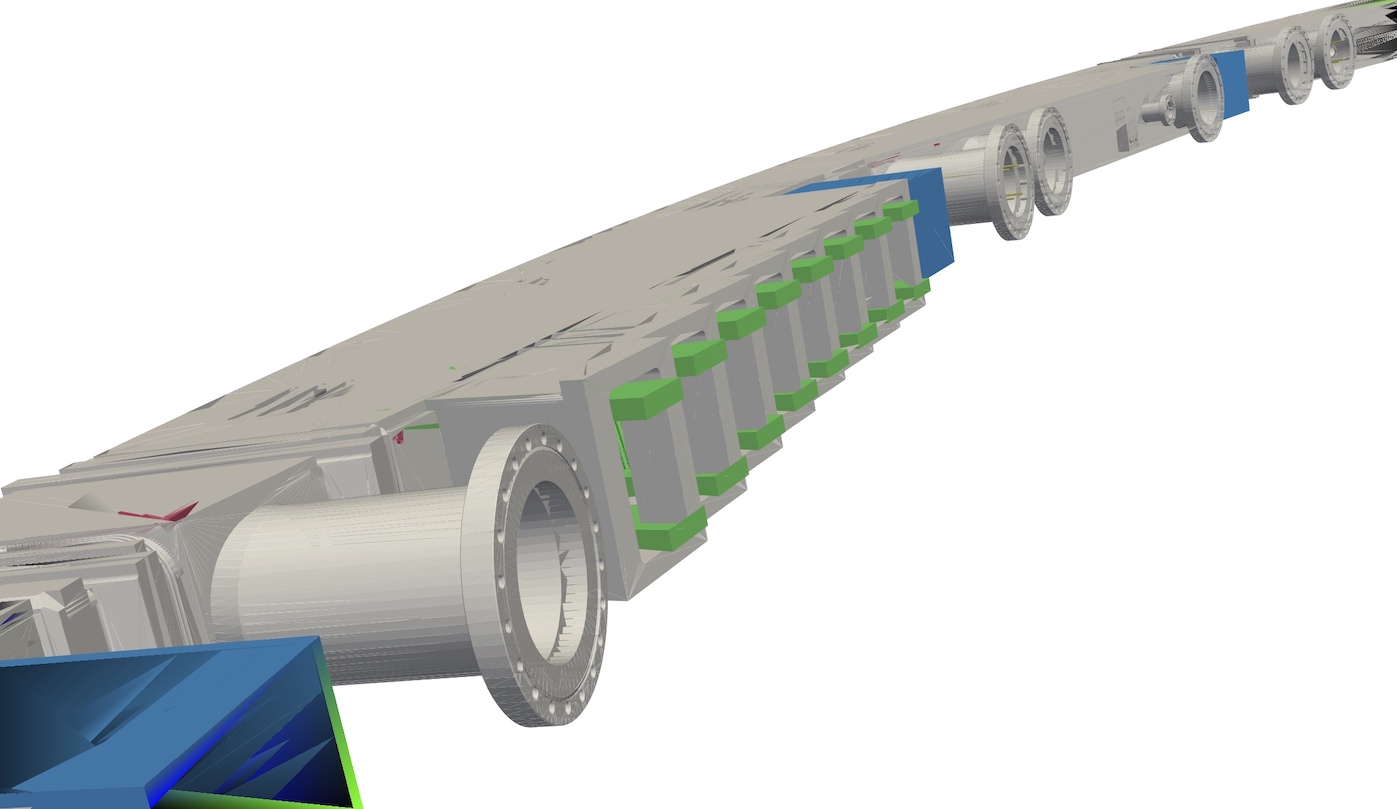}
    \caption[The storage ring in the simulation]{A portion of the storage ring in the simulation. Rendering courtesy of L. Welty-Rieger~\cite{Leah_2}.}
    \label{fig:geant4}
\end{figure}

A simulation framework is essential in understanding the performance of the experiment. This framework is a \verb!C++! code-base, that relies on Fermilab's central software infrastructure, \textit{art}~\cite{art}, which distributes external software products (e.g. \verb!ROOT!~\cite{root} and \verb!Geant4!~\cite{geant4}). The geometry of the detectors, in \verb!Geant4!, is used both during particle generation and during data reconstruction, for both simulated and non-simulated data. The geometry of the storage ring in \verb!Geant4! is shown in \cref{fig:geant4}.
\vspace{-0.3cm}

\graphicspath{{fig/}}

\chapter{Data acquisition system}
\label{ch:daq}

This chapter describes the DAQ system of the \gm2 experiment in \cref{sec:daq_overview}, while a detailed description of the tracking detector readout electronics is given in \cref{sec:trackerDAQ}.

\section{DAQ overview}\label{sec:daq_overview}
\begin{figure}[htpb]
    \centering
    \subfloat[]{\includegraphics[width=0.48\linewidth]{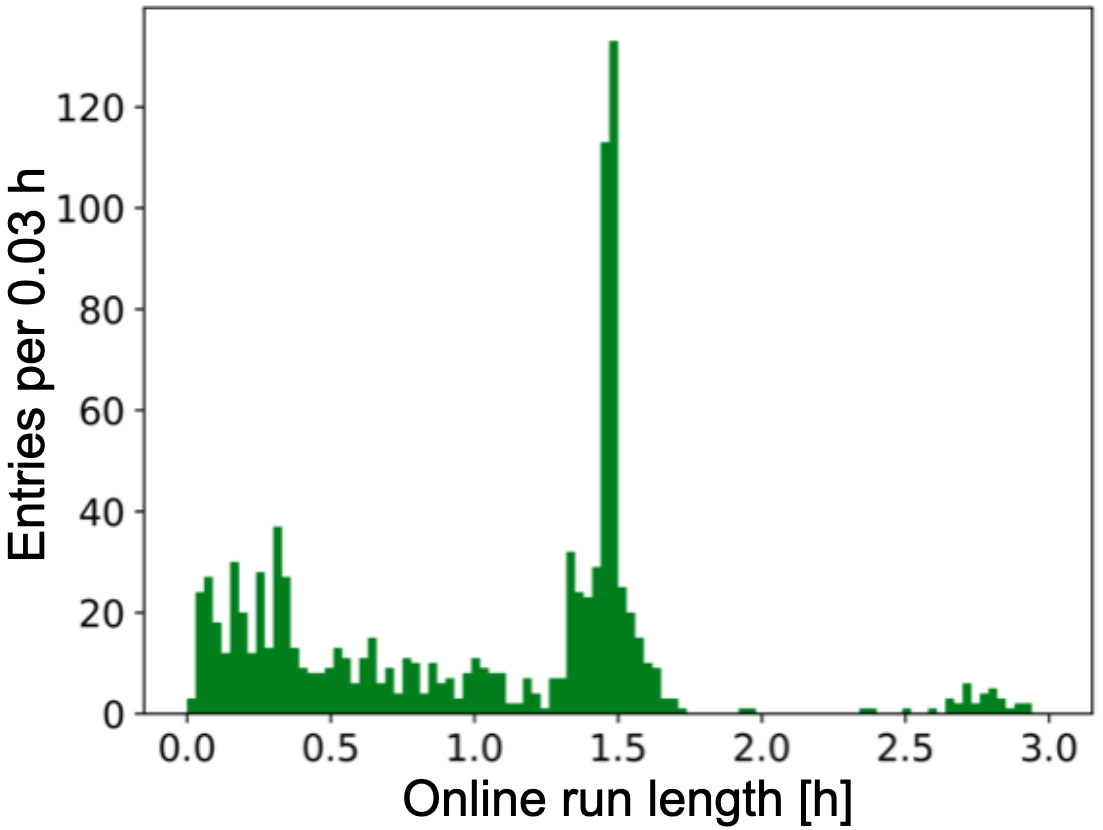}}
    \subfloat[]{\hspace*{2mm}\includegraphics[width=0.48\linewidth]{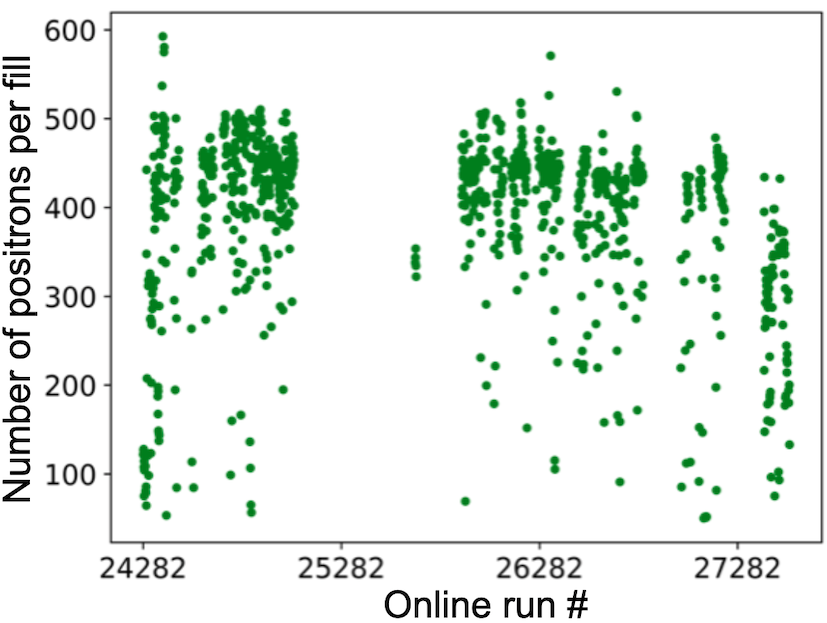}}
    \caption[\R2 DAQ performance]{\R2 DAQ performance. (a) The distribution of online run lengths (including systematic runs). (b) The raw number of recorded positrons per fill.}
    \label{fig:r2_daq}
\end{figure}

An essential component of the experiment is the \ac{DAQ} system~\cite{daq}, which manages the data flow from the detector electronics. In addition, the DAQ interfaces to the accelerator signals (triggers) and distributes a clock. The experiment is acquiring raw data at a rate of 20 GB/s. This is accomplished by employing a parallel data-processing architecture using 28 high-speed GPUs (NVIDIA \textit{Tesla K40}) to process data from 12-bit waveform digitisers, reducing the recorded (to tape) data rate to 200 MB/s. The system processes data from 1296 calorimeter channels (54 channels per calorimeter), two straw tracker stations, auxiliary detectors, the kicker, \ac{ESQ} and \ac{NMR} probes. The total data output of the experiment to tape during \R1 was 2 PB. Additionally, a PCI-based GPS synchronisation card is used to timestamp the digitised data to facilitate subsequent matching between the detector system readout and the magnetic field readout. The \ac{DAQ} system achieved $>90\%$ uptime in \R1, with nearly twice the amount of collected data compared to the E821 experiment, as shown in \cref{fig:pot}. The DAQ performance in \R2, in terms of the online run length and the number of recorded $e^+$ per fill, is shown in \cref{fig:r2_daq}.

\clearpage

\begin{figure}[htpb]
    \centering
    \includegraphics[width=\linewidth]{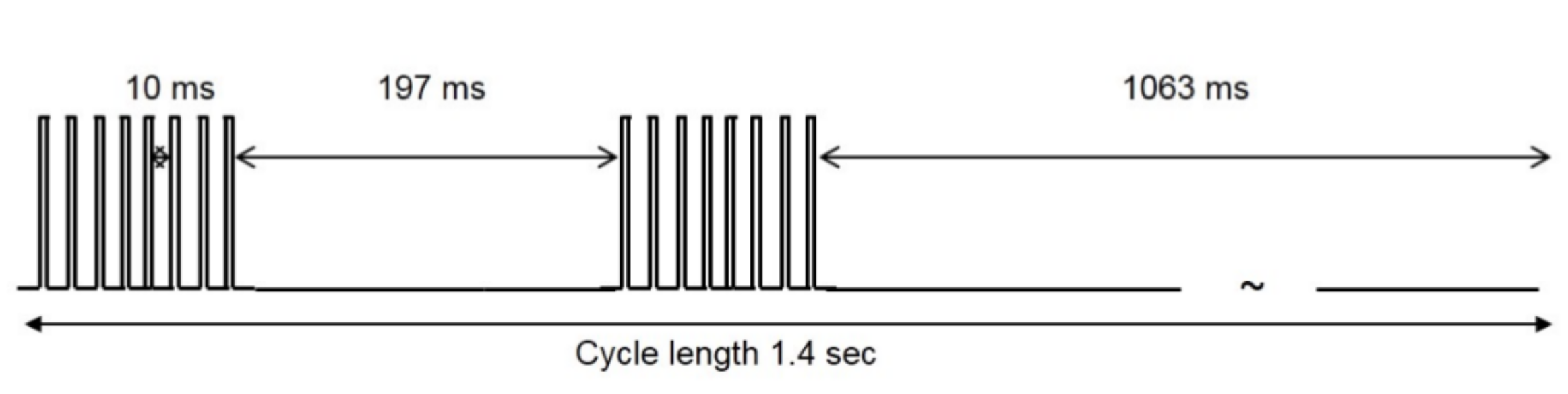}
    \caption[Fill structure]{Fill structure of the experiment: 1 ms muon fills are separated by 10 ms gaps, which set the frequency of the data readout. The larger gaps (with no fills) are the consequence of sharing the accelerator with other experiments at Fermilab. These time gaps are used by the DAQ system to write buffered data to disk. Image courtesy of the \gm2 collaboration~\cite{FNAL_TDR}.}
    \label{fig:fill}
\end{figure}

Each proton bunch delivered by the accelerator complex (see \cref{sc:beamline}) to the pion production target corresponds to a muon fill in the \gm2 storage ring. One muon fill corresponds to a single accumulation (i.e. triggered time window). DDR3 memory blocks allow the data from multiple triggers to be buffered in hardware, which allows the exploitation of the periodic long gaps between muon fills for asynchronous readout. The readout is decoupled from the trigger sequence (in time) to avoid buffer overfill, thus preventing back-pressure on the upstream components. The fill structure is shown in \cref{fig:fill}.

\clearpage
During the data-taking periods, two shifters are always present in the control room of the \gm2 experiment. The shifters control and monitor the data flow from the detector systems. This control is implemented via a state machine interface implemented in \texttt{MIDAS}~\cite{MIDAS}. In addition to shifters, 24/7 on-call \gm2 experts for the DAQ, as well as the other subsystems, are available to ensure any software or hardware issues are solved in a timely fashion.

\subsection{Data quality monitoring}
\begin{figure}[htpb]
    \centering
    \subfloat[]{\includegraphics[width=\linewidth]{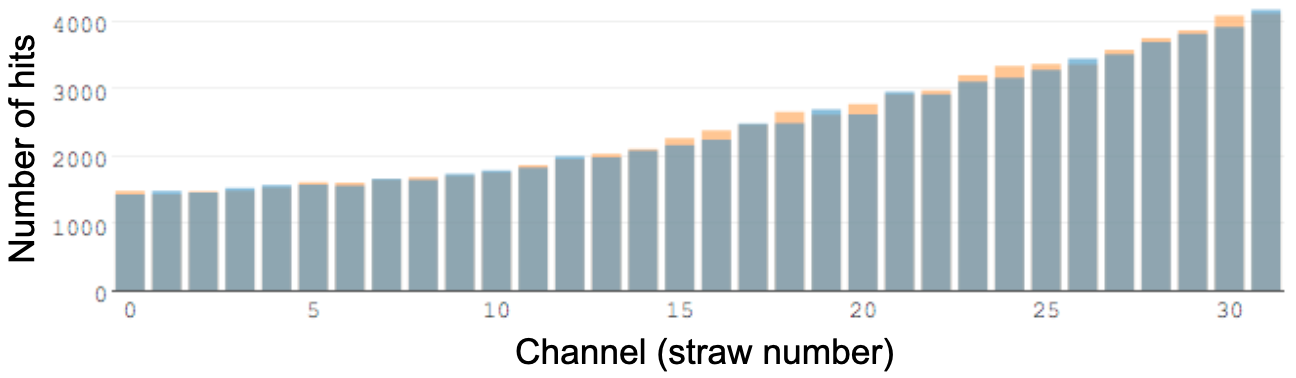}} \\
    \subfloat[]{\includegraphics[width=\linewidth]{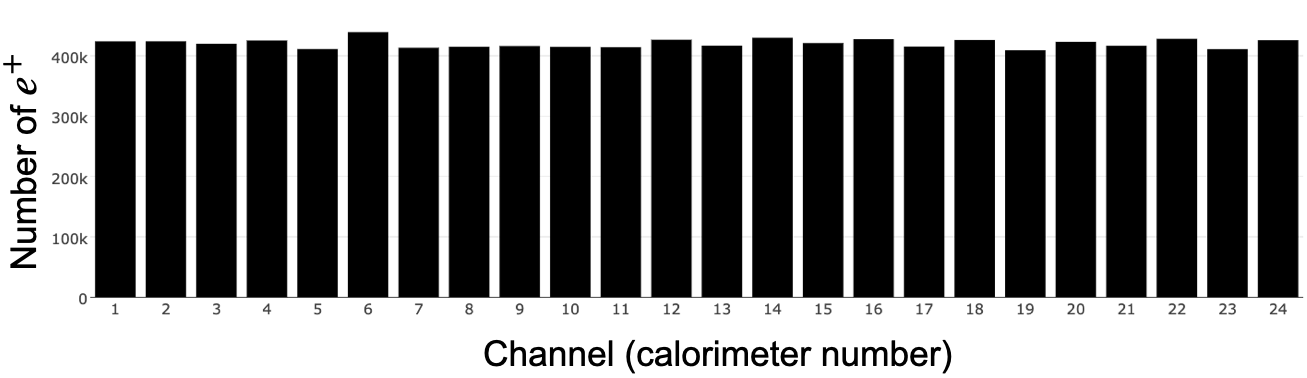}}
    \caption[Data quality monitoring plots]{Examples of data quality monitoring plots. (a) The recorded number of hits in a tracker module during a data-taking period. The number of hits is higher in the channels (straws) located closer to the muon beam.  (b) The recoded number of positrons in the 24 calorimeters.}
    \label{fig:dqm}
\end{figure}

To ensure that there are no issues during the data-taking, it is desirable to reconstruct and analyse a fraction of newly-recorded data. A data quality monitoring system~\cite{Aaron} assesses the performance of the experimental subsystems in real-time, as shown in \cref{fig:dqm}. 

\vspace{1cm}

\clearpage
\subsection{Clock blinding}\label{sec:blind}
\begin{figure}[htpb]
\centering
   \includegraphics[width=0.4\linewidth]{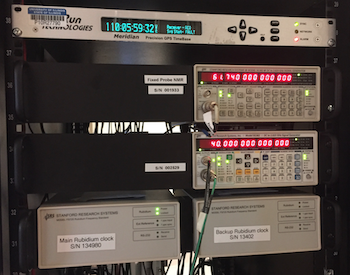}
   \vspace{-0.2cm}
   \caption[The master clock crate]{After the blinding was implemented, the master clock crate was covered and locked to the collaboration, and monitored weekly by Fermilab scientists independent of \gm2. Image courtesy of L. Gibbons~\cite{Lawrence}.}
  \label{fig:ccc}
\end{figure}

A common practice in scientific investigations is to try and reduce cognitive bias by \say{blinding} the data. In the \gm2 experiment, this is achieved in hardware by offsetting the $40$~MHz master clock, shown in \cref{fig:ccc}.
\vspace{-0.2cm}

\vspace{-0.34cm}
The clock is offset by a frequency $\zeta$, which shifts the measured value of $\omega_a$ (and $\omega_p$) by as much as $75$~ppm. This process is known as absolute blinding. $\zeta$ is monitored to prevent any drifts in the blinding signal, as described in~\cite{David}. The absolute blinding is augmented by an additional software blinding, shifting $\omega_a$ in the uniform range of $\pm24$~ppm with an additional $\pm1$~ppm Gaussian tail. The net shift in the value of $\omega_a$ due to this double-blinding is by design larger than the previous $\omega_a$ measurement's discrepancy with the theory of $2.4$~ppm (see \cref{sc:mma}). 

In fitting the periodic variation of the number of decay positrons in the so-called \say{wiggle plot} (e.g. \cref{fig:1a_wiggle}) to determine $\omega_a$, a parameter $R$ is defined
\small
\begin{equation}
    R = \frac{\omega_a-\omega_a^0}{\omega_a^0},
\end{equation}
\normalsize
where $\omega_a^0=1.439$~\SI{}{\micro\radian}$\mathrm{s}^{-1}$, and $R$ is expressed in ppm. The value of $R$ returned in fitting has the additional software blinding, $\Delta R^{i}_{\mathrm{s}}$, which is different for each analysis team, such that
\small
\begin{equation}
    \omega_a^i = 2\pi\cdot0.2291 \ \mathrm{MHz} \cdot (1+[R+\Delta R^{i}_{\mathrm{s}}]\times10^{-6}).
    \label{eq:blind}
\end{equation} 
\normalsize
Different analyses can be compared by setting $\Delta R^{i}_{\mathrm{s}}=0$, whilst still retaining the hardware blinding.
\clearpage

\section{Tracking detector DAQ} \label{sec:trackerDAQ}
This section describes the readout electronics of the tracking detector, as well as the testing and commissioning of the tracking detector DAQ at Fermilab. A more in-depth description of the development, testing and characterisation of the DAQ for the tracking detector is given by T. Stuttard~\cite{Tom}. 

\subsection{Readout electronics}\label{sec:elec}
\subsubsection{Front-end electronics}
\begin{figure}[htpb]
\centering
   \subfloat[]{\includegraphics[width=0.85\linewidth]{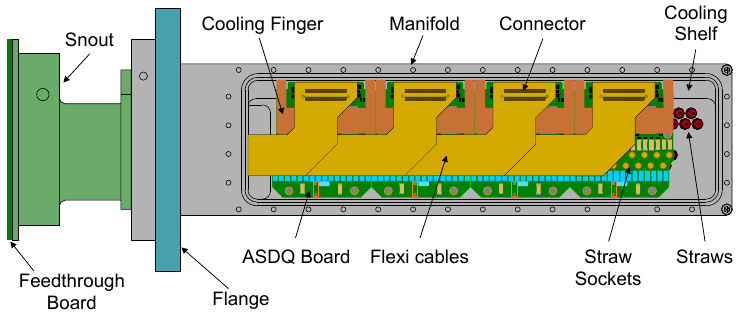}} \\ 
   \subfloat[]{\includegraphics[width=0.8\linewidth]{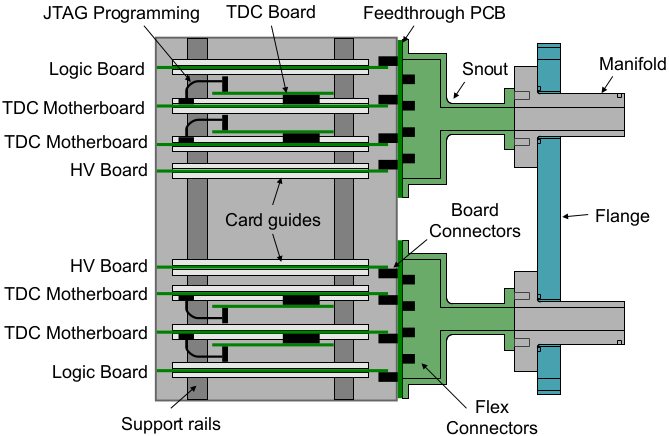}}
   \vspace{-0.2cm}
   \caption[Manifolds and electronics]{(a) Schematic showing the top view of the manifold and the detector electronics. (b) The side view of the logic boards are shown in the FLOBBER. Images courtesy of the \gm2 collaboration~\cite{FNAL_TDR}}
  \label{fig:tracker_electronics}
\end{figure}

The schematic of the \ac{FE} electronics is given in \cref{fig:tracker_electronics}.
\vspace{-0.3cm}

\clearpage
The \ac{ASDQ}~\cite{ASDQ} cards are located inside the top and bottom manifolds. The \ac{ASDQ} processes an analogue signal from a straw, and sends a digital signal to the \ac{TDC} board via a \textit{Flexicable}. Moreover, the \ac{ASDQ} filters the \ac{HV} from the \ac{HV} boards to the straws. A logic board serves as an interface between the \ac{TDC} and the \ac{BE} electronics. The electronic boards external to the gas volume (i.e. the manifold) reside in the \ac{FLOBBER}.

\subsubsection{Back-end electronics}
\begin{figure}[htpb]
    \centering
    \includegraphics[width=.62\linewidth]{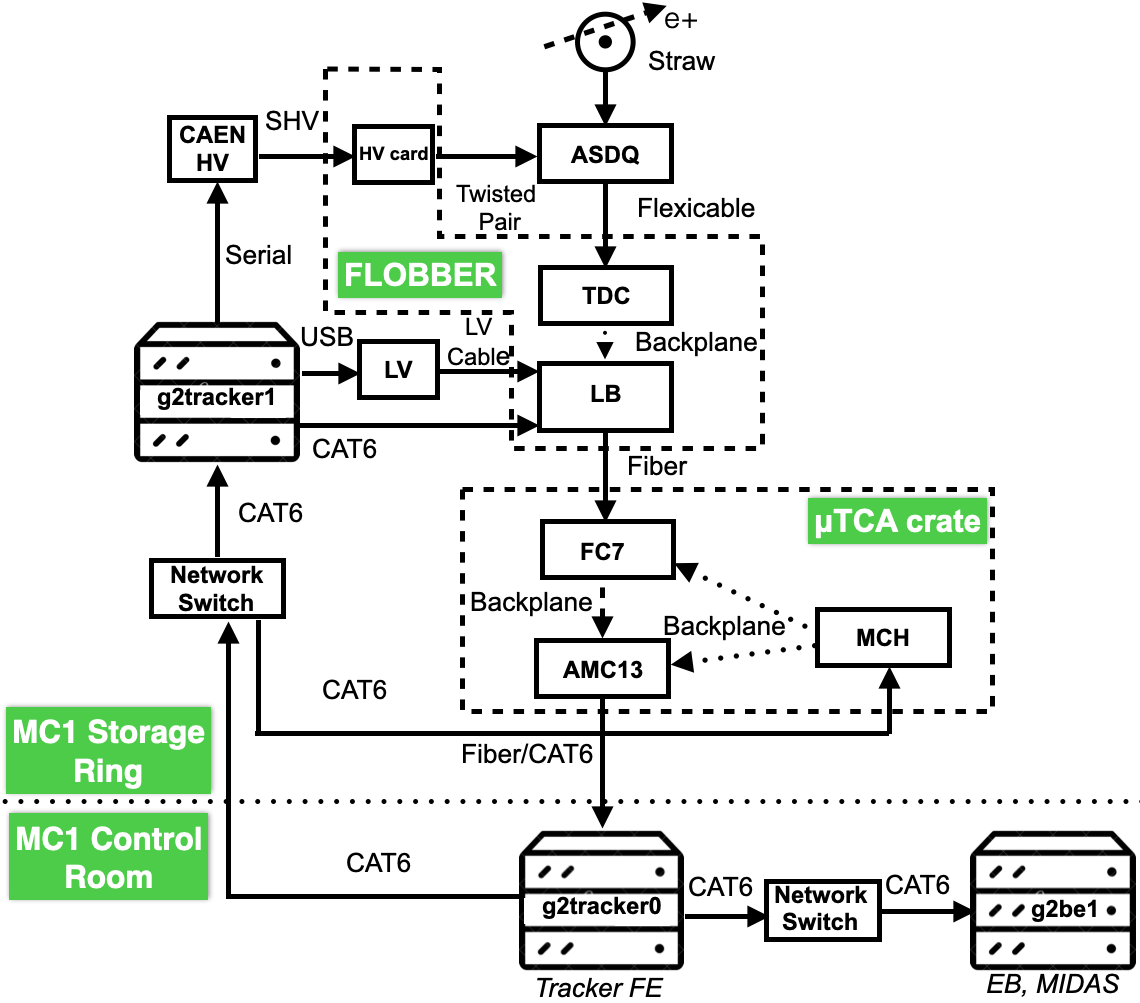}
    \vspace{-0.2cm}
    \caption[Tracker DAQ]{The schematic of the Tracker DAQ from the perspective of a straw.}
    \label{fig:daq1}
\end{figure}

A fibre cable, connected to a logic board, passes digitised data from all the tracking detectors in a station to the FC7~\cite{FC7} card in the $\mathrm{\upmu}$TCA~\cite{MicroTCA} crate (\textit{Vadatech VT891}). An AMC13~\cite{AMC13} card receives the data from the two FC7 cards (one per station), via a backplane, and transmits data to a local PC, which in turn passes it to the Event Builder on a single \ac{BE} machine, on which a central run-control framework -- \texttt{MIDAS} -- is run. The AMC13 is also connected to the 40 MHz external \textit{master clock}. A second tracker PC provides control of the \ac{HV}, \ac{LV} and \ac{SC} via USB serial interfaces. The MCH~\cite{MicroTCA} in the $\mathrm{\upmu}$TCA crate monitors and controls the FC7, AMC13, and manages the crate cooling. This readout chain is depicted in \cref{fig:daq1}.
\vspace{-0.2cm}
\clearpage

\subsection{DAQ development at UCL}
\begin{figure}[htpb]
    \centering
    \subfloat[]{\includegraphics[width=.4\linewidth]{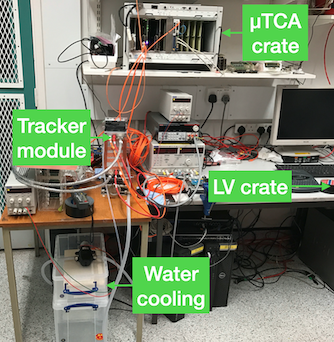}}
    \subfloat[]{\hspace*{0.2cm}\raisebox{4mm}{\includegraphics[height=5cm]{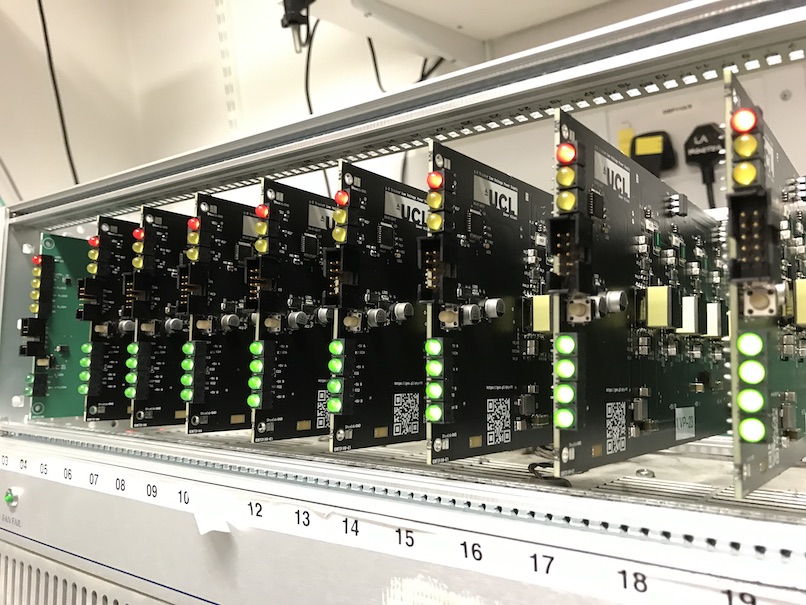}}}
    \vspace{-0.25cm}
    \caption[Test-station at UCL]{(a) Tracker test-station at UCL. A single tracker module with a LV crate, readout DAQ chain, as well as water cooling is shown. (b) LV boards inside the crate during testing.}
    \label{fig:test_station}
\end{figure}

\begin{figure}[htpb]
    \centering
    \subfloat[]{\includegraphics[height=5.2cm]{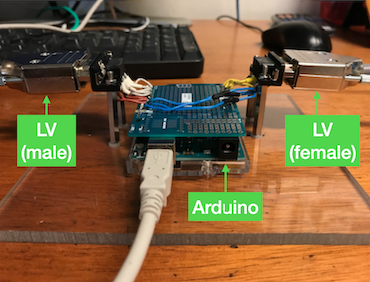}\label{fig:test_lv_a}}
    \subfloat[]{\hspace*{0.2cm}\raisebox{0mm}{\includegraphics[width=.42\linewidth]{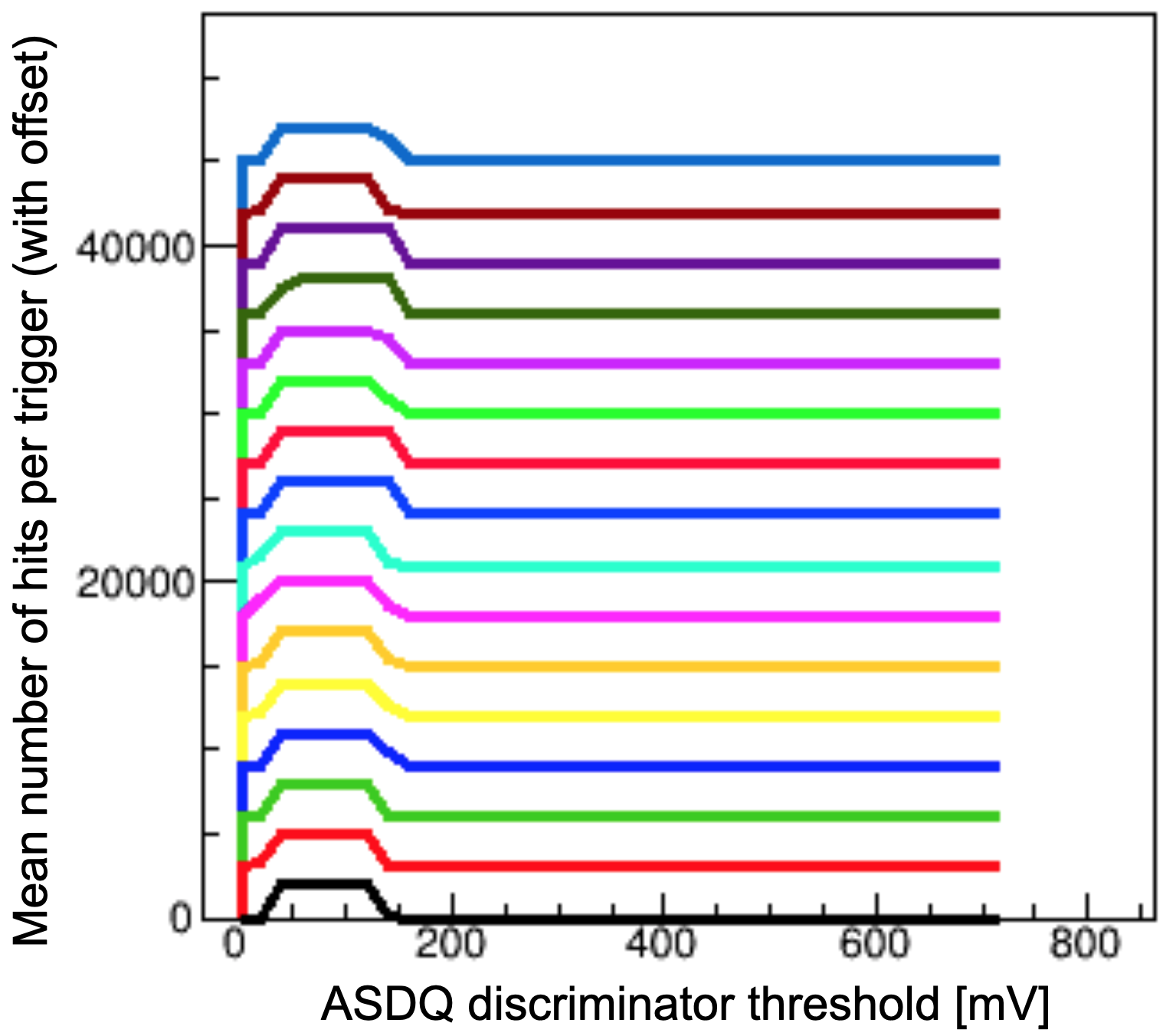}\label{fig:test_lv_b}}}
    \vspace{-0.25cm}
    \caption[LV test-stand]{ (a) \textit{Arduino} LV test-stand. Each of the 9-pins on the male end receives a signal, which is checked at the female connector.  (b) The number of noise hits in 16 ASDQ channels as a function of threshold. All channels display satisfactory noise levels.}
\end{figure}

The \ac{FE} electronics require a steady input of $+5$~V. The LV system was developed by engineers at UCL, and was integrated into the tracker test-station, as shown in \cref{fig:test_station}. Connections between the LV system and logic boards are made via cables with D-sub (9-pin) connectors, which were installed at UCL. To ensure the integrity of cables and connectors, an \textit{Arduino} LV test-stand was developed, as shown in \cref{fig:test_lv_a}. The tracker test-station was used to integrate and test DAQ components. One of the most common tests performed is an ASDQ threshold scan shown in \cref{fig:test_lv_b}.
\vspace{-0.1cm}
\vspace{-1.1cm}
\clearpage

\subsection{DAQ commissioning at Fermilab}
\begin{figure}[htpb]
    \centering
    \subfloat[]{\includegraphics[height=12cm]{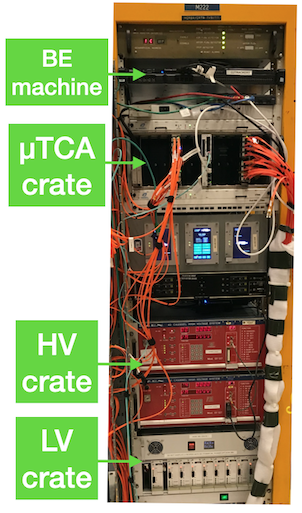}}
    \subfloat[]{\hspace*{0.2cm}\raisebox{2.8cm}{\includegraphics[height=6.5cm]{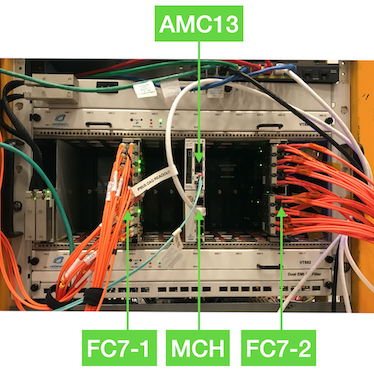}}}
    \caption[Tracker rack]{(a) The tracker rack inside the \gm2 storage ring. (b) The zoom-in on the $\mathrm{\upmu}$TCA crate: the orange fibre cables are carrying the data from the tracker modules to the two FC7 cards.}
    \label{fig:crate}
\end{figure}

The \ac{BE} electronics was assembled and tested at Fermilab as shown in \cref{fig:crate}. The commissioning of the \ac{FE} electronics and tracking detector itself is described in~\cref{sec:hardware}. The first data from the stored muon beam was acquired on 2 June 2017. 

\graphicspath{{fig/}}

\chapter{Tracking detector}\label{ch:tracker}

This chapter describes the hardware and software infrastructures of the tracking detector. The methodology of the \ac{EDM} measurement using data from the tracking detector is given in \cref{sec:edm_track}.

\section{Overview}
\begin{figure}[htpb]
    \centering
    \includegraphics[width=.85\linewidth]{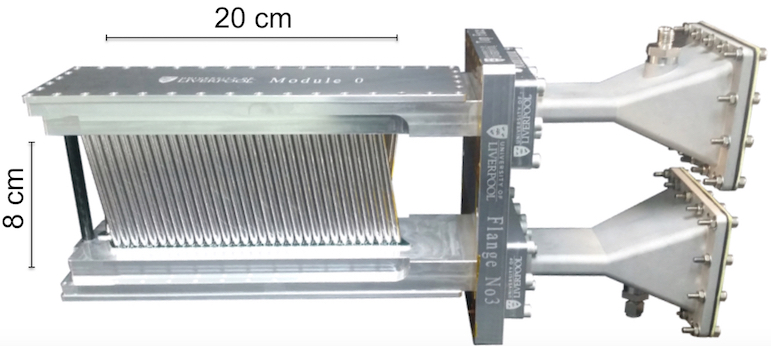} 
     \caption[Tracking detector module]{A single tracker module with 128 straws, which are held between two manifolds. A straw with a diameter of 5 mm is an ionisation chamber filled with 50:50 Ar:Ethane, with a central anode wire at $+1.65$~kV. The active tracking region is inside the storage ring vacuum of $10^{-9}$~atm. A carbon fibre post (black) supports the weight of the top manifold. Images courtesy of the \gm2 collaboration~\cite{FNAL_TDR}.}
    \label{fig:tracker_photo}
\end{figure}

\begin{figure}[htpb]
    \centering
    \includegraphics[width=.8\linewidth]{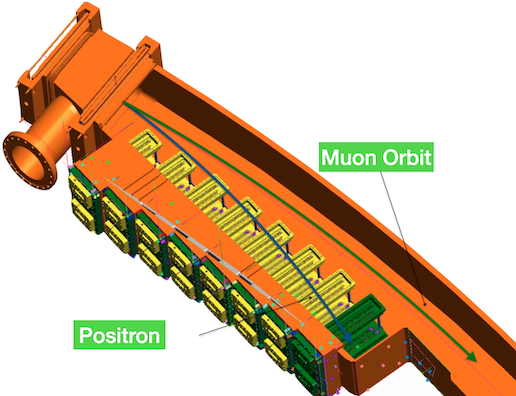} 
     \vspace{-0.2cm}
     \caption[Tracking detector]{Rendering of a positron (from a muon decay) trajectory passing through the tracker station before hitting the calorimeter. The modules are mounted into the vacuum chamber, with the module closest to the calorimeter indicated in green. Image courtesy of the \gm2 collaboration~\cite{FNAL_TDR}.}
    \label{fig:station}
\end{figure}

\begin{figure}[htpb]    
    \centering  
    \includegraphics[width=\linewidth]{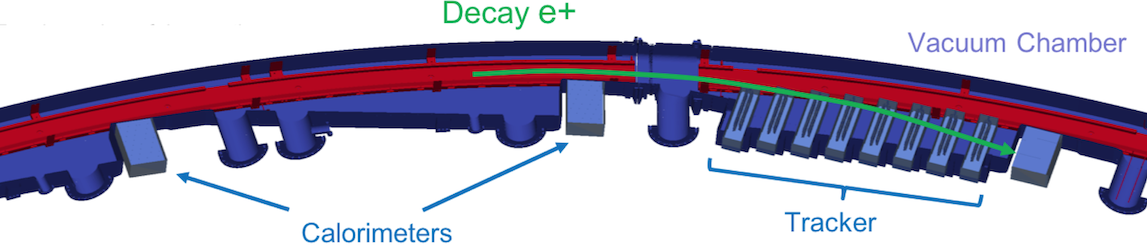}    
    \vspace{-0.3cm}
    \caption[Tracker in the vacuum chamber]{The tracker station is shown with respect to nearby calorimeter detectors on the inside of the storage ring. Rendering courtesy of the \gm2 collaboration~\cite{FNAL_TDR}.} 
    \label{fig:TrackerInChamber}    
\end{figure}

The primary aim of the tracking detector is to reduce the systematic uncertainty on $\boldsymbol{\omega_a}$ via a measurement of the muon beam profile and its evolution with time.  The geometry and the material of the straws are designed to minimise the energy loss and scatter of positrons. The tracking detector, shown in \cref{fig:tracker_photo}, measures the trajectory of the positron from the $\mu^+$ decay in the storage ring. Each tracker module consists of four layers of 32 straws oriented at $\pm7.5^{\circ}$ to the vertical. Each aluminised mylar straw is \SI{15}{\micro\metre} thick, and is held at 1~atm pressure. Each straw is filled with a 50:50 Ar:Ethane mixture and contains a central wire that is held at a $+1.65$~kV potential. The modules are inside the vacuum of $10^{-9}$~atm and experience a predominantly vertical magnetic field. The magnetic field in the tracker region varies radially from $1.45$~T -- at the closest point to the stored muon beam -- to $1.0$~T. Eight tracker modules make up a tracker station, as shown in \cref{fig:station,fig:TrackerInChamber}, with the two stations, labelled S12 and S18, located in front of two calorimeters, as indicated in \cref{fig:ring_pic}. 

\clearpage
\vspace{-0.5cm}
\clearpage

\vspace{-0.2cm}
\section{Hardware}\label{sec:hardware}
The design, manufacture and testing of the tracker modules were done by a team of engineers and scientists at the University of Liverpool, with a detailed description of the process available in~\cite{Tabitha,Will,Talal}.

\subsection{Commissioning at Fermilab}
\begin{figure}[htpb]
    \centering
    \subfloat[]{\includegraphics[height=6cm]{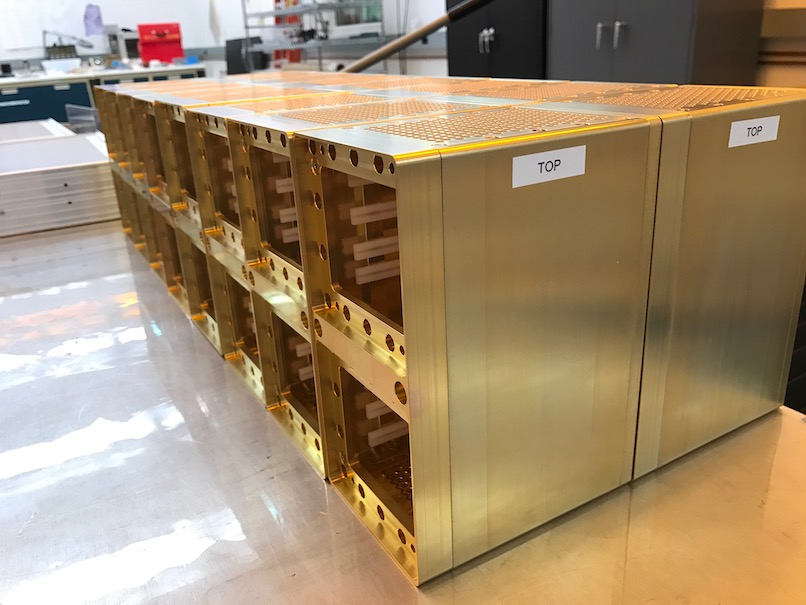}} 
    \subfloat[]{\hspace*{0.1cm}\includegraphics[height=6cm]{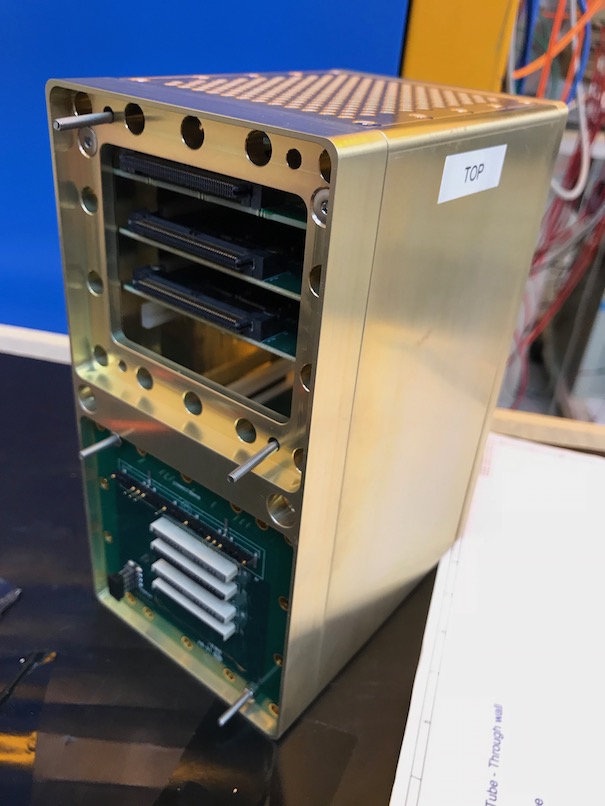}}
    \vspace{-0.2cm}
     \caption[FLOBBER assembly]{Assembly: a) FLOBBER units, and b) FE electronics inside a unit.}
     \label{fig:flobber_assembly}
\end{figure}

The assembly and commissioning of the detector components took place at Fermilab, with the assembled detector system ready for the June 2017 commissioning run. The installation of \ac{FE} electronics into the \ac{FLOBBER} units are shown in \cref{fig:flobber_assembly}. 
\vspace{-0.3cm}

\begin{figure}[htpb]
    \centering
    \includegraphics[width=0.95\linewidth]{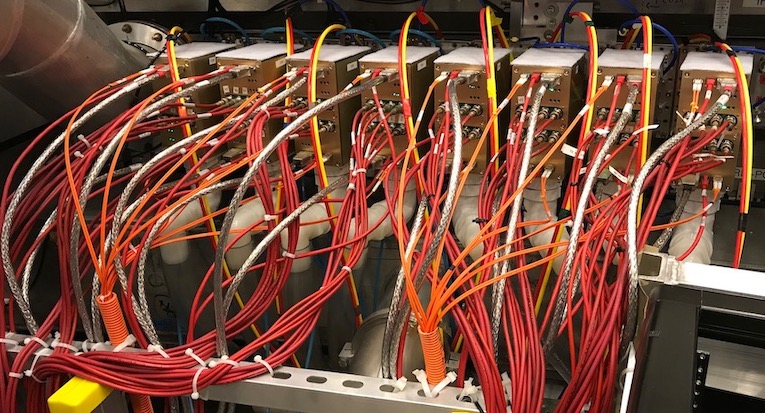}
    \caption[Tracker station]{Assembled tracker station S12 during \R1. The \ac{FLOBBER} units are seen on the outside. The readout fibre cables (orange) bring the data upstream to the DAQ system, while HV (red) and LV (silver) cables provide power to the FE electronics.}
    \label{fig:TrackerStation}
\end{figure}

\begin{figure}[htpb]
    \centering
    \includegraphics[width=0.95\linewidth]{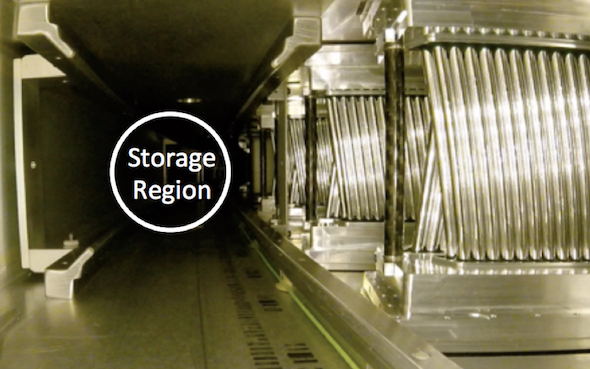}
    \caption[The view of the straws from within the storage region]{The view of the straws and the carbon fibre post from within the vacuum chamber. The muon storage region is indicated. Image courtesy of J. Grange~\cite{Joe_tracker}.} 
    \label{fig:storage}
\end{figure}

\vspace{-0.2cm}
\cref{fig:TrackerStation} shows the FE electronics outside of the vacuum chamber, while \cref{fig:storage} provides a view of the active tracking region inside of the vacuum chamber. 
\clearpage

  
\clearpage
\section{Software} \label{sc:track_soft}
The actual measurement that is being made by the tracker is the time a charged particle traversed a straw -- the hit time. This time needs to be correlated with the mean hit time across multiple straws to calculate the so-called drift-time in a straw. The drift-time is then used to calculate the distance from the central anode wire that the charged particle passed in a given straw. A combination of multiple hit distances allows a reconstruction of the trajectory of that particle. The reconstructed trajectory allows for the measurement of the particle's momentum, as well as -- via \say{track extrapolation} -- an estimation of where the $\mu^+$ decay producing the $e^+$ took place.

\subsection{Hit formation} \label{sub:hit_formation}
\begin{figure}[htpb]
    \centering
    \includegraphics[width=\linewidth]{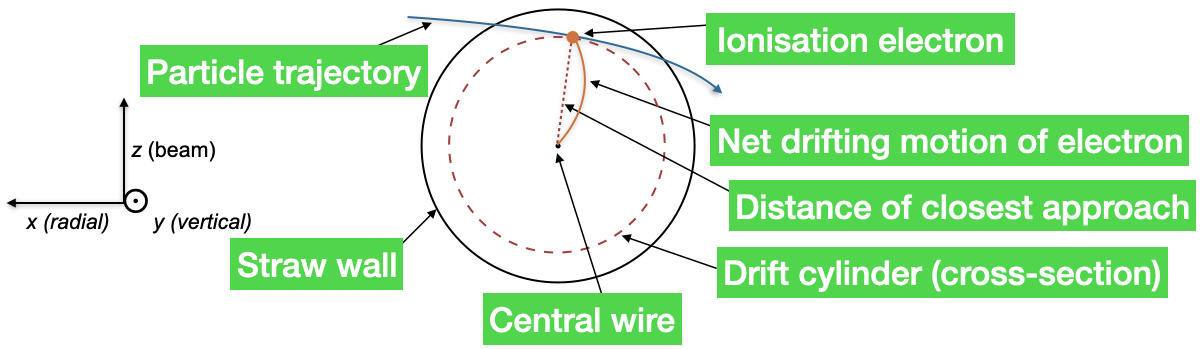}
    \caption[Drift circle]{The drift circle, as a cross-section of the drift cylinder, is indicated. The magnetic field, $B_y=1.45$~T, is along the vertical ($y$).}
    \label{fig:drift_cirlce}
\end{figure}

\begin{figure}[htpb]
\centering
\subfloat[]{\includegraphics[width=0.4\linewidth]{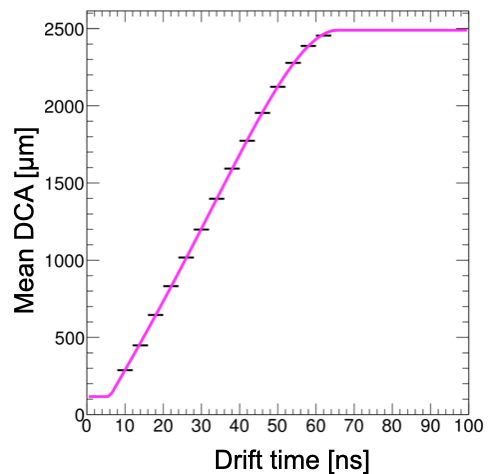}\label{fig:drift}}
\subfloat[]{\includegraphics[width=0.4\linewidth]{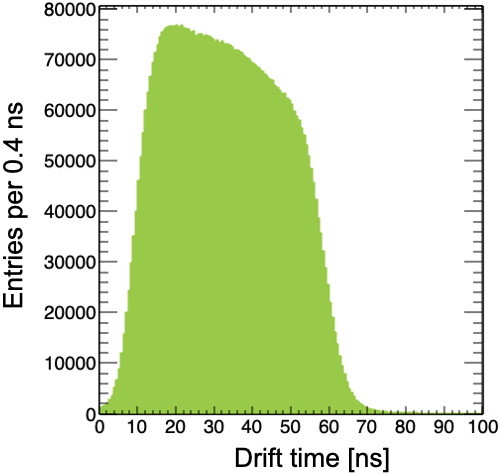}\label{fig:drift_hist}}
\vspace{-0.2cm}
\caption[Drift-time]{ (a) Mean track \ac{DCA} versus drift time. Plot courtesy of G. Hesketh~\cite{Gavin_rt}. (b) The histogram of drift-times in an online run 15922 ($\sim1$~h).}
\end{figure}

The detected signal in the tracker comes from the collected charge on the central wire in the straw. This charge is due to the ionisation electrons, liberated from Ar atoms by a charged particle traversing the straw. The liberated electrons drift towards the wire, which is held at $+1.65$~kV. The net drifting motion of the liberated electrons is curved due to the presence of the magnetic field. The produced charge signal on the wire will be digitised and the drift-time recorded, if the pulse passes the threshold set by the discriminator. The absolute hit position in the straw is not known, only a \textit{drift circle} is determined, a radius of which is given by the \ac{DCA}, as illustrated in \cref{fig:drift_cirlce} in a 2D representation through a cross-section of a straw.
\clearpage
The radius of the drift circle is inferred from the time-to-distance relationship shown in \cref{fig:drift}. The shape of the drift-time distribution is shown in \cref{fig:drift_hist}, where three efficiency regimes are apparent: there are fewer hits with small and large drift-times, corresponding to the hits close and far away from the wire, respectively. The closer tracks produce fewer secondary ionisations, while tracks close to the straw wall produce fewer primary ionisations.
\vspace{-0.3cm}

\begin{figure}[htpb]
    \centering
   \includegraphics[width=0.58\linewidth]{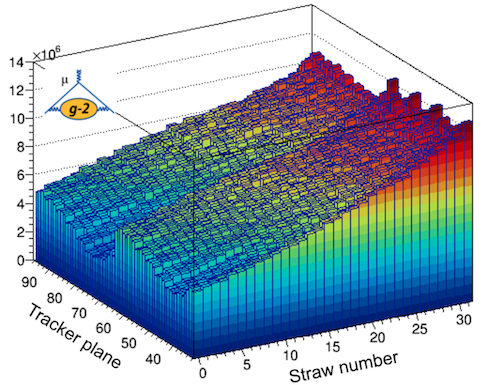}
    \vspace{-0.2cm}
    \caption[Hit occupancy]{Distribution of recorded hits in the two stations. The hit density peaks closer to the storage region on the right. Plot courtesy of the \gm2 collaboration~\cite{FNAL_TDR}.}
    \label{fig:occupancy}
\end{figure}

\vspace{-0.2cm}
The hit occupancy distribution in both stations is shown in \cref{fig:occupancy}, where no cuts are applied, demonstrating that there are no noisy straws. 
\vspace{-0.2cm}
\clearpage

\subsection{Straw coordinate system} \label{sc:cs}
\begin{figure}[htpb]
    \centering
    \includegraphics[width=0.45\linewidth]{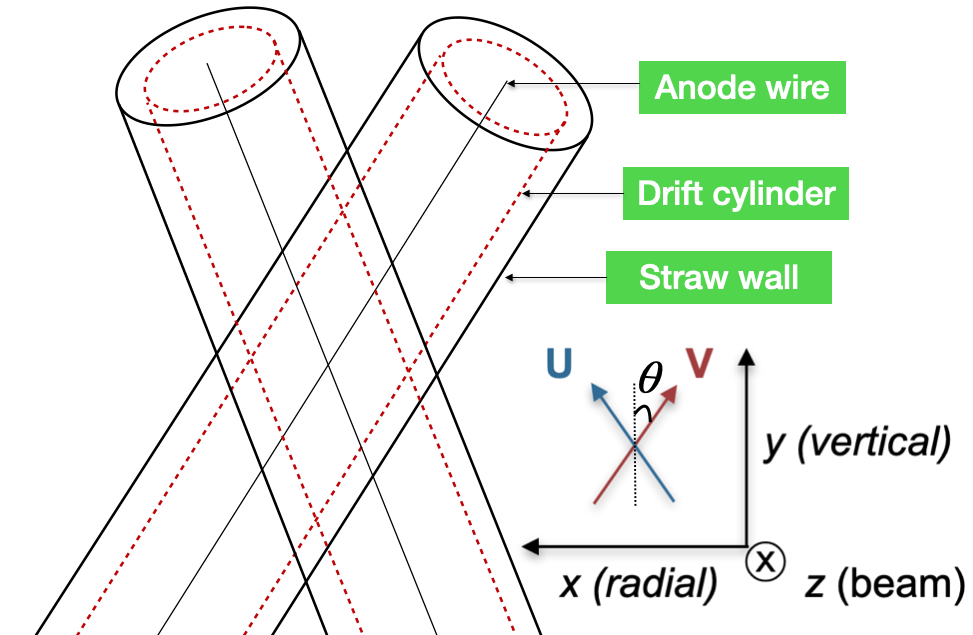}
    \vspace{-0.2cm}
    \caption[Straw coordinate system]{Straw coordinate system.}
    \label{fig:doublet}
\end{figure}

The straws are split into two stereo orientations -- U and V -- each at $\theta=\pm7.5^{\circ}$ to the vertical as shown in \cref{fig:doublet}.
\vspace{-0.2cm}
\vspace{-0.4cm}

A transformation between the global coordinate system (XY) and the local (UV) coordinate system is done via 
\small
\begin{equation}
\begin{pmatrix}u\\v\\\end{pmatrix}=\begin{pmatrix}\cos \theta &-\sin \theta \\\sin \theta &\cos \theta \\\end{pmatrix} \begin{pmatrix}x\\y\\\end{pmatrix}.
\end{equation}
\normalsize

\subsection{Left-Right ambiguity} \label{sub:left_right_ambiguity}
At the point of generation (in simulation) of a hit in a straw, or fitting of a track (in data or simulation), a \ac{LR} \say{sign} is assigned. For example, if the hit is generated on the left side of the straw centre relative to the beam, it will be an \say{L} hit. Hits generated close to the straw centre can be smeared to the other side. This effect can be further magnified if a given hit is also displaced in the same direction (i.e. away from the truth hit position) by the effect of misalignment. 

\subsection{Track reconstruction}\label{sc:track_param}
\begin{figure}[htpb]
    \centering
    \includegraphics[width=.58\linewidth]{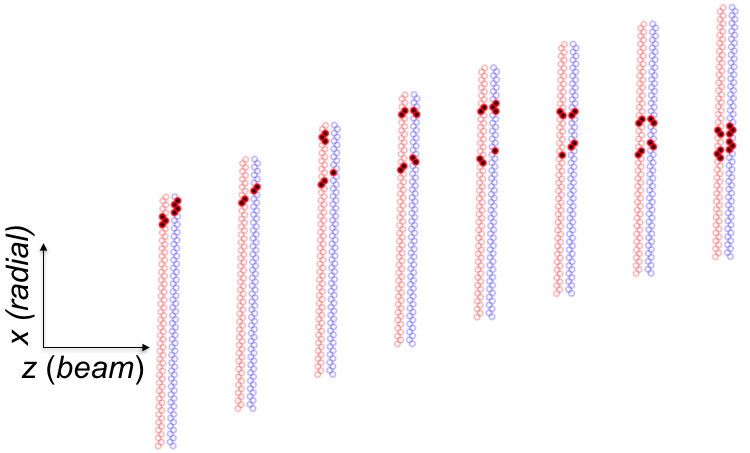} 
    \vspace{-0.2cm}
    \caption[Hits in the straws]{Hits in the straws from two tracks close in time.}
    \label{fig:hits}
\end{figure}

The \gm2 track reconstruction framework was developed by N. Kinnaird~\cite{Nick}, and is implemented with the \verb!GEANE! package~\cite{Geane}. The framework incorporates a model of the tracker's geometry and material, as well as a model of the magnetic field, and utilises transport and error matrices for particle propagation though the straws. \verb!GEANE! is be able to reconstruct tracks when the candidate hits are close together in space and time, as shown in \cref{fig:hits}.
\clearpage
\vspace{-0.2cm}

\begin{figure}[htpb]
\centering
\includegraphics[width=0.93\textwidth]{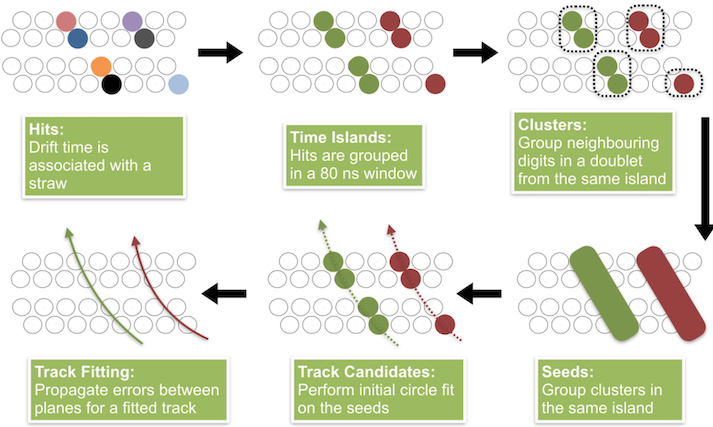}
\vspace{-0.2cm}
\caption[The tracking path]{The steps leading to reconstructed charged particle track trajectories: the originally recorded drift-time data goes through multiple algorithms to form a track. The \texttt{GEANE} package is used in the last step of track fitting.}
\label{fig:Tracking-Path}
\end{figure}

\begin{figure}[htpb]
\centering
    \includegraphics[width = 0.6\linewidth]{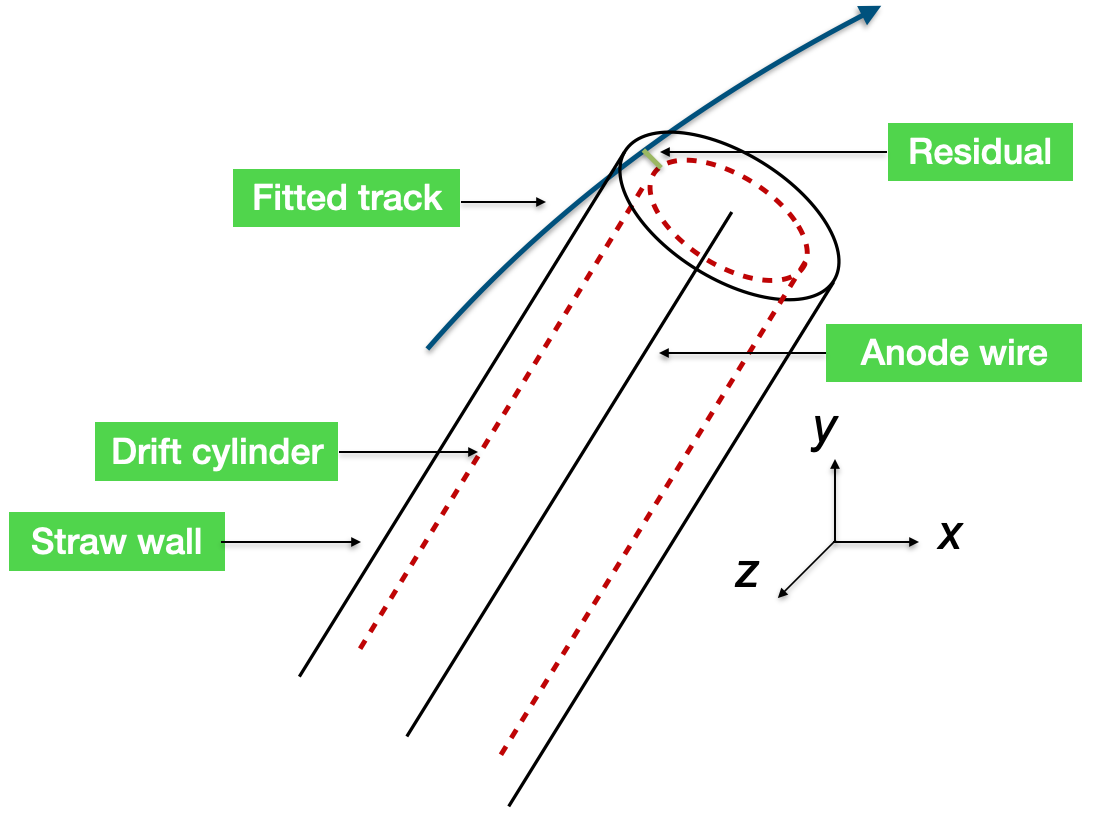}  
    \vspace{-0.2cm}
    \caption[A fitted track though a single straw]{A fitted track passing though a single straw. A residual between the fitted track and the measurement is indicated.}
\label{fig:DriftCylinder} 
\end{figure}

The full tracking algorithm is schematically represented in \cref{fig:Tracking-Path}, while the formation of a residual between the measurement (i.e. drift circle) and the prediction (i.e. fitted track) is depicted in \cref{fig:DriftCylinder}. 
\vspace{-0.2cm}
\vspace{-0.2cm}

\begin{figure}[htpb]
    \centering
    \includegraphics[width=0.6\linewidth]{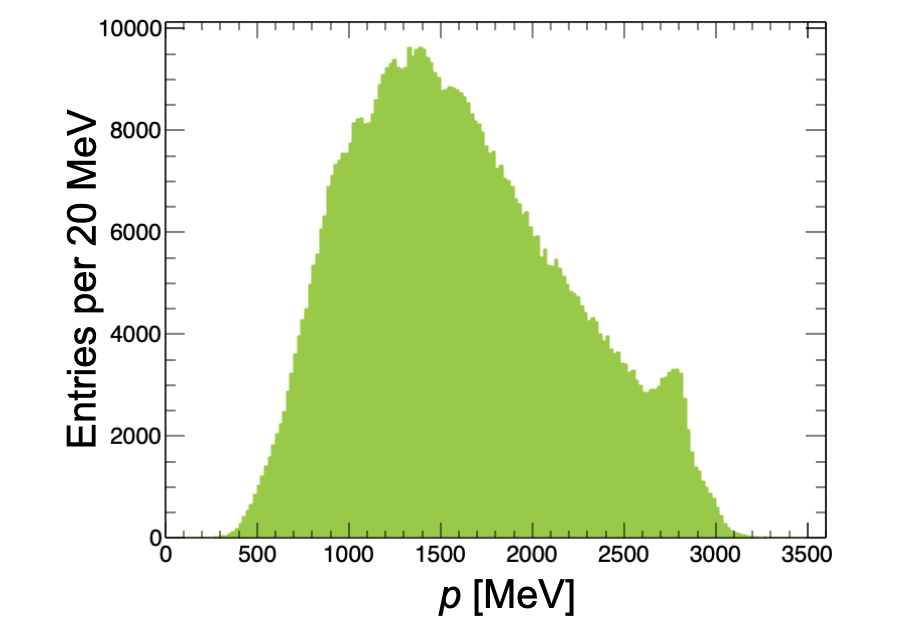}
    \vspace{-0.2cm}
    \caption[Measured track momentum]{Distribution of track momentum ($p$) in an online run 15922. The peak at 2700 \MeV is due to lost-muons (see \cref{sc:lost_muons}).}
    \label{fig:Tracker-Momentum}
\end{figure}

The \ac{DCA} resolution of hits within the straws is approximately \SI{120}{\micro\metre} to \SI{150}{\micro\metre}~\cite{Nick}, which surpasses the design goal of \SI{240}{\micro\metre}. Similarly, a momentum resolution was determined to be at $2\%$~\cite{Nick}, which corresponds to $\sim30$~MeV for the mid-momentum tracks, and is in-line with the design goal. The distribution of measured track momenta is shown in \cref{fig:Tracker-Momentum}.
\clearpage
\vspace{-0.5cm}
\vspace{-0.2cm}

\subsection{Track extrapolation}
\begin{figure}[htpb]
    \centering
    \includegraphics[width=.85\linewidth]{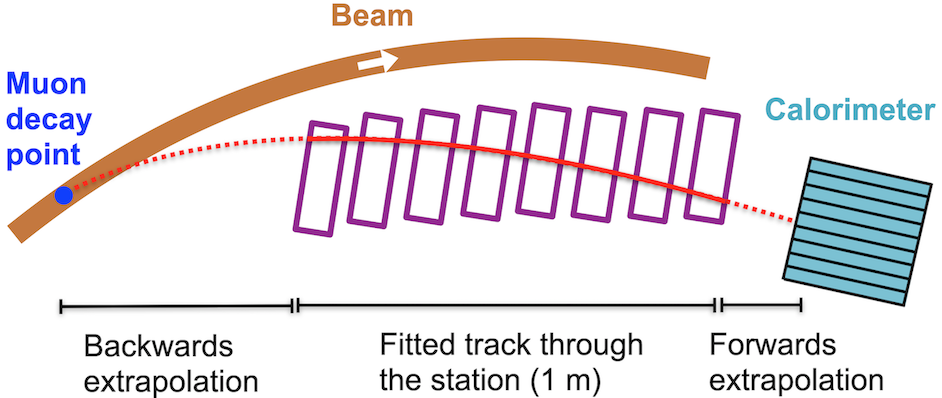} 
    \vspace{-0.2cm}
    \caption[The track extrapolation technique]{A fitted track is extrapolated backwards to the muon decay point, and forwards into the calorimeter.}
    \label{fig:extrap}
\end{figure}

\begin{figure}[htpb]
    \centering
    \subfloat[]{\raisebox{3.5mm}{\includegraphics[width=.52\linewidth]{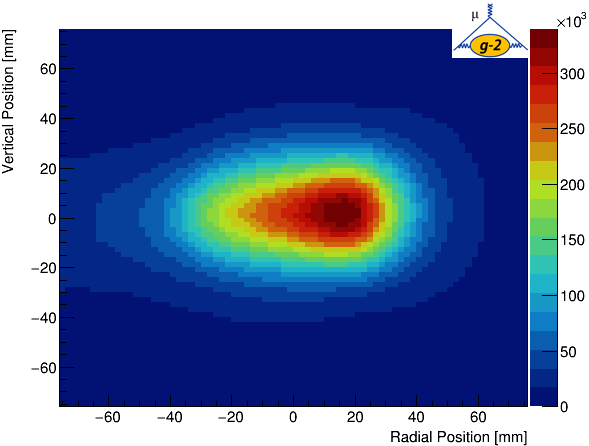} \label{fig:beam}}}
    \subfloat[]{\includegraphics[width=.45\linewidth]{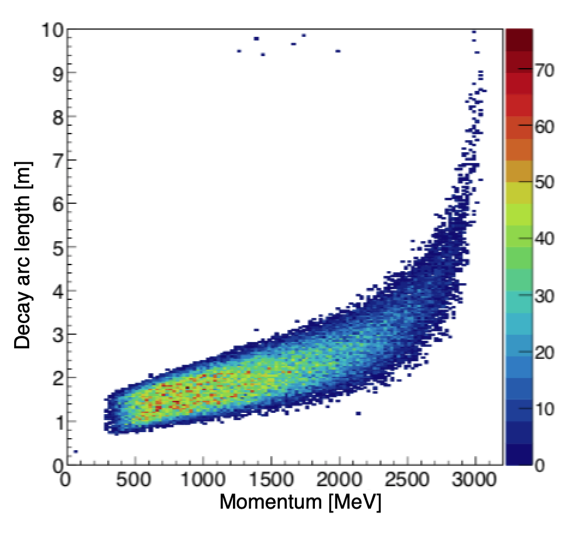} \label{fig:arc}}
    \vspace{-0.2cm}
    \caption[Backward extrapolation of tracks]{Backward extrapolation of tracks. (a) Reconstructed beam profile from tracks that have been extrapolated back to their decay position. (b) Reconstructed decay arc length as a function of track momentum. Plots courtesy of S. Charity~\cite{Saskia}.}
\end{figure}

The \gm2 track extrapolation framework was developed by S. Charity~\cite{Saskia}. The fitted tracks are extrapolated back to the most probable muon decay point, as shown in \cref{fig:extrap}, using a \textit{Runge-Kutta}~\cite{Runge} algorithm that propagates the tracks through the varying magnetic field, until the point of radial tangency is reached. The extrapolated tracks are used to measure the muon beam profile, as shown in \cref{fig:beam}, as well as to calculate the decay arc length, as shown in \cref{fig:arc} where it is clear that higher momentum tracks originate further away from the detector.
\clearpage
\vspace{-0.7cm}
\vspace{-0.3cm}

\begin{figure}[htpb]
    \centering
    \subfloat[]{\includegraphics[width=.48\linewidth]{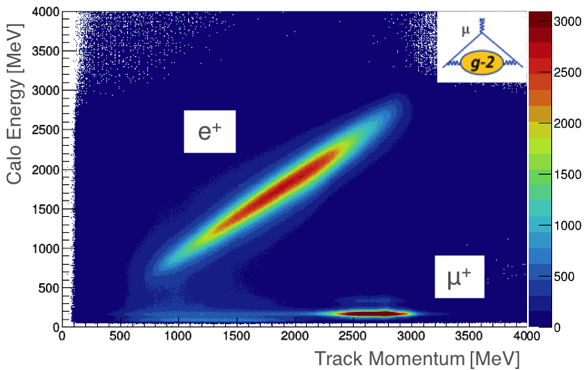} \label{fig:ep}}
    \subfloat[]{\includegraphics[width=.48\linewidth]{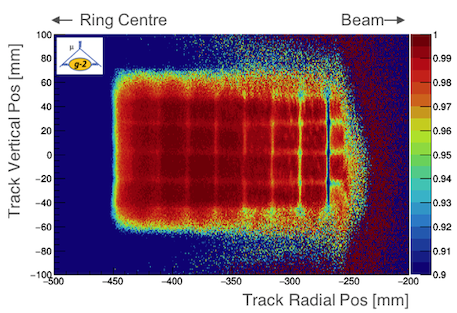} \label{fig:face}}
    \vspace{-0.2cm}
    \caption[Forward extrapolation of tracks]{Forward extrapolation of tracks. Tracks and calorimeter clusters are matched based on their time proximity. (a) Comparing the measured momentum from the tracking detectors with the energy from the calorimeters shows two distinct populations -- positrons and high momentum \say{lost-muons}. (b) Tracks extrapolated to the front face of the calorimeter. The number of matched tracks is divided by the total number of tracks to show the efficiency as a function of position. Nearly all the missing calorimeter hits resemble \say{lost-muons}. Plots courtesy of the \gm2 collaboration~\cite{FNAL_TDR}.}
\end{figure}

The tracks can also be extrapolated forward to the calorimeter, enabling particle identification, as shown in \cref{fig:ep}, as well as to investigate the efficiency of matched calorimeter clusters and tracks, as shown in \cref{fig:face}. The efficiency decreases in the gaps between the crystals where a \say{lost-muon} might split its small energy deposition between the two crystals, neither of which goes above the threshold. The data in \cref{fig:face} contain early times in the fill (\textless~\SI{30}{\micro\second}), so there are more \say{lost-muons}, which artificially lowers the efficiency, but brings out the crystal structure (c.f. \cref{fig:calo_det}). Matching via $E/p$ can also be used to independently monitor the calorimeter gain, and identify the rate of pileup in the calorimeter.
\clearpage

\vspace{-0.5cm}
\subsection{Track quality cuts} \label{sc:track_quality}
Analogous to the data quality control described in \cref{sec:dqc}, the track reconstruction has its own set of criteria to define the quality of a track. These criteria~\cite{James_TQ} are then used to define a sample of tracks appropriate for a particular analysis. Of particular interest to the analysis in this thesis are the cuts imposing:
\begin{itemize} \itemsep -7pt 
\item Extrapolated track did not pass through a significant amount of material (e.g. the wall of a vacuum chamber)
\item Track passed through at least 12 straws
\item Track has a good fit quality with a \pval$>5\%$ 
\end{itemize}
Imposing these and other criteria remove $\sim60\%$ of the tracks, which is an acceptable reduction in statistics given the gain in the data quality. 

\subsection{Magnetic field convolution} \label{sec:field_convolution}
\begin{figure}[htpb]
    \centering
    \includegraphics[width=.6\linewidth]{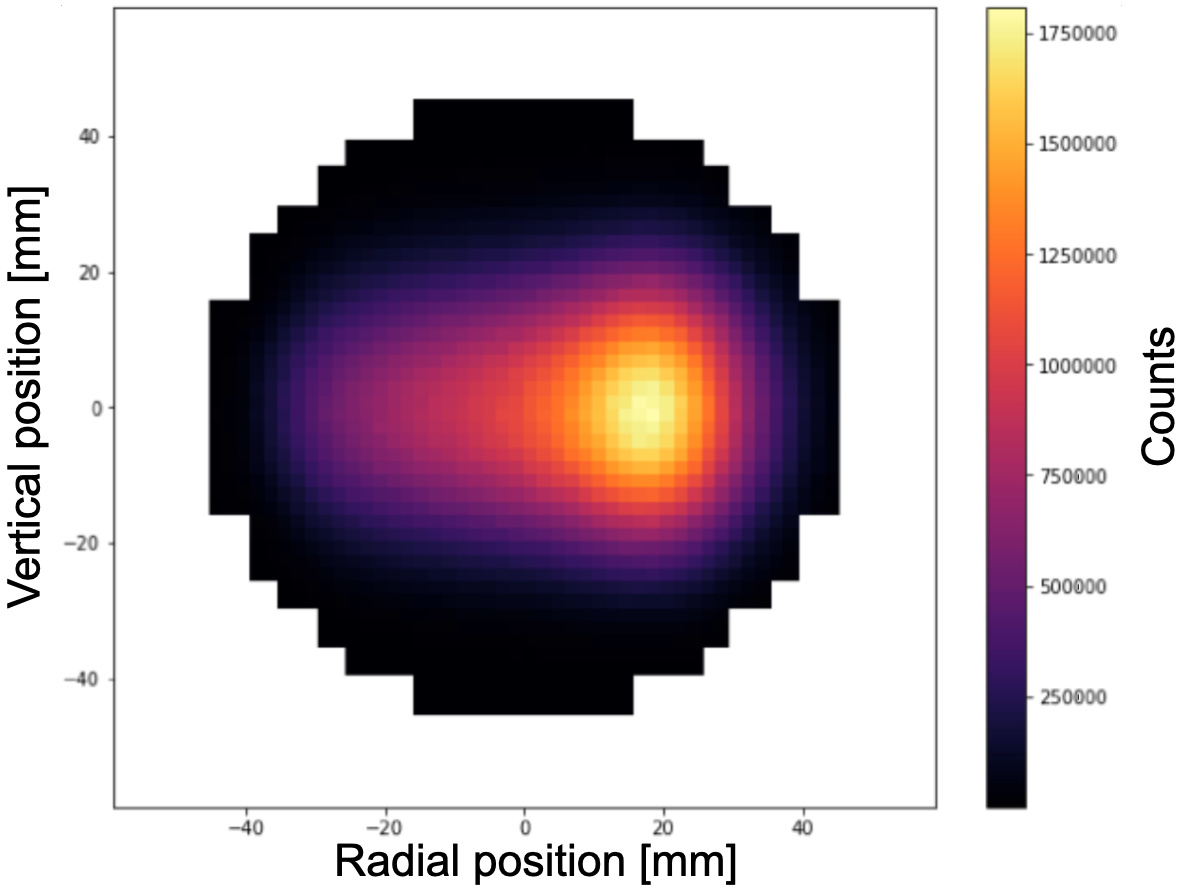}
    \caption[The acceptance-corrected beam profile]{The acceptance-corrected muon profile. Plot courtesy of J. Bono and S. Charity~\cite{JasonSaskia}.}
    \label{fig:beam_acc}
\end{figure}

The magnetic field (\cref{fig:field}) measured by the trolley is convoluted with the beam profile (\cref{fig:beam}) measured by the trackers to find the average field, $\langle B \rangle$, experienced by the muons before decay. Additionally, the calorimeter acceptance (i.e. a fraction of events where a positron is detected by the calorimeters) must be taken into account to estimate the distribution of the muons that are used in the $\omega_a$ analysis. An example of the acceptance-corrected beam profile is shown in \cref{fig:beam_acc}.

\subsection{Tracker calibration and efficiency}\label{sub:tracker_calibration_and_track_refinement}
The track-reconstruction efficiency is defined as the fraction of recorded hits that are associated with tracks. The work to improve the tracking efficiency is ongoing and has so far been able to improve it by a factor of four~\cite{Gavin}. Several factors determine the efficiency including the efficacy of the time-island and clustering algorithms (see \cref{fig:Tracking-Path}), the time-to-distance calibration, the efficacy of making the correct \ac{LR} assignment, and the alignment of the detectors. 

The alignment is an important aspect in improving both the track-reconstruction efficiency and the tracker resolution, with implications for reducing the uncertainty in the tracker-based $\omega_a$ analysis (see \cref{ch:wiggle}) and the \ac{EDM} analysis (see \cref{ch:edm}). The alignment is discussed in detail, and the results of the alignment procedure presented, in \cref{ch:align_error,ch:align,ch:align_curvature}.

\graphicspath{{fig/}}

\chapter{Systematic contribution of the alignment to the beam extrapolation}
\label{ch:align_error}

\section{Introduction}

A precise calibration of the tracking detector is required to reduce the systematic uncertainty on the $a_{\mu}$ measurement and improve the sensitivity to a muon \ac{EDM}. The calibrations are the time-to-distance relationship and the alignment. The alignment has two components: internal and global. The internal alignment considers the positions of the tracking modules within a station, while the global alignment considers the absolute position of the station relative to the rest of the experiment. The internal position of the tracking modules must be known to a high level of precision. Therefore, a physics-level (i.e.~track-based) alignment was implemented with data from \R1 using the \mpt framework~\cite{mp2}. The global alignment was implemented using laser survey measurements of the tracking station chambers. The results and methodology of the global and internal alignments are presented in \cref{ch:align}. 

It is imperative, for the beam measurements made with the tracking detectors, to have an estimate of the systematic uncertainty that comes from an internally misaligned detector. One way to produce such an estimate is to add sets of known misalignment offsets to the positions of the tracker modules and reconstruct data with these offsets. A comparison can then be performed between the nominal case and a case with the added offsets.

\section{Methodology}  

\begin{figure}[htpb]
    \centering
    \subfloat[]{\includegraphics[width = 0.49\linewidth]{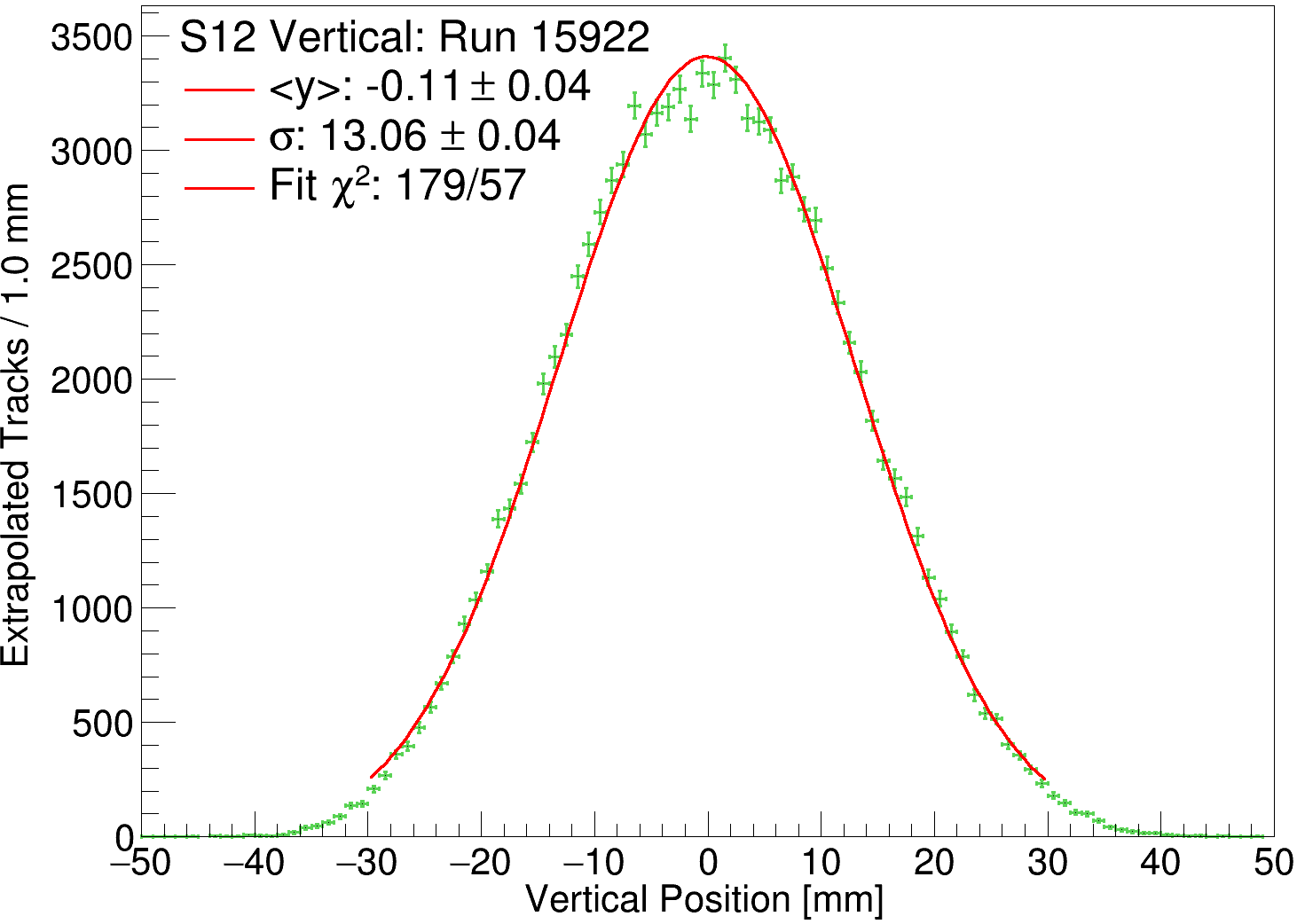}}   
    \subfloat[]{\includegraphics[width = 0.49\linewidth]{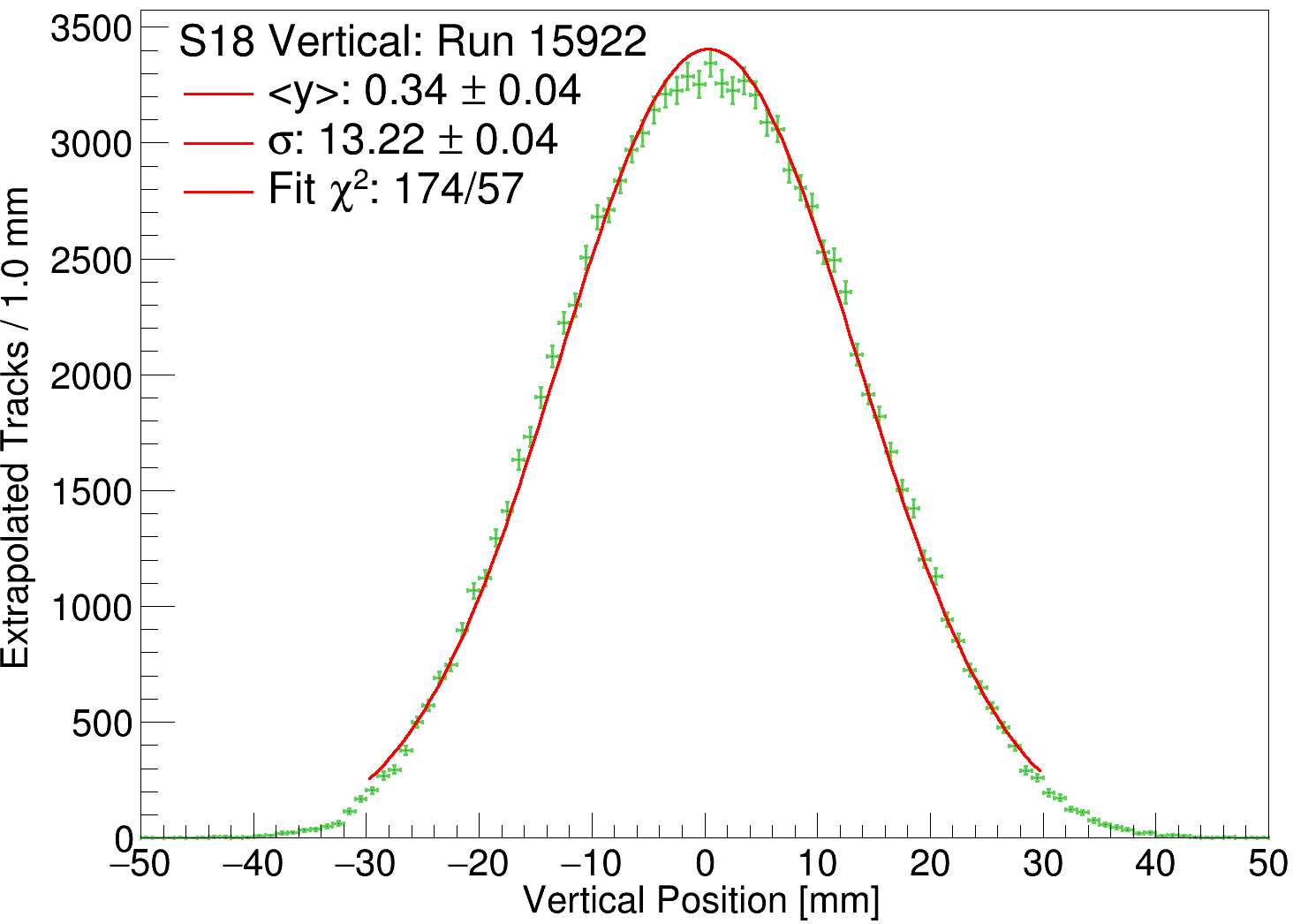}}  \\
    \subfloat[]{\includegraphics[width = 0.49\linewidth]{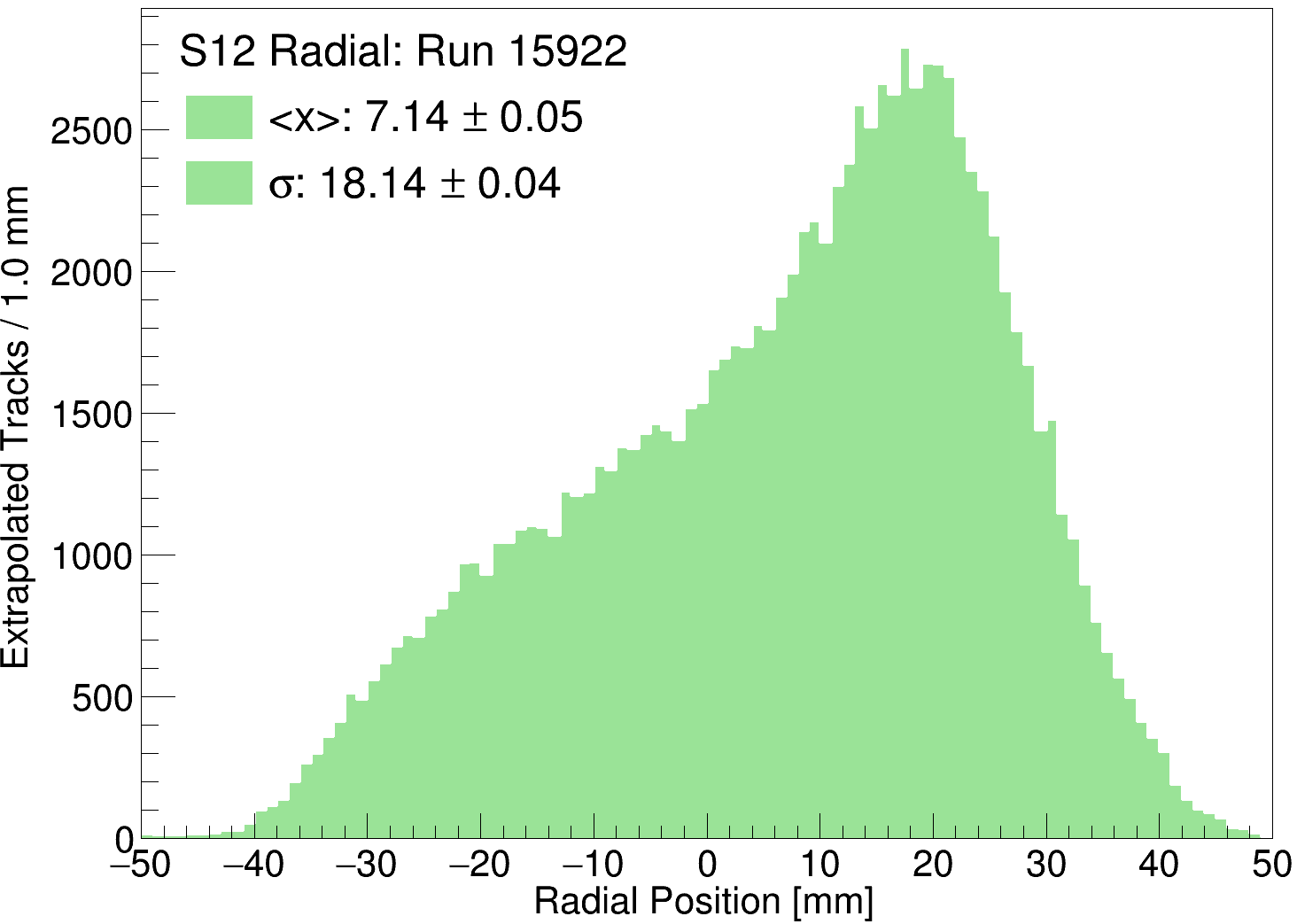}}  
    \subfloat[]{\includegraphics[width = 0.49\linewidth]{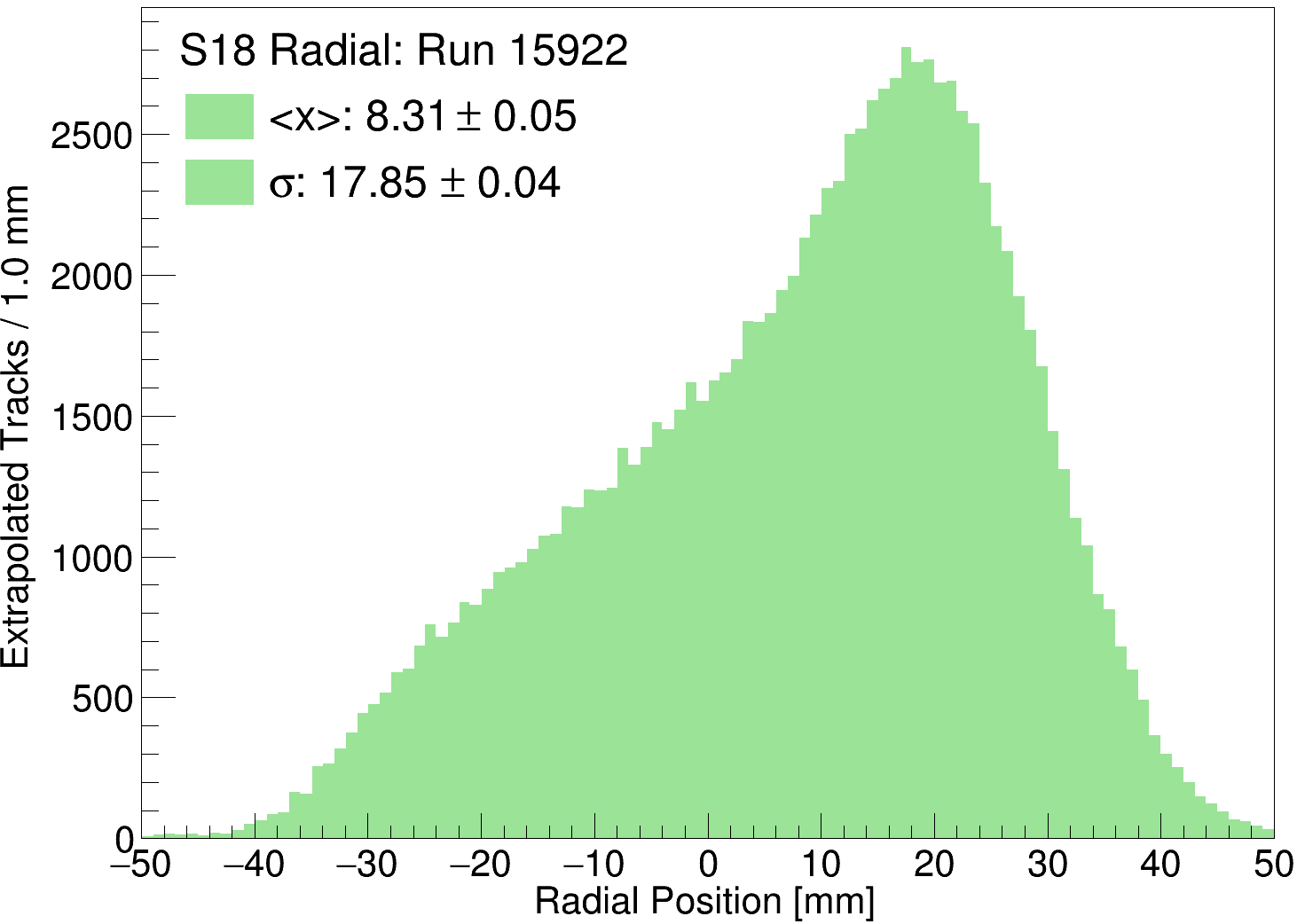}}
    \caption[Run 15922 beam extrapolation]{Run 15922. Nominal radial and vertical beam position reconstructed from $\sim1.5\times10^5$ tracks in the two stations: (a) S12 vertical, (b) S18 vertical, (c) S12 radial and (d) S18 radial. The mean and the width of each distribution are shown.}
    \label{fig:nominalBeam}
\end{figure}

The focus of this study was to quantify the effect that the internal misalignments have on the radial and vertical estimate of the beam's position. The chosen run (nominal case) for this study was run 15922 (22 April 2018) that contained one hour of data, and the nominal extrapolated radial and vertical beam means and widths for both stations are shown in \cref{fig:nominalBeam}. 

\clearpage
All extrapolations were done after applying the track quality cuts. Additionally, a further cut removing tracks that extrapolate back more than \SI{50}{\milli\metre} from the beam centre was implemented. This selects tracks that have come from a uniform field region of the storage ring (see \cref{sc:field}). Moreover, only tracks with a time more than \tms were considered. Such tracks have originated from the decay muons that have already undergone scraping (see \cref{sc:scraping}) and have a stable orbit around the ring. 

To extract the mean and width of the beam, the vertical distribution was fitted with a Gaussian function between \SI{\pm30}{\milli\metre} using the $\chi^2\textnormal{-minimisation}$ method. The radial distribution does not have a Gaussian shape, and so the mean and the width are extracted directly from data. The reason for the non-Gaussian shape of the radial distribution is discussed in \cref{sc:kicker}.

\begin{figure}[htpb]
    \centering
   \subfloat[]{\includegraphics[width=0.89\linewidth]{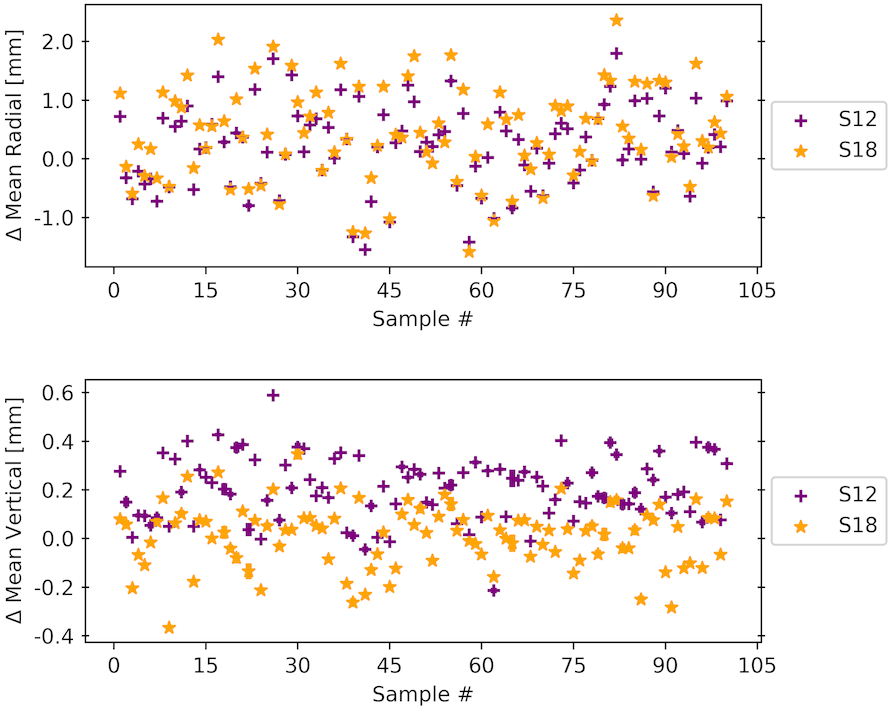}}  \\ 
    \subfloat[]{\includegraphics[width=0.89\linewidth]{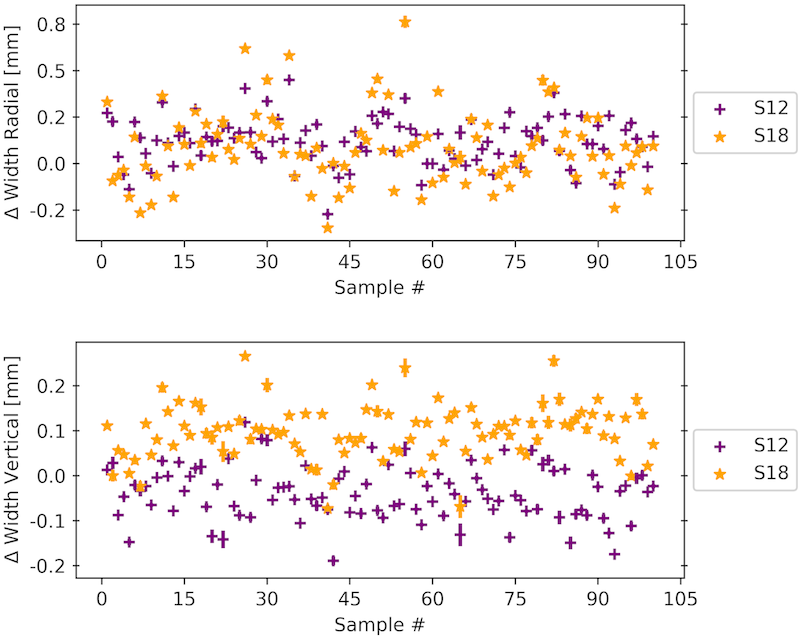}}
    \caption[The difference in the the extrapolated beam positions]{The difference in (a) the mean and (b) the width (standard deviation) of the extrapolated radial and vertical beam positions compared to the nominal case in S12 and S18.}
    \label{fig:P0_100}
\end{figure}

Each of the eight modules in a station was misaligned independently with a misalignment in the range of \SI{-100}{\micro\metre} to $+$\SI{100}{\micro\metre}. In this study, 100 samples of random offsets were used, as shown in \cref{fig:P0_100}. In \cref{ch:align}, it is shown that the mean measured misalignment per module is \SI{31}{\micro\metre} and \SI{82}{\micro\metre} radially and vertically, respectively. Therefore, the estimated systematic error in this study overestimates internal misalignment contributions to the beam extrapolation. 

The task consisted of running the track reconstruction 100 times on a single data run. Each of the two tracker stations yields approximately $\sim4\times10^5$ tracks per run ($\sim$ one hour), which is reduced to $\sim1.5\times10^5$ tracks after the track quality cuts are applied, as described in \cref{sc:track_quality}. 

\clearpage

\clearpage
\section{Results}
\begin{figure}[htpb]
    \centering
    \subfloat[]{\includegraphics[width = 0.49\linewidth]{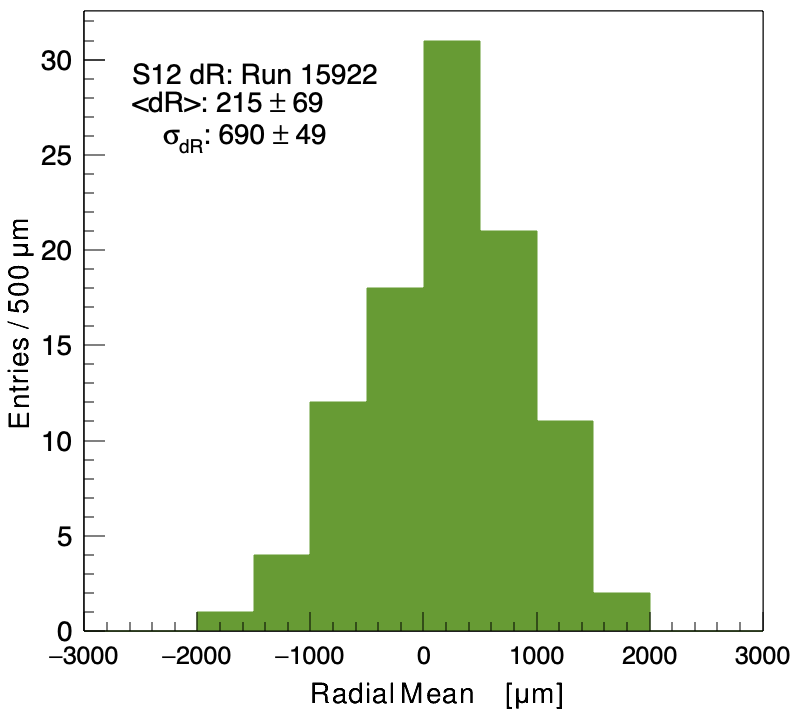}}   
    \subfloat[]{\includegraphics[width = 0.49\linewidth]{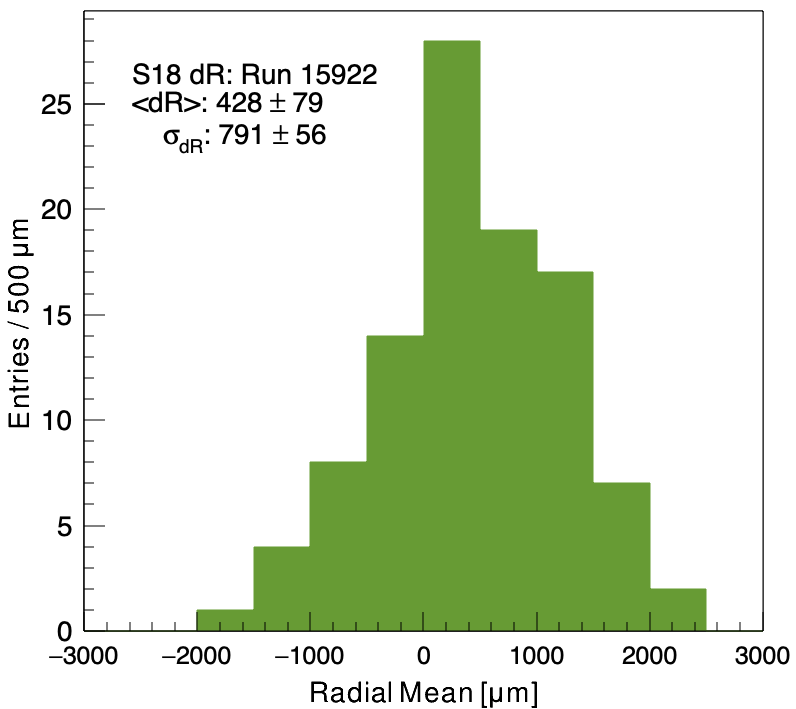}}  \\
    \subfloat[]{\includegraphics[width = 0.49\linewidth]{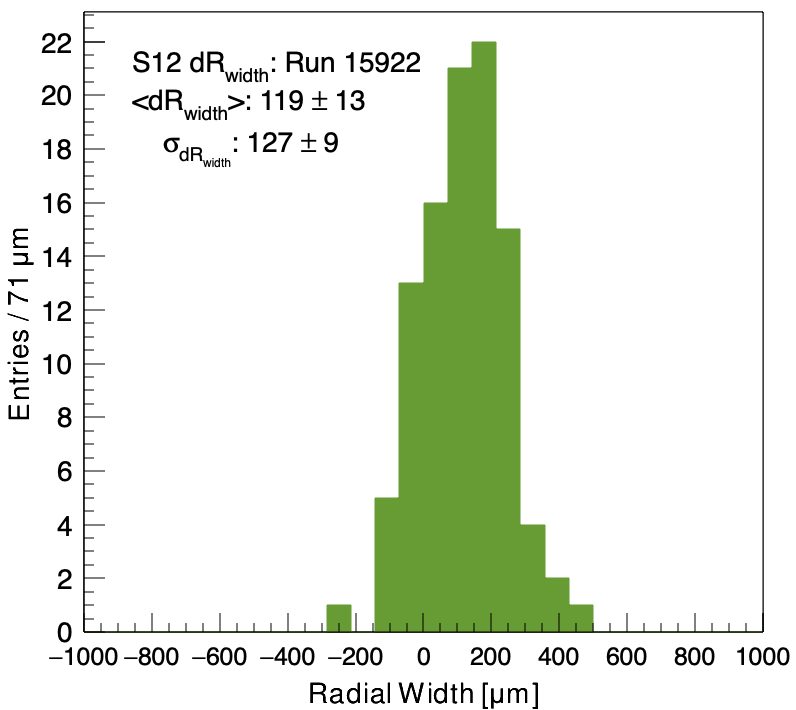}}  
    \subfloat[]{\includegraphics[width = 0.49\linewidth]{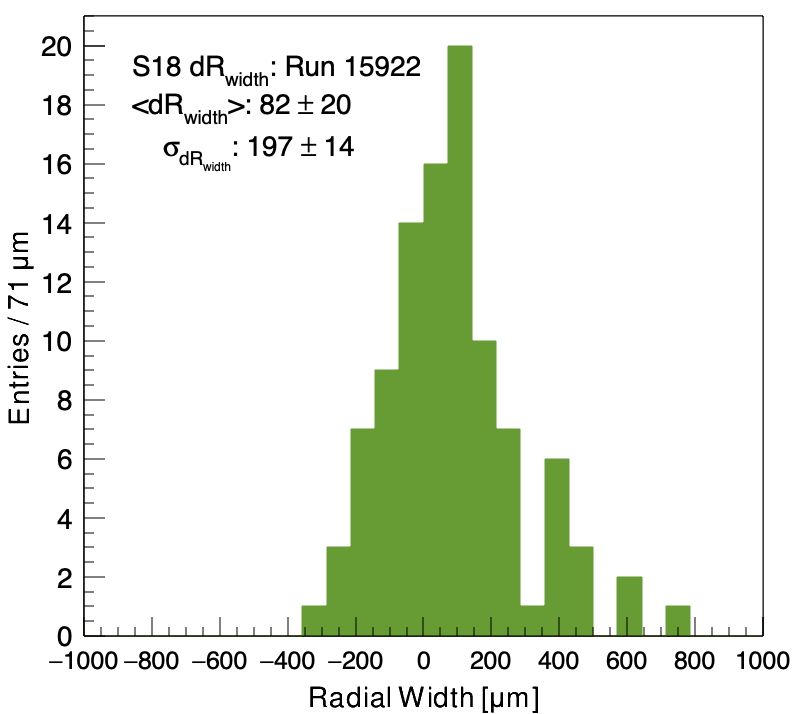}}
    \caption[Nominal case versus 100 randomised cases radially]{Difference between the nominal case and 100 samples for the radial beam extrapolation: (a) S12 radial mean, (b) S18 radial mean, (c) S12 radial width and (d) S18 radial width.}
    \label{fig:AUE_rad}
\end{figure}

\begin{figure}[htpb]
    \centering
    \subfloat[]{\includegraphics[width = 0.49\linewidth]{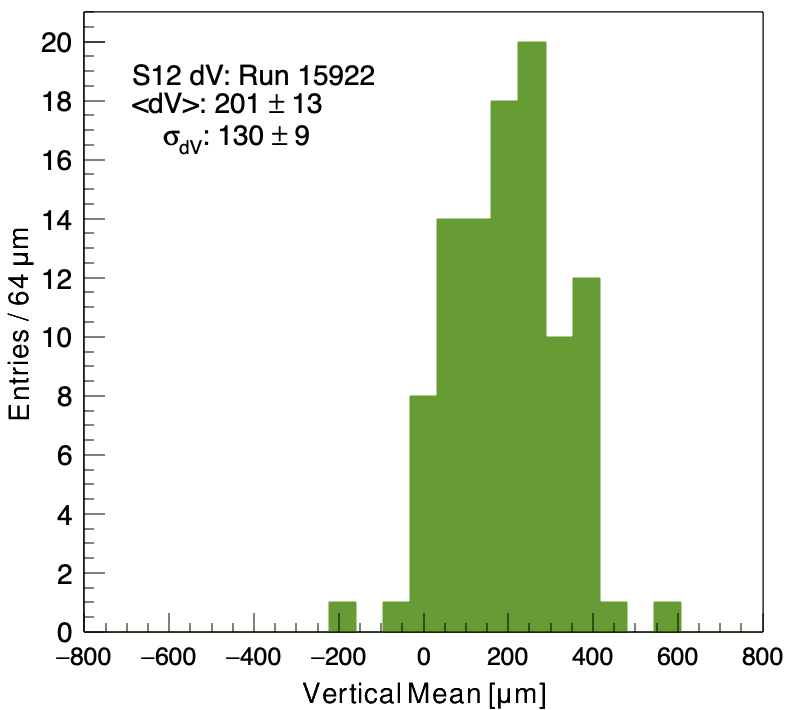}}   
    \subfloat[]{\includegraphics[width = 0.49\linewidth]{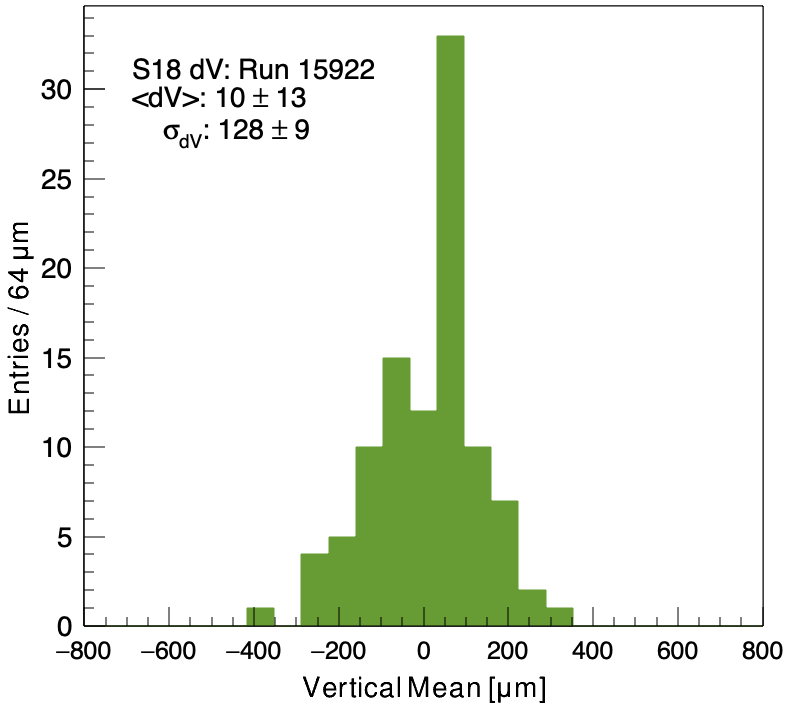}}  \\
    \subfloat[]{\includegraphics[width = 0.49\linewidth]{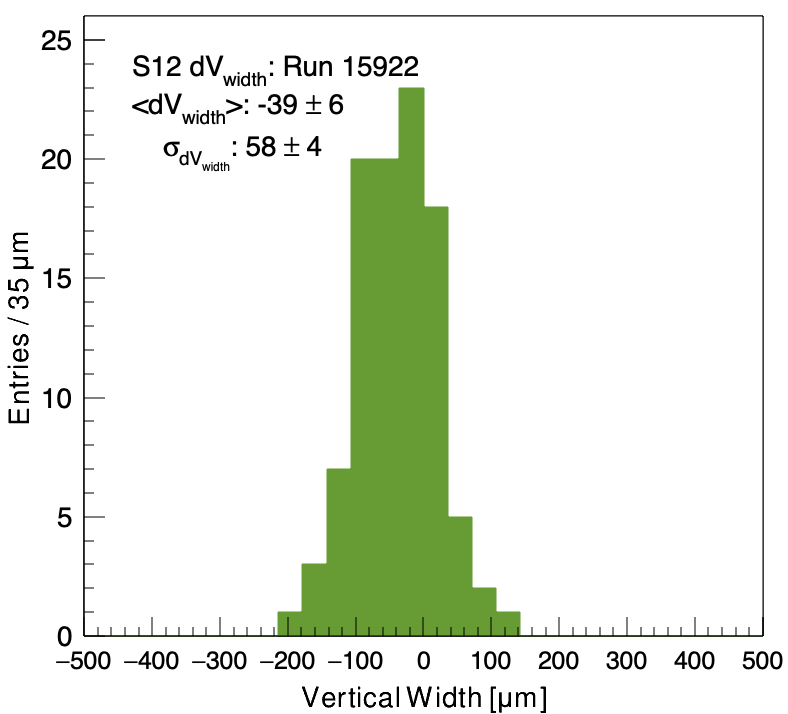}}  
    \subfloat[]{\includegraphics[width = 0.49\linewidth]{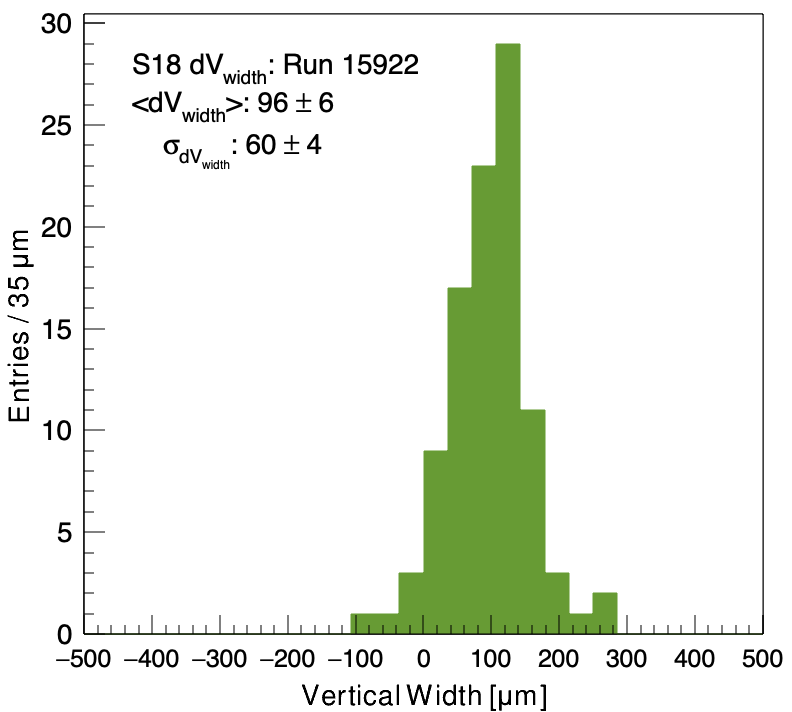}}
    \caption[Nominal case versus 100 randomised cases vertically]{Difference between the nominal case and 100 samples for the vertical beam extrapolation: (a) S12 vertical mean, (b) S18 vertical mean, (c) S12 vertical width and (d) S18 vertical width.}
    \label{fig:AUE_ver}
\end{figure}

The results from the impact of the misalignment on the radial and vertical beam extrapolation are shown in \cref{tab:radAUE} and \cref{tab:verAUE}, respectively. The extracted quantities of interest are: (1) the difference in the mean extrapolated radial position, $\langle\mathrm{dR}\rangle$, and (2) its standard deviation, $\mathrm{\sigma_{dR}}$, as well as the difference in the width of the beam, $\langle\mathrm{dR_{\mathrm{width}}}\rangle$, and its standard deviation, $\mathrm{\sigma_{dR_{\mathrm{width}}}}$. This difference is defined with respect to the nominal value. $\mathrm{R}_0$, extrapolated from run 15922, as shown below for the radial position
\begin{equation}
    \mathrm{dR}_i = \mathrm{R}_{0} - \mathrm{R}_i,
    \label{eq_dR}
\end{equation}
where $\mathrm{R}_i$ is one of the randomised samples. The results from \cref{fig:P0_100} were accumulated in \cref{fig:AUE_rad} and \cref{fig:AUE_ver}, where a comparison between a nominal and a randomised sample was performed according to \cref{eq_dR}. The mean is then extracted from the ensemble as 
\begin{equation}
    \langle\mathrm{dR}\rangle = \frac{1}{N}\sum_{i=1}^{N=100}(\mathrm{dR}_i),
\end{equation}
and similarly for $\mathrm{\sigma_{dR}}$, $\langle\mathrm{dR_{\mathrm{width}}}\rangle$, and $\mathrm{\sigma_{dR_{\mathrm{width}}}}$, and the four counterpart vertical ($\mathrm{V}$) quantities. 

The uncertainty on the beam extrapolation from the internal misalignment was taken as the mean from the two stations, with the uncertainty computed in quadrature. The final four systematics that were used in propagating the internal alignment contribution to the beam extrapolation are summarised in \cref{tab:finalAUE}.
\clearpage
\begin{table}[htpb]  
  \centering
  \begin{tabular}{lrrrr}
    \toprule
           & $\langle\mathrm{dR}\rangle$ [\SI{}{\micro\meter}] & $\mathrm{\sigma_{dR}}$ [\SI{}{\micro\meter}]& $\langle\mathrm{dR_{\mathrm{width}}}\rangle$ [\SI{}{\micro\meter}] & $\mathrm{\sigma_{dR_{\mathrm{width}}}}$ [\SI{}{\micro\meter}]\\ \midrule
    S12 $\oplus$ S18 & 322 $\pm$ 105  & 741 $\pm$ 74  & 101 $\pm$ 24 & 162 $\pm$ 17 \\ \bottomrule
  \end{tabular}
  \caption[Uncertainties on the radial beam extrapolation]{Combined uncertainties, in both stations, on the radial beam extrapolation arising from a \SI{100}{\micro\metre} misalignment.}
  \label{tab:radAUE}
\end{table}
\clearpage

\begin{table}[htpb]  
  \centering
  \begin{tabular}{lrrrr}
    \toprule
           & $\langle\mathrm{dV}\rangle$ [\SI{}{\micro\meter}] & $\mathrm{\sigma_{dV}}$ [\SI{}{\micro\meter}]& $\langle\mathrm{dV_{\mathrm{width}}}\rangle$ [\SI{}{\micro\meter}] & $\mathrm{\sigma_{dV_{\mathrm{width}}}}$ [\SI{}{\micro\meter}] \\ \midrule
    S12 $\oplus$ S18  & 106 $\pm$ 18  & 129 $\pm$ 13   & 29 $\pm$ 8 & 59 $\pm$ 6  \\ \bottomrule
  \end{tabular}
  \caption[Uncertainties on the vertical beam extrapolation]{Combined uncertainties, in both stations, on the vertical beam extrapolation arising from a \SI{100}{\micro\metre} misalignment.}
  \label{tab:verAUE}
  \end{table}
\clearpage
\section{Internal alignment contribution to the beam extrapolation}\label{sec:propag_pitch}
The beam extrapolation determines the radial and vertical positions of the muon beam in the storage ring. If the internal misalignment of the tracking detectors is not determined and corrected for, the $1\sigma$ uncertainty on the mean radial and vertical extrapolated beam positions correspond to 0.741 mm and 0.129 mm, respectively. 

The pitch correction (see \cref{sec:pitch}) is given by \cref{eq:pitch}. The $\mathrm{\sigma_{dV_{\mathrm{width}}}}$ from \cref{tab:finalAUE} potentially yields a 1.5 ppb uncertainty on the pitch correction arising from the internal misalignment, if the misalignment is not accounted for. Once the misalignment is corrected for, there is a negligible uncertainty in $C_{\mathrm{pitch}}$.
\begin{table}[htpb]  
  \centering
  \begin{tabular}{rrrr}
    \toprule
            $\mathrm{\sigma_{dR}}$ [\SI{}{\milli\meter}] & $\mathrm{\sigma_{dR_{\mathrm{width}}}}$ [\SI{}{\milli\meter}] & $\mathrm{\sigma_{dV}}$ [\SI{}{\milli\meter}] & $\mathrm{\sigma_{dV_{\mathrm{width}}}}$ [\SI{}{\milli\meter}]  \\ \midrule
    
      0.741 & 0.162 & 0.129 & 0.059  \\ \bottomrule
  \end{tabular}
  \caption[The contribution of a randomised misalignment to the beam extrapolation uncertainty]{The final contribution of a randomised internal misalignment to the beam extrapolation uncertainty.}
  \label{tab:finalAUE}
\end{table}

\graphicspath{{fig/}}

\chapter{Internal alignment of the \texorpdfstring{\gm2}~straw tracking detectors}
\label{ch:align}

\section{Introduction}
\begin{figure}[htpb]
    \centering
    \includegraphics[width=0.7\linewidth]{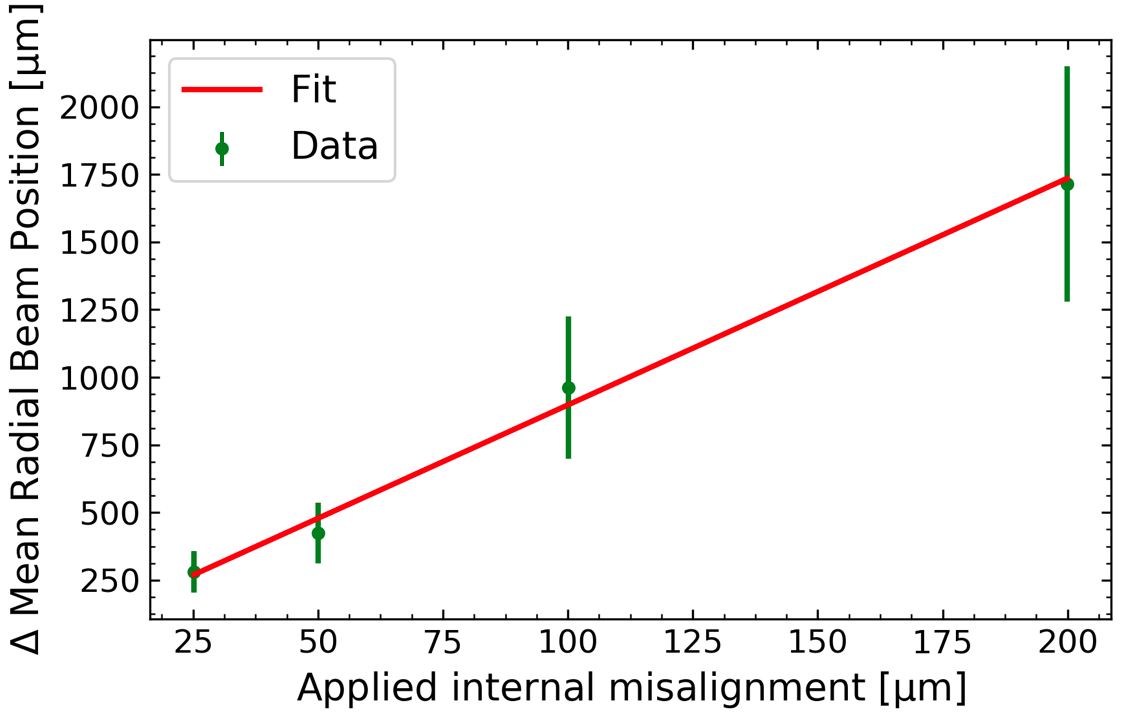}
    \caption[The change in the extrapolated radial beam position]{The change in the extrapolated radial beam position, relative to the non-misaligned case, resulting from a randomised radial misalignment of individual tracker modules at four scales of misalignment.}
    \label{fig:rad_inter}
\end{figure}

A track-based internal alignment of the two tracking stations was implemented using data from \R1. A Monte Carlo simulation was also developed to understand how the detector geometry affects how well the alignment can be determined, as well as to test the alignment procedure itself. A high-precision internal alignment of the tracking system is motivated by the need to minimise the uncertainty on the extrapolated beam position. This is illustrated in \cref{fig:rad_inter}, which highlights that even a relatively small-scale misalignment can have a large impact on the extrapolated radial beam position. Additionally, a precise alignment is essential for performing a search for the muon \ac{EDM}, as described in \cref{sec:edm_track}. Broadly speaking, an internal misalignment of an element of a tracking detector results in a residual between a hit position (e.g. \ac{DCA} of a hit to a wire) and the fitted track. This residual arises from the fact that the assumed detector position, used in the fitting of the track, is not the actual position of that detector. The alignment procedure aims to establish the corrections to the assumed detector position, and hence, minimise the residuals.
\clearpage
This chapter describes the internal alignment procedures. \cref{sc:mp2} describes the chosen framework for the alignment, as well as the general methodology of the internal alignment of a straw tracking detector. \cref{sc:align_simple} describes the integrity tests of the alignment procedure in a simplified 2D case with straight tracks. \cref{sec:align_sim} describes the results of the alignment using the full 3D geometry of the real tracking system in an inhomogeneous magnetic dipole field with curved tracks. At each stage, a thorough comparison was made between simulation, data, and where available, analytical predictions. Finally, in \cref{sc:align_real} alignment results with data are presented. An overview of the global (external) alignment, which established an absolute position of the tracker stations relative to the rest of the experiment, is described in \cref{sec:align_global}. 

\section{Alignment methodology} \label{sc:mp2}
The chosen alignment framework was \mpt~\cite{mp2_3}, which simultaneously fits many parameters describing the detector geometry and the input data. The framework accounts for the correlations between different alignment elements. This framework is widely used in particle physics: the inner tracker of LHCb~\cite{LHCb}, and the Belle II~\cite{Belle2} vertex detector have both been aligned using \mpt. Moreover, a track-based alignment with 50,000 parameters was successfully implemented by CMS~\cite{CMS_1, CMS_2}. Outside of particle physics, \mpt has been used in medical physics for the alignment of Positron Emission Tomography Scanners~\cite{PET}.

\begin{figure}[htpb]
    \centering
    \includegraphics[width=0.75\linewidth]{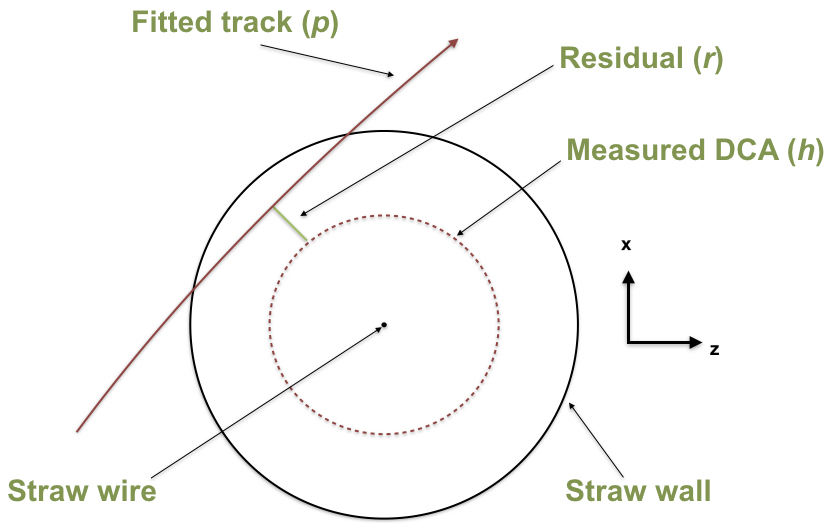}
    \caption[A vertical cross-section of a straw]{A vertical cross-section of a straw showing the DCA of a single hit ($h$) and the residual ($r$) with respect to the fitted track ($p$).}
    \label{fig:residual}
\end{figure}

The alignment is essentially a \ac{LSR} with a large number of parameters. These parameters can be divided into two classes: global and local parameters. Global parameters (i.e. \textit{alignment parameters} or \textit{geometry parameters}) affect all tracks (e.g. the radial position of a detector). Local parameters (i.e. \textit{track parameters}) are associated with individual fitted tracks. For example, a straight line-fit in 2D has two local parameters: a slope and an intercept. \mpt performs a \ac{LSR}, using both global and local parameters, minimising a linearised function of a sum of residuals. A residual, $r_i$, is defined as the difference between a fitted (predicted) position, $p_i$, and a measured hit position, $h_i$, as follows
\begin{equation}
    r_i = p_i - h_i,
\end{equation}
where the fitted track is described by some parametrisation. This is demonstrated in \cref{fig:residual}.
\clearpage

This parametrisation can be written in terms of the aforementioned global, $\boldsymbol{a}$, and local, $\boldsymbol{b}_j$ parameters, and the dependence of these parameters on the residual is
\begin{equation} \label{eq:residual}
    r_i(\boldsymbol{a},\boldsymbol{b}_j) = f_j(\boldsymbol{a},\boldsymbol{b}_j) - h_i,
\end{equation}
where $f_j$ is the fit parametrisation, and subscript $j$ defines the association of the local parameters to a particular track. The $\chi^2$ of many residuals from a single track is given by
\begin{equation}
    \chi^2_j(\boldsymbol{a},\boldsymbol{b}_j) = \sum_{i=1}^{\mathrm{hits}}{\frac{\big(r_i(\boldsymbol{a},\boldsymbol{b}_j)\big)^2}{(\sigma_i^{\mathrm{det}})^2}},
\end{equation}
where $\sigma_i^{\mathrm{det}}$ is the estimate of the uncertainty in the hit position. 
The total $\chi^2$ for many tracks is then
\begin{equation}
\chi^2(\boldsymbol{a}, \boldsymbol{b}) = \sum_{j}^{\mathrm{tracks}}\sum_{i}^{\mathrm{hits}} \frac{\big(r_{i,j}(\boldsymbol{a},\boldsymbol{b}_j)\big)^2}{(\sigma^{\mathrm{det}}_{i,j})^2}.
\end{equation}
A function $F$ is then minimised with respect to a variation of the global and local parameters
\small
\begin{equation}
F(\boldsymbol{a},\boldsymbol{b})=\frac{\partial\chi^2(\boldsymbol{a},\boldsymbol{b})}{\partial(\boldsymbol{a},\boldsymbol{b})}=\sum_{j}^{\mathrm{tracks}}\sum_{i}^{\mathrm{hits}} \frac{1}{(\sigma^{\mathrm{det}}_{i,j})^2} \frac{\big(r_{i,j}(\boldsymbol{a_0},\boldsymbol{b_0}_{,j})+\frac{\partial r_{i,j}}{\partial \boldsymbol{a}}
\delta \boldsymbol{a} +\frac{\partial r_{i,j}}{\partial \boldsymbol{b}_j}\delta \boldsymbol{b}_j\big)^2}{\partial(\boldsymbol{a},\boldsymbol{b})}=0,
\label{eq:obj_fun}
\end{equation}
\normalsize
where $\boldsymbol{a_0}$ and $\boldsymbol{b_0}$ are the initial geometry and track parameters, respectively. The corrections to the global parameters, $\delta\boldsymbol{a}$, which minimise $F$, are added to the assumed geometry of the detector to improve the relative alignment and the overall quality of the fitted tracks.

\subsection{Generalisation of the alignment methods}
\begin{figure}[htpb]
\centering
\includegraphics[width=0.9\linewidth]{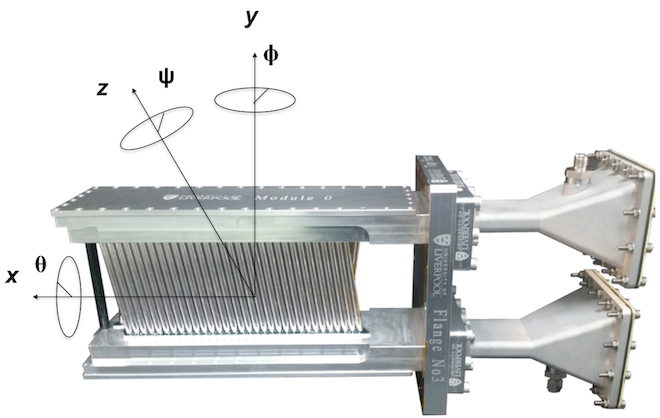}
\caption[The translation ($x$, $y$, $z$) and rotation ($\theta$, $\phi$, $\psi$) coordinates]{The translation ($x$, $y$, $z$) and rotation ($\theta$, $\phi$, $\psi$) coordinates for a single module.}
\label{fig:Rotation}
\end{figure}

The alignment parameter basis, $\delta\boldsymbol{a}$, is defined by six \ac{DoF} in the Euclidean 3D space, as shown in \cref{fig:Rotation},
\begin{equation} \label{eq:align_par}
\delta\boldsymbol{a} = \begin{pmatrix} \delta x  \\ \delta y \\ \delta z \\  \delta \theta \\ \delta \phi\\ \delta \psi \end{pmatrix},
\end{equation}
where $\theta$, $\phi$ and $\psi$ are the Euler angles. These \ac{DoF} describe three translations and three rotations of a detector.

In the minimisation from \cref{eq:obj_fun}, a vector containing the derivatives of the residuals with respect to the alignment parameters, $\frac{\partial r}{\partial \boldsymbol{a}}$, is used. The individual derivatives depend on the functional form of the residual in \cref{eq:residual}.
\clearpage

A line in 3D can be parametrised with a variable parameter $t$ as follows
\begin{equation}
\boldsymbol{s}(t) = \begin{pmatrix}x\\y\\z\\\end{pmatrix} = \begin{pmatrix}x_L + m_x  t\\y_L + m_y t\\t\\\end{pmatrix},
\end{equation}
as one needs only four parameters to specify a line in 3D. Assuming the track is not parallel to the $xy$ plane, as will be the case for all the tracks of interest in this study, the parametrised track equations is 
\begin{equation}
\boldsymbol{s}_T(t) =  \begin{pmatrix}x_T + m_x  t\\y_T + m_y  t\\  t\\\end{pmatrix} =  \boldsymbol{v}t+\boldsymbol{r}_T^0=\begin{pmatrix}m_x\\ m_y \\  1\\\end{pmatrix}t + \begin{pmatrix}x_T \\y_T  \\  0\\\end{pmatrix}.
\label{eq:track}
\end{equation}
This yields directly the four local parameters $\boldsymbol{b}$
\begin{equation}
\boldsymbol{b} = \begin{pmatrix}x_T \\y_T \\m_x \\ m_y\end{pmatrix},
\end{equation} 
where the two intercept points $x_T$ and $y_T$ are at $z=0$, and two slopes $m_i$, where $i=x, y$, are given by
\begin{equation}
m_i = \frac{p_i}{p_z}=\tan\theta_i, 
\label{eq:m_i}    
\end{equation}
where $p_i$ is the momentum in the $i$ direction, and $\theta_i$ is the corresponding angle. Therefore, the vector containing derivatives of the residuals with respect to the fitted track parameters is given by
\small
\begin{equation}
\frac{\partial r}{\partial \boldsymbol{b}} = \begin{pmatrix} \frac{\partial r}{\partial x_T} \\ \frac{\partial r}{\partial y_T} \\ \frac{\partial r}{\partial m_x} \\ \frac{\partial r}{\partial m_y}  \end{pmatrix}. 
\end{equation}
\normalsize
\clearpage
\subsection{Coordinate system transformations}
\begin{figure}[htpb]
\centering
\includegraphics[width=0.9\linewidth]{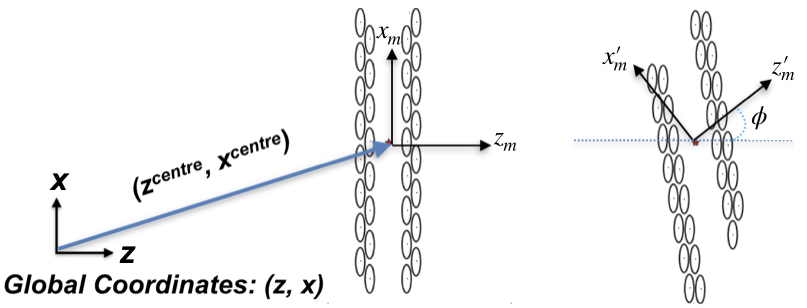}
\caption[The coordinate system under a 2D rotation]{The coordinate system used to define an anticlockwise rotation in 2D.}
\label{fig:CS}
\end{figure}

The coordinate systems used in the alignment are defined below. It is important to note, that due to the geometry of the straw tracker, the alignment is not sensitive to shifts in the $z$ direction, the \say{beam direction}, and corrections to translations along $z$ will not be derived. The alignment corrections for the other global parameters will be derived directly in the station coordinate system. The coordinate systems are:\\
1) Global (ring) coordinates $\boldsymbol{g}=(x_g, y_g, z_g)$. \\
2) Detector (tracker station) coordinates $\boldsymbol{d}=(x_d, y_d, z_d)$, with the origin of $\boldsymbol{d}$ at the centre of the first of the eight modules. \\
3) Local (tracker module) coordinates $\boldsymbol{m}=(x_m, y_m, z_m)$, with the origin of $\boldsymbol{m}$ at the centre of rotation of a module.  \\
In the case of 3D translations and rotations, a tracker module is connected to the station coordinate system via the relation
\begin{equation}
\boldsymbol{d}=\boldsymbol{R}^{T}\boldsymbol{m} + \boldsymbol{d}_0,
\end{equation}
where $\boldsymbol{d}_{0}$ is the module position in the station coordinates, and $\boldsymbol{R}^T$ is the 3D rotation matrix (see \cref{eq:rot_matrix}). An example of such a transformation in 2D is illustrated in \cref{fig:CS}. With the above definition, the alignment corrections, as given in the tracker station frame, can be represented by
\begin{equation}
\boldsymbol{d}=\boldsymbol{R}^{T}\delta\boldsymbol{R}(\boldsymbol{m}+\delta\boldsymbol{m}) + \boldsymbol{d}_{0},
\end{equation}
where corrections for translations, $\delta\boldsymbol{m}$, and rotations, $\delta\boldsymbol{R}$ form the alignment parameter vector described in \cref{eq:align_par}.

\subsection{Parameter optimisation methods}
The number of all the parameters in \cref{eq:obj_fun} could be large, hence, the differential equation is first linearised, to allow for a subsequent matrix inversion, and then reorganised into a block matrix equation~\cite{mp2_2} of the form
\begin{equation}
\left(\boldsymbol{C}\right)\left(\boldsymbol{a}\right) = - \left(\boldsymbol{g}\right),
\end{equation}
where $\boldsymbol{C}$ is a matrix containing the connection between the local and global parameters, $\boldsymbol{a}$ is the correction vector (for global parameters), and $\boldsymbol{g}$ a vector of the normal equations~\cite{mp2}. Hence, a matrix inversion, whose dimension is given by the number of global and local parameters, is necessary to obtain the corrections to the global parameters that simultaneously minimises the $\chi^2$. However, the size of the matrix can be reduced, as the derivatives with respect to the global parameters are the only parameters of interest, as contained in the sub-matrix $\boldsymbol{C}_{22}$
\begin{equation}
\left(
\begin{array}{c|c}
\boldsymbol{C}_{11} & \boldsymbol{C}_{21}^{T} \\ 
 \hline
\boldsymbol{C}_{21} & \boldsymbol{C}_{22} \\    
\end{array}
\right)
\begin{pmatrix} 
\boldsymbol{a}_{1} \\ 
 \hline
\boldsymbol{a}_{2} \\   
 \end{pmatrix}
=-
\begin{pmatrix} 
\boldsymbol{g}_{1} \\ 
 \hline
\boldsymbol{g}_{2} \\    
\end{pmatrix}.
\end{equation}
Advanced matrix algebra techniques and the assumption that $\boldsymbol{C}_{22}$ is invertible are used~\cite{mp2_3} to remove the unnecessary parameters, such that only the inverse of $\boldsymbol{C}_{22}$ is needed to establish the corrections to the alignment parameters. 

\subsection{Alignment software framework}
\begin{figure}[htpb]
    \centering
    \includegraphics[width = 0.9\linewidth]{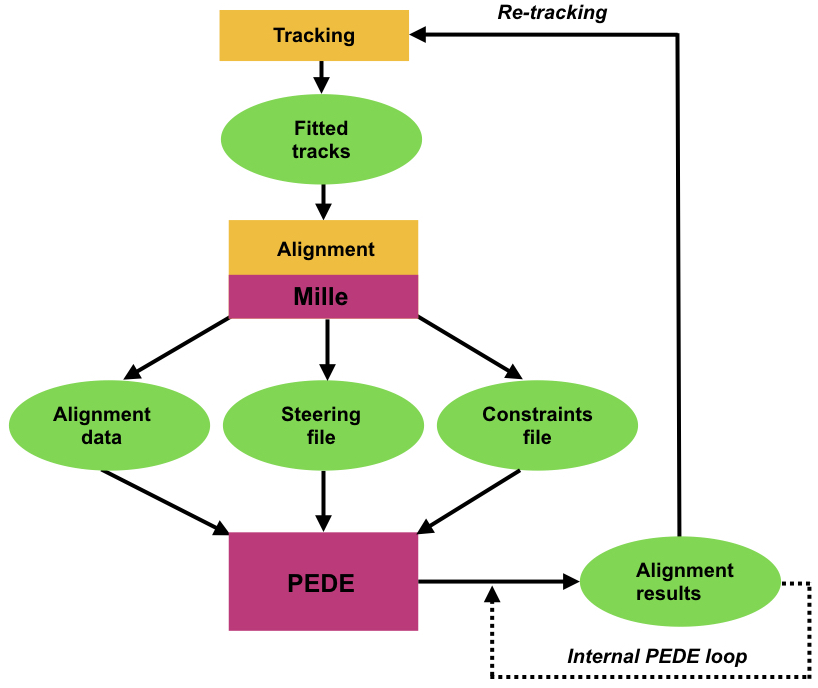}
    \caption[An overview of the software components]{An overview of the software components used in the alignment. The external packages \mille and \pede are shown in purple, while the \gm2 \art modules for tracking and alignment are shown in orange. The outputs of each of the three stages (tracking, alignment, residual minimisation) are indicated in green.}
    \label{fig:mp2}
\end{figure}

An alignment software module, written in \cpp, has been developed as part of the \gm2 \art software framework~\cite{art}. This module is run after the track-fitting (see \cref{sc:track_soft}), and it can be used on data or simulation. The geometry is defined at the tracking stage and is passed to the alignment module, which also calculates the residuals of the selected tracks, as well as computes the local and global derivatives. A \cpp module (\mille) is provided~\cite{mp2_3} to write this information into a binary file which is passed to \pede (a \texttt{Fortran} executable), which performs a simultaneous fit via the matrix inversion described above. The alignment module is also responsible for writing a constraints file (specifying the redundant \ac{DoF}), and a steering file (specifying the mathematical methods used by \pede). The final outputs of the \pede algorithm are labels and the corresponding fitted values of the global parameters and their errors. The components of the algorithm are shown in \cref{fig:mp2}.
\clearpage
Given that the number of total alignment parameters for the tracking detector is relatively small ($<\mathcal{O}(100)$), then the \pede inversion method with 10 iterations (internal \pede loop) and a 0.01 convergence rate on $F$ in \cref{eq:obj_fun} is adequate for the minimisation of $F$. The advantage of this inversion method is that it also provides uncertainties on the corrections to the global parameters. The memory space requirement for the matrix inversion and the execution time are acceptable. For an entire run of data with $\sim 5\times10^5$ tracks the execution time of \pede is $\mathcal{O}(1~\mathrm{min})$ and the memory footprint is below $200$~MB. The initial choice of convergence rate and number of iterations is empirical and is comparable to the values deployed by \mpt developers~\cite{GBL}. The final choice of steering and constraints methods is discussed in more detail in \cref{app:steer}.

\section{Alignment validation in simulation}\label{sc:align_simple}
To validate the alignment software independently of the \gm2 \art framework, a standalone framework was developed~\cite{Gleb_git_1}. This framework, written in \cpp and \python, interfaces with \mpt, and also incorporates a simplified detector geometry. It generates straight tracks and reconstructs them in 2D. This is described in \cref{sc:2d_geometry}. \cref{sc:3d_geometry} describes the progression to a 3D geometry with a simplified \art tracking chain with straight tracks, without a magnetic field and no detector material. The next implementation using a uniform magnetic field, curved tracks, and detector material is described in \cref{sec:UMF}. The final implementation with an inhomogeneous magnetic field in the full \gm2 simulation is presented in \cref{sec:IMF}. 

\subsection{2D geometry} \label{sc:2d_geometry}
Translational and rotational misalignments (as defined in \cref{fig:Rotation}) in the middle modules were considered, with the first and the last modules \textit{fixed} with no misalignment. An outline of the standalone simulation and fitting procedure to validate and test the alignment using four tracker modules (with four layers per module) with straight tracks in 2D is given below: 
\begin{enumerate} \itemsep -2pt
    \item Define the ideal geometry (i.e. assumed straw coordinates).
    \item Define the misaligned geometry (i.e. actual (\textit{truth}) straw coordinates).
    \item Generate straight tracks of the form $x = mz+c$ through the misaligned geometry. 
    \item Calculate the measurement \ac{DCA}, $h_i$, from the truth track and a straw hit in each layer (see \cref{fig:residual}).
    \item Smear the \ac{DCA} by the detector resolution, $\sigma_p^{\mathrm{det}}$.
    \item Reconstruct the \ac{DCA} in the ideal geometry.
    \item Fit a line to the reconstructed hit points (or reconstructed \textit{drift circles}, as described in the next section), and calculate the residual between the fit and the hit point.
    \item Pass residuals, as well as global and local derivatives, to \pede to perform the minimisation of the residuals using the matrix inversion method.
\end{enumerate}

\subsubsection{Translational misalignment in $x$ with a straight line-fit}
\begin{figure}[htpb]
    \centering
    \subfloat[]{\includegraphics[width = 0.49\linewidth]{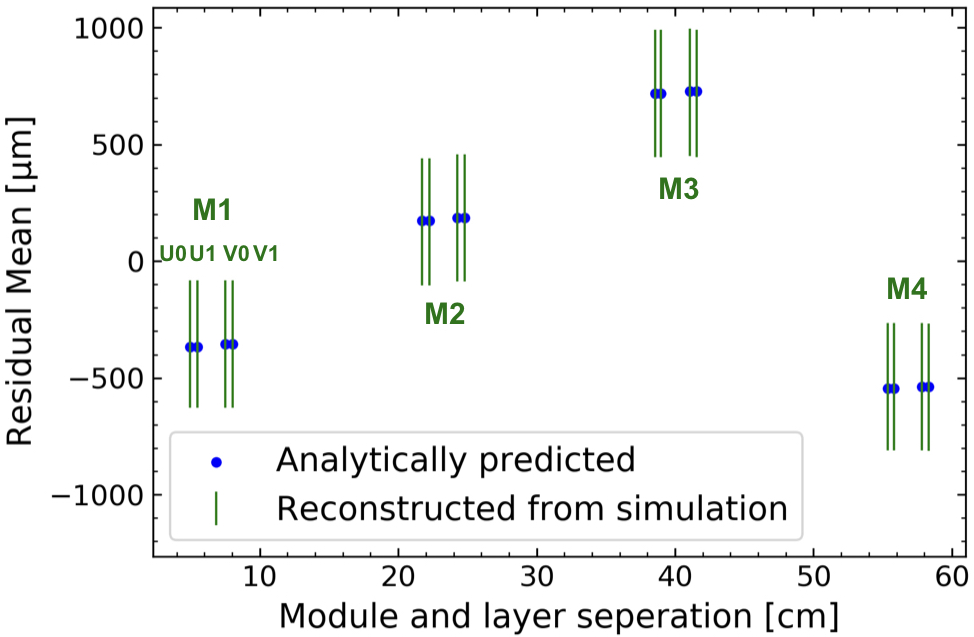}}
    \subfloat[]{\includegraphics[width = 0.49\linewidth]{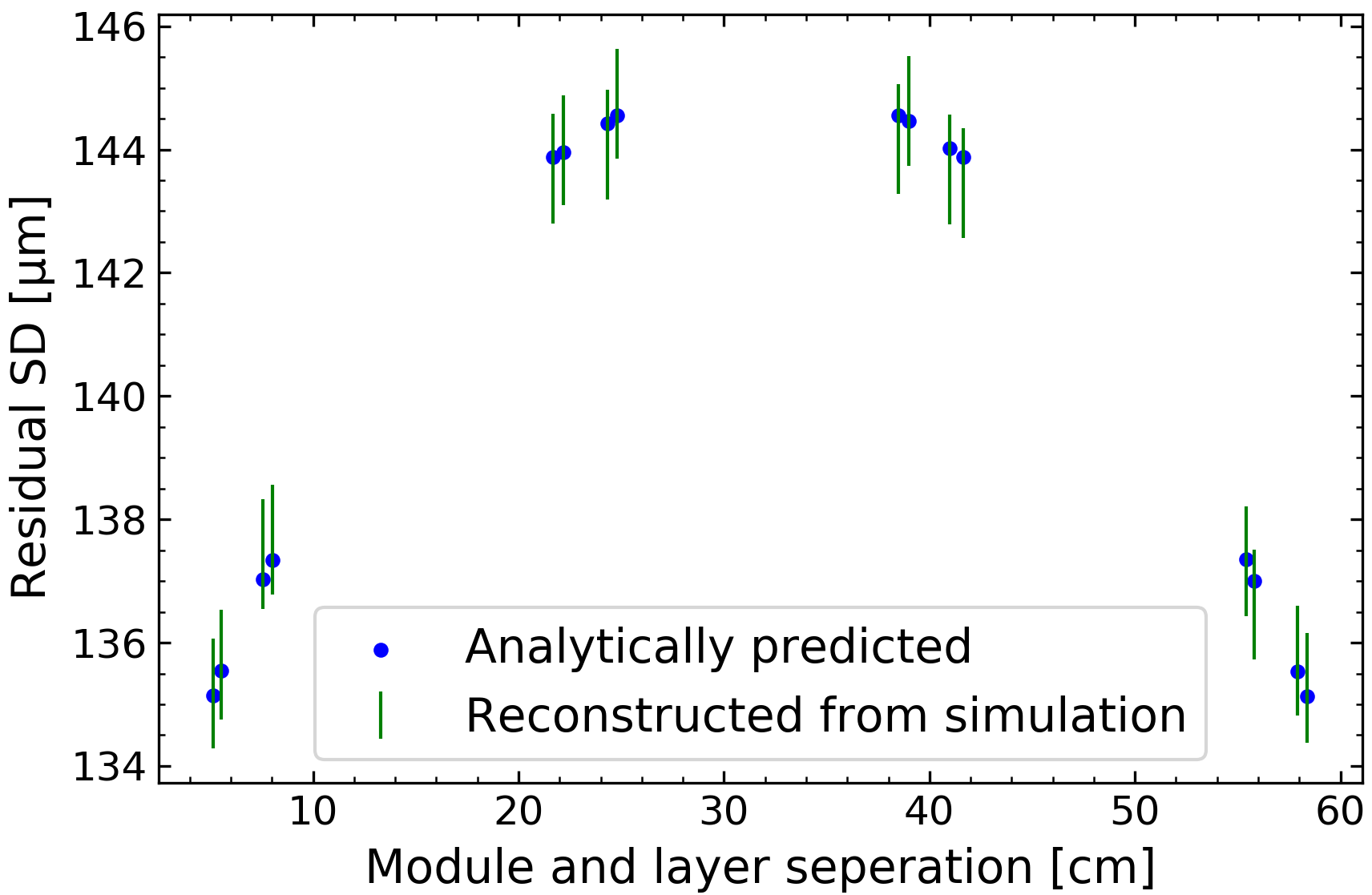}}
    \caption[Residual parameters from simulation]{Analytically predicted and reconstructed residual parameters from simulation. (a) Residual mean with modules 2 and 3 misaligned by \SI{500}{\micro\metre} and \SI{1000}{\micro\metre}, respectively. The net effect of the misalignment shifts the mean position of the residuals in each layer (U0, U1, V0, V1) in each of the four modules according to \cref{eq:shear}. (b) Residual SD with no misalignment. The SD is smaller on the outer modules (M1 and M4), as the \textit{lever-arm effect} produces better fits away from the mean $z$ position, according to \cref{eq:res}.}
    \label{fig:res}
\end{figure}

\begin{figure}[htpb]
    \centering
    \includegraphics[width = 0.65\linewidth]{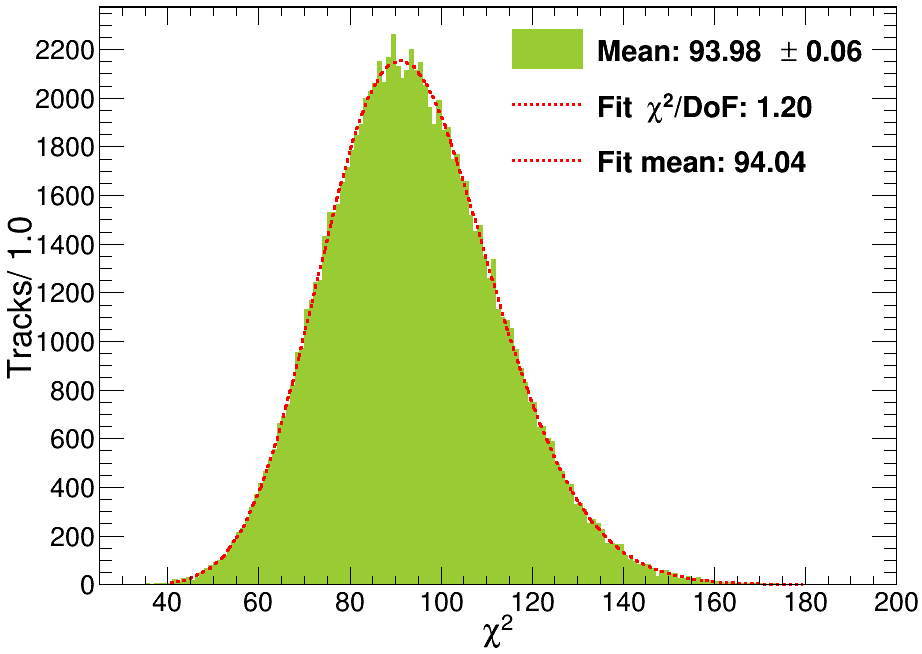}
    \caption[$\chi^2$ distribution from the simulation]{$\chi^2$ distribution from the simulation, and a fit of a non-central $\chi^2$ function ($\mathrm{mean}=94.04$, $\mathrm{DoF}=14$). The expected mean, $\langle \chi^2_{\mathrm{Mis}} \rangle$, is predicted from \cref{eq:chi2}, with M2 and M3 misaligned by \SI{-500}{\micro\metre} and $+$\SI{500}{\micro\metre}, respectively.} 
    \label{fig:chi2}
\end{figure}

In the simple 2D geometry it is possible  to define three distinct manifestations of misalignment: \\
1) $M^c_p$ is the \textit{characteristic} misalignment, which is specific for each individual straw layer $p$. $M^c_p$ is set in the simulation as the \textit{truth} misalignment. The aim of the alignment procedure is to then recover this input misalignment as close as possible to the truth. The recovered misalignment is called the \textit{reconstructed} misalignment. \\
2) $M_0$ is the \textit{overall} misalignment given by
\begin{equation}
M_0 = \frac{\sum_{i=1}^{P} M^c_i}{P},
\end{equation}
where $P$ is the total number of detector layers.\\
3) $M^{s}_p$ is the \textit{shear} misalignment which corresponds to the mean of the residual, $\langle r_p \rangle$, for a layer $p$
\begin{equation}
\langle r_p \rangle = \frac{\sum_{j=1}^{N}{r_{j,p}}}{N},
\end{equation}
where $N$ is the total number of tracks. With these definition, $M^{s}_p$ is then given by
\begin{equation}
M^{s}_p = \langle r_p \rangle =  M^c_p  - M_0 - \frac{z_p\sum_{i=1}^{P}{ M^c_{i}z_i}}{\sum_{i=1}^{P}{z_i^2}},
\label{eq:shear}
\end{equation}
where $z_i$ is the horizontal position (along the beam) of a layer $i$. The \ac{SD} for a distribution of residuals in layer $p$ is given by
\begin{equation}
\sigma_{p}^2 = \frac{\sum_{j=1}^{N}{(r_{j,p} - \langle r_p \rangle)^2}}{N}.
\end{equation}
Using the above equations, it is possible to show (see \cref{app:2D}) that the expected mean $\chi^2$ for a given set of characteristic misalignments is given by
\begin{equation}
\langle \chi^2_{\mathrm{Mis}} \rangle = P - 2 +  \frac{
    \sum_{i=1}^{P}(M^{c}_i)^2-\frac{\sum_{i=1}^{P}{(M^{c}_i)^2}}{P} - 
    \frac{2\sum_{i=1}^{P}M^{c}_i z_i}{\sum_{i=1}^{P}(z_i)^2}}
{\sum_{i=1}^{P}(\sigma_i^{\mathrm{det}})^2},
\label{eq:chi2}
\end{equation}
where the summation is over all layers. For a straight track-fit in 2D, there are two redundant \ac{DoF}, which are removed from the above equation. Any misalignment will shift the mean of the residuals for the misaligned and the non-misaligned modules. Moreover, for a non-misaligned scenario, the expected \ac{SD} of the residuals at layer $p$ is given by
\begin{equation}
\sigma_p=\sigma^\mathrm{det}\sqrt{\frac{P-1}{P}-\frac{z_i^2}{\sum_{i=1}^{P} z_i^2}}.
\label{eq:res}
\end{equation}
The analytically derived equations for $\langle \chi^2_{\mathrm{Mis}} \rangle$, $\sigma_p$ and $\langle r_p \rangle$ are compared to the predictions from the simulation in \Crefrange{fig:res}{fig:chi2}, and are seen to agree excellently.  

\begin{figure}[htpb]
    \centering
    \includegraphics[width = 0.99\linewidth]{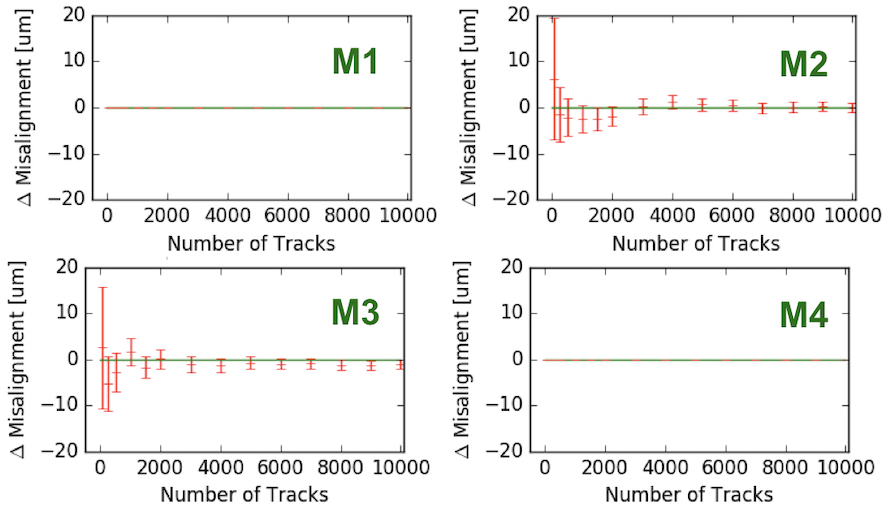}
    \caption[Misalignment comparison between truth and prediction]{The difference in misalignment between the simulation input and the \pede prediction versus the number of generated tracks. M1 and M4 were not misaligned and fixed by constraints.}
    \label{fig:pede}        
\end{figure}

Having established the integrity of the analytical predictions, \pede was then used to study the number of tracks required for the alignment parameters to converge on acceptable values. The inputs to \pede are summarised below. The residual was defined according to \cref{eq:residual}
\begin{equation}
r = \mathrm{Track}(x,m,c) - \mathrm{Hit} = x - h = mz+c - h,
\end{equation}
where $h$ is the closest vertical distance to a wire, in this simplified case. A constant detector resolution (i.e. hit smearing) was defined for all hits of 
\begin{equation}    
\sigma^{\mathrm{det}} = \SI{150}{\micro\metre}.
\end{equation}
Two local parameters defining the slope ($m$) and intercept ($c$) of the track were used and only a single global (alignment) parameter ($x$) was used such that the derivatives are:
\begin{equation}
\frac{ \partial r}{\partial c} = 1, \ \frac{ \partial r}{\partial m} = z,\ \mathrm{and} \ \frac{ \partial r}{\partial x} = \frac{ \partial r}{\partial c} = 1.
\end{equation}
Two modules were misaligned by \SI{500}{\micro\metre} and \SI{1000}{\micro\metre}, and the difference between the known input misalignment and that determined from \pede is plotted in \cref{fig:pede} as a function of the number of tracks. It is seen that acceptable alignment parameters are determined after $\mathcal{O}(5000)$ tracks. 

\subsubsection{Translational misalignment in $x$ with a circle-fit}
In the previous study, a simplified straight line-fit to the hit positions was used. In reality this is not possible, due to the so-called \ac{LR} ambiguity (see \cref{sub:left_right_ambiguity}), and the fact that the measurement in the straw is a drift circle (see \cref{sub:hit_formation}). In the case of a misalignment, the LR ambiguity can cause the alignment minimisation to fail. For this reason, only hits with $h$ greater than \SI{500}{\micro\metre} are used. Moreover, this requirement is also necessary to avoid a discontinuity in the residual derivative (see \cref{eq:2D_DLC1}). 

\begin{figure}[htpb]
    \centering
    \subfloat[]{\includegraphics[width = \linewidth]{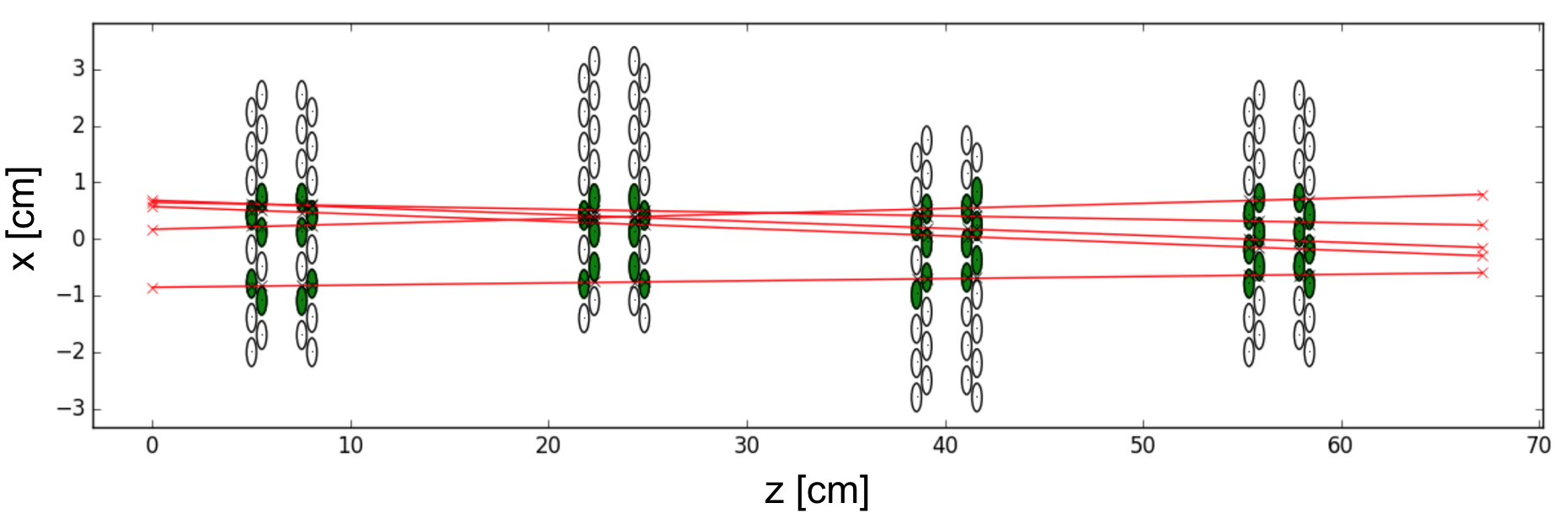}} \\
    \subfloat[]{\includegraphics[width = \linewidth]{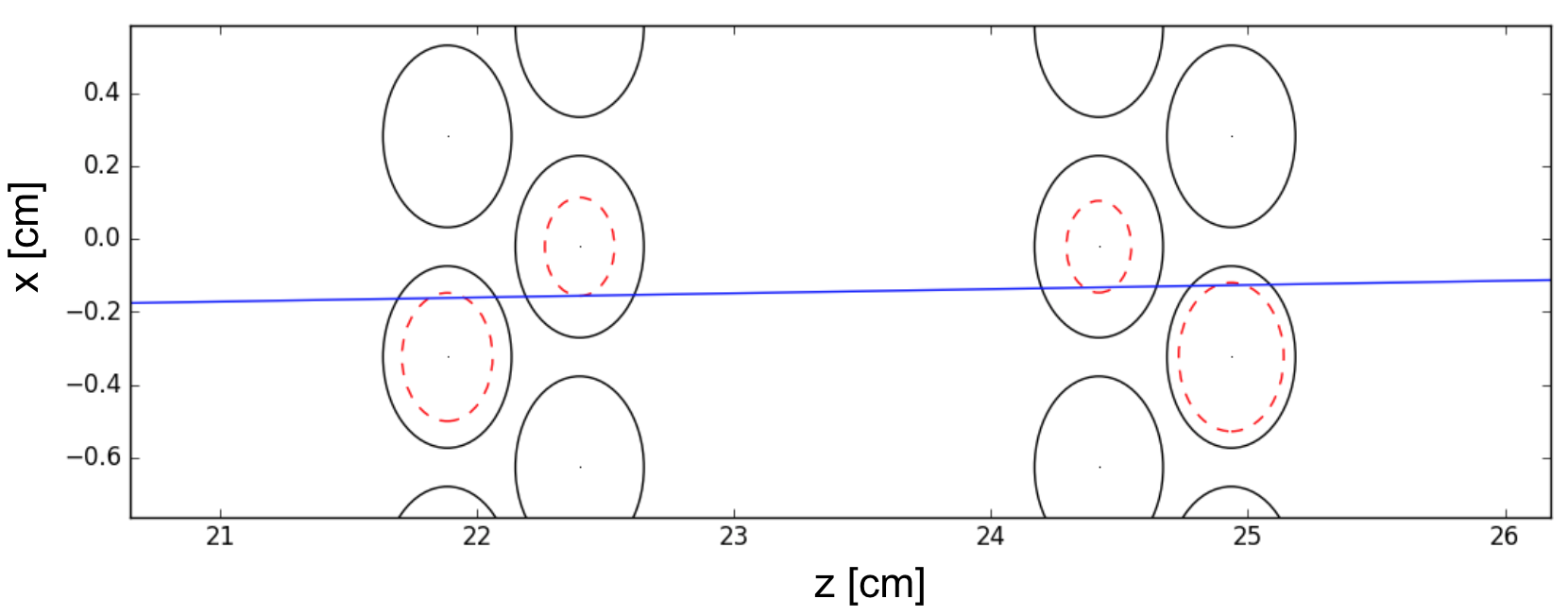}}
    \caption[Simulated tracks through a misaligned tracker station]{(a) Simulated tracks through a misaligned tracker station. Straws with recorded hits are indicated in green. (b) Reconstructed tracks through the assumed geometry. The recorded measurements from (a) are shown as drift circles within the straws, and the fitted track is shown in blue.}
    \label{fig:tracks}
\end{figure}

A simulation framework incorporating 2D misalignments using a \textit{circle-fit} was implemented. An example of five generated tracks in four tracker modules is shown in \cref{fig:tracks}. Due to the empirical nature of the circle-fit (which is performed by finding the minima numerically), the exact analytical predictions for $\sigma_p$, $\langle r_p \rangle$, and $\langle \chi^2_{\mathrm{Mis}} \rangle$ cannot be derived. However, other tools exist to check the solutions, as described in the rest of this chapter. 

\begin{figure}[htpb]
    \centering
    \subfloat[]{\includegraphics[width = 0.5\linewidth]{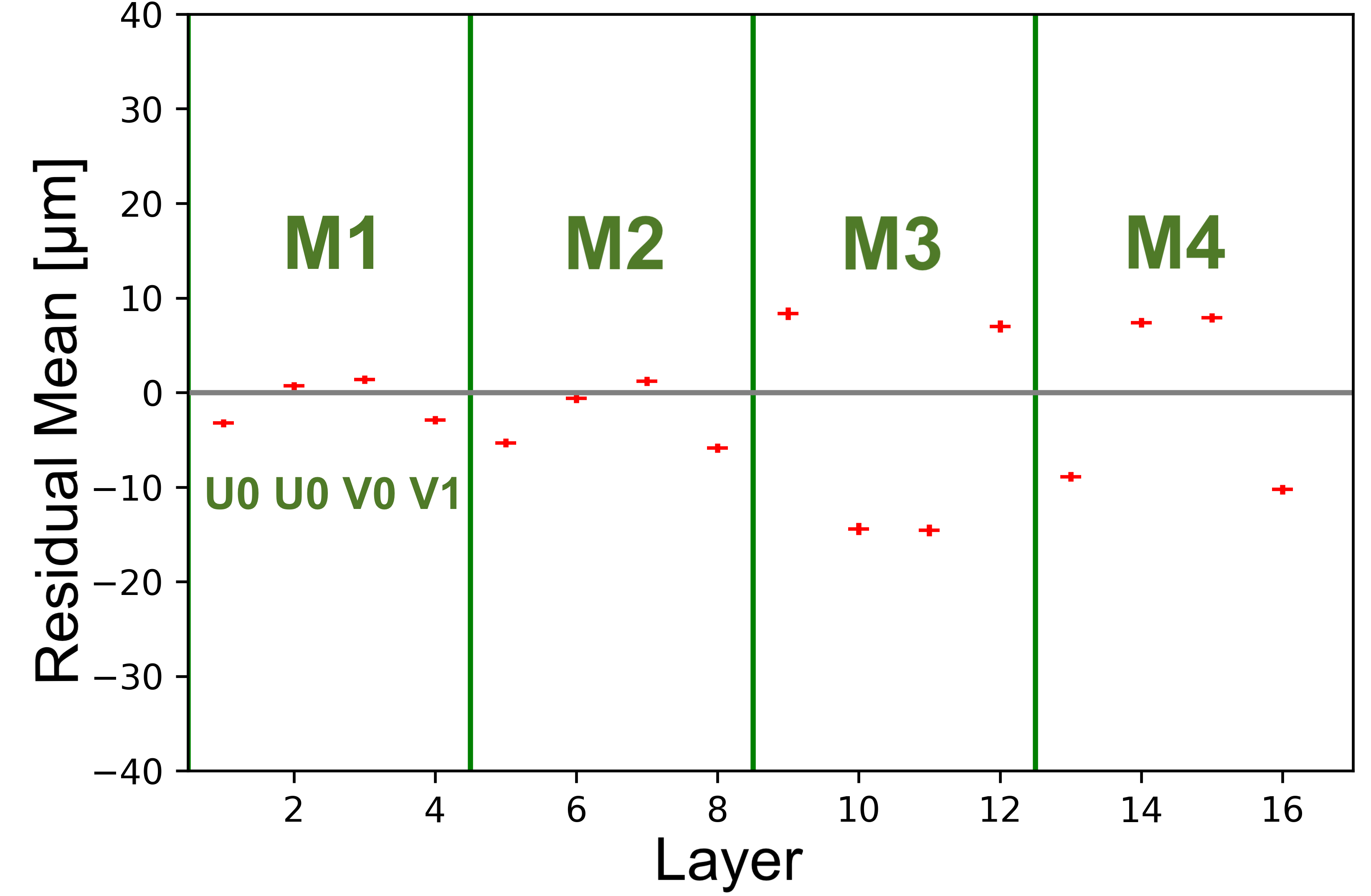}} 
    \subfloat[]{\includegraphics[width = 0.5\linewidth]{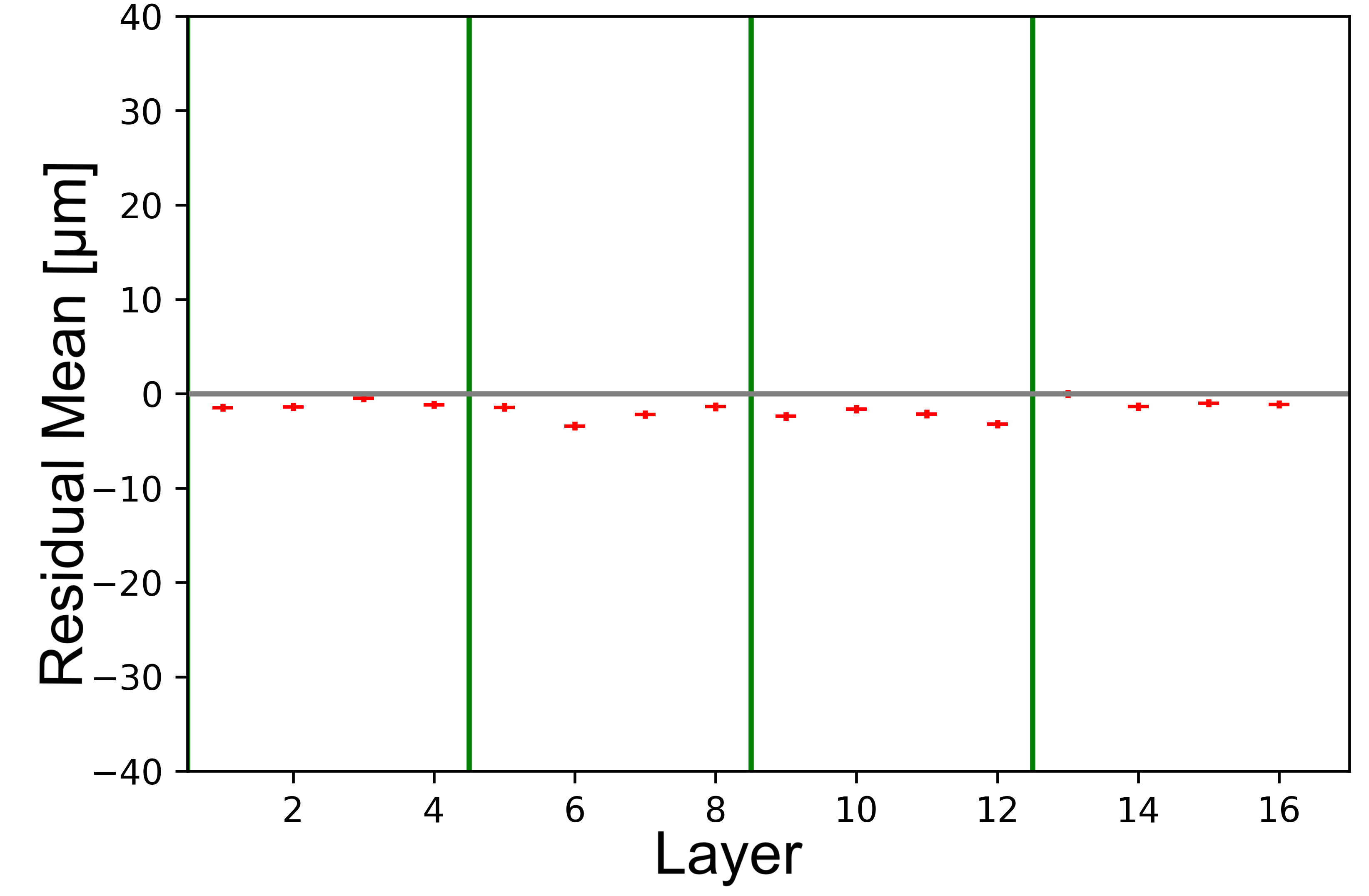}}
    \caption[Residual mean per layer]{Residual mean per layer (U0, U1, V0, V1) in the four modules. (a) Before alignment with M3 offset by \SI{100}{\micro\metre}. The residuals in neighbouring modules, which were not misaligned, are also affected. (b) After alignment. The truth misalignment is recovered within \SI{5}{\micro\metre}, after the first iteration. The distribution of residuals around $0$ is indicative of an aligned detector.}
    \label{fig:resMean_2D}
\end{figure}

\begin{figure}[htpb]
    \centering
    \subfloat[]{\includegraphics[width = 0.52\linewidth]{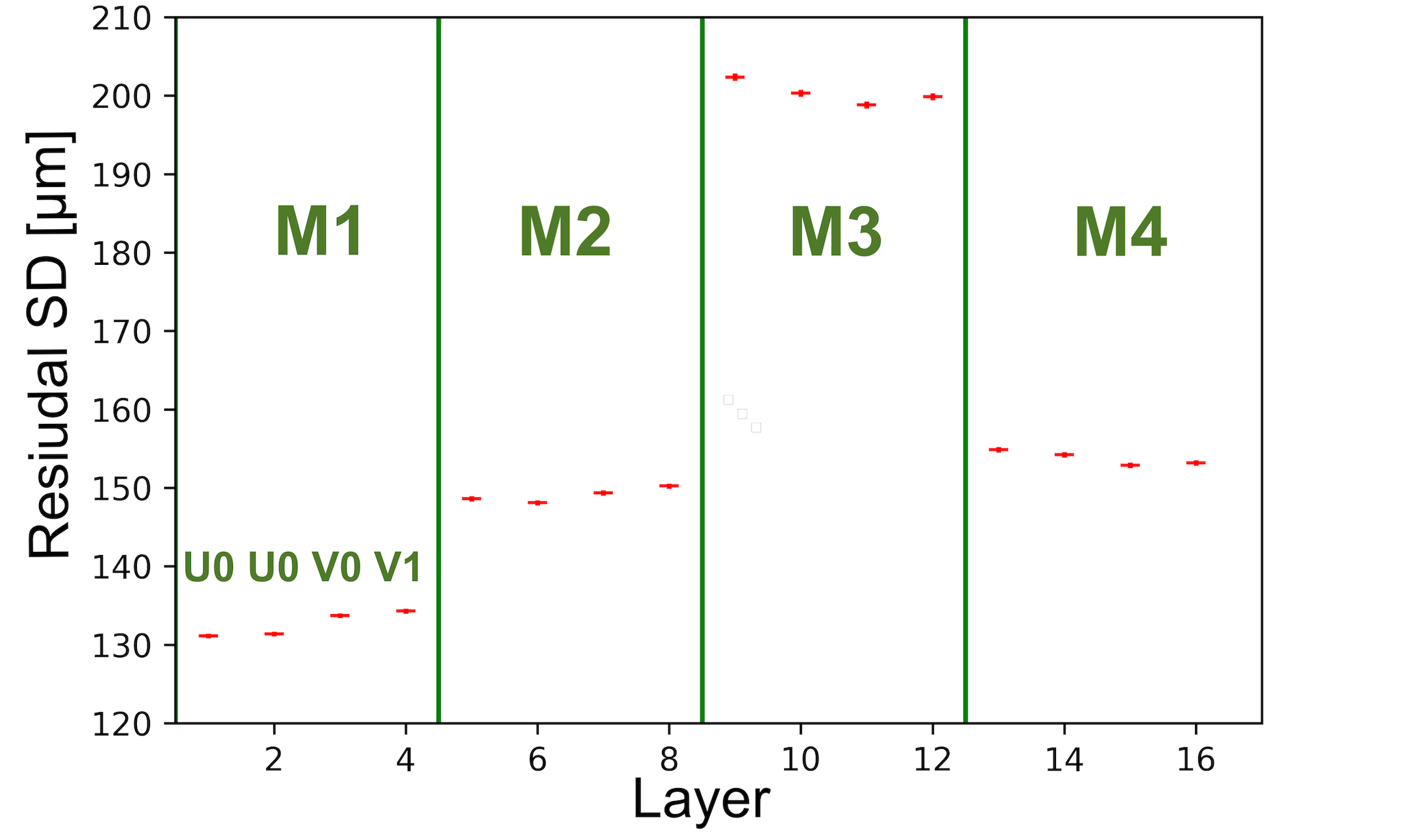}}
    \subfloat[]{\includegraphics[width = 0.48\linewidth]{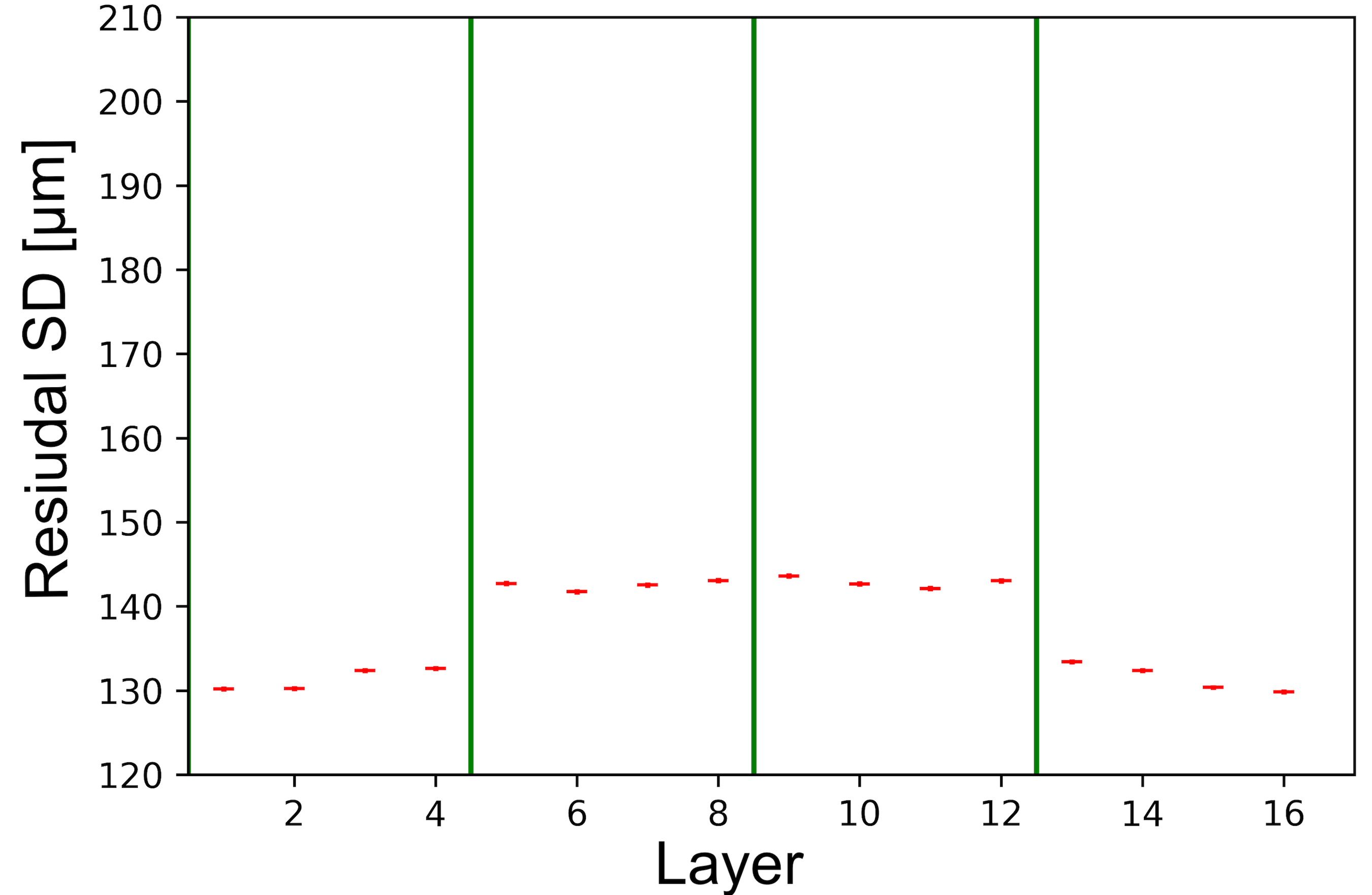}}
    \caption[Residual SD per layer]{Residual SD. (a) Before alignment with M3 offset by \SI{100}{\micro\metre}. The presence of relatively large misalignment in M3 is apparent. The residuals in neighbouring modules are also affected. (b) After alignment. The residual \ac{SD} is at the nominal level as in \cref{fig:res}, indicative of an aligned detector. }
    \label{fig:resSD_2D}
\end{figure}

\begin{figure}[htpb]
    \centering
    \subfloat[]{\includegraphics[width = 0.5\linewidth]{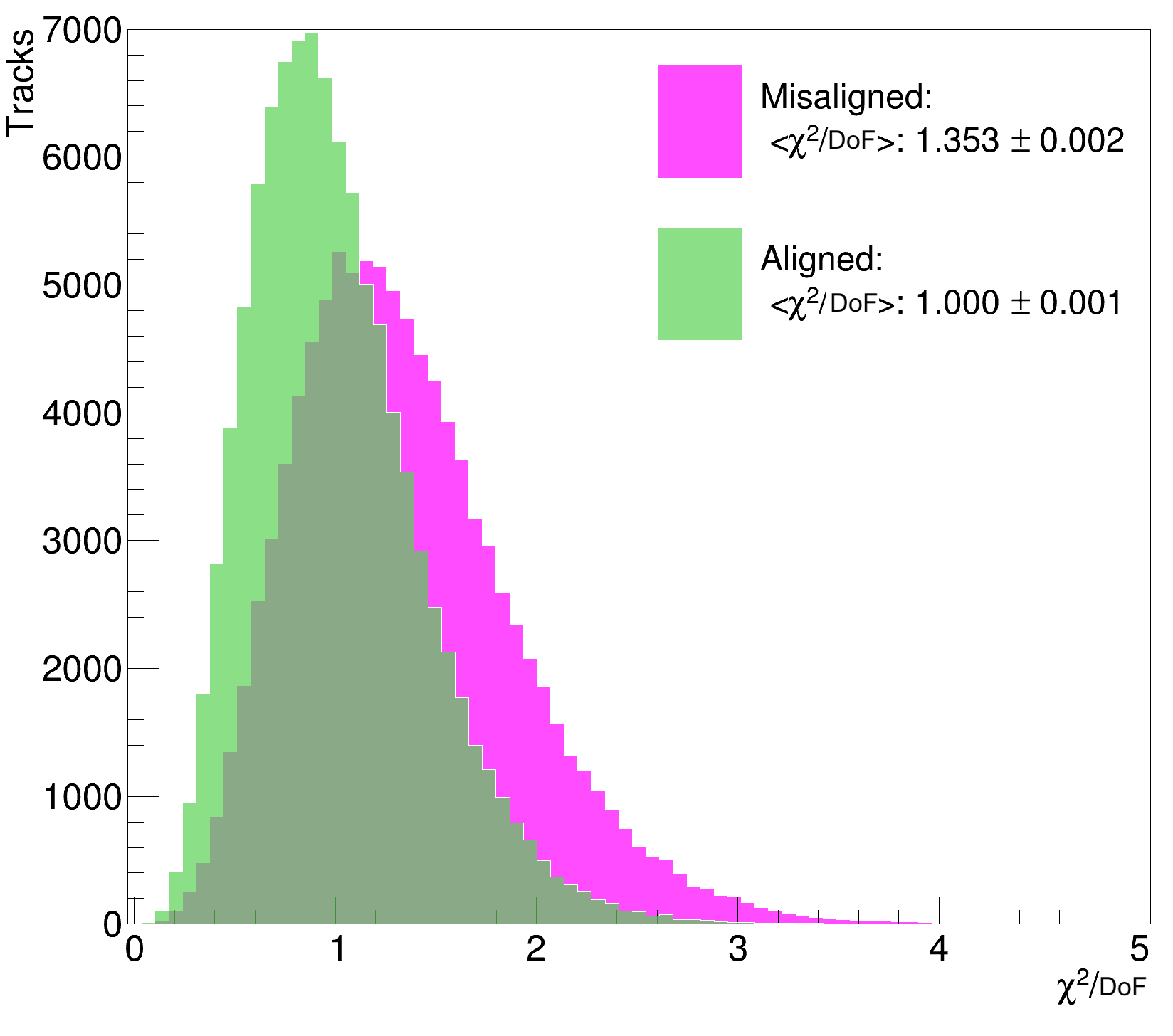}}
    \subfloat[]{\includegraphics[width = 0.5\linewidth]{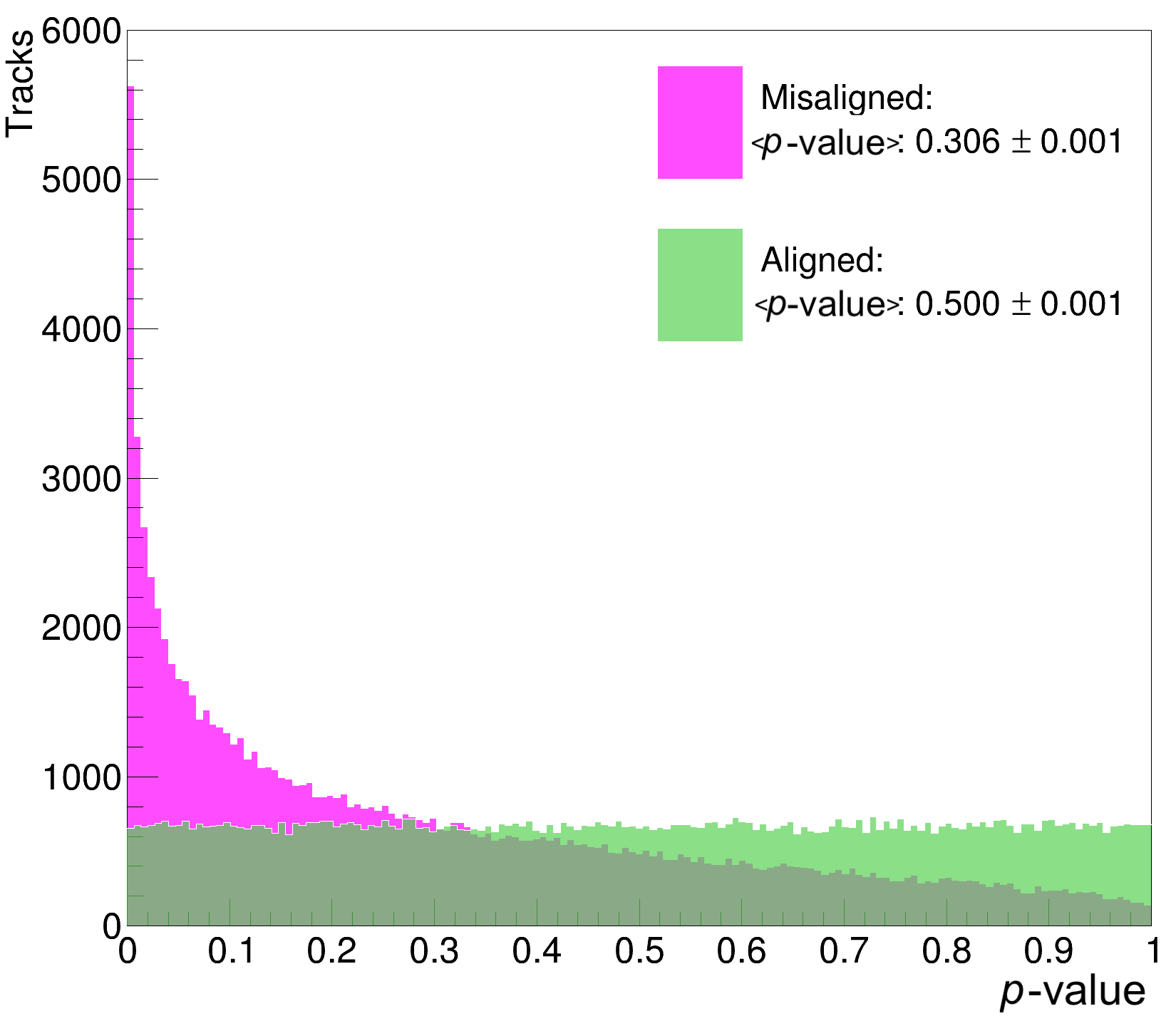}}
    \caption[Alignment \fom]{Alignment \fom: (a) the $\nicefrac{\chi^2}{\mathrm{DoF}}$ after alignment is indicative of a correct detector position, (b) the \pval distribution for the fitted tracks before alignment clearly highlights a poor fit for a misaligned detector.}
    \label{fig:chi_p_2D}
\end{figure}

The inputs to \pede in the case of a circle-fit using straight tracks are given below. In the circle-fit, the closest point on the drift circle to the track is determined first, with the residual, $r$, given by
\begin{equation}    
r = \mathrm{DCA}(x,m,c) - h = \frac{ |c+mz-x| }  { \sqrt{m^2+1} } - h,
\end{equation}
with two local derivatives, $\frac{\partial r}{\partial c}$ and $\frac{ \partial r}{\partial m}$ given in \cref{eq:2D_DLC1,eq:2D_DLC2}, and a single global derivative
\begin{equation}    
\frac{\partial r}{\partial x} = \frac{\partial r}{\partial c}.
\end{equation}
\clearpage
A simple \SI{100}{\micro\metre} misalignment of M3 was analysed. The improvement in the distribution of residuals is apparent as seen in \cref{fig:resMean_2D} and \cref{fig:resSD_2D}; with the final \fom for alignment given in \cref{fig:chi_p_2D}. To obtain the results after the alignment, the predicted global parameters by \pede were used as the correctional offsets to the assumed geometry. 

\subsubsection{Translational and Rotational misalignment in the $xz$ plane with a circle-fit}
The aim of this section is to evaluate the circle-fit residuals with a misalignment induced by an anticlockwise rotation ($\phi$) in the $xz$ plane through the detector centre. 

There are three coordinate systems defining straw positions (see \cref{fig:CS}): \\
1) Global coordinates ($z$, $x$) relative to the \say{global} (0, 0). The derivatives of interest are given in these coordinates. \\ 
2) Local module coordinates ($z_m$, $x_m$) in the un-rotated frame relative to the centre of rotation of a module, which is given by ($z^{\mathrm{centre}}$, $x^{\mathrm{centre}}$).\\ 
3) Local module coordinates ($z'_m$, $x'_m$) in the rotated frame. Such that, the centre of rotation ($z^{\mathrm{centre}}_m$, $x^{\mathrm{centre}}_m$) = ($z^{'\mathrm{centre}}_{m}$, $x^{'\mathrm{centre}}_m$) = (0,0), in the local coordinates.
With $z_m'=z_m(\phi)$ and $x'_m=x_m(\phi)$ transformations, the action of an anticlockwise rotation is given by the reduced 2D form of the rotation matrix (see \cref{eq:rot_matrix}), and the equation for the residual is derived in \cref{app:2D_rot}. 

\begin{figure}[htpb]
\centering
\includegraphics[width = 0.95\linewidth]{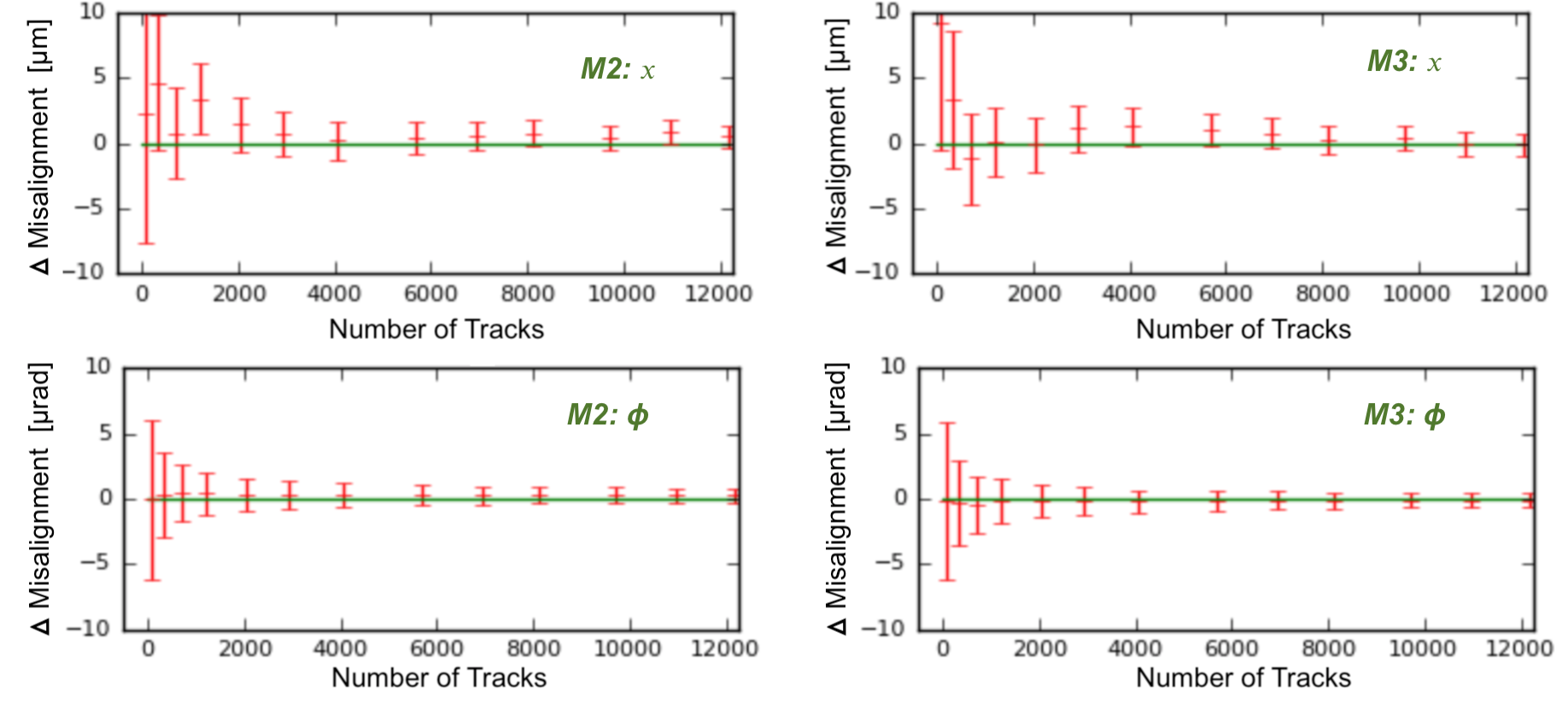}  
    \caption[Comparison between truth and prediction with rotations]{The difference in misalignment between the simulation input and the \pede prediction versus the number of generated tracks.}
\label{fig:PEDEFOM} 
\end{figure}

The result of the \pede minimisation for modules 2 and 3 in $x$ and $\phi$ is shown in \cref{fig:PEDEFOM}. Again, the correct alignment is realised after $\mathcal{O}(5000)$ tracks are fitted. 

\subsection{3D geometry} \label{sc:3d_geometry}
\begin{figure}[htpb]
\centering
\includegraphics[width = 0.56\linewidth]{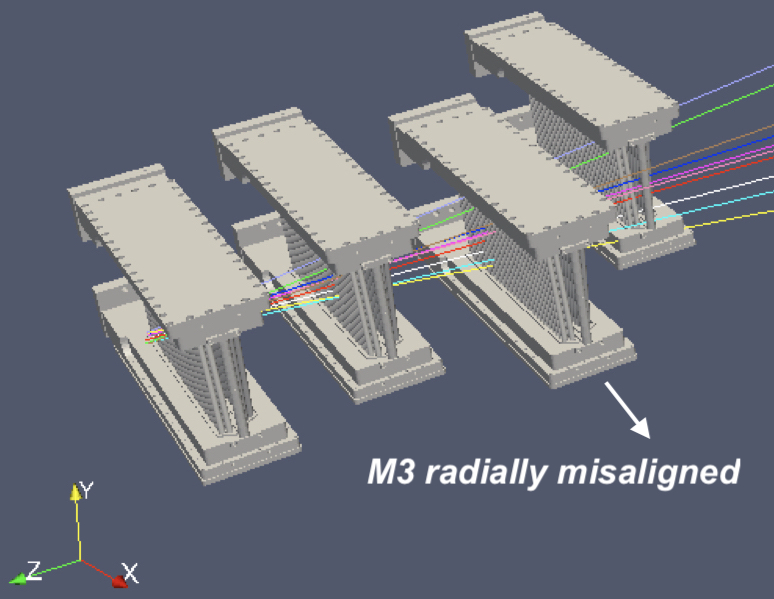}  
\caption[Straight truth tracks through a radially misaligned station]{Straight truth tracks through a radially misaligned station of four tracker modules.}
\label{fig:4M} 
\end{figure}

\begin{figure}[htpb]
\centering
\includegraphics[width=0.7\linewidth]{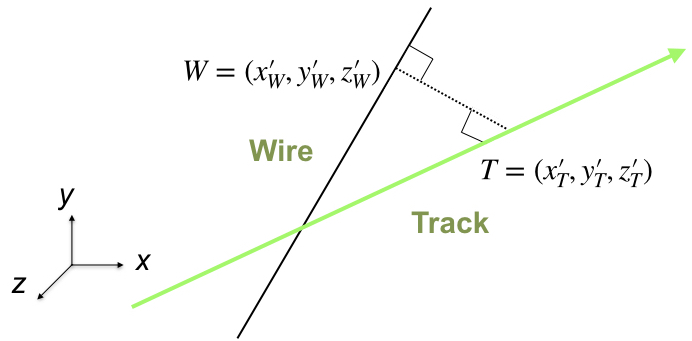}
\caption[Track-to-wire DCA in 3D]{Track-to-wire \ac{DCA} in 3D.}
\label{fig:WireLine}
\end{figure}

In this section we will consider the case of a misalignment in 3D, with the tracker geometry coming from \textit{art}. The residual of interest, between the reconstructed track through the detector and the \say{drift cylinder} (as shown in \cref{fig:DriftCylinder}), is now a function of the \ac{DCA} between two skew lines: the track and the straw wire. The assumption of the two lines being non-parallel will hold for all physical tracks of interest; the track and the wire can intersect, however. The visualisation of this case study with four modules is given in \cref{fig:4M}. In this instance the $x$ misalignment is \textit{radial} and the $y$ is \textit{vertical}.
\vspace*{-0.3cm}
\clearpage
The \ac{DCA} between the track and the wire is given by
\footnotesize
\begin{equation}
\mathrm{DCA} = | \boldsymbol{W} - \boldsymbol{T} | =  | \begin{pmatrix}x'_W - x'_T\\y'_W - y'_T\\z'_W - z'_T\\\end{pmatrix} | = \sqrt{ (x'_W - x'_T)^2 + (y'_W - y'_T)^2 + (z'_W - z'_T)^2}, 
\label{eq:DCA_geom_req}
\end{equation}
\normalsize
where \textbf{W} and \textbf{T} are the points of closest approach between the wire and the track, respectively -- as indicated by the $'$ notation on the coordinates. The vector $\boldsymbol{WT}$ is perpendicular to both the wire and the track, as shown in \cref{fig:WireLine}. 

In order to form the required derivatives with respect to local and global parameters, one needs the functional form of the \ac{DCA} expressed in terms of the reconstructed track parameters and the wire parameters. These derivations are given in \cref{app:3D}.

There are three possible rotations (see \cref{fig:Rotation}) around the centre of the tracker module. A rotation around the $y$-axis is the same as considered previously in \cref{eq:2D_drdphi}, but needs to be extended to a 3D geometry. With the constraint that a point along the straw will have the same vertical height ($y$) before and after the above rotation, the derivative for the anticlockwise rotation $\phi$ along the $y$-axis given by
\begin{equation}
   \frac{\partial r}{\partial \phi} =  \frac{\partial r}{\partial z_W} (-x_W + x^{\mathrm{centre}}) + \frac{\partial r}{\partial x_W} (z_W - z^{\mathrm{centre}}),    
\end{equation}
and similarly for the other two rotations, as shown in \cref{app:3D}.

\subsection{Uniform field} \label{sec:UMF}
\begin{figure}[htpb]
\centering
\includegraphics[width=0.6\linewidth]{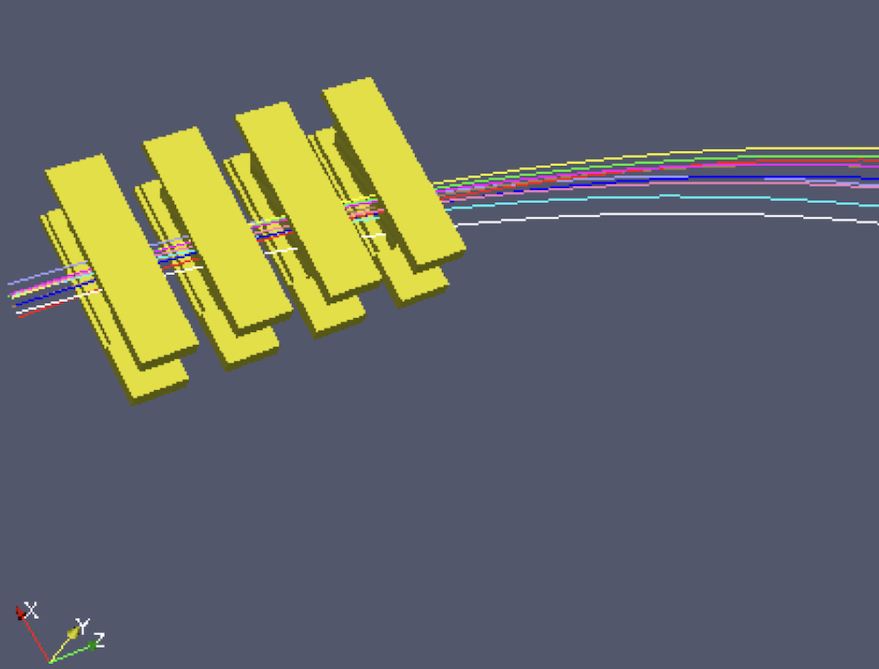}
\caption[Curved tracks in simulation]{A uniform vertical magnetic field of \SI{1.5}{\tesla} bends the particles as they traverse the tracker station of four modules.}
\label{fig:MF_tracker}
\end{figure}

A uniform field throughout the tracker region of \SI{1.5}{\tesla} was implemented in the simulation as shown in \cref{fig:MF_tracker}.

In a magnetic field, a fifth local parameter can be added to describe the track-curvature, $\kappa$, where 
\begin{equation}
\kappa = \frac{1}{p},
\end{equation}
and $p$ is the momentum of the particle, and the approximation $p \approx p_z$ is used to calculate the fifth local derivative, $\frac{ \partial r}{\partial \kappa}$, as derived in \cref{eq:dlc5}.

\begin{figure}[htpb]
    \centering
    \includegraphics[width=0.8\linewidth]{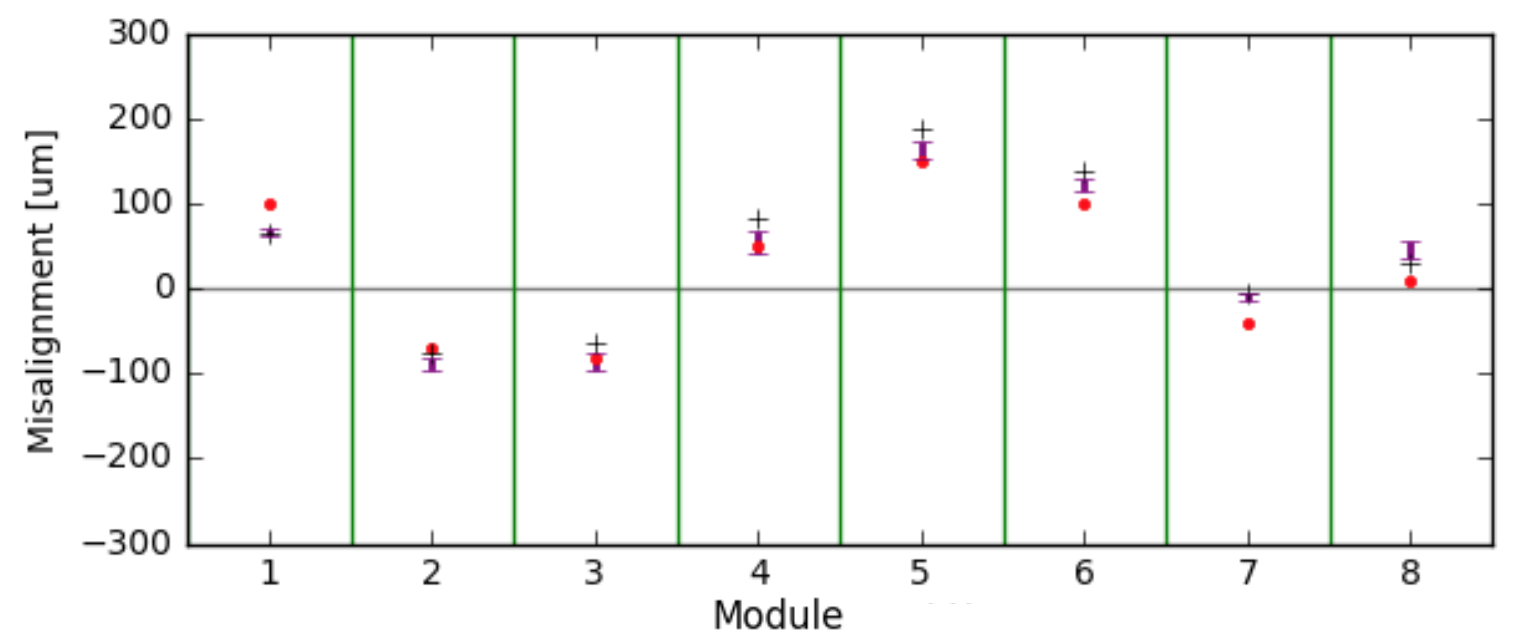}
    \caption[The radial alignment results]{The radial alignment results without constraining the global DoF. The results were obtained with eight modules in 3D with curved tracks after two iterations. The truth misalignment is indicated by a circle (\textbf{$\bullet$}), the first iteration is indicated by a cross (\textbf{+}), with no error, and the second iteration is indicated by a line (\textbf{I}) with an error bar.}
    \label{fig:UMF}
\end{figure}

Results of the radial alignment with eight modules in 3D with curved tracks are shown in \cref{fig:UMF}. The non-convergence of the recovered alignment, as compared to the truth misalignment in some modules, is indicative of not constraining the global \ac{DoF}, as described in \cref{sc:align_constr}.

\subsection{Constraining global parameters} \label{sc:align_constr}
In the particular case of the radial and vertical alignment of the tracking detectors with curved tracks in a magnetic field, constraints to five global \ac{DoF} are applied. There are two overall translations, radially and vertically, that must sum to 0, and two global rotation angles, also fixed at 0. The global rotation is defined through the centre of the station. This is done to minimise the \say{lever-arm effect}. The fifth constraint addresses a radial \textit{detector curvature} (sensor curvature)~\cite{CMS_3} due to the radially curving tracks in the magnetic field. The constraints can be summarised as follows: the radial constraints equation
\begin{equation}
    0=a(x-x_0)^2+b(x-x_0)+c,
    \label{eq:curve} 
\end{equation}
and the vertical constraints equation  
\begin{equation}
    0=b(x-x_0)+c,
\end{equation}
where $x_0=\SI{482.394}{\milli\metre}$ is the centre of the station, and $a$, $b$, and $c$ are the curvature, rotation, and translation parameters, respectively. The summary of the steering and constraint inputs to \pede is given in \cref{app:steer}.

The motivation behind constraining the overall translations and rotations is simple: internal alignment should not return the module offsets related to the global movements. The global alignment should independently measure these global movements, as described in \cref{sec:align_global}. The constraint on the radial detector curvature deserves a special mention, as it is essentially a \say{local minima problem}. The internal alignment can place the tracker modules along a curved path, as the residuals will be unchanged. This radial detector curvature, however, will change the measurement of the momentum of the tracks, and the extrapolated beam position. The detector curvature, therefore, must be measured and eliminated by another means, as described in \cref{ch:align_curvature}.

\subsection{Inhomogeneous magnetic field.} \label{sec:IMF}
\begin{figure}[htpb]
    \centering
    \includegraphics[width=0.55\linewidth]{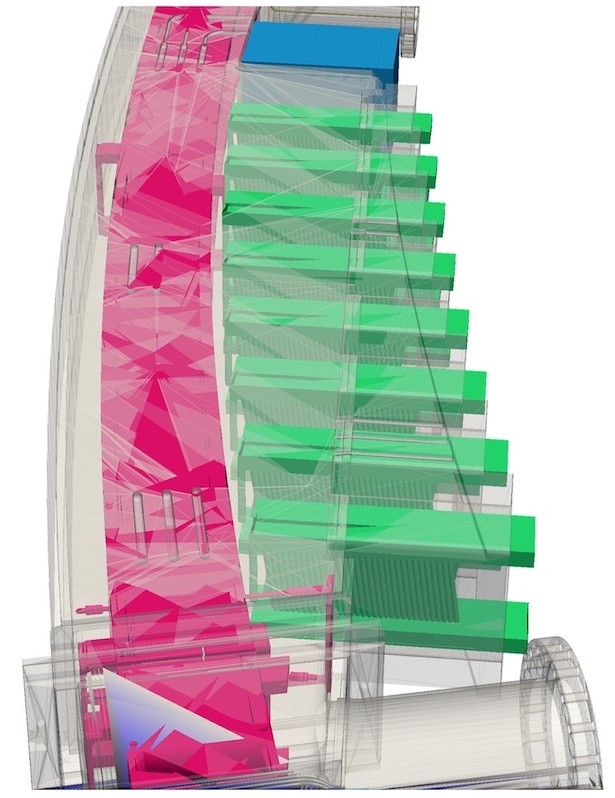}
    \caption[The geometry of tracker modules inside the vacuum chamber]{The geometry of tracker modules inside the vacuum chamber in simulation.}
    \label{fig:VacChamModules}
\end{figure}

Finally, to make the simulation as realistic as possible, the modules were placed in the vacuum chamber as shown in \cref{fig:VacChamModules}.

\begin{figure}[htpb]
    \centering
    \includegraphics[width=0.75\linewidth]{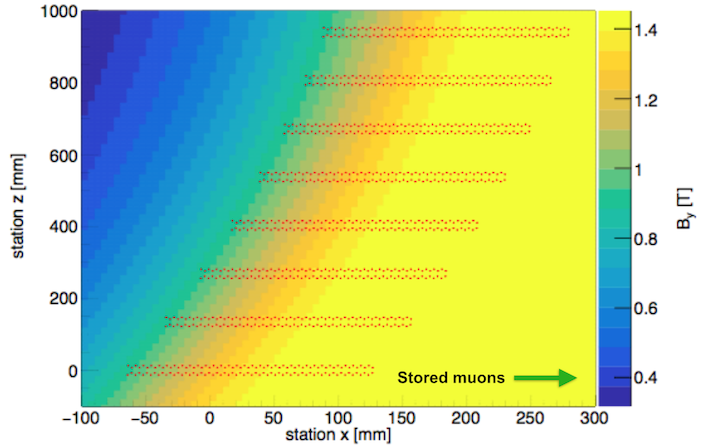}
    \caption[The magnetic field in the tracker region ]{The magnetic field in the tracker region varies radially from 1.45 T to 1.0 T across a distance of 10 cm. The position of the stored beam relative to the station is indicated by an arrow. Plot courtesy of N. Kinnaird~\cite{Nick}.}
    \label{fig:IMF}
\end{figure}

The full experimental simulation framework (\verb!gm2ringsim!), as described in \cref{sec:simulation_framework}, is now used. This also adds an additional complication of a radially non-uniform magnetic field in the tracker region as shown in \cref{fig:IMF}, and multiple scattering due to the presence of material (e.g. a vacuum chamber).  

\subsection{Tracking cuts for alignment}\label{sc:tracking_cuts_for_alignment}
The following cuts were used to select tracks, from simulation or data, to determine the final alignment:
\small
\begin{itemize} \itemsep -2pt 
    \item \textbf{Maximum layers with multiple hits = 0.} Only select tracks that have an unambiguous set of hits in a single straw per layer. 
    \item \textbf{DCA~\textgreater~$500~\mathrm{\mu m}$.} A hit with a small \ac{DCA} is not used in the alignment. This is required to prevent a residual discontinuity as we cross the LR boundary (\cref{eq:dlc1}-\cref{eq:dlc4}).
    \item \textbf{$p$-value~\textgreater~0.005.} Only tracks that have a reasonable fit quality are used. 
    \item \textbf{Hits~\textgreater~9.} Only tracks that have hits in at least three modules are used. 
    \item \textbf{\nicefrac{$p_z$}{$p$}~\textgreater~0.93.} Tracks that have large curvature are removed, required by the approximations used in \cref{eq:dlc5}.
\end{itemize}
\normalsize

\subsection{Iterative alignment}\label{sc:iteration}
\begin{figure}[htpb]
    \centering
    \includegraphics[width =\linewidth]{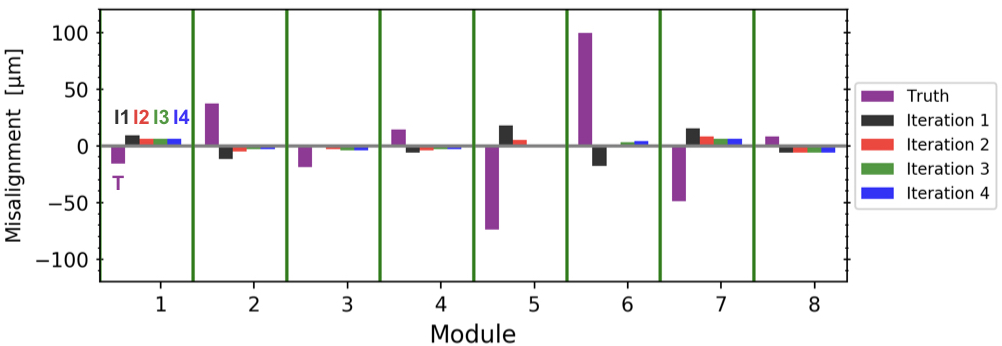}
    \caption[The radial alignment results versus iteration]{The radial alignment results per module in a tracker station in the simulation are shown. The results of the reconstructed misalignment, from the four iterations (I1-I4), are shown as the difference between the truth (T) and a reconstructed misalignment. The reconstructed misalignment from iterations 3 and 4 overlap, indicating convergence.}
    \label{fig:residuum}
\end{figure}

The process of iterative alignment uses the initial module offsets from \pede in the subsequent re-tracking. In iteration two onwards, the tracks are formed through a corrected detector geometry, and only small alignment corrections are returned in the subsequent \pede alignment, as shown in \cref{fig:residuum}. The iterative alignment is considered to converge when the returned alignment results agree with a previous iteration to within the \pede error (few \SI{}{\micro\metre}).

\clearpage
\vspace{-0.4cm}
\section{Alignment results in simulation}\label{sec:align_sim}
\begin{figure}[htpb]
    \centering
    \includegraphics[width =\linewidth]{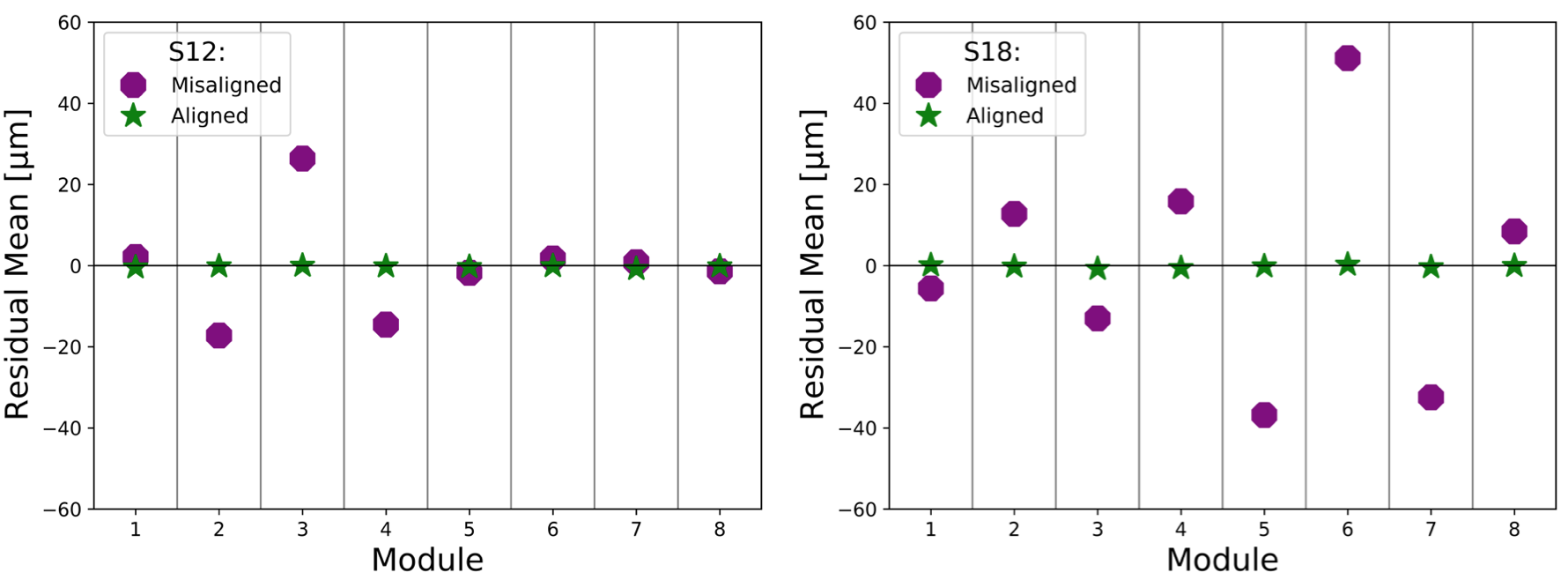}
    \vspace{-0.1cm}
    \caption[The mean values of the residuals per module]{The mean values of the residuals per module before and after the alignment in the simulation. The mean residual is expected to be at $0$ for an aligned detector.}
    \label{fig:Res_Sim}
\end{figure}

\begin{figure}[htpb]
    \centering
    \includegraphics[width = \linewidth]{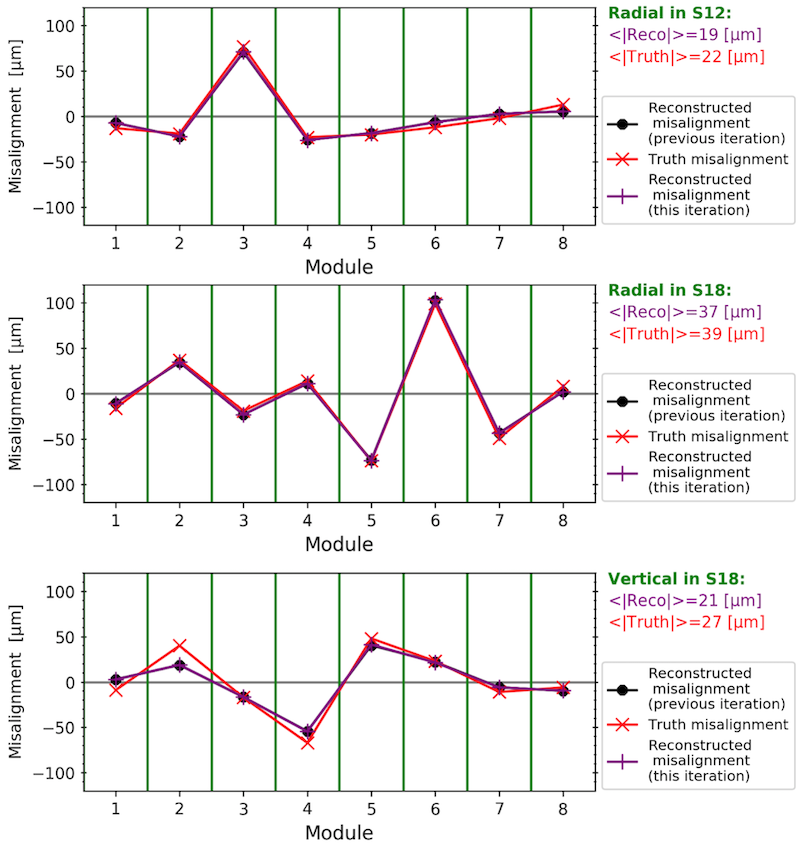}
    \caption[The final simulation alignment results]{The final simulation alignment results showing that alignment stability was reached. S12 was not misaligned vertically. The alignment results from the previous iteration and the current iteration overlap. The absolute mean reconstructed ($\langle|\mathrm{Reco}|\rangle$) and the absolute mean truth ($\langle|\mathrm{Truth}|\rangle$) misalignments per module are indicated. The alignment convergence, with respect to the truth, was established within \SI{3}{\micro\metre} and \SI{6}{\micro\metre} radially and vertically, respectively.}
    \label{fig:Align_Sim}
\end{figure}

\begin{figure}[htpb]
    \centering
    \includegraphics[width =0.9\linewidth]{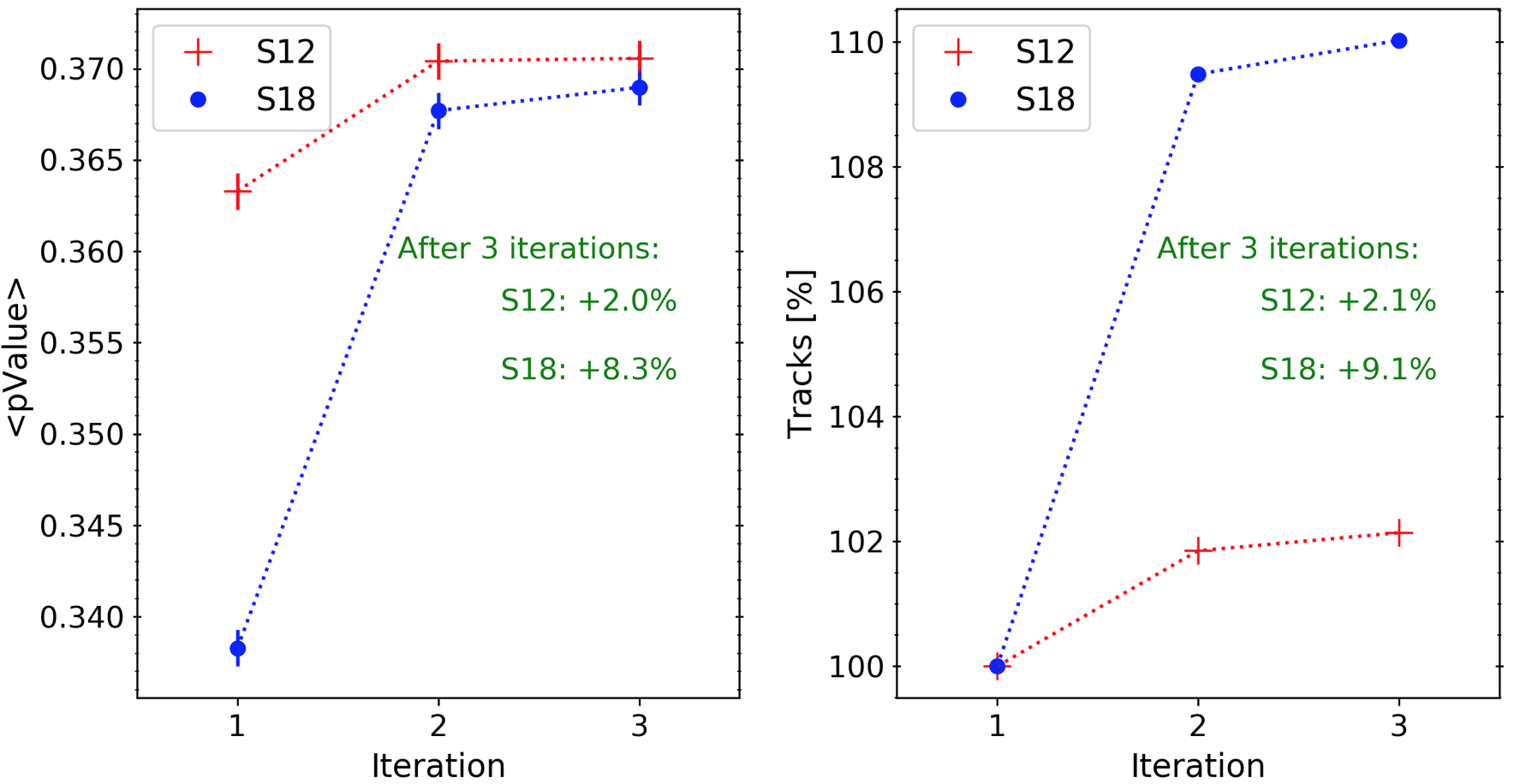}
    \caption[The improvement in the mean \pval in simulation]{The improvement in the mean \pval and the number of tracks as a function of the alignment iteration in simulation. The overall input misalignment in station 18 was greater, hence, after the alignment, the improvement in station 18 is more significant than in station 12.}
    \label{fig:Iter_Sim}
\end{figure}

The mean values of the residuals per module before and after the alignment are shown in \cref{fig:Res_Sim}. The alignment results in both stations after three iterations are shown in \cref{fig:Align_Sim}. The improvement in the mean \pval and number of reconstructed tracks as a function of number of iterations is shown in \cref{fig:Iter_Sim}.
\vspace{-0.2cm}
\clearpage
The overall input (i.e. truth) misalignment in station 18 was greater; hence, after the alignment, the improvement in station 18 is more significant than in station 12. The alignment stability was reached, and the alignment convergence was established within \SI{3}{\micro\metre} and \SI{6}{\micro\metre} radially and vertically, respectively, in simulation.
\clearpage
\vspace{-0.2cm}
\section{Alignment results with data}\label{sc:align_real}
\begin{figure}[htpb]
    \centering
    \includegraphics[width = \linewidth]{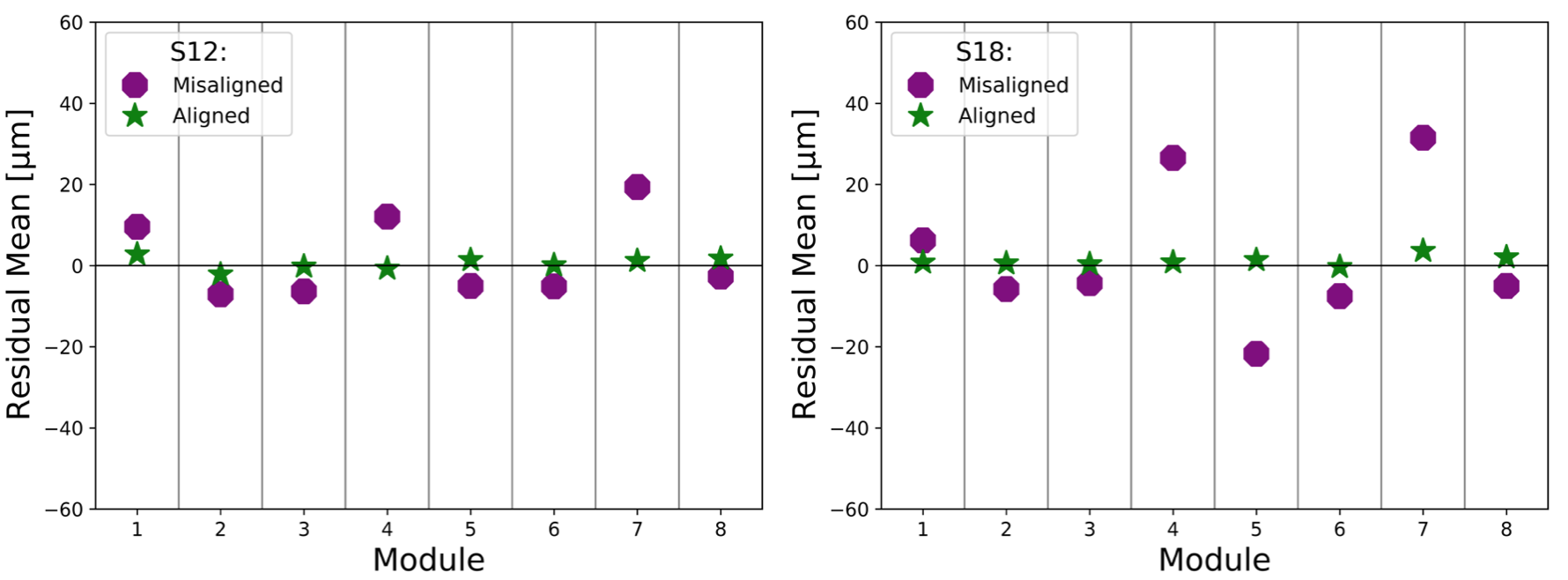}
    \vspace{-0.1cm}
    \caption[The mean values of the residuals per module]{The mean values of the residuals per module before and after the alignment in data. The mean residual is expected to be at $0$ for an aligned detector.}
    \label{fig:Res_Data}
\end{figure}

\begin{figure}[htpb]
    \centering
    \includegraphics[width = 0.9\linewidth]{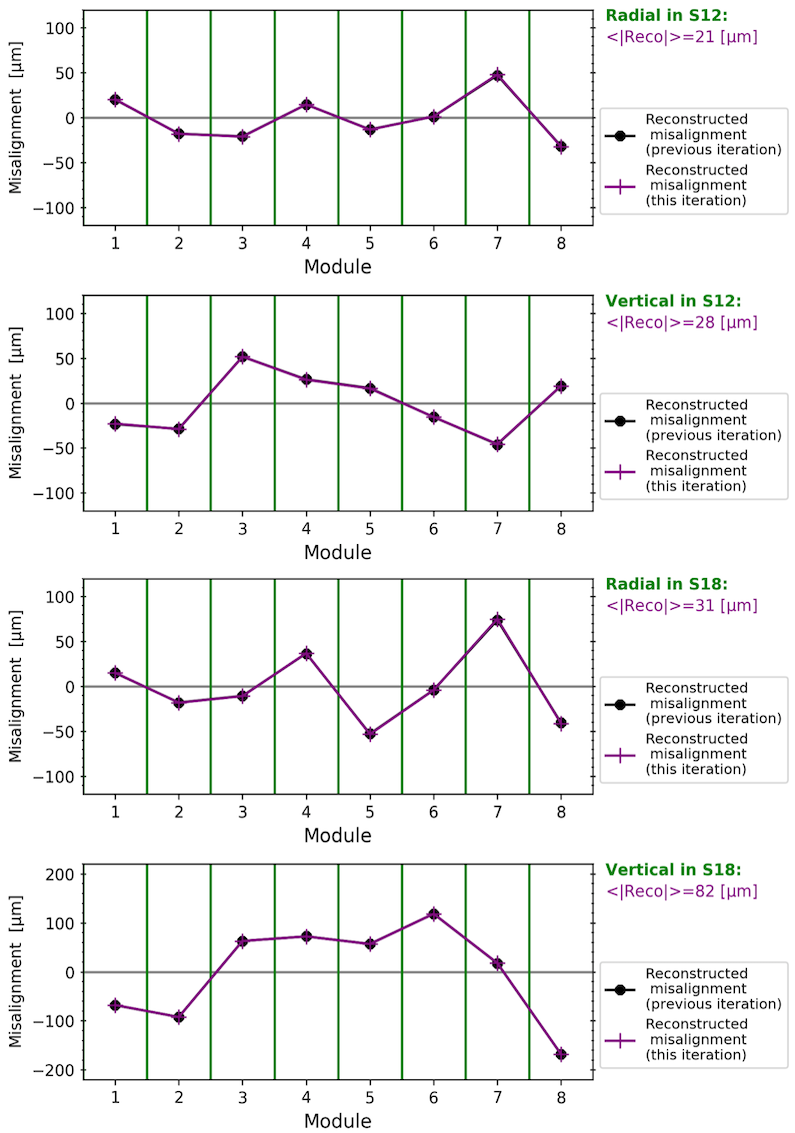}
    \caption[The alignment results with data]{The alignment results with data. The absolute mean reconstructed ($\langle|\mathrm{Reco}|\rangle$) misalignment per module is indicated. The alignment stability was reached, as seen by the alignment results from the previous iteration and the current iteration completely overlapping.}
    \label{fig:Align_Data}
\end{figure}

\begin{figure}[htpb]
    \centering
    \includegraphics[width =0.78\linewidth]{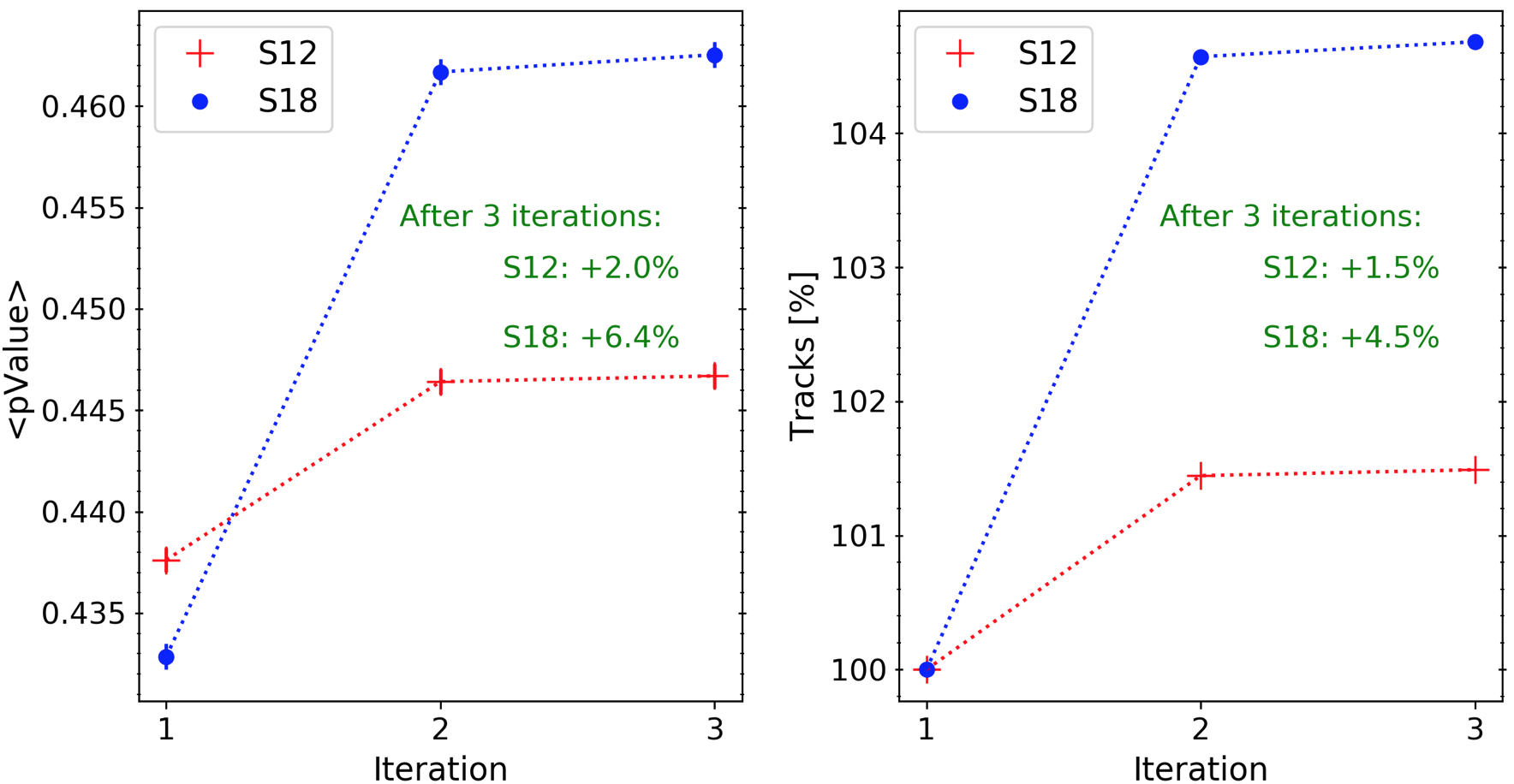}
    \vspace{-0.25cm}
    \caption[The improvement in the mean \pval in data.]{The improvement in the mean \pval and the number of tracks as a function of the alignment iteration.}
    \label{fig:Iter_Data}
\end{figure}

\begin{figure}[htpb]
    \centering
    \includegraphics[width = 0.90\linewidth]{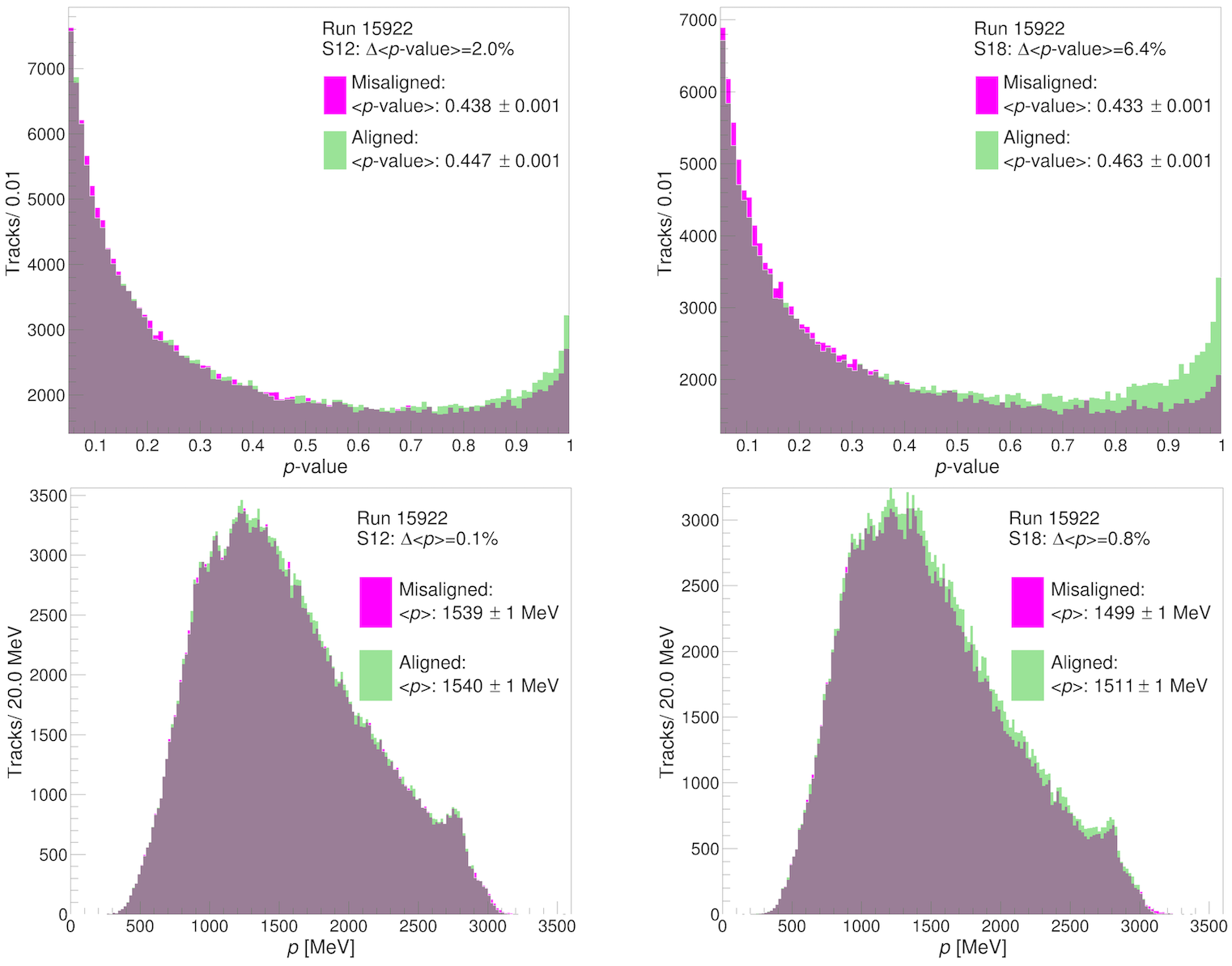}
    \vspace{-0.25cm}
    \caption[The distribution of track momentum after the alignment]{The distribution of track \pvals and momentum in both stations before and after the alignment. Stations 12 and 18 have 2.0\% and 6.4\% improvement in the mean track \pval, respectively, with only a small change in the mean track momentum. The peak at 2700 \MeV is due to lost-muons (see \cref{sc:lost_muons}), as the alignment considers all tracks, not just $e^+$ from $\mu^+$ decays, that have passed the alignment cuts defined in \cref{sc:tracking_cuts_for_alignment}.}
    \label{fig:Align_Pval}
\end{figure}

\begin{figure}[htpb]
    \centering
    \includegraphics[width =\linewidth]{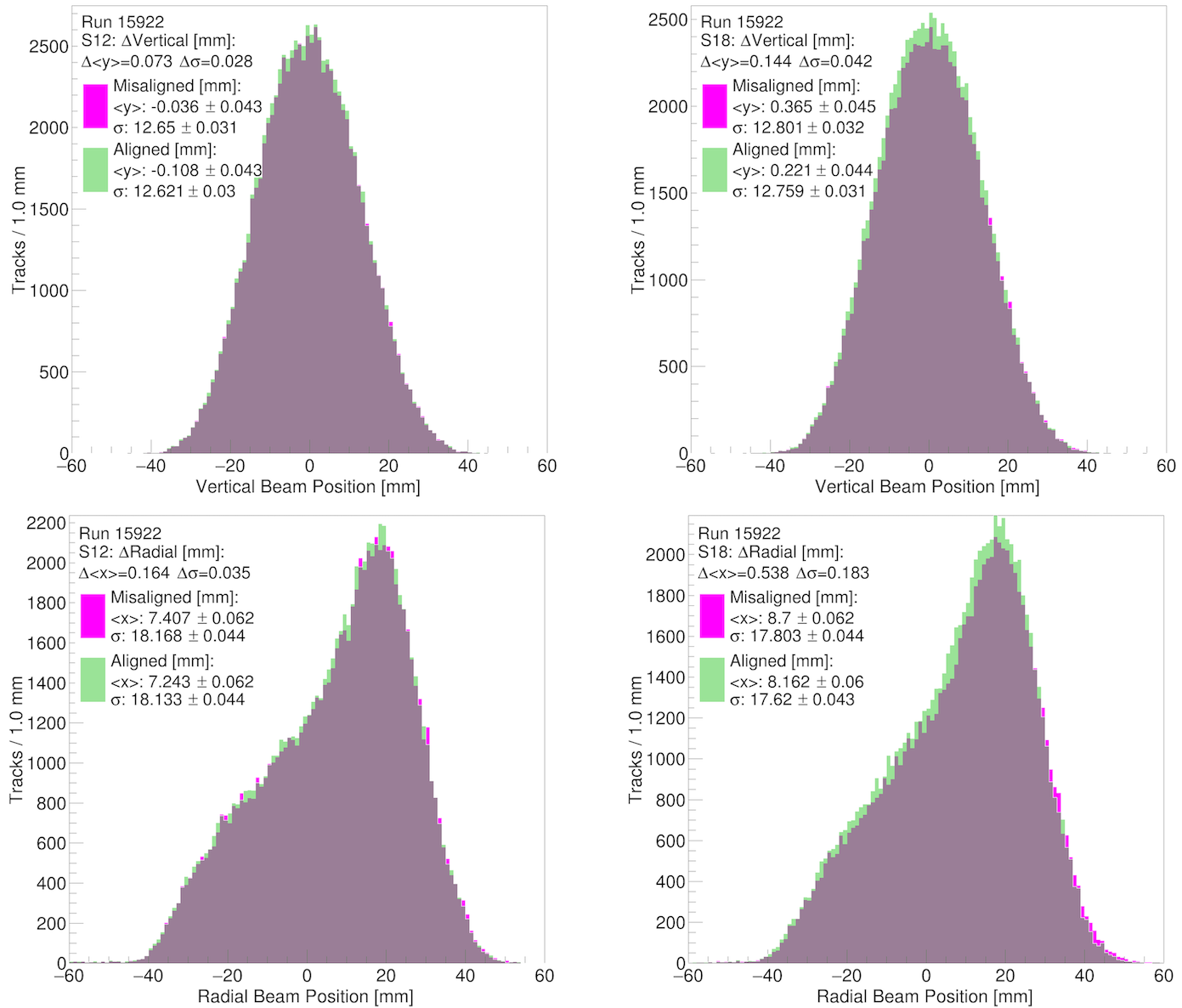}
    \caption[The extrapolated beam positions after the alignment]{The vertical and radial distributions of the extrapolated beam positions in both stations before and after the alignment.}
    \label{fig:Align_Beam}
\end{figure}

The mean values of the residuals per module before and after the alignment, using run 15922, are shown in \cref{fig:Res_Data}. The alignment results in both stations after three iterations are shown in \cref{fig:Align_Data}. The results were obtained with $\mathcal{O}(10^5)$ tracks, and were then used in the track reconstruction.  
\vspace{-0.2cm}
\clearpage
\vspace{-0.5cm}
The improvement in the mean \pval and number of reconstructed tracks as a function of number of iterations is shown in \cref{fig:Iter_Data}. The change in the distribution of \pvals and momentum, as well as in the beam extrapolation are shown in \cref{fig:Align_Pval} and \cref{fig:Align_Beam}, respectively. 
\clearpage
\vspace{-0.7cm}
\clearpage
Stations 12 and 18 have a 2.0\% and 6.4\% improvement in the mean track \pval, respectively. This implies that after the alignment the quality of the reconstructed tracks is higher. Moreover, due to the improved alignment, more tracks are reconstructed from the same hits in station 12 and 18, by 1.5\% and 4.5\%, respectively. 

The extrapolated tracks (for $t>$~\tms) from stations 12 and 18 have a radial shift towards the centre of the ring of \SI{164}{\micro\metre} and \SI{538}{\micro\metre}, respectively. The extrapolated tracks have a vertical shift downwards of \SI{73}{\micro\metre} and \SI{144}{\micro\metre} for stations 12 and 18, respectively.

\subsection{Alignment monitoring system and database}
\begin{figure}[htpb]
    \centering
    \includegraphics[width = 0.6\linewidth]{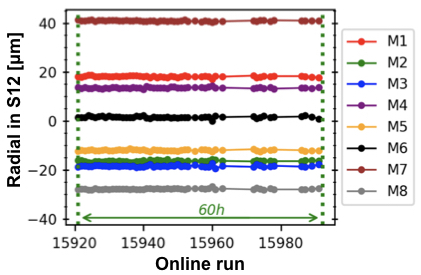}
    \caption[The radial alignment results (Run-1a dataset)]{The radial alignment results in the eight modules of station 12 across 60 hours of \gm2 data (Run-1a dataset).}
    \label{fig:60h}
\end{figure}

As seen from \cref{fig:Iter_Data}, the most significant alignment corrections come after the first iteration. Subsequent iterations yield corrections comparable to the \pede uncertainty. Hence, an assessment of the alignment stability across all \gm2 runs is possible, by running the alignment directly on the reconstructed tracks from all runs without an iterative alignment. The radial alignment results per module in station 12 over a 60-hour period (Run-1a dataset) are shown in \cref{fig:60h}.
\vspace{-0.1cm}
\vspace{-0.1cm}

\begin{figure}[htpb]
    \centering
    \includegraphics[width = 0.95\linewidth]{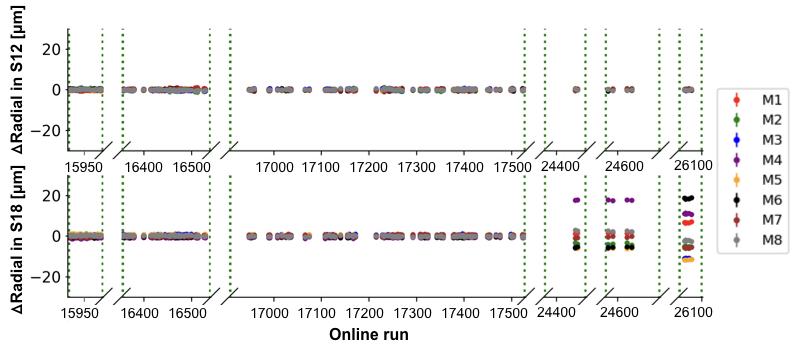}
    \caption[The alignment stability results (\R1 and \R2)]{The alignment stability, radially and vertically, across \R1 (online run \textgreater 10000) and \R2 (online run \textgreater 20000) in both stations. The alignment stability is displayed as the difference between the mean alignment for a given run and the alignment from a fixed reference run (run 15922). S12 displays stable alignment throughout the two Runs. The deviations in modules 4 (M4) and 6 (M6) in S18 are due to physical module swaps~\cite{Swaps} before \R2, and after run 25086.}
    \label{fig:Monitor}
\end{figure}

The alignment stability in both stations, radially and vertically, across \R1 and \R2 is shown in \cref{fig:Monitor}.
\vspace{-0.1cm}
\clearpage
The change in the internal alignment of the two modules is seen in \cref{fig:Monitor}: before the start of \R2 in module 4, and after run 25086 in module 6. These modules were replaced~\cite{Swaps} with spare modules, since the existing modules had developed a slightly higher gas leak rate into the storage ring vacuum than the specification. New alignment constants for the replaced modules were derived. The alignment constants were written into a \texttt{PostgreSQL} database, where each set of constants is associated to a given range of runs.
\vspace{-0.06cm}
\subsection{Outlook}
\begin{figure}[htpb]
    \centering
    \subfloat[]{\raisebox{7mm}{\includegraphics[width = 0.32\linewidth]{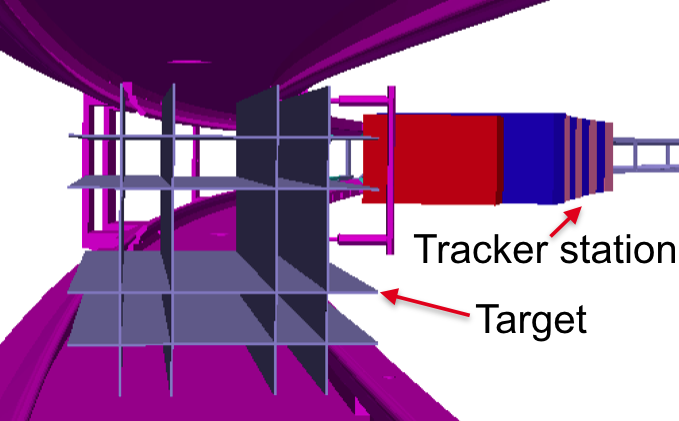}}}
    \subfloat[]{\includegraphics[width = 0.30\linewidth]{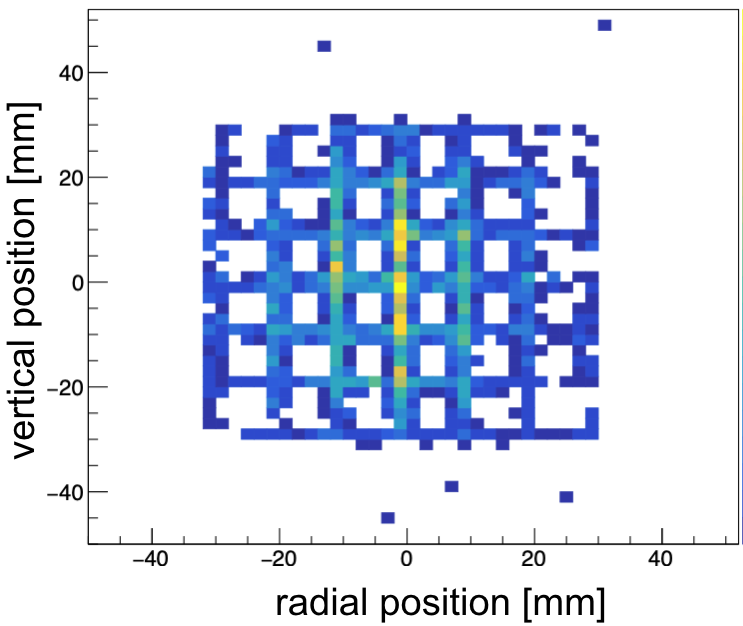}}
    \vspace{-0.2cm}
    \caption[The implementation of a tungsten scattering target]{The implementation of a tungsten scattering target in the simulation~\cite{Sean}. (a) Target location inside of the \gm2 ring relative to the tracker station. (b) Vertical and radial extrapolated beam positions. The target's grid structure in seen in the tracker.}
    \label{fig:scatTar}
\end{figure}

The internal alignment, radially and vertically, of the two tracker stations in \R1 and \R2 has been successfully determined. Moreover, the alignment manual~\cite{Gleb_manual}, to monitor the alignment in the future, has been produced. There are, however, a few more activities that are planned to further refine and improve the alignment:
\vspace{-0.2cm}
\small
\begin{itemize} \itemsep -5pt 
    \item \textbf{Scattering target}. A placement of a scattering target, as shown in \cref{fig:scatTar}, in a known position in the \gm2 storage ring will give a precise location of the starting position of the tracks. This project is being developed by S. Foster~\cite{Sean}.
    \item \textbf{Internal alignment with rotations}. The current tracking algorithm (see \cref{sc:track_soft}) assumes that the tracker planes are parallel to each other. The use of a Kalman filter algorithm, currently in development by A. Luca~\cite{Alessandra}, will allow for rotations of tracker planes, and hence, will improve the alignment. 
\end{itemize}
\normalsize
\vspace{-0.47cm}
\clearpage

\subsection{Alignment impact on the systematic uncertainties}
\subsubsection{Pitch correction (\texorpdfstring{$\omega_a$}~)}
The beam extrapolation determines the radial and vertical positions of the muon beam in the storage ring. The accuracy of this determination is affected by detector calibrations, such as the alignment. The internal alignment of the tracking detectors is now established, with the $1\sigma$ uncertainties on the mean and width of the beam given in \cref{tab:align_pitch_final}. In the table, the combined uncertainty, in both stations, from \cref{fig:Align_Beam} was taken.
\begin{table}[htpb]  
  \centering
  \begin{tabular}{lrrrr}
    \toprule
            & $\mathrm{\sigma_{dR}}$ [\SI{}{\milli\meter}] & $\mathrm{\sigma_{dR_{\mathrm{width}}}}$ [\SI{}{\milli\meter}] & $\mathrm{\sigma_{dV}}$ [\SI{}{\milli\meter}] & $\mathrm{\sigma_{dV_{\mathrm{width}}}}$ [\SI{}{\milli\meter}]  \\ \midrule
    
      S12 $\oplus$ S18 & 0.351 & 0.109 & 0.109 & 0.035  \\ \bottomrule
  \end{tabular}
  \caption[The contribution of the internal alignment to the beam extrapolation uncertainty]{The contribution of the internal alignment to the beam extrapolation uncertainty in both stations.}
  \label{tab:align_pitch_final}
\end{table}

However, the alignment is now accounted for, as the correct module positions are now used in the track reconstruction. For example, in \cref{ch:align_error}, the uncertainty on the pitch correction, $\Delta C_{\mathrm{pitch}}$, from an averaged randomised internal misalignment was estimated to be 1.5 ppb. After the implementation of the internal alignment, this uncertainty was eliminated. This has reduced the total contribution to the uncertainty on the pitch correction from tracking (see \cref{sec:pitch}) from 8.6 ppb to 8.4 ppb. 

\subsubsection{Field convolution (\texorpdfstring{$\omega_p$}~)}
The magnetic field is convoluted with the beam profile measured by the tracking detectors to find the average field experienced by the muons before decay, as described in \cref{sec:field_convolution}. As previously discussed, the measurement of the beam profile from the tracking detectors is affected by the internal alignment. A study of an impact of translations on the extrapolated beam profile from the tracking detectors on the field measurements was performed by Jason Bono and Saskia Charity~\cite{JasonSaskia}. Given the scale of translations induced by the alignment in \cref{fig:Align_Beam}, it is possible to estimate the corresponding uncertainties on $\langle B \rangle$, and hence $\omega_p$ (see \cref{eq:omega_p}).
\clearpage
The results for both stations are given in \cref{tab:b_align}, with the average vertical and radial uncertainties on $\langle B \rangle$ of 3.8 ppb and 0.7 ppb, respectively. After the implementation of the internal alignment, these uncertainties were eliminated. This compares to the overall goal of a 70 ppb uncertainty on the determination of $\omega_p$ at the end of the experiment's data taking.
\begin{table}[htpb]  
  \centering
  \begin{tabular}{rrrr}
    \toprule
           S12 radial & S18 radial & S12 vertical & S18 vertical \\ \midrule
    
      1.5 ppb & 6.1 ppb & 0.4 ppb & 0.9 ppb  \\ \bottomrule
  \end{tabular}
  \caption[The alignment contribution to the uncertainty on $\langle B \rangle$]{The alignment contribution to the uncertainty on $\langle B \rangle$.}
  \label{tab:b_align}
\end{table}

\graphicspath{{fig/}}

\chapter{Detector curvature}
\label{ch:align_curvature}

\section{Introduction}
\begin{figure}[htpb]
    \centering
    \includegraphics[width = \linewidth]{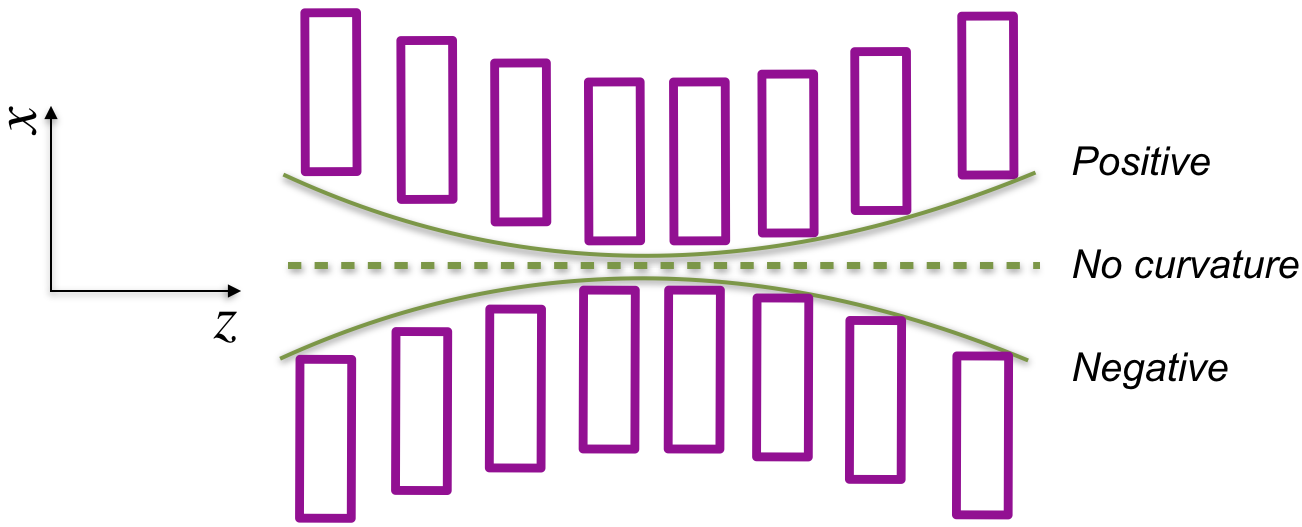}
    \caption[The positions of tracker modules with curvatures]{An illustration of radially misaligned detectors that cause an erroneous momentum measurement. The positions of the eight tracker modules are shown with positive and negative curvatures. Positive curvature is defined as away from the centre of the \gm2 storage ring.}
    \label{fig:curveEx}
\end{figure}

A radial curvature in the positions of the tracker modules, as shown in \cref{fig:curveEx}, changes the measurement of the momentum of the tracks, and the extrapolated beam position. As described in \cref{sc:align_constr}, this detector curvature must be measured and constrained by a method that is independent of the internal alignment. One such method is described here, along with the derivations of the tracking systematic uncertainties arising from the curvature.
\clearpage

\section{Impact on alignment and tracking}
\begin{figure}[htpb]
    \centering
    \subfloat[]{\includegraphics[width = 0.62\linewidth]{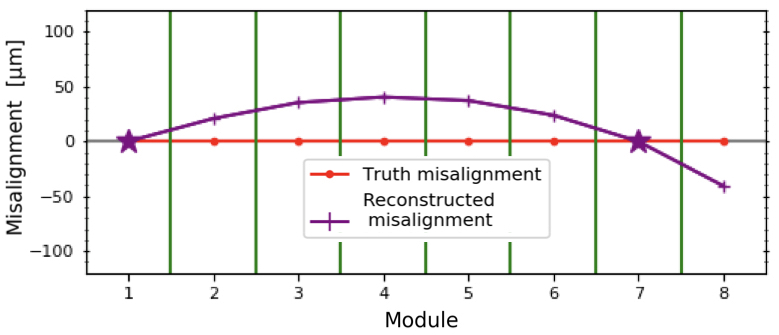}} 
    \subfloat[]{\raisebox{2.5mm}{\includegraphics[width = 0.38\linewidth]{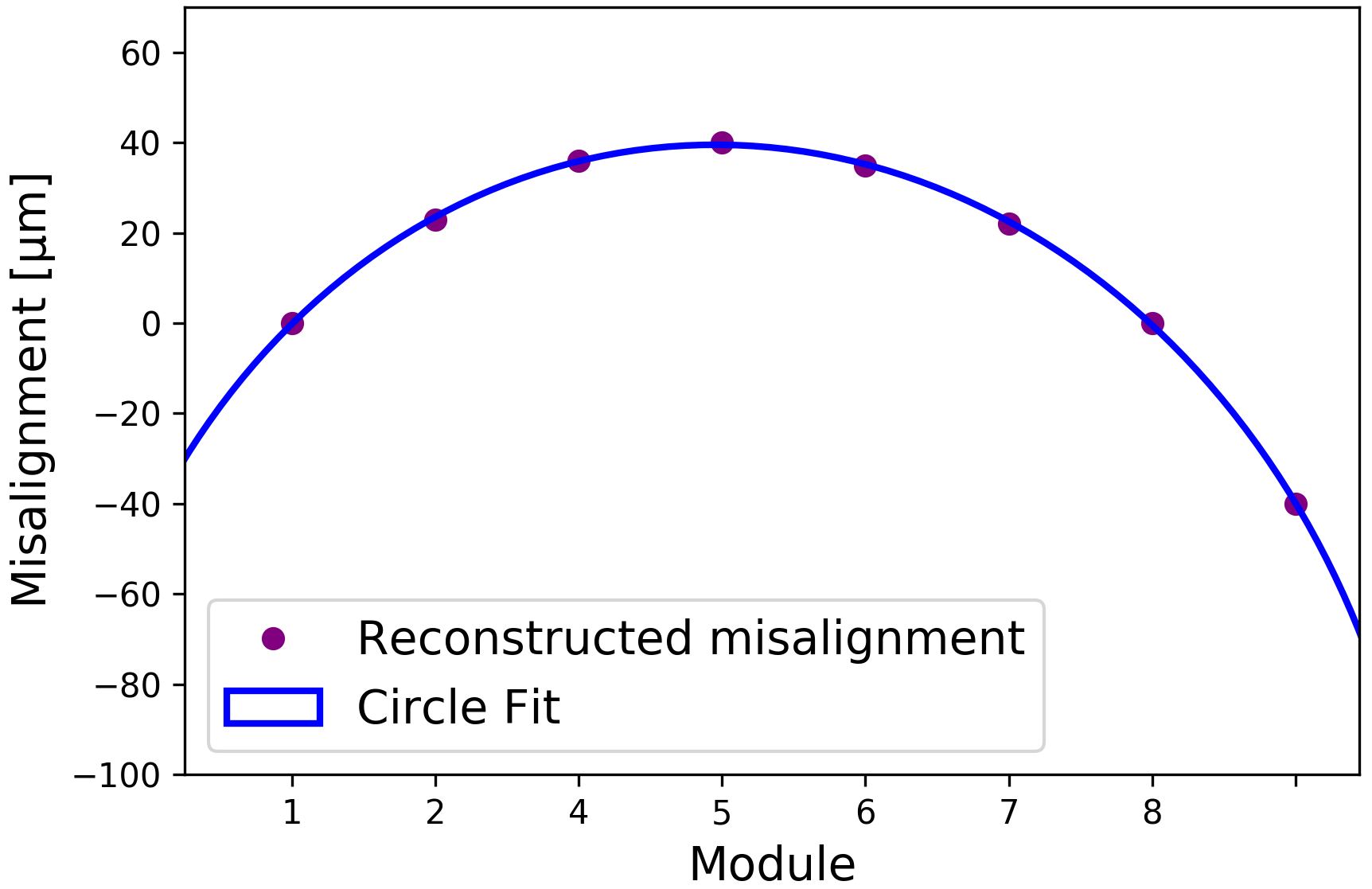}}}
    \vspace{-0.1cm}
    \caption[An unconstrained detector curvature]{An unconstrained detector curvature yields an erroneous result. (a) Truth misalignment and reconstructed alignment radially. (b) A circle fit to the reconstructed module positions from (a). }
    \label{fig:Curve1}
\end{figure}

If the internal alignment is established without constraining for the radial detector curvature, the recovered alignment results will contain an erroneous curvature. Even for a case of no input misalignment in simulation, the absence of the correct constraints will introduce this erroneous curvature, as shown in \cref{fig:Curve1}.
\vspace{-0.45cm}

\begin{figure}[htpb]
    \centering
    \subfloat[]{\includegraphics[width = 0.495\linewidth]{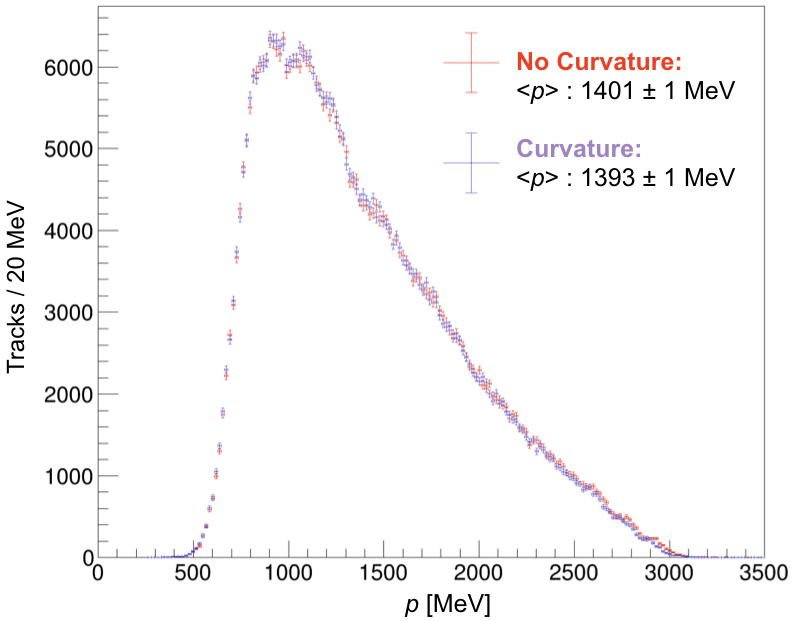}}
    \subfloat[]{\includegraphics[width = 0.505\linewidth]{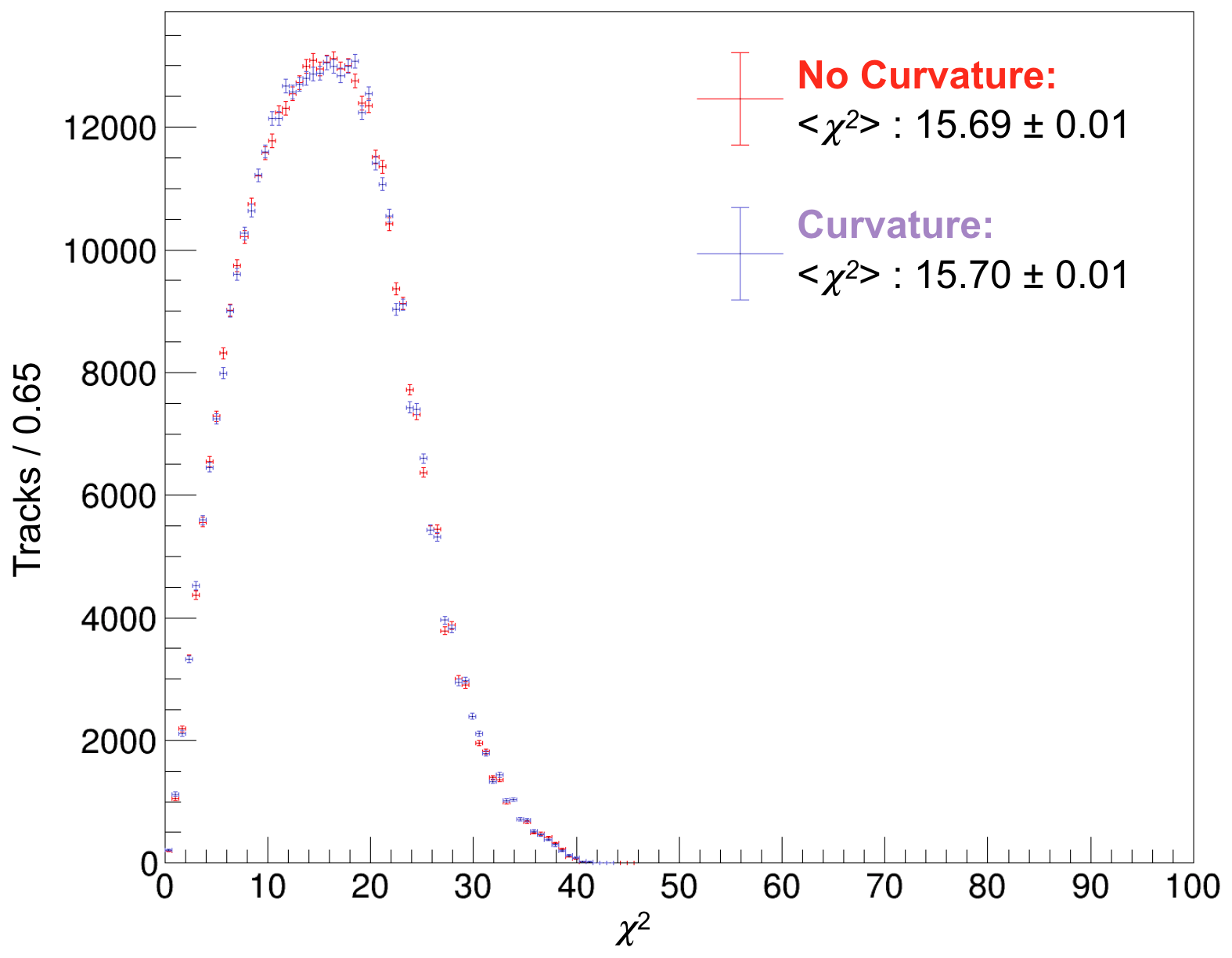}}
    \vspace{-0.1cm}
    \caption[The detector curvature effect on the reconstructed tracks]{The detector curvature effect on the reconstructed tracks: (a) change in the mean momentum due to the detector curvature, (b) the $\chi^2$ distribution of tracks is unaffected.}
    \label{fig:MomCurve}
\end{figure}

\begin{figure}[htpb]
    \centering
    \subfloat[]{\includegraphics[width = 0.62\linewidth]{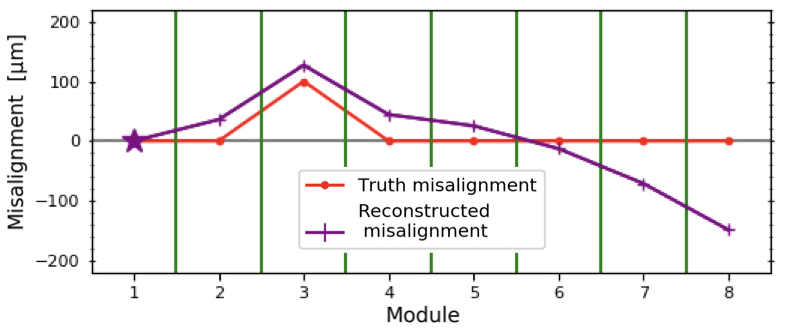}} 
    \subfloat[]{\raisebox{2.5mm}{\includegraphics[width = 0.38\linewidth]{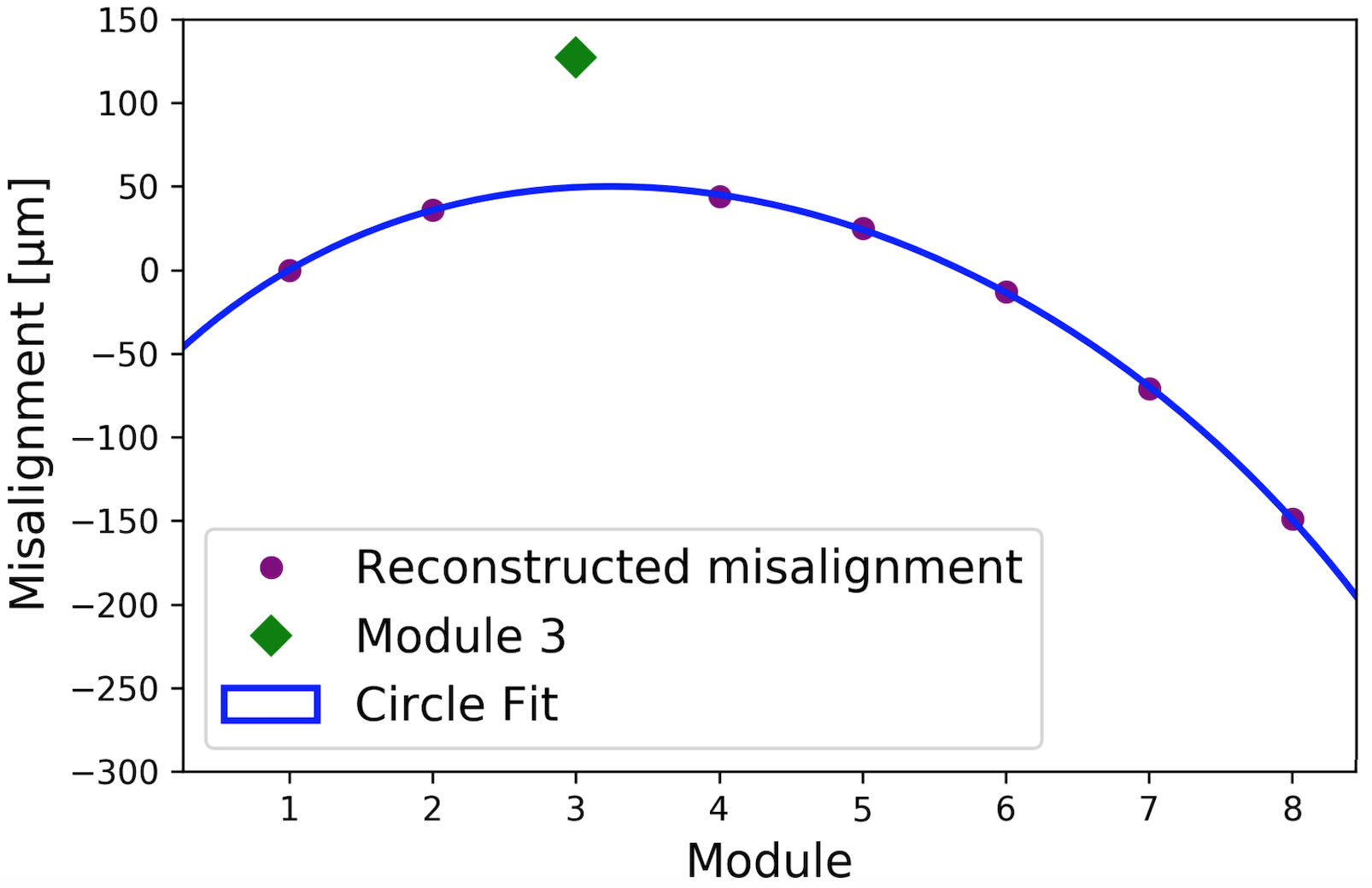}}}
    \caption[An unconstrained detector curvature with misalignment]{An unconstrained detector curvature with an additional internal misalignment in one module. (a) Truth misalignment and reconstructed alignment radially. (b) A circle fit to the reconstructed module positions from (a).}
    \label{fig:Curve2}
\end{figure}

\vspace{-0.25cm}
Re-tracking with the modules placed in the suggested position from \cref{fig:Curve1} changes the momentum distribution of the tracks, but not their $\chi^2$ distribution, as shown in \cref{fig:MomCurve}. This is suggestive of the \say{local minima problem} -- \mpt minimises the $\chi^2$ function, and without providing more information, this wrong solution could be a \say{valid} solution.
\vspace{-0.35cm}
\clearpage
Moreover, if a module is misaligned internally, this is manifested as a deviation from the overall radial detector curvature in that module, as shown in \cref{fig:Curve2}.

\begin{figure}[htpb]
    \centering
    \subfloat[]{\includegraphics[width = 0.65\linewidth]{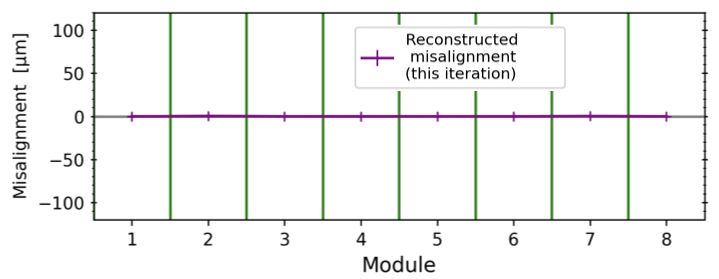}} \\
    \subfloat[]{\includegraphics[width = 0.65\linewidth]{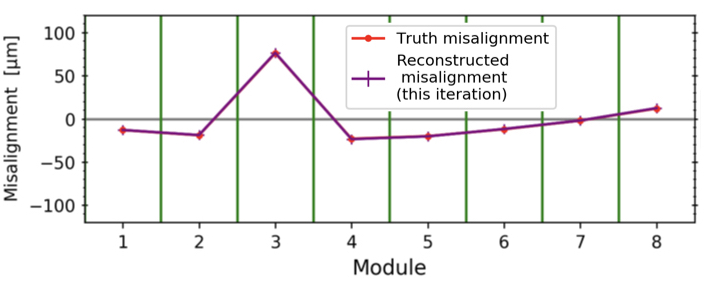}}
    \caption[A constrained radial detector curvature]{Constrained radial detector curvature with: (a) no misalignment, and (b) internal misalignment.}
    \label{fig:Curve3}
\end{figure}

However, if the correct constraints from \cref{eq:bowing} are used, the radial detector curvature effect is removed completely as shown in \cref{fig:Curve3}.
\clearpage

\section{Methodology}
\begin{figure}[htpb]
\centering
\subfloat[]{\includegraphics[width=0.49\linewidth]{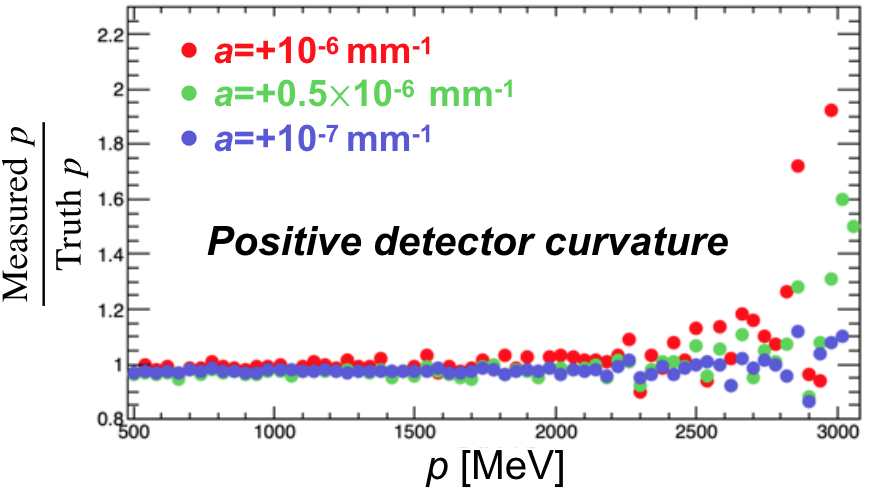}}
\subfloat[]{\includegraphics[width=0.49\linewidth]{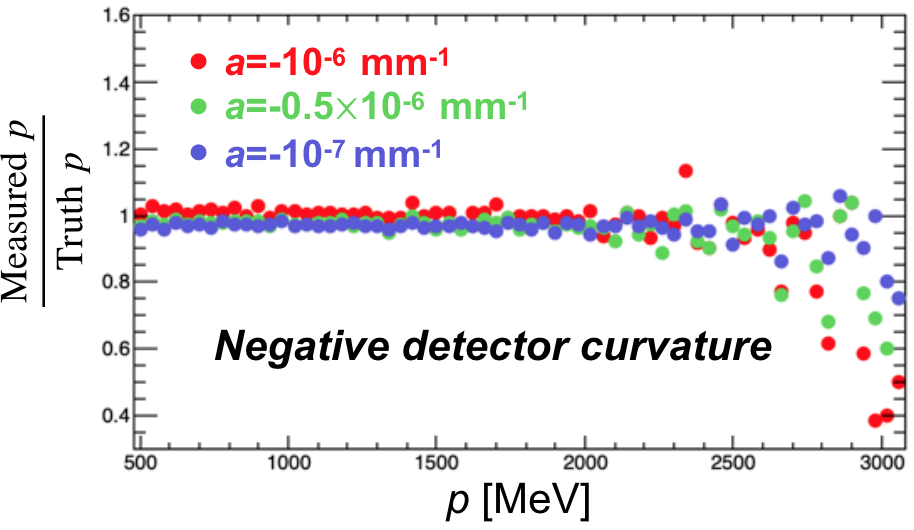}}\\
\subfloat[]{\includegraphics[width=0.55\linewidth]{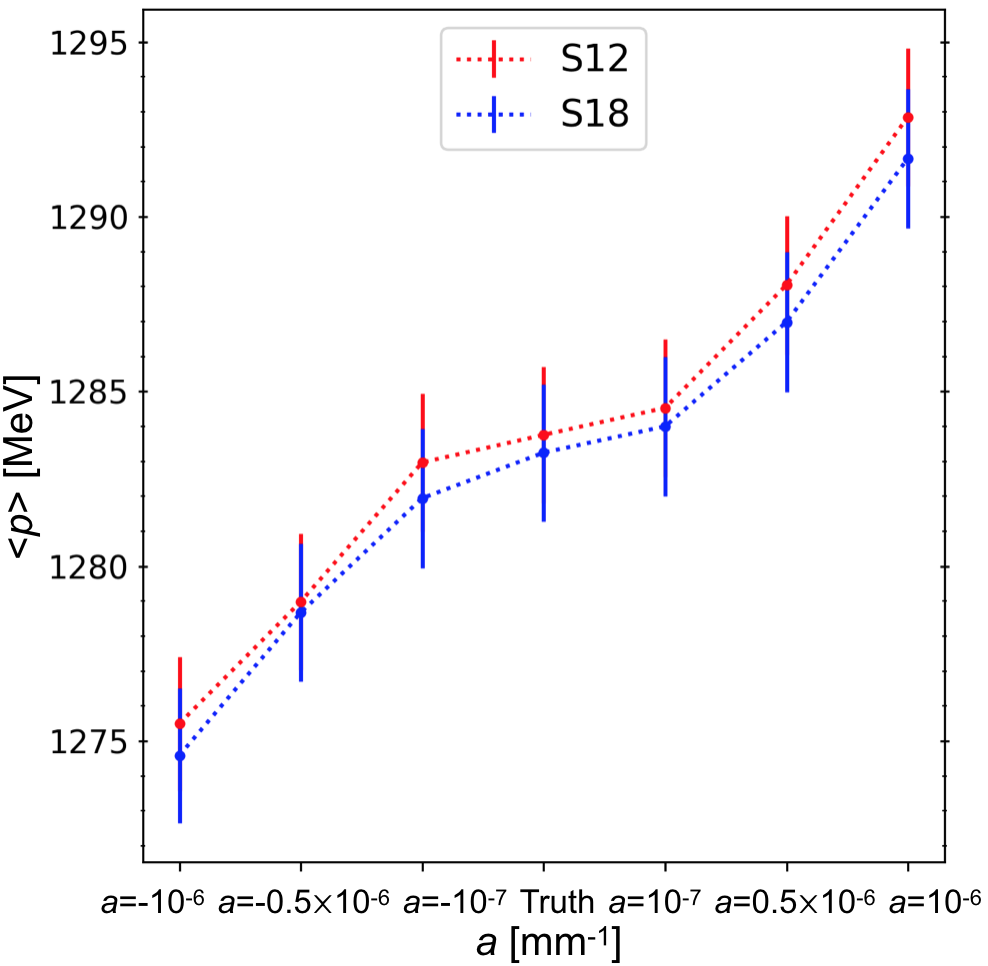}}
\caption[The radial detector curvature effect on the track momentum]{The high momentum, $p$, tracks have the largest response to the input detector curvature, $a$, as seen for the positive detector curvature in (a) and for the negative detector curvature in (b), shown here as a fraction of the measured momentum over the truth (no curvature) momentum. The overall mean momentum of the tracks changes depending on the applied curvature as seen in (c).}
\label{fig:MomDep}
\end{figure}

This potential radial detector curvature, $a$ (see \cref{eq:curve}), needs to be measured independently of the internal alignment, analogous to the global alignment measurements, described in \cref{sec:align_global}. One property that makes this measurement possible is the momentum-dependence, $p$, of the extrapolated tracks with respect to the input detector curvature, as shown in \cref{fig:MomDep}.

\begin{figure}[htpb]
    \centering
    \includegraphics[width = 0.7\linewidth]{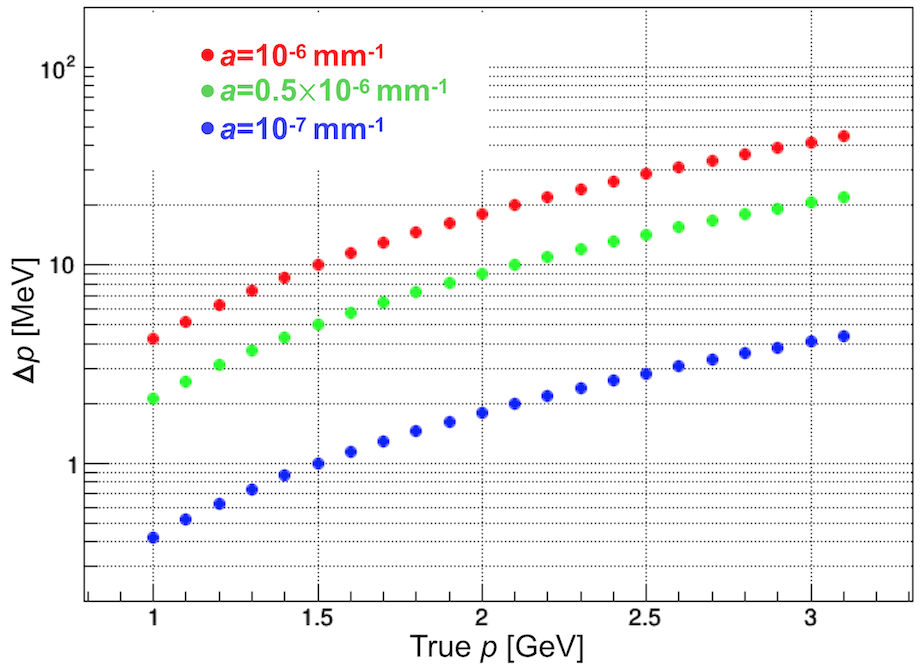}
    \caption[The momentum change caused by a change in the curvature]{The momentum change ($\Delta p$) caused by a change in the detector curvature, $a$, as a function of momentum.}
    \label{fig:MomDep2}
\end{figure}

\clearpage
A standalone toy-model for this momentum-dependence of the curvature as a function of momentum is shown in \cref{fig:MomDep2}.

The previously seen effect is confirmed: larger curvatures have a larger impact on the track momentum, and high momentum tracks have the largest response to a given curvature. To further study this effect, laser-survey data (see \cref{sec:align_global}) of the modules was used to define a reasonable curvature range of the modules.

\begin{figure}[htpb]
    \centering
    \includegraphics[width = 0.7\linewidth]{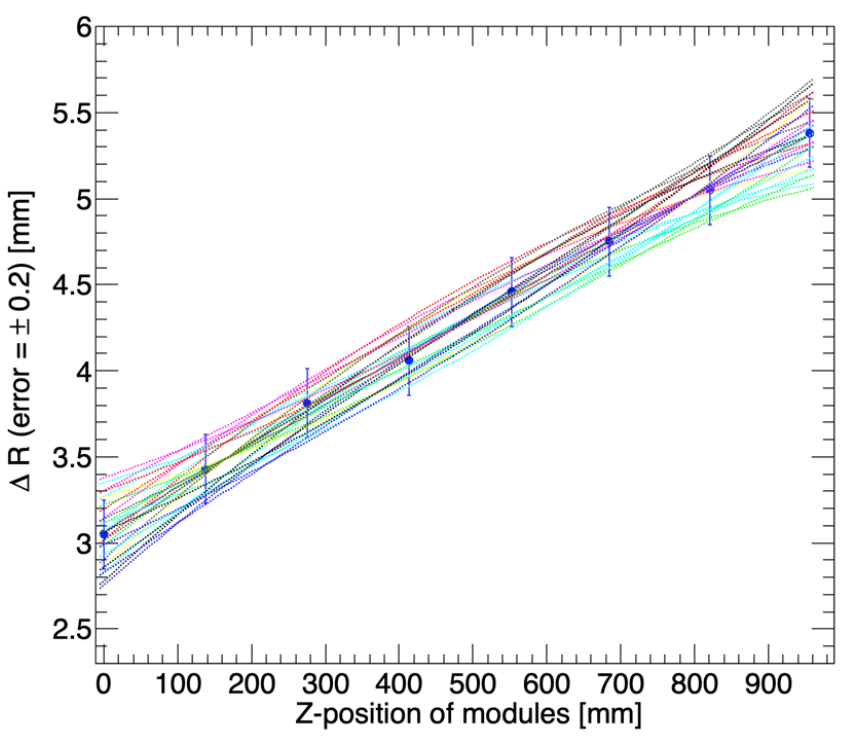}
    \caption[The 27 Mahalanobis fits to the survey measurements]{The 27 Mahalanobis fits to the survey measurements of station 18. $\Delta R$ is defined as the difference between the assumed module position and the position from the survey. The range of these fits encompasses $\pm1\sigma$ around the nominal curvature.}
    \label{fig:Mah1}
\end{figure}

\begin{figure}[htpb]
    \centering
    \includegraphics[width = 0.6\linewidth]{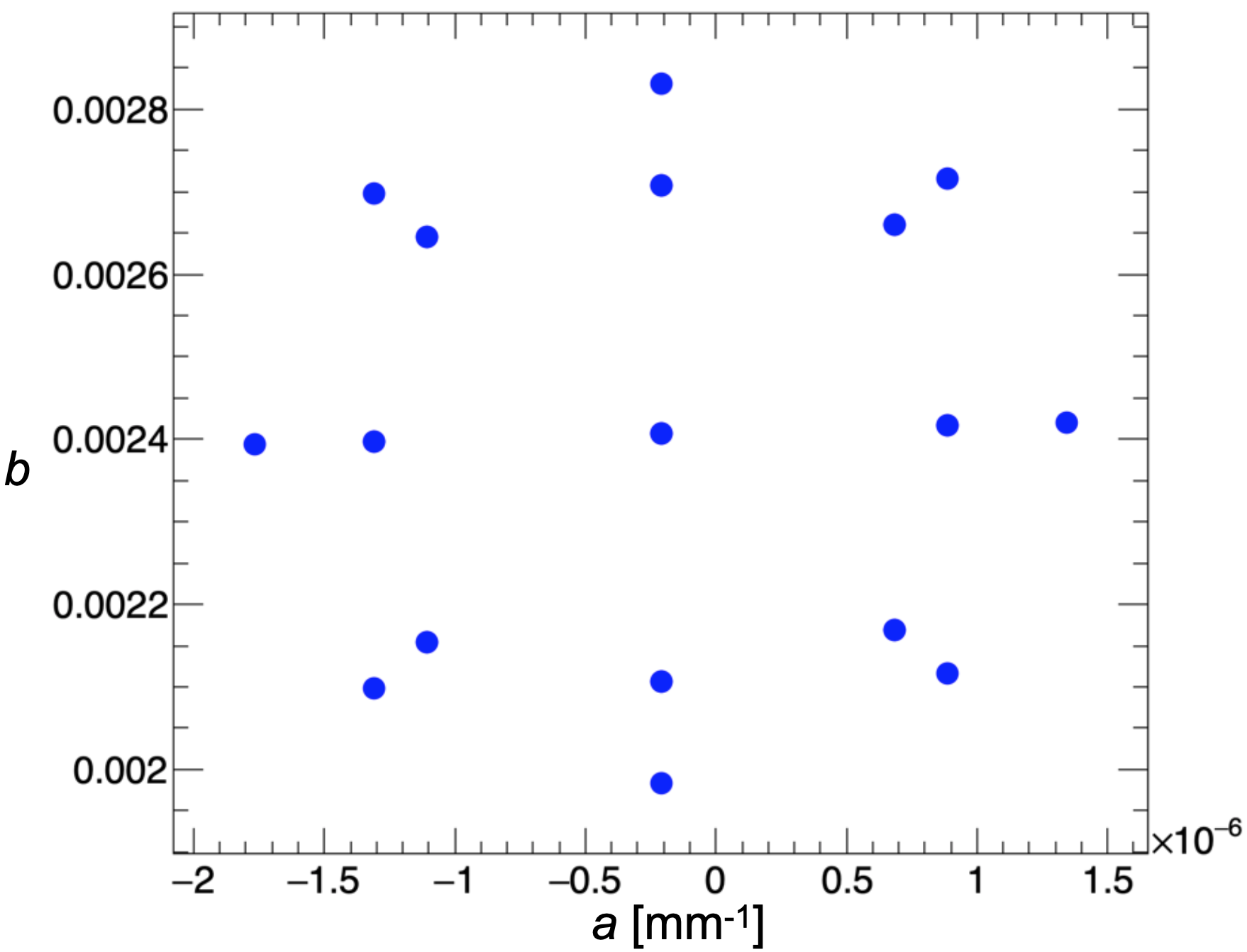}
    \caption[The 27 parameters from the Mahalanobis fits]{The 27 parameters from the Mahalanobis 1$\sigma$ band fits and the associated curvature ($a$) values; $b$ was defined in \cref{eq:curve} as the rotation parameter.}
    \label{fig:Mah2}
\end{figure}

The laser-survey alignment has an error of \SI{200}{\micro\metre} on an individual module position. This error can be used to fit a nominal (best fit) curve to the module positions. The curvature parameter ($a$) is extracted directly from a fit to \cref{eq:curve}, along with the curvature parameters that lie within a 1$\sigma$ band of the nominal curvature. These are obtained using the Mahalanobis method~\cite{Mah}. For a three parameter fit, there are 27 possible Mahalanobis fits. These fits are shown in \cref{fig:Mah1}, and the $a$ parameter for these fits are shown in \cref{fig:Mah2}.
\clearpage

\subsection{Data-simulation comparison} \label{sub:data_simulation_comparison}
\begin{figure}[htpb]
    \centering
    \includegraphics[width = 0.8\linewidth]{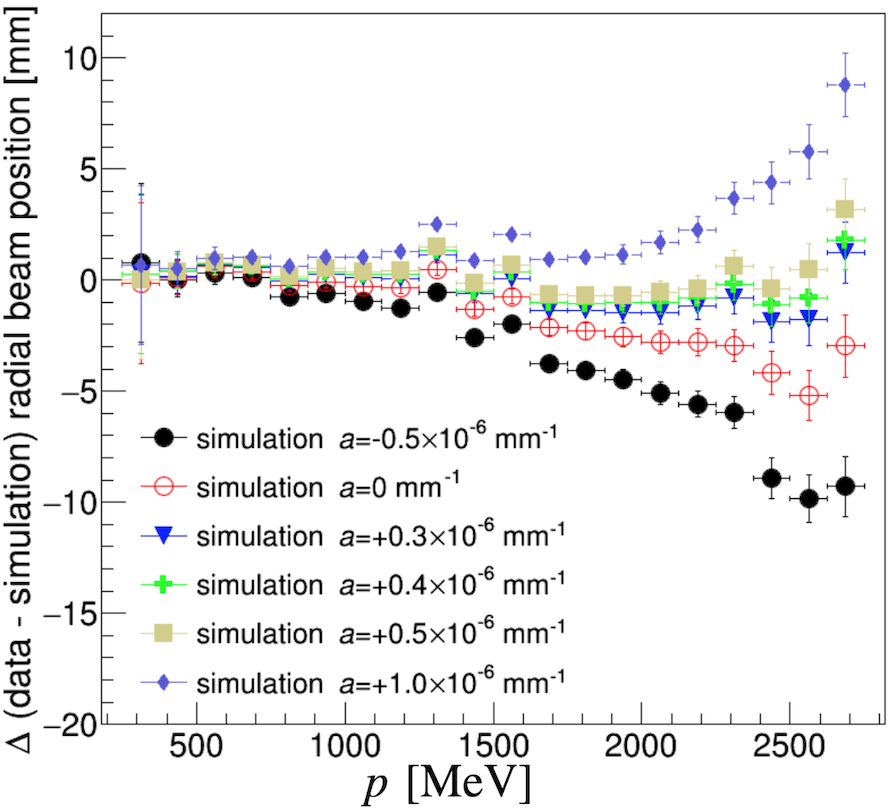}
    \caption[Extrapolated radial beam position with data]{Extrapolated radial beam position with data in station 18, compared to a simulation at various levels of detector curvature ($a$), as a function of track momentum. Negative curvature in data can be ruled out. It is also clear that the curvature in data is no larger than $+10^{-6} \ \mathrm{mm}^{-1}$.}
    \label{fig:Det1}
\end{figure}

\cref{eq:curve} and the estimated curvature parameters from \cref{fig:Mah2} were used to misalign tracker modules in a curved position in the simulation. A comparison of the various curvatures with data, where this additional curvature was not applied, was then performed. The assumption is that the simulated curvature that best matched the data is the corresponding curvature in data. The comparison criteria was the extrapolated radial beam position, as shown in \cref{fig:Det1}, for a selection of the most closely matched curvatures in station 18. The extrapolation on data was performed after internal and external alignments as described in \cref{sc:align_real} and \cref{sec:align_global}, respectively. Moreover, the simulation's beam distribution was tuned to match the data in \R1 (see \cref{sub:improved_simulation_for_r1}). 

\clearpage
\subsection{Comparison methods} \label{sub:comparison_methods}
\begin{figure}[htpb]
    \centering
    \includegraphics[width = 0.45\linewidth]{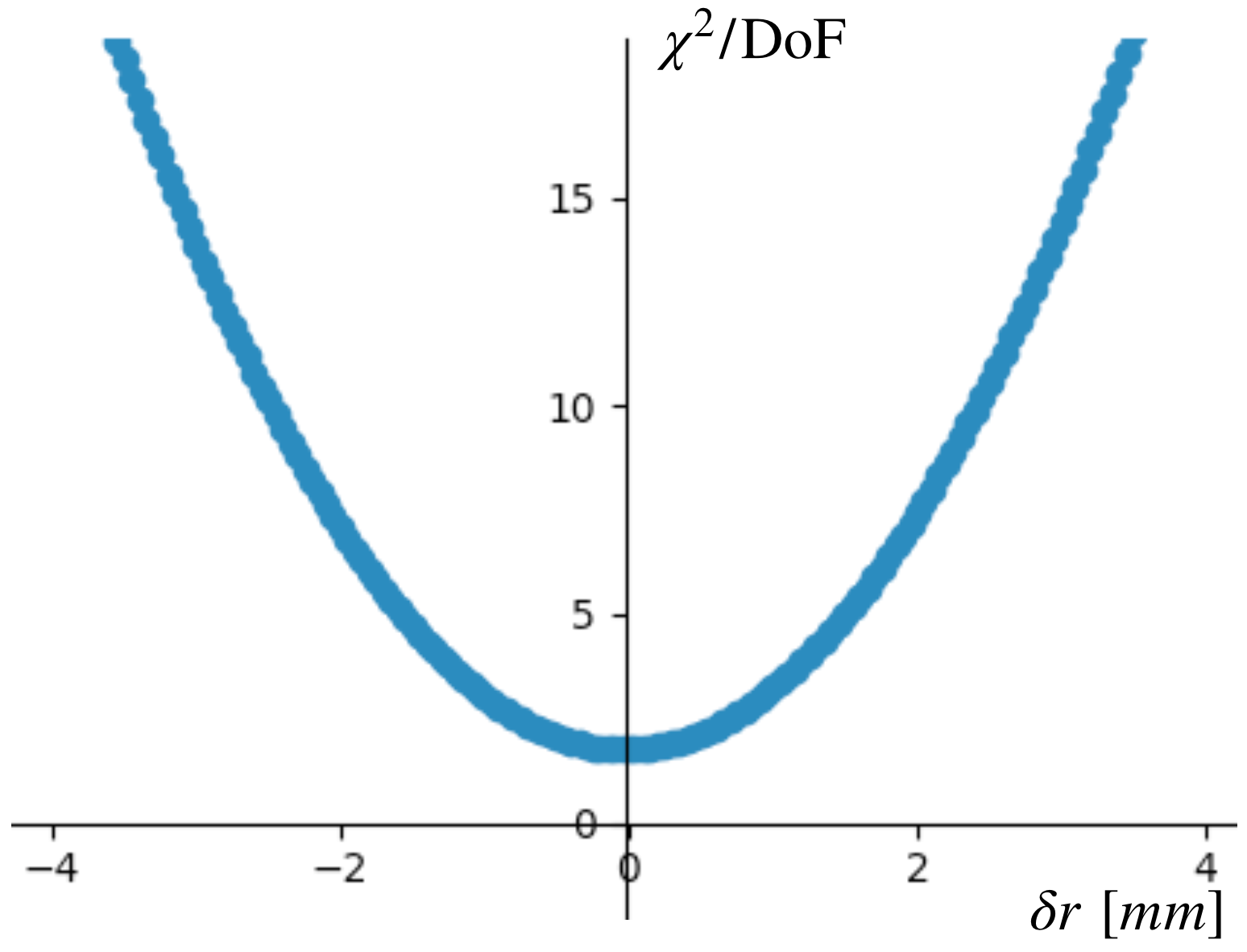}
    \caption[The $\chi^2 / \mathrm{DoF}$ change for a curvature of $a=0.35\times10^{-6} \ \mathrm{mm}^{-1}$]{The $\chi^2 / \mathrm{DoF}$ change for a curvature of $a=0.35\times10^{-6} \ \mathrm{mm}^{-1}$ as a function of the radial shift ($\delta r$).}
    \label{fig:Det2_1}
\end{figure}

The results in \cref{fig:Det1} can now be compared using the $\chi^2$ method. This is done by calculating the $\chi^2$ per momentum bin for a given simulation curvature 
\begin{equation}
   \chi^2/\mathrm{DoF} = \frac{1}{\mathrm{N_{bins}}}\sum_{i}^{\mathrm{N_{bins}}} \frac{(\mathrm{data_i-simulation_i}+\delta r)^2}{(\Delta \mathrm{data_i})^2+(\Delta \mathrm{simulation_i})^2},
\end{equation} 
as the function of $\delta r$, the amount of radial shift (for all momentum bins) needed in simulation to align the extrapolated curve with data, such that
\begin{equation}
    \frac{\langle\chi^2\rangle}{\delta r} = 0.
\end{equation}
\vspace{-0.25cm}
The assumption is that the simulation result with the smallest corresponding $\chi^2/\mathrm{DoF}$ is the best representation of the curvature in the data. An example scan over $\delta r$ for a curvature of $a=0.35\times10^{-6} \ \mathrm{mm}^{-1}$ is shown in \cref{fig:Det2_1}.
 \vspace{-0.25cm}
 
\begin{figure}[htpb]
    \centering
    \subfloat[]{\includegraphics[width = 0.64\linewidth]{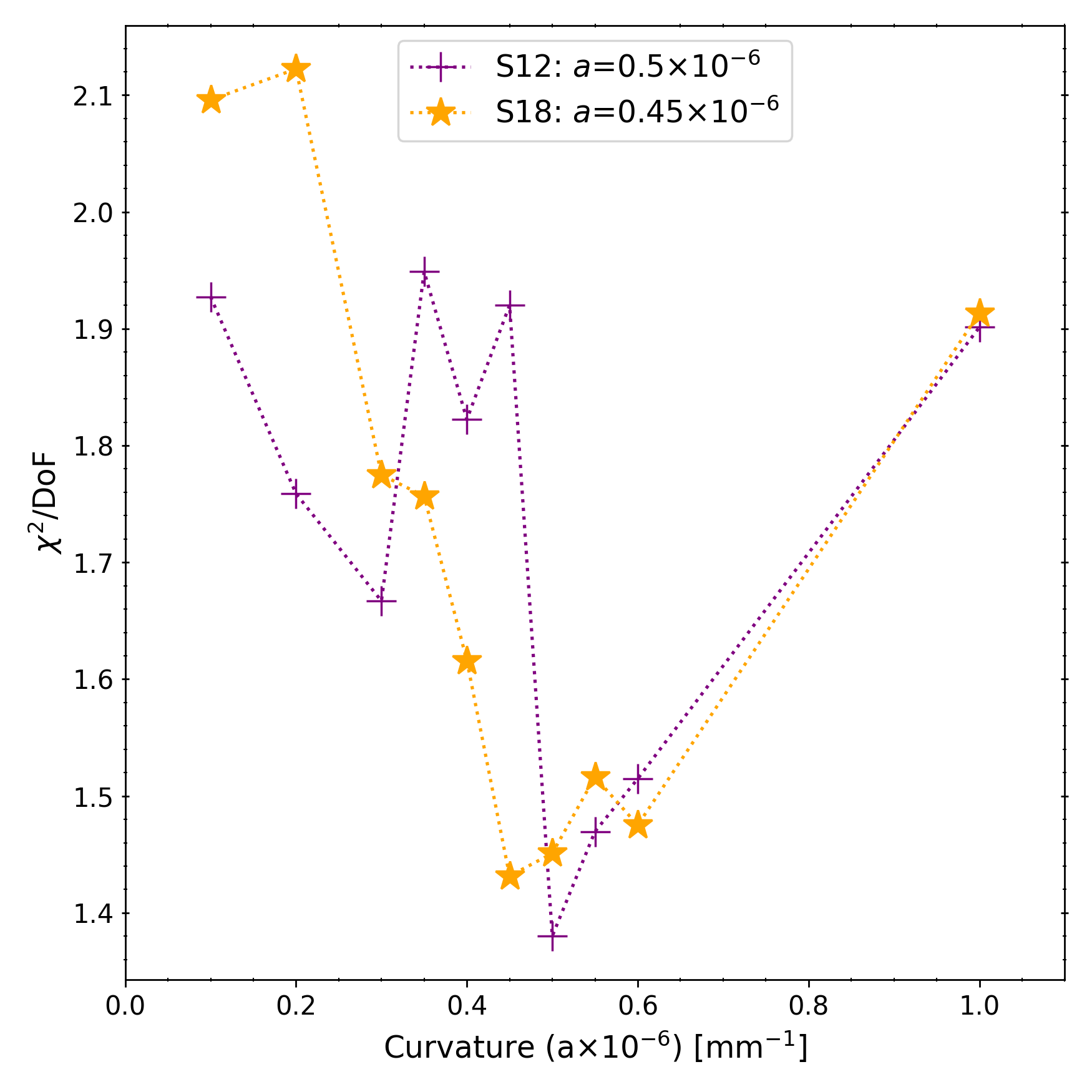} \label{fig:Det2}} \\
    \subfloat[]{\includegraphics[width = 0.64\linewidth]{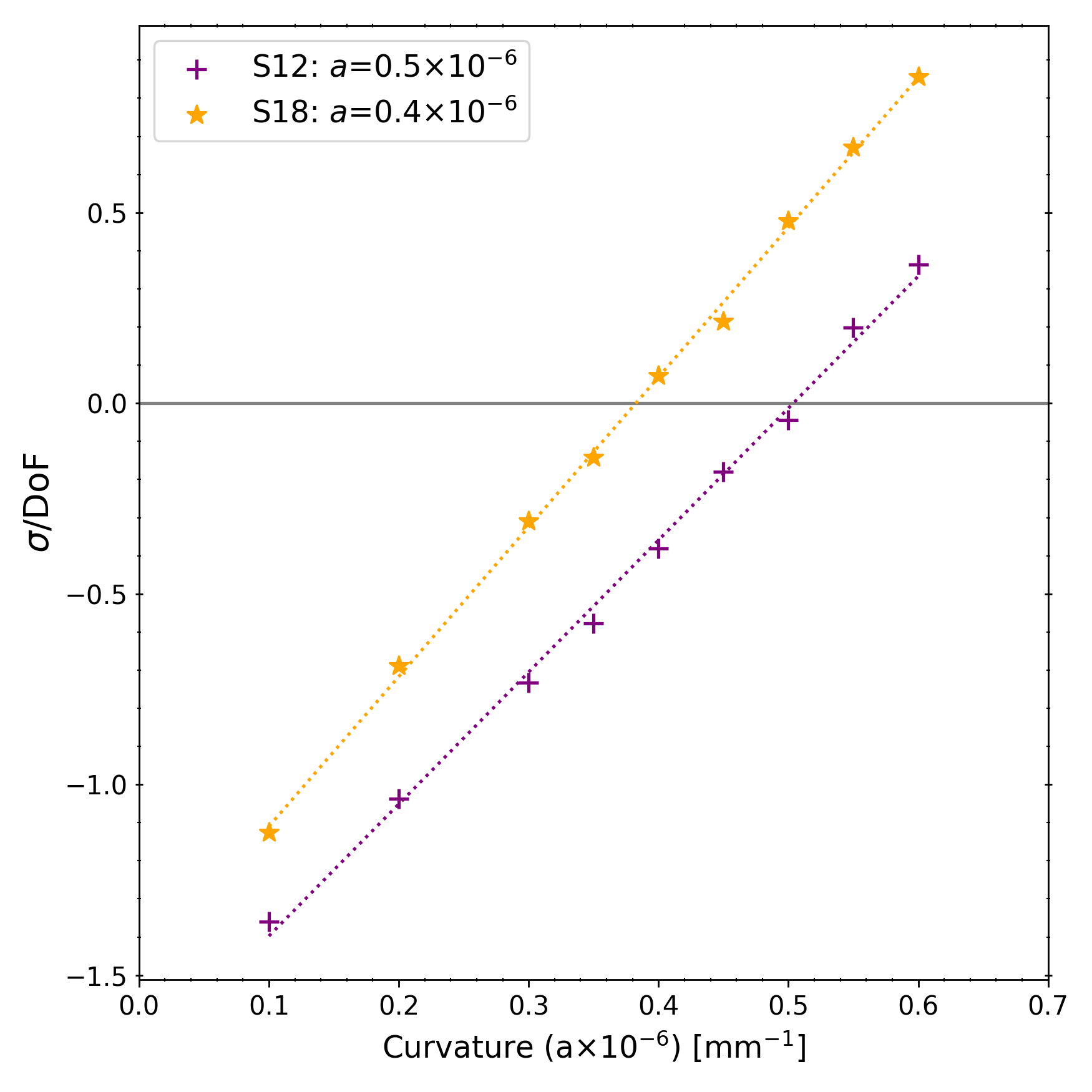}\label{fig:Det3}}
    \vspace{-0.2cm}
    \caption[The results of the two comparison methods]{The results of the two comparison methods. (a) The $\chi^2$ method. The most likely curvature in data, as compared to simulation, has the smallest value of $\chi^2 / \mathrm{DoF}$: $a=0.5\times10^{-6} \ \mathrm{mm}^{-1}$ for station 12, and $a=0.45\times10^{-6} \ \mathrm{mm}^{-1}$ for station 18.
    (b) The residual method. The most likely curvature in data is represented by the simulation with the lowest $|\sigma/\mathrm{DoF}|$: $a=0.5\times10^{-6} \ \mathrm{mm}^{-1}$ for station 12, and $a=0.4\times10^{-6} \ \mathrm{mm}^{-1}$ for station 18.}
\end{figure}

\vspace{-0.25cm}
The final results of the $\chi^2$ method are summarised in \cref{fig:Det2}. Alternatively, as a cross-check, a residual method can be used. This method calculates $\sigma$ for each momentum bin, and extracts the lowest absolute value of $\sigma/\mathrm{DoF}$ over many curvatures from
 \vspace{-0.20cm}
\begin{equation}
    \sigma/\mathrm{DoF} = \frac{1}{\mathrm{N_{bins}}}\sum_{i}^{\mathrm{N_{bins}}}\frac{\mathrm{data_i}-\mathrm{simulation_i}}{\sqrt{(\Delta \mathrm{data_i})^2+(\Delta \mathrm{simulation_i})^2}}.
\end{equation}
\vspace{-0.2cm}
The results from the residual method are summarised in \cref{fig:Det3}.
\clearpage
\clearpage
\subsection{Extrapolation with detector curvatures}
\begin{figure}[htpb]
    \centering
    \includegraphics[width = \linewidth]{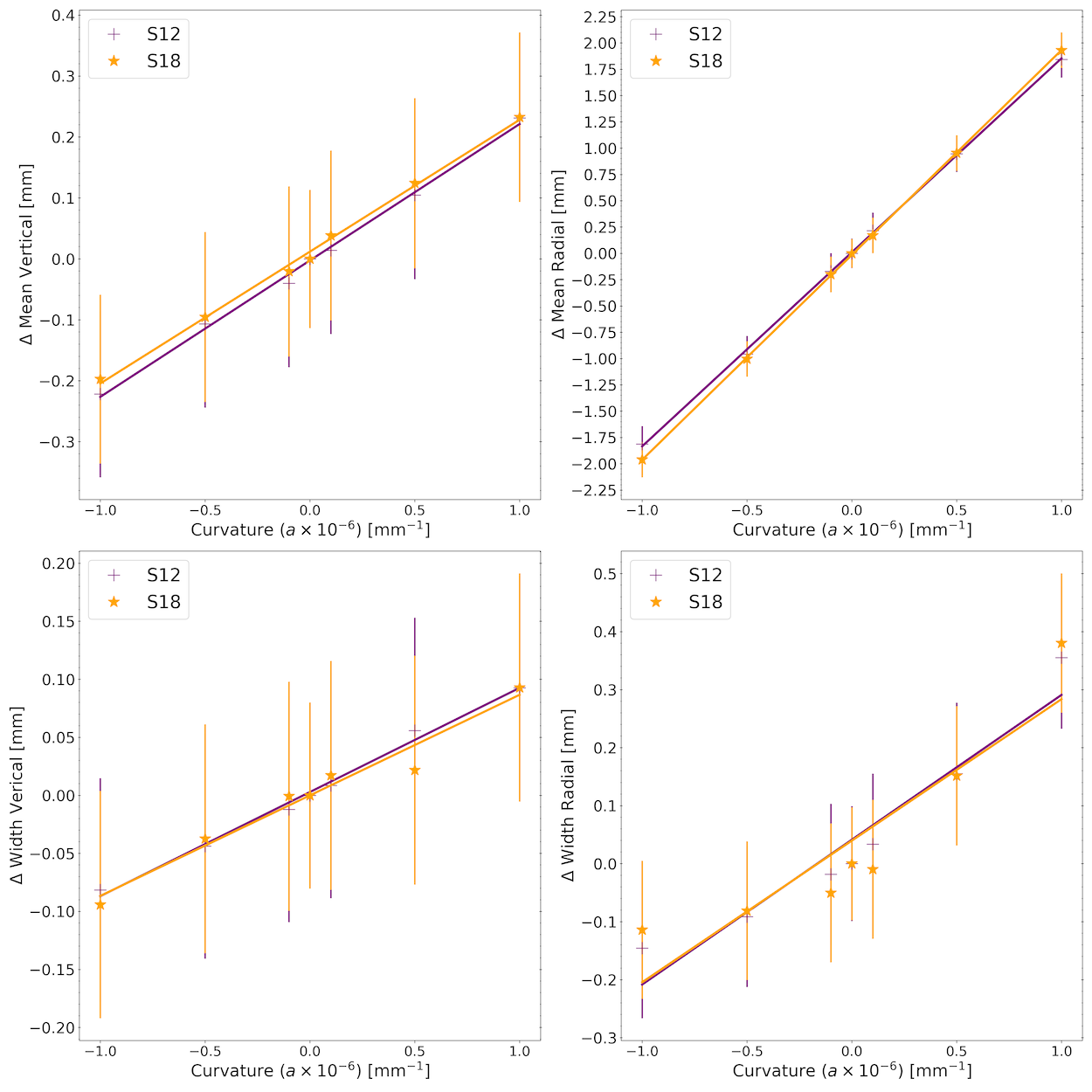}
    \caption[Beam extrapolation with applied curvature in simulation]{The difference in the extrapolated mean and width of the radial and vertical beam position for six values of curvature and the nominal value in the simulation. Due to the tracker UV straw geometry (see \cref{sc:cs}), a change in the radial position also induces a change in the vertical position.}
    \label{fig:SimExtMom}
\end{figure}

To examine the effect of these curvatures, $\sim4\times10^5$ tracks were re-fitted with a curvature offset and the tracks were extrapolated back to determine a beam position. This shift of this beam position, both radially and vertically, from the nominal, is shown in \cref{fig:SimExtMom}.

\begin{figure}[htpb]
    \centering
    \includegraphics[width = \linewidth]{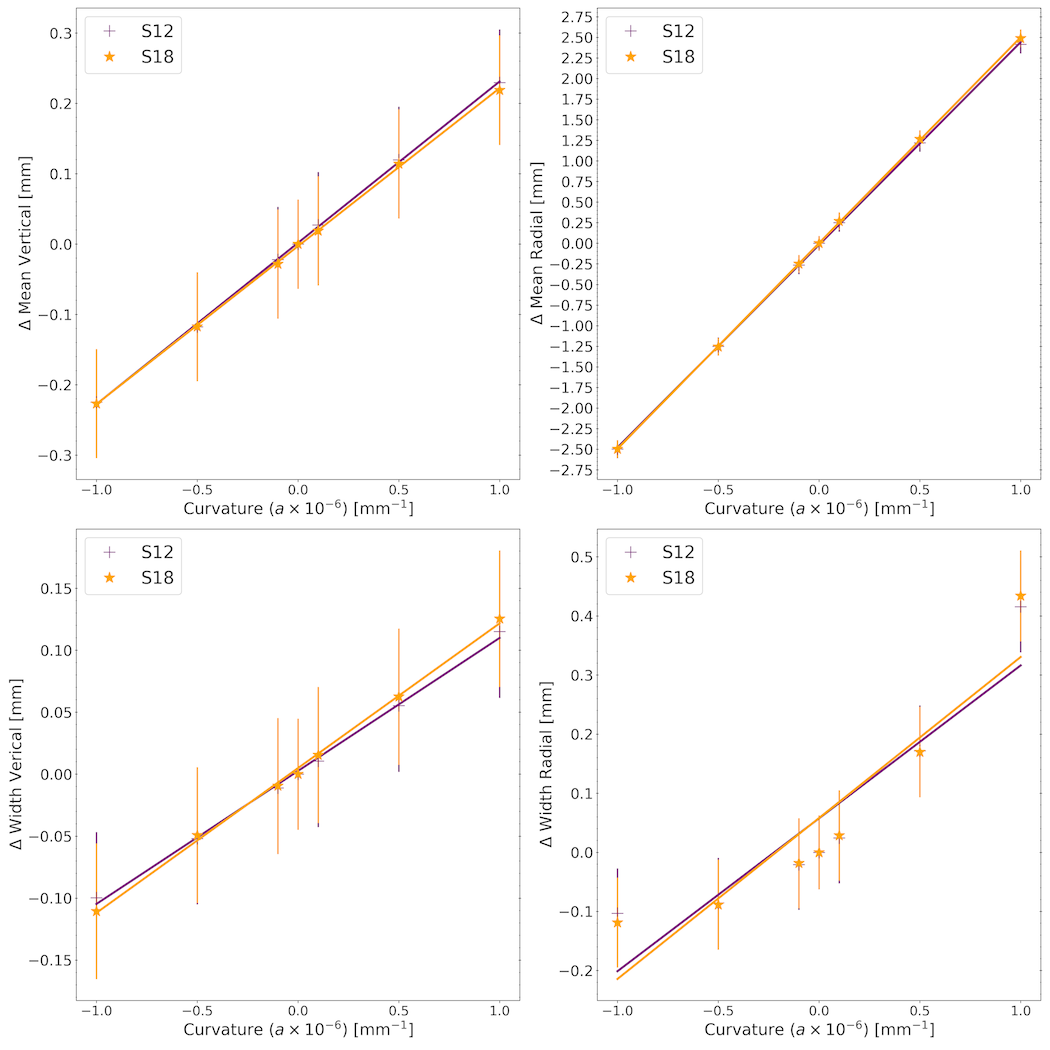}
    \caption[Beam extrapolation with applied curvature in data]{The difference in the extrapolated mean and width of the radial and vertical beam position for six values of curvature and the nominal value in the data.}
    \label{fig:ExtMom}
\end{figure}

\clearpage
Finally, a similar scale of curvatures can be applied to data, as shown in \cref{fig:ExtMom}. The uncertainty coming from the detector curvature estimated by comparison with simulation can now be directly extracted from these plots.
\clearpage

\section{Results and outlook}
The beam extrapolation determines the radial and vertical positions of the muon beam in the storage ring. The accuracy of this determination is affected by detector effects, such as the radial detector curvature. To estimate the contribution of the detector curvature to beam extrapolation, the largest of the values from the two methods in \cref{sub:comparison_methods} was used, for a conservative estimate. This value is $a=0.5\times10^{-6} \ \mathrm{mm}^{-1}$, and the corresponding beam extrapolation uncertainties, extracted from \cref{fig:ExtMom}, are given in \cref{tab:align_curvature_final}.
\begin{table}[htpb]  
  \centering
  \begin{tabular}{lrrrr}
    \toprule
            & $\mathrm{\sigma_{dR}}$ [\SI{}{\milli\meter}] & $\mathrm{\sigma_{dR_{\mathrm{width}}}}$ [\SI{}{\milli\meter}] & $\mathrm{\sigma_{dV}}$ [\SI{}{\milli\meter}] & $\mathrm{\sigma_{dV_{\mathrm{width}}}}$ [\SI{}{\milli\meter}]  \\ \midrule
    
      S12 $\oplus$ S18 & 1.25 & 0.20 & 0.10 & 0.05  \\ \bottomrule
  \end{tabular}
  \caption[The contribution of the detector curvature to the beam extrapolation uncertainty]{The contribution of the detector curvature ($a$) to the beam extrapolation uncertainty in both stations, given $a=0.5\times10^{-6} \ \mathrm{mm}^{-1}$.}
  \label{tab:align_curvature_final}
\end{table}

However, there is a motivation to improve the detector curvature estimation, given it is one of the larger tracking systematics (see \cref{sec:pitch}). Some of the possible improvements are given below:
\small
\begin{itemize} \itemsep -2pt
    \item \textbf{Improved tracking and simulation}. An improvement to the tracking algorithms will allow for a more precise limit on the curvature to be placed. Such improvements include LR assignment, hit selection, and improving track quality, and are being developed by G. Sweetmore~\cite{George}, S. Grant~\cite{Sam}, and A. Luca~\cite{Alessandra}, respectively. Moreover, an improved simulation, that is more representative of real data will also benefit the curvature study. 
    \item \textbf{Cosmic muons}. An alternative estimation of the curvature is possible with cosmic ray muons during periods when the magnetic field in the ring if off. However, there are numerous challenges associated with this study, such as maintaining long periods without the magnet on, developing a precise straight-track fitting algorithm, and estimating the residual magnetic field. 
\end{itemize}
\normalsize

\clearpage

\section{Impact on the systematic uncertainties}
\subsubsection{Pitch correction (\texorpdfstring{$\omega_a$}~)}
Using \cref{eq:pitch}, and an upper bound on $a=0.5 \times 10^{-6} \ \mathrm{mm}^{-1}$ from \cref{sub:comparison_methods}, with the corresponding $\mathrm{\sigma_{dV_{\mathrm{width}}}}=0.05$~mm (see \cref{tab:align_curvature_final}), yields $\Delta C_{\mathrm{pitch}} = 0.7$~ppb. 

\subsubsection{Field convolution (\texorpdfstring{$\omega_p$}~)}
A study of an impact of translations on the extrapolated beam profile from the tracking detectors on the field measurements was performed by Jason Bono and Saskia Charity~\cite{JasonSaskia}. Given the scale of translations induced by the curvature of $a=0.5\times10^{-6} \ \mathrm{mm}^{-1}$, of 1.25 mm and 0.1 mm radially and vertically, respectively (see \cref{tab:align_curvature_final}). It is possible to estimate the corresponding uncertainties on $\langle B \rangle$, and hence $\omega_p$ (see \cref{eq:omega_p}). The results for both stations are given in \cref{tab:curve_align}, with the average vertical and radial uncertainties in both stations on $\langle B \rangle$ of 12.9 ppb and 0.6 ppb, respectively.
\begin{table}[htpb]  
  \centering
  \begin{tabular}{rrrr}
    \toprule
            S12 radial   & S18 radial & S12 vertical & S18 vertical \\ \midrule
    
      11.6 ppb & 14.2 ppb & 0.6 ppb & 0.6 ppb  \\ \bottomrule
  \end{tabular}
  \caption[The detector curvature contribution to the uncertainty on $\langle B \rangle$]{The detector curvature contribution to the uncertainty on $\langle B \rangle$.}
  \label{tab:curve_align}
\end{table}

\graphicspath{{fig/}}

\chapter{A preliminary track-based \texorpdfstring{$\omega_a$} ~ analysis}\label{ch:wiggle}

This chapter presents preliminary results from the first attempt at an $\omega_a$ analysis with data from the tracking detector. This type of analysis has the advantage of having a different set of systematic uncertainties (e.g. pile-up, \say{lost-muons}), as compared to the calorimeter-based analysis. The available statistics in the tracker-based analysis is significantly lower. However, given the previous $\omega_a$ measurement's discrepancy with the theory of $2.4$~ppm (see \cref{sc:mma}) and if the tracking detector measures the same value as the BNL, then to exclude the theoretical prediction at $95\%$~CL requires a measurement with a precision of $1.4$~ppm. In this chapter, an estimate is made of when this threshold will be reached with the acquisition of more data, based on the preliminary analysis of \R1 data. 

\section{Analysis overview}
The analysis work in this chapter follows closely the methodology of the $\omega_a$ analysis using calorimeter data~\cite{Aaron,David,Nick}. The first statistical assessment of \R1 data, applying time cuts (see~\cref{sc:scraping}) and momentum cuts (see~\cref{sub:momentum_cuts_wiggle}), is summarised in~\cref{tab:DS_track}. The number of tracks (above the $p$ threshold) is $\sim 130$ times less than the number of $e^+$ (above the $E$ threshold) recorded by the calorimeters (c.f. \cref{tab:DS} with \cref{tab:DS_track}).

\clearpage
\begin{table}[htpb]  
  \centering
  \small
  \begin{tabular}{ccccccc}
    \toprule
     Dataset & All quality & Quality tracks & Quality  tracks\\
     name & tracks &  ($t>\SI{30}{\micro\second}$) & ($t>\SI{30}{\micro\second}$, $p>1.8$~GeV) \\ \toprule
    1a & $3.51\times10^7$ &  $2.27\times10^7$ & $7.25\times10^6$ \\ \midrule
    1b & $4.83\times10^7$  &  $3.12\times10^7$ &  $9.93\times10^6$ \\ \midrule
    1c & $7.23\times10^7$ &   $4.60\times10^7$  &  $1.46\times10^7$  \\ \midrule
    1d & $1.40\times10^8$ &  $8.63\times10^7$  &  $2.74\times10^7$   \\ \toprule \toprule
    Total & $2.96\times10^8$ & $1.86\times10^8$ & $5.92\times10^7$  \\ \bottomrule 
  \end{tabular}
  \normalsize
  \caption[\R1 datasets for track-based analysis]{For each dataset in \R1, the total number of quality tracks (see \cref{sc:track_quality}), as well as the number of tracks above certain time and momentum cuts, are shown. Applying the time cut gives an estimate of the number of useful tracks for the EDM analysis, while the additional momentum cut gives an estimate of the number for the $\omega_a$ analysis.}
  \label{tab:DS_track}
\end{table}

\vspace{-0.7cm}
\subsection{Five-parameter fit}
\begin{figure}[htpb]
    \centering
    \includegraphics[width=0.75\linewidth]{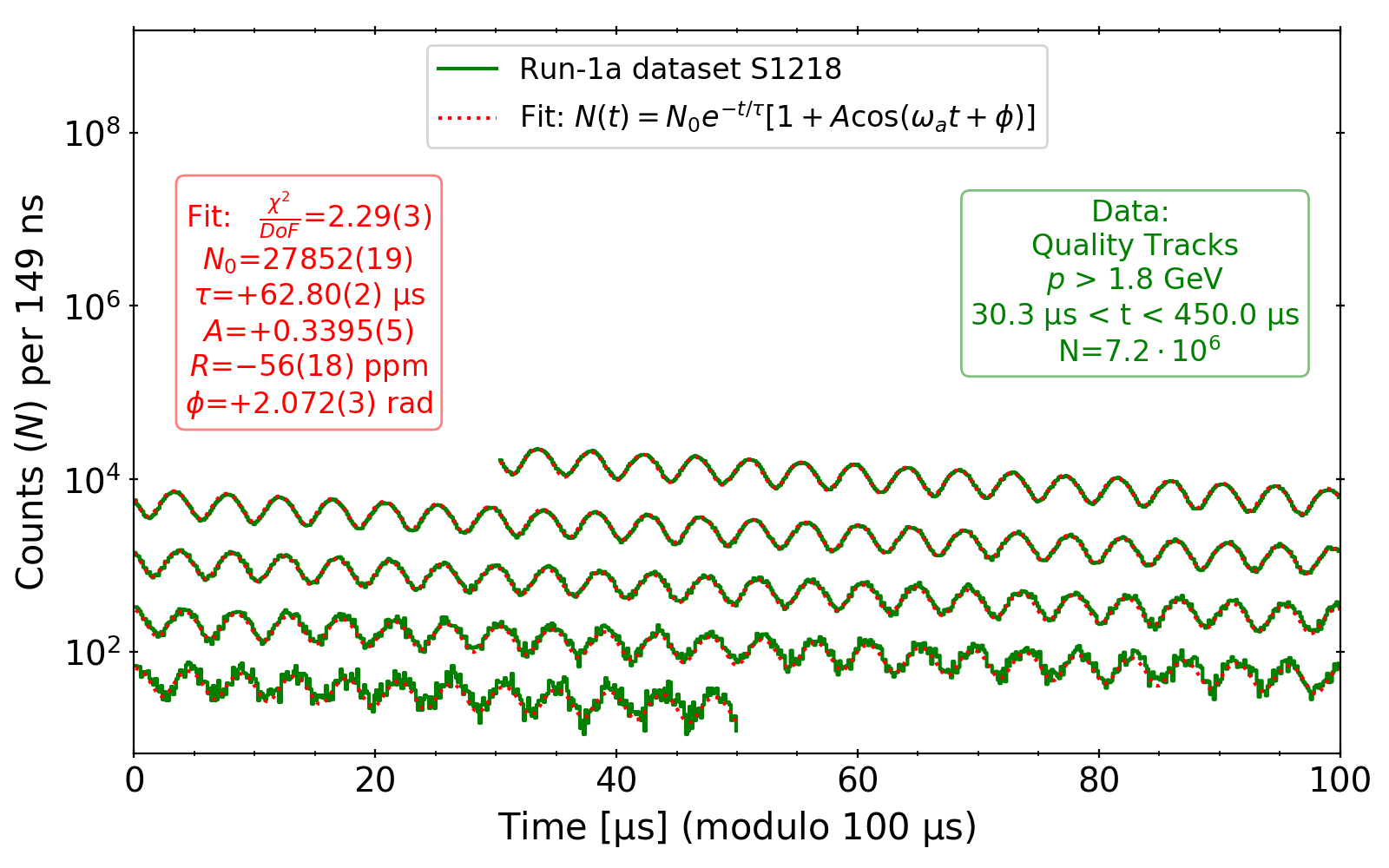} 
    \vspace{-0.2cm}
    \caption[Five-parameter wiggle fit]{Five-parameter fit to data from the Run-1a dataset in both tracker stations (S12 and S18).}
    \label{fig:S1218_5par}
\end{figure}

The first step in the analysis was a simple five-parameter fit to the number of quality tracks as a function of time 
\vspace{-0.1cm}
\begin{equation}
    N(t)=N_{0}e^{-t/\tau}[1+A\cos(\omega_at+\phi)],
    \label{eq:5_par}
\end{equation}
where $N_0$ is the overall normalisation, $\tau$ is the time-dilated muon lifetime, $A$ is the asymmetry, and $\phi$ is the phase. The value of $\omega_a$ returned from the fit is blinded in software and hardware (see \cref{sec:blind}), so that a value of $R$ (see \cref{eq:blind}) is reported instead. The result of the fitting procedure is shown in~\cref{fig:S1218_5par}. Such a plot is known colloquially  as a \say{wiggle plot}. The choice of the width of a bin in this histogram is $149.2$~ns, the cyclotron period. It is important to note that the start-time of the fit should be at the edge of a bin; otherwise, the fit quality is negatively affected.

\begin{figure}[htpb]
    \centering
    \subfloat[]{\raisebox{2mm}{\includegraphics[width=0.5\linewidth]{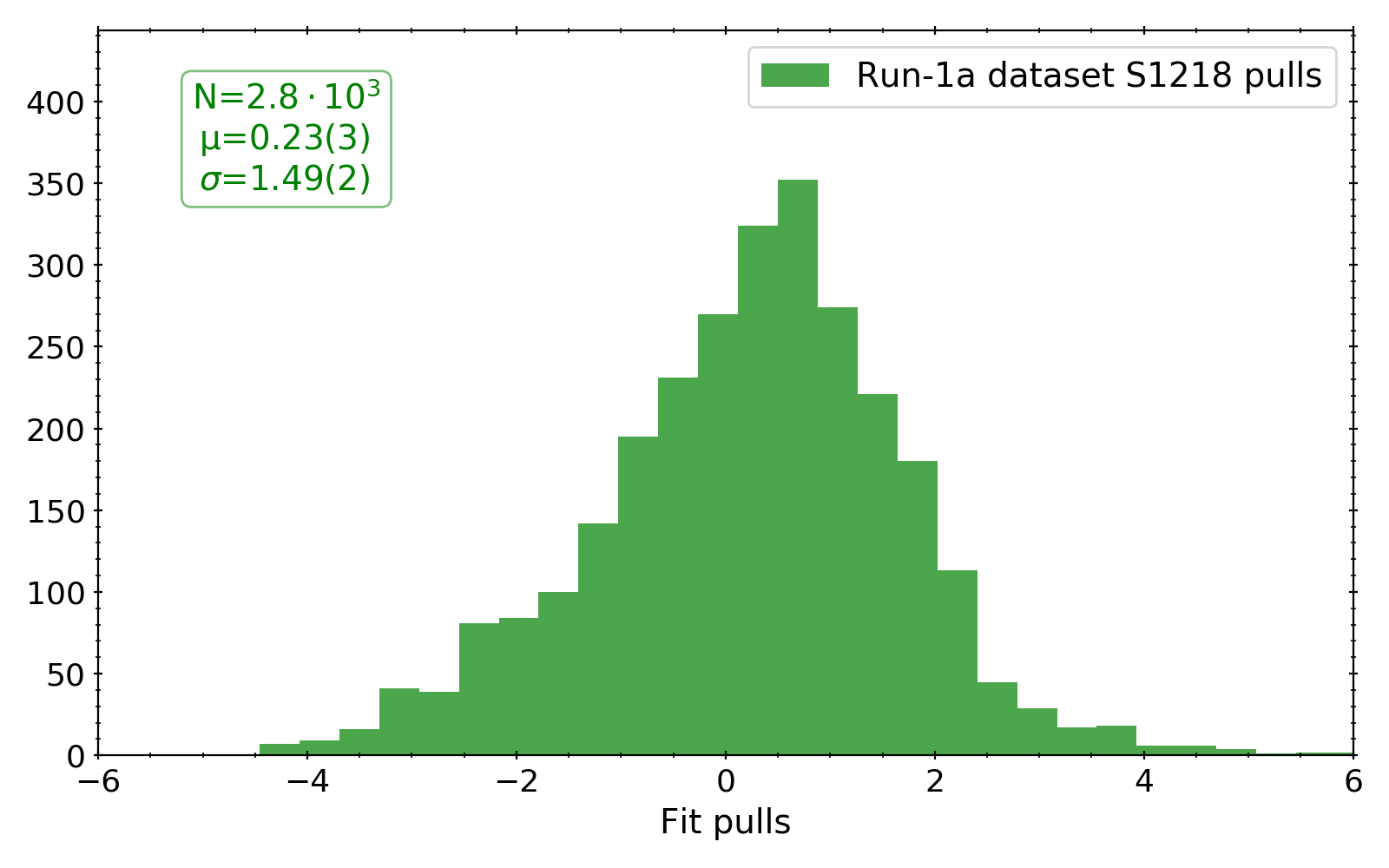} \label{fig:S1218_5par_pull}}}
    \subfloat[]{\includegraphics[width=0.49\linewidth]{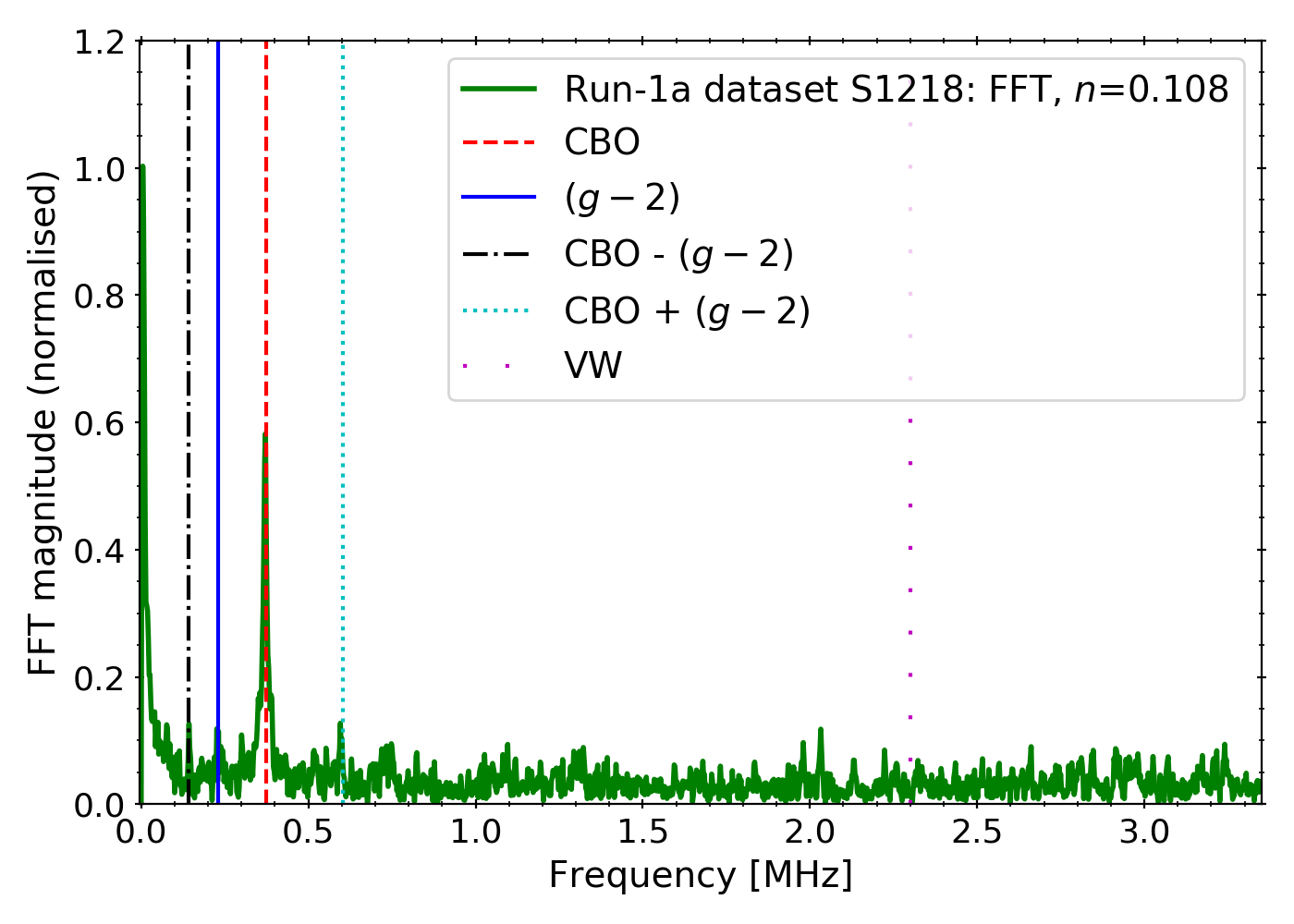} \label{fig:S1218_5par_fft}}
    \vspace{-0.2cm}
    \caption[Five-parameter residuals and FFT]{Five-parameter fit (a) pulls and (b) FFT from the Run-1a dataset.}
    \label{fig:S18_5par_res_fft}
\end{figure}

The quality of the above fit can be assessed in terms of the distribution of the fit pulls (i.e. normalised residual between the fit function in a given time bin and the data), as shown in~\cref{fig:S1218_5par_pull}.

\clearpage

\vspace{-0.8cm}

\vspace{-0.2cm} 
The non-centred distribution of fit pulls with $\sigma$ significantly larger than $1$ suggests that the chosen function doesn't describe the data well. This can be further verified by transforming the residuals from the time to the frequency domain, using a \ac{FFT}~\cite{FFT}. A large peak at $0$~MHz is observed, which is likely due to effects from lost-muons or pile-up. Moreover, a peak at the \ac{CBO} frequency (see \cref{sec:CBO}) is also seen, which also couples to the \gm2 frequency,~$f_a$,
\begin{equation}
    f_a = \frac{2\pi}{\omega_a} \approx 0.23 \ \mathrm{MHz}.
\end{equation}
The expected frequencies of the peaks in~\cref{fig:S1218_5par_fft} depend on the field-index, $n$ (see \cref{sc:scraping} and \cref{tab:DS}.).

\clearpage

\subsection{Nine-parameter fit}
The presence of the large \ac{FFT} peak at the \ac{CBO} frequency motivates incorporating additional terms in the fit function, which describe the impact of the CBO on the number oscillation, such that 
\begin{equation}
        N(t)=N_0e^{-t/\tau}[1+A\cos(\omega_at+\phi)]\cdot C(t),
\end{equation}
where $C(t)$ is the CBO function~\cite{Joe_CBO} given by
\begin{equation}
     C(t) = 1.0 + e^{-t / \mathrm{T_{CBO}}}\mathrm{A_{CBO}} \cos(\omega_{\mathrm{CBO}} t + \phi_{\mathrm{CBO}}).
     \label{eq:cbo_fit}
\end{equation}

\begin{figure}[htpb]
    \centering
    \subfloat[]{\raisebox{3mm}{\includegraphics[width=0.5\linewidth]{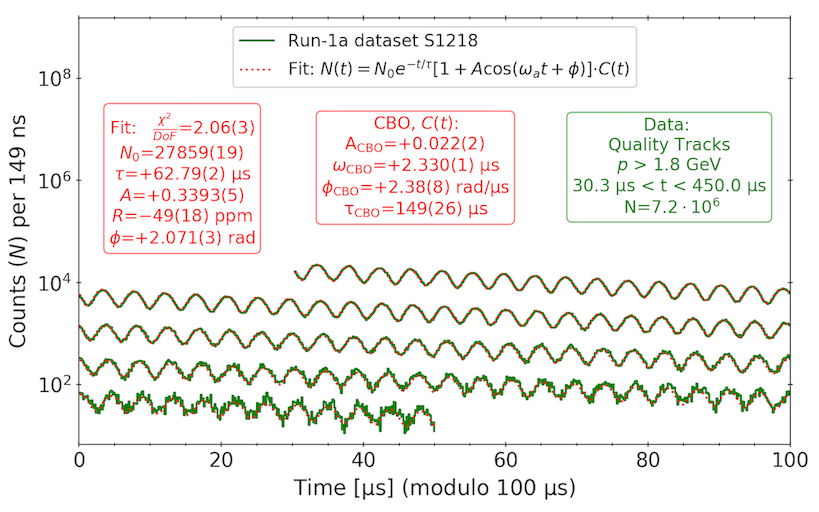} \label{fig:S1218_9par_fit}}}
    \subfloat[]{\includegraphics[width=0.48\linewidth]{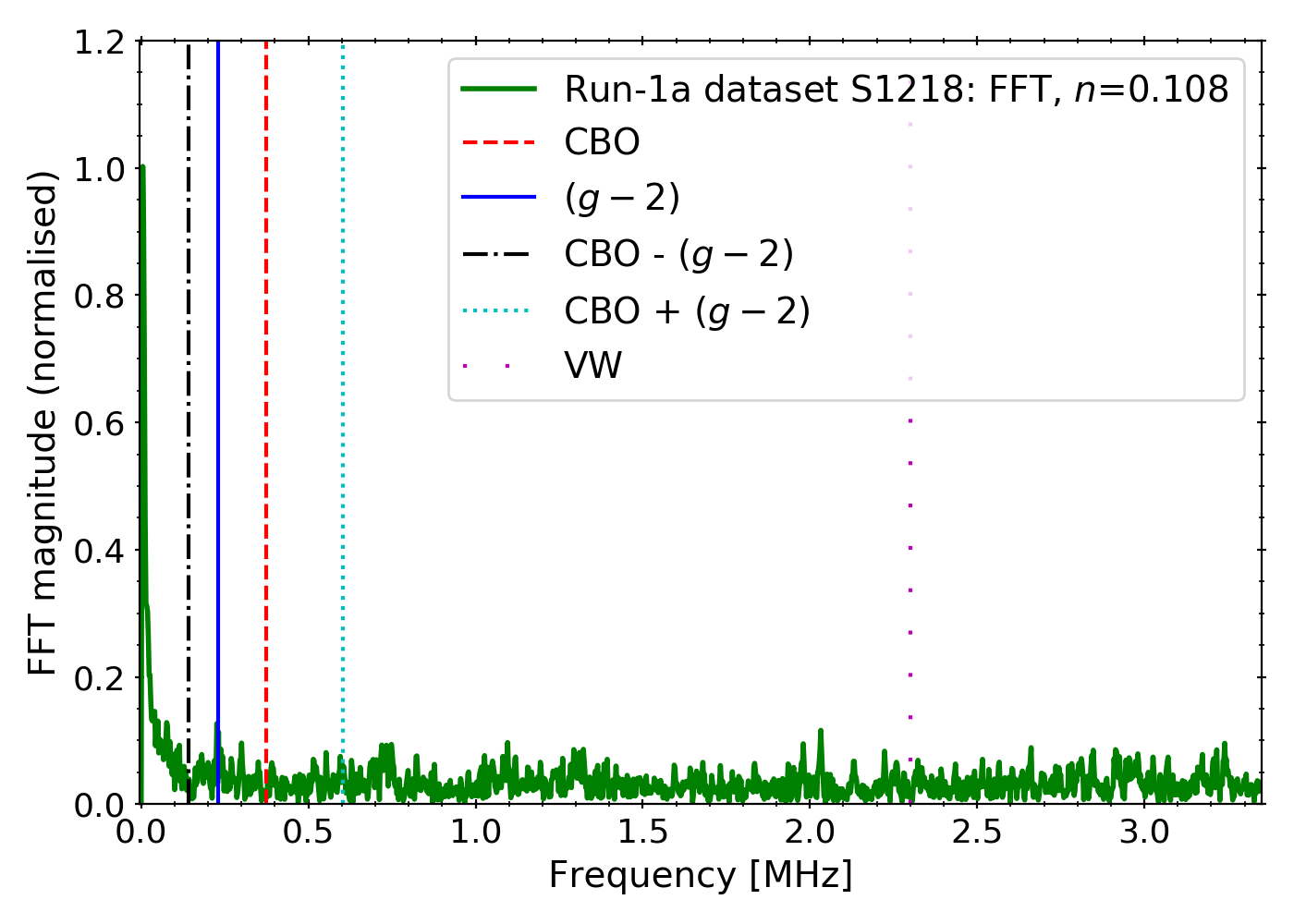} \label{fig:S1218_9par_fft}} 
    \vspace{-0.2cm}
    \caption[Nine-parameter fit and FFT]{Nine-parameter (a) fit and (b) FFT from the Run-1a dataset.}
    \label{fig:S1218_9par}
\end{figure}

\cref{eq:cbo_fit} is analogous to the original five-parameter function, but with each term now describing the CBO parameters: lifetime ($\mathrm{T_{CBO}}$), amplitude ($\mathrm{A_{CBO}}$), angular frequency ($\omega_{\mathrm{CBO}}$), and phase ($\phi_{\mathrm{CBO}}$). This fit is shown in~\cref{fig:S1218_9par_fit}. With the CBO terms in the fit, the $\frac{\chi^2}{\mathrm{DoF}}$ is improved. To verify that the correct model was used to describe the CBO's influence on the oscillation, the \ac{FFT} analysis was repeated as shown in~\cref{fig:S1218_9par_fft}. The CBO peak is no longer present. 
\vspace{-0.3cm}

\subsection{Ten-parameter fit}
\begin{figure}[htpb]
    \centering
    \includegraphics[width=0.8\linewidth]{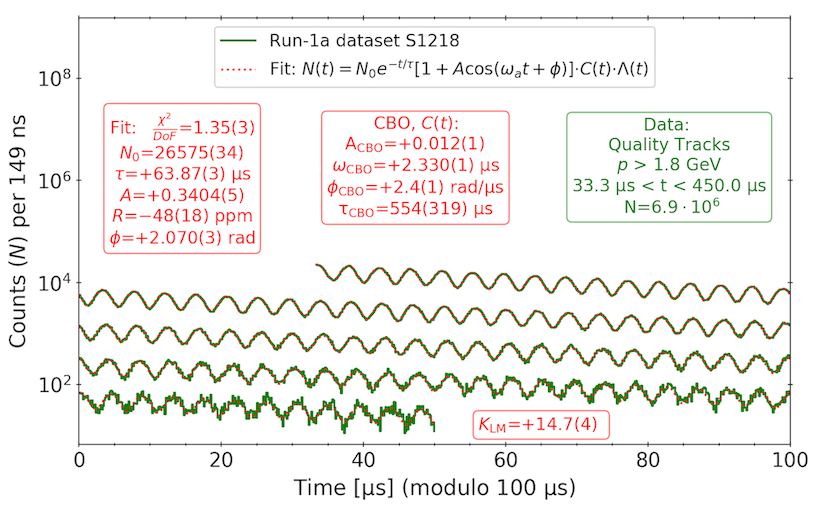} 
    \vspace{-0.3cm}
    \caption[Ten-parameter wiggle fit]{Ten-parameter fit to data from the Run-1a dataset.}
    \label{fig:S1218_10par}
\end{figure}

The remaining feature in the FFT spectrum in~\cref{fig:S1218_9par_fft} that needs to be accounted for is the peak at $0$~MHz. The physical origin of this feature are lost-muons (see \cref{sc:lost_muons}). This can be incorporated by adding a term accounting for the lost-muons in the fit, such that the fit equation becomes
\begin{equation}
    N(t)=N_0e^{-t/\tau}[1+A\cos(\omega_at+\phi)]\cdot{C(t)}\cdot \Lambda(t),
\end{equation}
where $\Lambda(t)$ is
\begin{equation}
    \Lambda(t) = 1 - K_{\mathrm{LM}} \int^t_{t_0} L(t') e^{t' / \tau}dt',
    \label{eq:lm}
\end{equation}
the lost-muon term, with an arbitrary fit parameter $K_{\mathrm{LM}}$ (i.e. normalisation). In this analysis, $\Lambda(t)$ was taken from~\cite{Nick}. It should be noted, however, that~\cite{Nick} defines the rate of lost-muons for a calorimeter-based determination of $\omega_a$. The different acceptance of the tracking detectors means it is at best an approximation of the lost-muon rate for a track-based $\omega_a$ analysis. Nevertheless, incorporating the lost-muon term yielded improved results, as compared to the nine-parameter fit, as shown in~\cref{fig:S1218_10par}. The fit $\frac{\chi^2}{\mathrm{DoF}}$ can be further improved by adding systematic effects to the fit (e.g. pile-up), or performing a dedicated lost-muon analysis in the tracking detector. However, the fit quality is sufficient for this preliminary analysis to be able to estimate the statistical precision on the tracker-based determination of $\omega_a$.
\vspace{-0.1cm}

\begin{figure}[htpb]
    \centering
    \subfloat[]{\raisebox{2mm}{\includegraphics[width=0.5\linewidth]{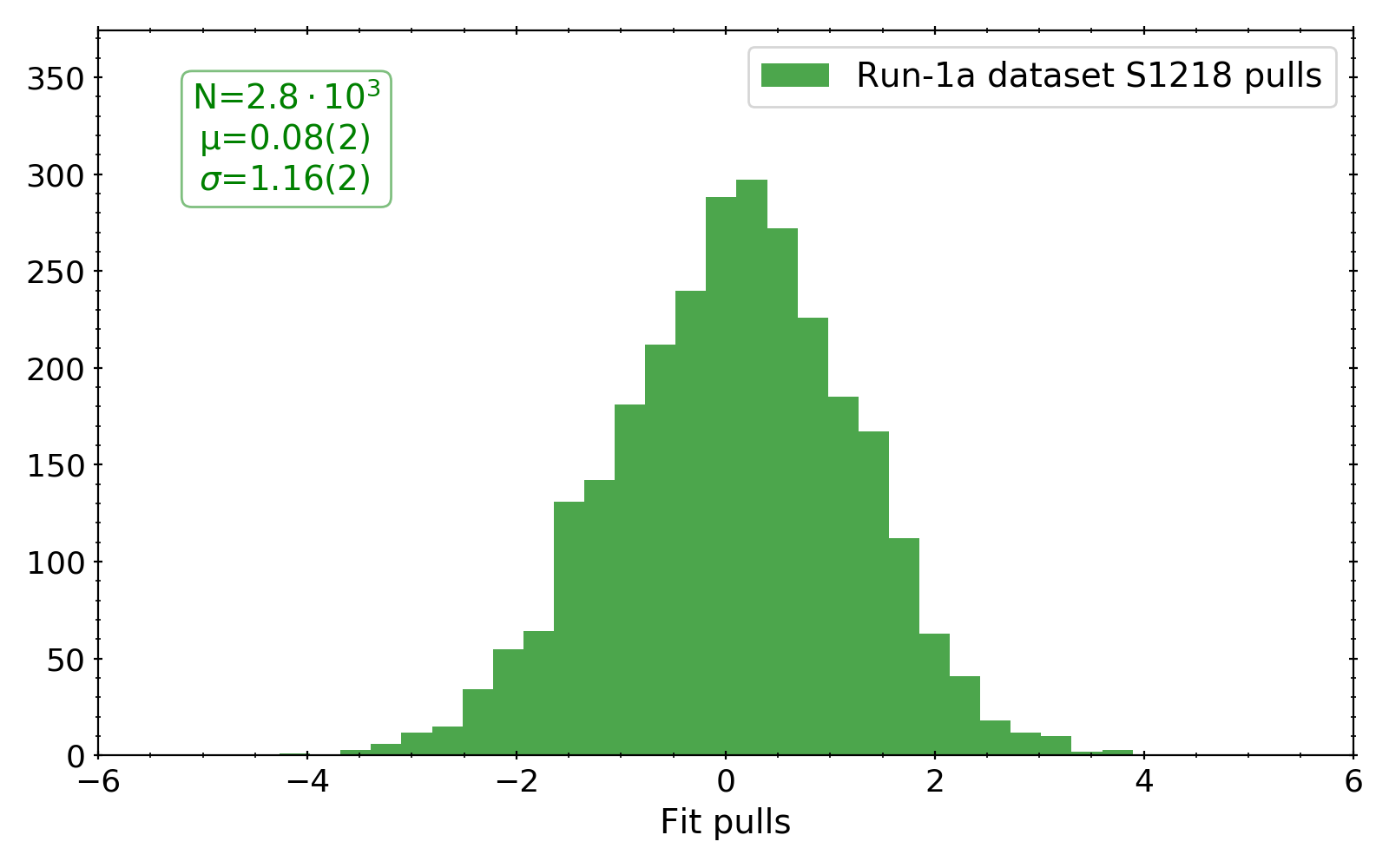} \label{fig:S1218_10par_pulls}}}
    \subfloat[]{\includegraphics[width=0.49\linewidth]{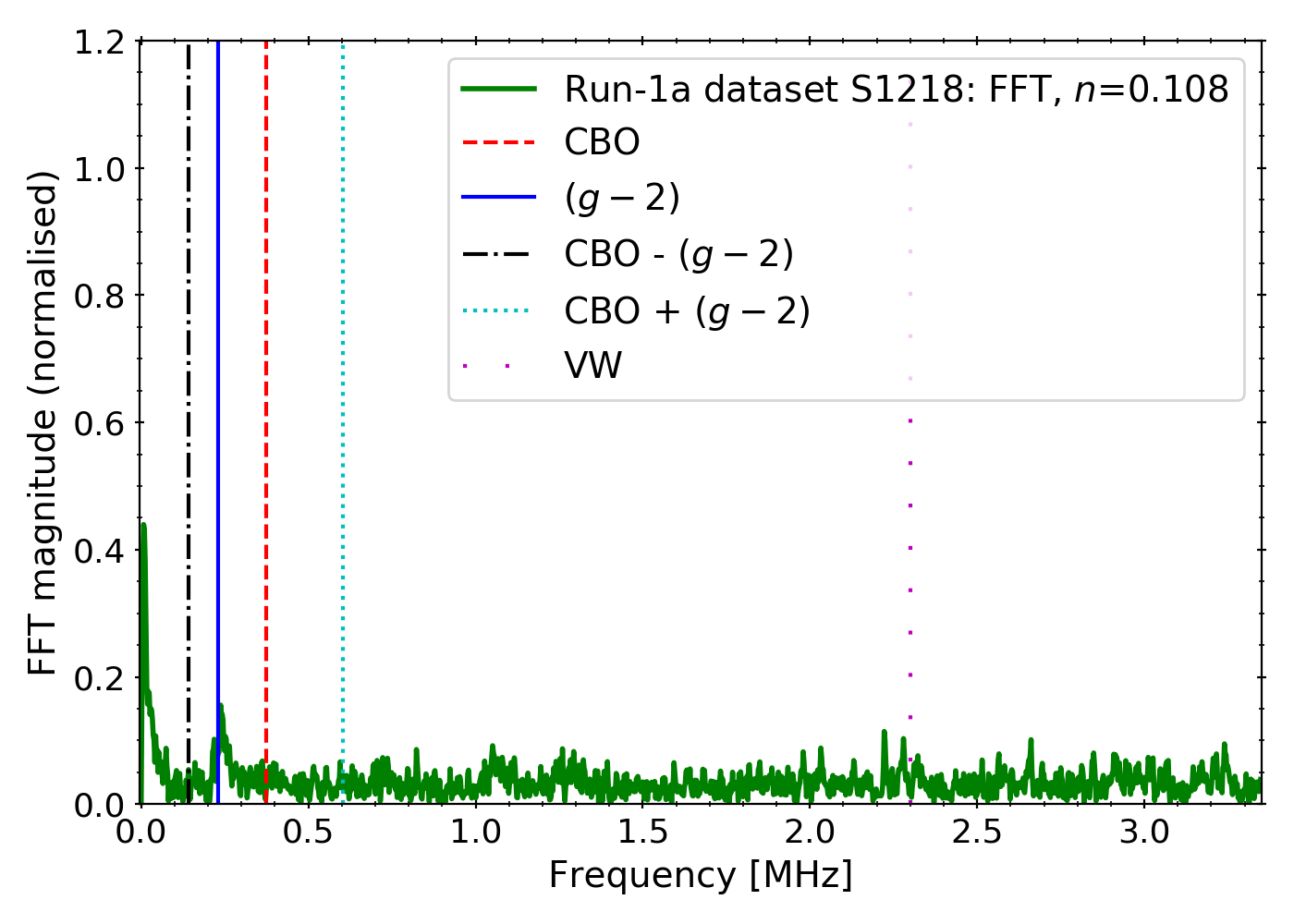} \label{fig:S1218_10par_fft}} 
    \vspace{-0.3cm}
    \caption[Ten-parameter pulls and FFT]{Ten-parameter fit (a) pulls and (b) FFT from the Run-1a dataset.}
    \label{fig:S1218_10par_res_fft}
\end{figure}

\vspace{-0.1cm}
With the ten-parameter fit, the slow effect is now diminished as seen in the improved distribution of fit pulls in~\cref{fig:S1218_10par_pulls}, and the prominent peak at $0$~MHz is significantly reduced, as seen in~\cref{fig:S1218_10par_fft}.

\clearpage

\vspace{-0.6cm}
\section{Results}
The statistical uncertainties on $R$, obtained from a ten-parameter fit, for all four \R1 datasets are summarised in \cref{tab:wiggle_fit}, while the corresponding plots are contained in~\cref{sec:ancillary_plots_wa}. 
\vspace{-0.1cm}
\begin{table}[htpb]  
  \centering
  \begin{tabular}{ccccc}
    \toprule
           Dataset & 1a & 1b & 1c & 1d  \\ \midrule
          $\delta R$ (\ac{ppm}) & $18$ &  $15$ &  $13$ &  $10$ \\ \midrule
        \end{tabular}
  \vspace{-0.1cm}
  \caption[$\delta R$ in \R1]{$\delta R$ from the fit to the tracker data across \R1 datasets.}
  \label{tab:wiggle_fit}
\end{table}

\vspace{-0.1cm}
The combined statistical uncertainty can be approximated~\cite{stat_error} via
\begin{equation}
    \delta \bar{R} = \frac{1}{\sqrt{\sum_i\delta R_i^{-2}}} = 7 \ \mathrm{ppm},
    \label{eq:wa_ppm_r1}
\end{equation}
which assumes a Gaussian distribution of errors and no correlations between the datasets. 
To verify this estimation, the statistical uncertainty on $R$ of 1.4 ppm from a fit of $930\times10^6$ calorimeter events~\cite{Nick} in the Run-1a dataset can be used to predict the expected precision on the tracker data of $59\times10^6$ events, yielding
\begin{equation}
  \delta R = \sqrt{\frac{930}{59}} \times 1.4 = 6 \ \mathrm{ppm}.
  \label{eq:wa_ppm_calo}
\end{equation}
The estimated uncertainties in \cref{eq:wa_ppm_r1,eq:wa_ppm_calo} compare well, given the precision on the calorimeter data is derived from a fit with better estimates of systematic effects.

\clearpage

It is now possible to calculate the number of tracks required to reach a defined precision on $R$, and hence $\omega_a$, given that $59.2\times10^6$ tracks yield a $\sim7$ ppm uncertainty. Therefore, to reach a precision on the $\omega_a$ measurement of 1.4 ppm, and thus confirm the BNL-SM discrepancy at $95\%$ CL, should the same value be measured, would require $1.3\times10^9$ tracks. While to reach a comparable precision of the final $\omega_a$ measurement at the BNL, of 0.54 ppm~\cite{BNL_AMM}, would require the experiment to accumulate $8.7\times10^{9}$ tracks. 

\section{Outlook} 
For the tracker-based determination of $\omega_a$ to become competitive, more tracker data must be acquired, and the tracking efficiency increased. This efficiency is described in~\cref{sub:tracker_calibration_and_track_refinement}, and will be applied to the already-collected data, yielding at least a factor of four increase in the number of tracks. 

\begin{figure}[htpb] 
    \centering
    \includegraphics[width=0.75\linewidth]{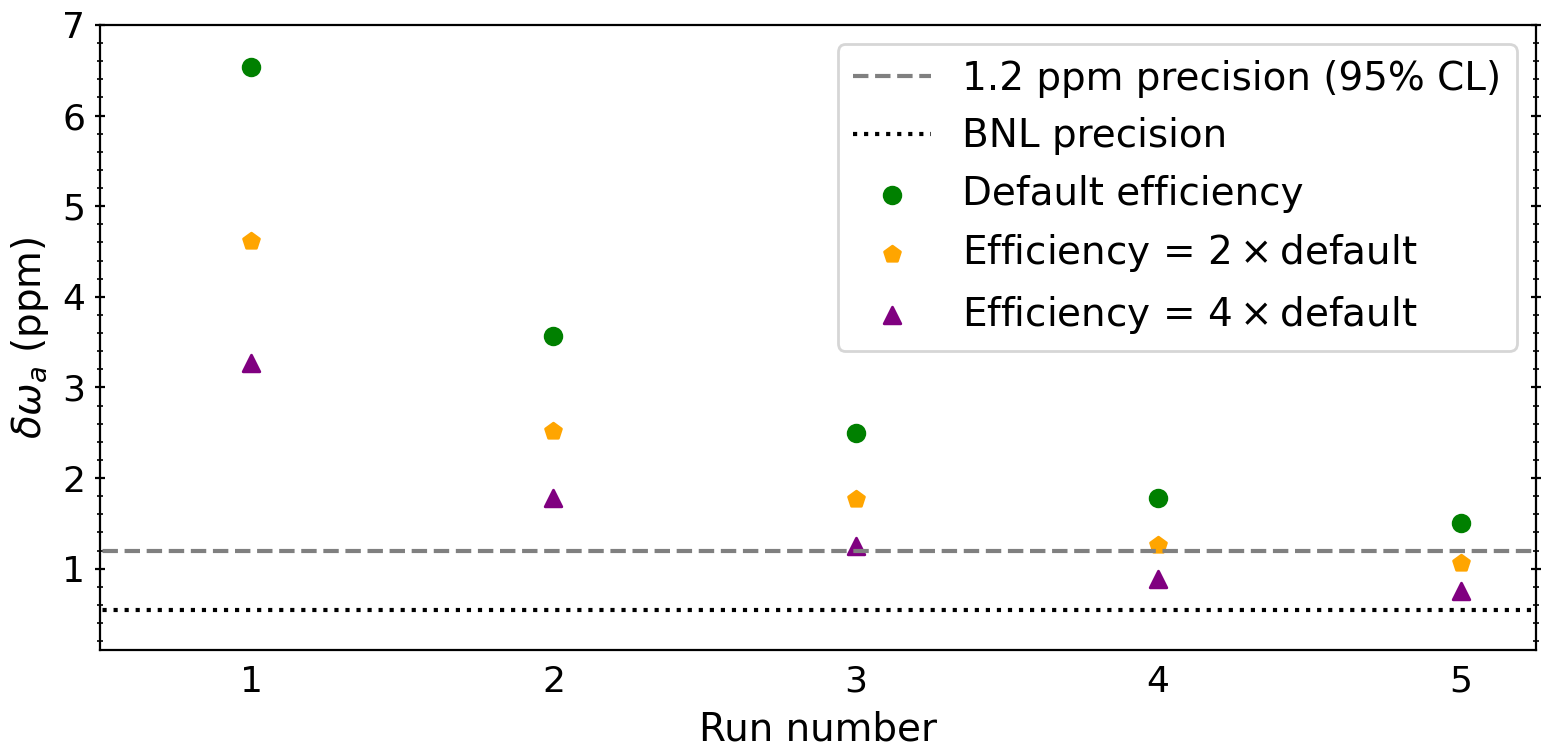} 
    \vspace{-0.25cm}
    \caption[Projected precision on $\omega_a$ with tracker data]{Projected precision on $\omega_a$ with tracker data.}
    \label{fig:money}
\end{figure}

Using \cref{fig:pot}, a relation can be made between the raw number of $e^+$ collected in \R1 and the number of tracks that were used in this preliminary $\omega_a$ analysis. A projection can than be made, using \cref{fig:pot_R3}, on the expected number of tracks that can be used in the determination of $\omega_a$. This projection, shown in \cref{fig:money}, is based on the assumption that the increase in the quality data (see \cref{sec:dqc}) will be between $60\%$ and $80\%$~\cite{Fred_R2,Mark_private}, as compared to \R1. Given the current improvements in tracking, a $95\%$ CL limit can be placed by \R3. With further improvements, a comparable precision to the one achieved at the BNL \gm2 (with calorimeter data) can be reached by \R5 with the Fermilab \gm2 tracking detectors alone.
 \vspace{-0.2cm}

\graphicspath{{fig/}}

\chapter{Systematic studies of the EDM}\label{ch:edm} 

This chapter describes the methodology of the EDM measurement using data from the tracking detectors.  \cref{sec:edm_track} contains the results of the EDM analysis in the simulation. The impact of radial and longitudinal magnetic fields on the EDM and $\omega_a$ determination is introduced in \cref{sec:rad_long}, while the results of their measurements are given in \cref{sec:rad} and \cref{sec:lon}, respectively. 

\section{Introduction}\label{sec:intro_ana}
\begin{figure}[htpb]
    \centering
    \subfloat[]{\includegraphics[width=0.4\linewidth]{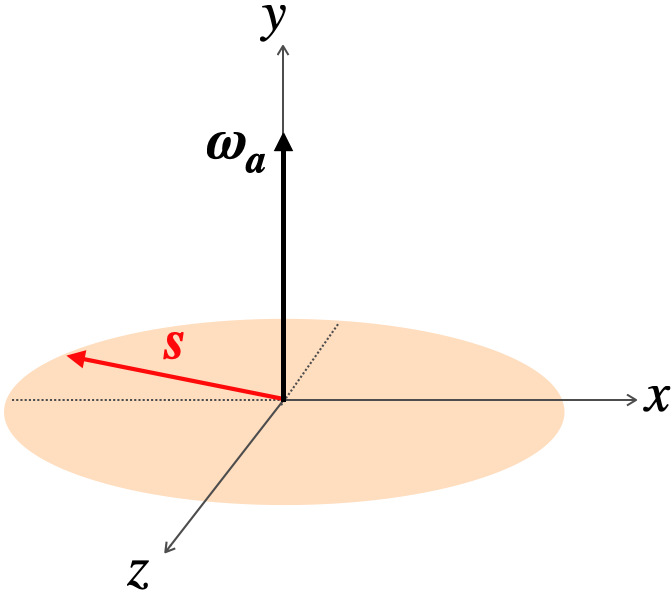}} 
    \subfloat[]{\hspace*{0.08\linewidth}\includegraphics[width=0.4\linewidth]{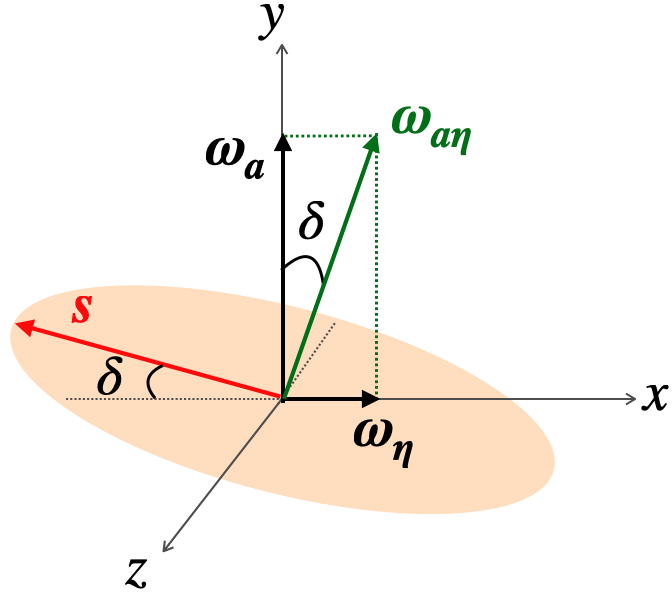}} 
    \vspace{-0.2cm}
    \caption[Precession-plane tilt due to an EDM]{The tilt in the precession-plane of the muon due to an EDM, in the muon rest-frame. The spin ($\boldsymbol{s}$) of the muon precesses in the circle that is orthogonal to the vertical magnetic field ($B_y$). The momentum vector ($\beta$) is along the $z$-axis. The centre of the storage ring is along the $x$-axis in the orientation. (a) A zero EDM produces no tilt.  b) The presence of an EDM tilts the plane by an angle $\delta$. The precession of $\boldsymbol{s}$ now has a vertical component.}
    \label{fig:edm_tilt}
\end{figure}

The effect of a potential \ac{EDM} of the muon, which was introduced in \cref{sec:edm}, would result in a tilt of the muon spin precession-plane as shown in \cref{fig:edm_tilt}. This tilt changes the observed precession frequency by an addition of the precession resulting from a muon EDM~\cite{LDM}, $\omega_{\eta}$,
\begin{equation}
    \boldsymbol{\omega_{\eta}} = \eta\frac{e}{2m_{\mu}}\left(\frac{\boldsymbol{E}}{c}+\boldsymbol{\beta}\times\boldsymbol{B}\right).
    \label{eq:omega_eta}
\end{equation}
\cref{eq:omega_eta} describes the muon experiencing a torque in the presence of the electric field produced by the \ac{ESQ}, as well as the apparent electric field in the muon rest-frame resulting from the Lorentz boost to the laboratory-frame magnetic field~\cite{BNL_EDM}. $\boldsymbol{\omega_{\eta}}$ is orthogonal to the vertically oriented $\boldsymbol{\omega_{a}}$, and is pointing radially inwards towards the centre of the \gm2 storage ring. This results in the spin precession-plane tilting out from the orbit plane towards the centre of the storage ring. 

\clearpage

\vspace{-0.5cm}
There are two important consequences of a non-zero muon EDM: an increase in the observed precession frequency, and an introduction of an oscillation of the average vertical angle of the emitted positrons from the muon decay. The magnitude of the observed precession frequency is now given by 
\begin{equation}
    \omega_{a\eta}=\sqrt{\omega_{a}^2 + \omega_{\eta}^2}.
\label{eq:mu_edm}
\end{equation}
The resultant tilt angle of the spin precession-plane in the muon rest-frame, $\delta$, is given by
\begin{equation}
    \delta =\tan^{-1}\left(\frac{\omega_{\eta}}{\omega_a}\right) = \tan^{-1}\left(\frac{\eta\beta}{2a_{\mu}}\right).
\label{eq:delta_mrf}
\end{equation}
As the positrons are preferentially emitted along the direction of muon spin (see \cref{fig:mu_decay}), the tilt will make the positrons on average downward going when the spin is pointing towards the centre of the \gm2 storage ring, and upward going when the spin points away from the centre of the ring. Therefore, the observed average vertical angle varies with the magnitude of $\omega_{a\eta}$, and this angle is the observable that is used in the \gm2 experiment to measure the EDM using the tracking detectors.

\clearpage
Finally, it is essential to emphasise that when the muon spin and momentum vectors are aligned, the maximum $\omega_a$ signal is observed. However, the vertical angle oscillation is at the maximum when the spin is pointing radially outwards, that is, $90^{\circ}$ out-of-phase with the momentum vector. This property is used to distinguish between an EDM and other effects, which are in-phase with $\omega_a$, that can also impact the vertical angle oscillation. This is discussed in \cref{sec:rad_long}, for potential non-zero radial and longitudinal magnetic field components. 

\section{EDM search with the tracking detectors}\label{sec:edm_track}
The goal of the Fermilab \gm2 experiment is to measure $\delta$ to within $\SI{0.4}{\micro\radian}$, resulting in a sensitivity to an \ac{EDM} of $\sim10^{-21}~e\cdot$cm~\cite{Becky}. The tracking detectors will realise an \ac{EDM} measurement through the direct detection of an oscillation in the average vertical angle of the decay positrons. The tracking detectors measure this vertical angle, $\theta_y$, through the momentum components of a track
\begin{equation}
  \theta_y = \tan^{-1}\left(\frac{p_y}{p_z}\right).
\end{equation}

\begin{figure}[htpb]
    \centering
    \includegraphics[width=0.65\linewidth]{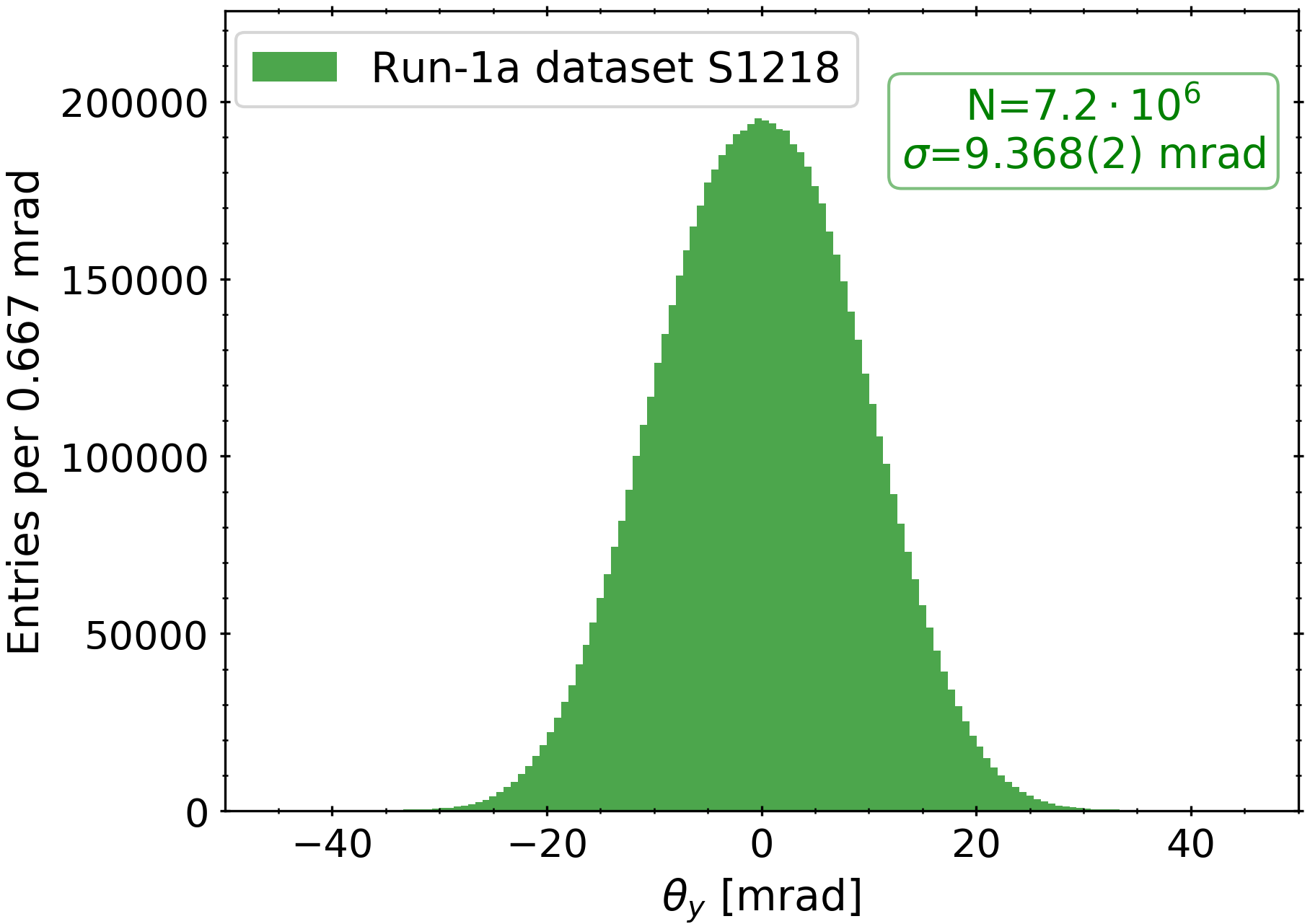}
    \caption[The distribution of measured vertical angles of tracks.]{The distribution of measured vertical angles ($\theta_y$) of tracks in the Run-1a dataset in tracking stations 12 and 18.}
    \label{fig:theta_y}
\end{figure}

For example, a distribution of measured vertical angles in the Run-1a dataset are shown in~\cref{fig:theta_y}, with a width of $9.368(2)$~mrad.
\clearpage

\subsection{Simulating a large EDM signal}
In order to test and develop the analysis techniques, a simulation with a known value of the EDM was used. The chosen value of the EDM was $d_{\mu} = 5.4\times10^{-18} \ e\cdot{\mathrm{cm}}$, which corresponds to a value $\sim 30$~times larger than the EDM limit set at the \ac{BNL} experiment~\cite{BNL_EDM}. This value was chosen to resolve the EDM signal with a relatively low number of available tracks from simulation and permitted a cross-check with previous simulation results~\cite{Saskia}. 

It is possible to analytically predict the observed oscillation amplitude, $A_{\mathrm{EDM}}$, of $\theta_y$
\begin{equation}
  A_{\mathrm{EDM}}=a_{\mathrm{EDM}}\delta ',
  \label{eq:a_edm}
\end{equation}
where $\delta '$ is the precession-plane tilt angle in the lab-frame, and $a_{\mathrm{EDM}}$ is the asymmetry of $\sim0.13$~\cite{Joe_Saskia}, which accounts for the fact that not all positrons are emitted in the direction of the polarisation vector. The tilt angle in the muon rest-frame, $\delta$, is accessible via a Lorentz boost (see \cref{sec:lorentz_boost})
\begin{equation}
  \delta ' = \tan^{-1}\left(\frac{\tan(\delta)}{\gamma}\right).
\end{equation}
Finally, substituting $\eta$ from \cref{eq:edm_1} into \cref{eq:delta_mrf} allows $A_{\mathrm{EDM}}$ to be directly expressed in terms of fundamental constants and the input EDM signal ($d_{\mu} = 5.4\times10^{-18} \ e\cdot{\mathrm{cm}}$)
\begin{equation}
  A_{\mathrm{EDM}}=a_{\mathrm{EDM}}\tan^{-1}\left(\frac{2d_{\mu}\beta m_{\mu}c}{a_{\mu}\gamma e \hbar}\right) = 0.22 \ \mathrm{mrad}.
  \label{eq:pred}
\end{equation}

\subsection{EDM measurement strategy in the simulation}\label{sub:edm_measurement_simulation}
\begin{figure}[htpb]
    \centering
    \subfloat[]{\includegraphics[width=0.5\linewidth]{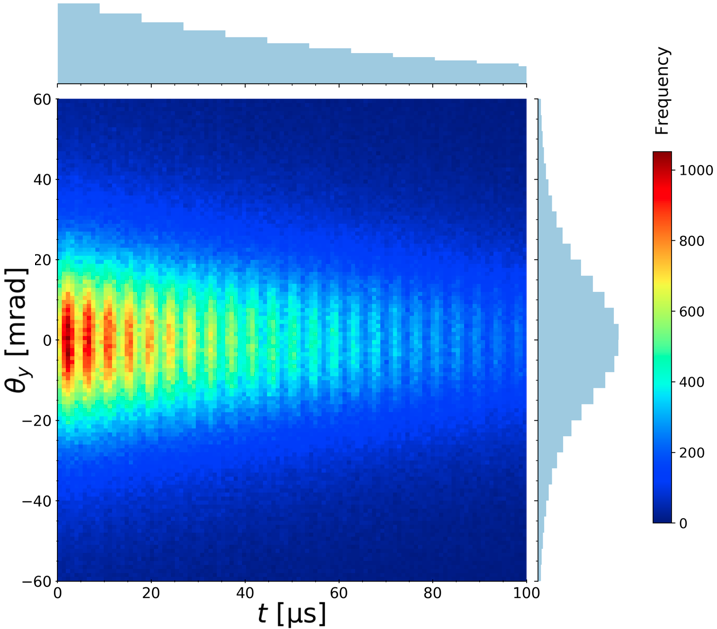}\label{fig:theta_time}} 
    \subfloat[]{\includegraphics[width=0.5\linewidth]{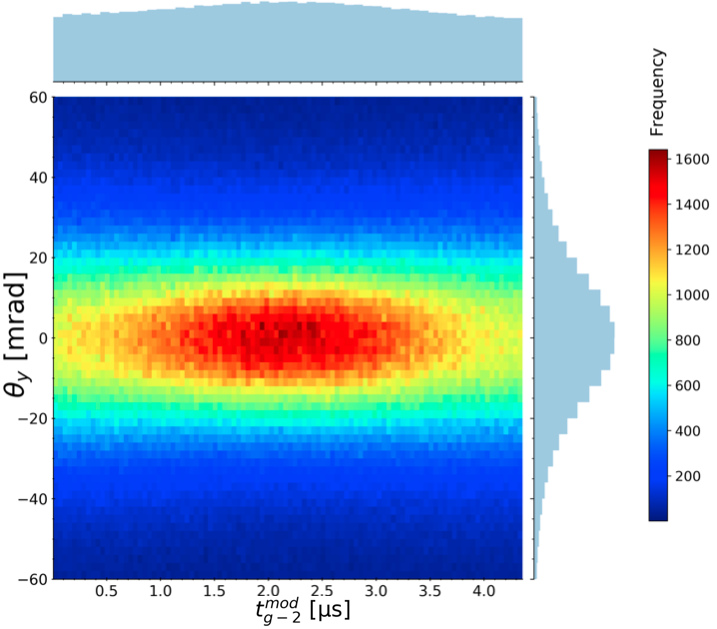}\label{fig:theta_mod}} 
    \vspace{-0.2cm}
    \caption[Vertical angle versus time in fill]{(a) $\theta_y$ versus time in fill. (b) $\theta_y$ versus modulated time in fill.}
    \label{fig:theta_fill}
\end{figure}

To observe the oscillation of the vertical angle, and hence measure $A_{\mathrm{EDM}}$, it is necessary to consider how $\theta_y$ changes with time, as shown \cref{fig:theta_time}, where $t$ is the track time in a fill. This time is then modulated by the $g-2$ period ($\sim$~\SI{4.4}{\micro\second}), $T_{g-2}$, 
\begin{equation}
  T_{g-2} = \frac{2\pi}{\omega_a}.
\end{equation}
\clearpage
This is analogous to the method used in the BNL EDM analysis~\cite{Sossong}, which is known as the period-binned analysis. This modulation method is useful for looking at periodic effects, such as the vertical angle oscillation. Periodic effects that do not have the same period as $T_{g-2}$ (e.g. betatron oscillations mentioned in~\cref{sec:CBO}) will generate a flat background when many periods are considered, while the signal of interest adds constructively. The time modulation, $t^{\mathrm{mod}}_{g-2}$, is given by
\begin{equation}
   t^{\mathrm{mod}}_{g-2} = \left(\frac{t}{T_{g-2}} - \mathrm{int}\left[\frac{t}{T_{g-2}}\right]\right)T_{g-2}.
   \label{eq:t_mod}
 \end{equation} 
\vspace{-1.3cm}

\begin{figure}[htpb]
    \centering
    \includegraphics[width=0.65\linewidth]{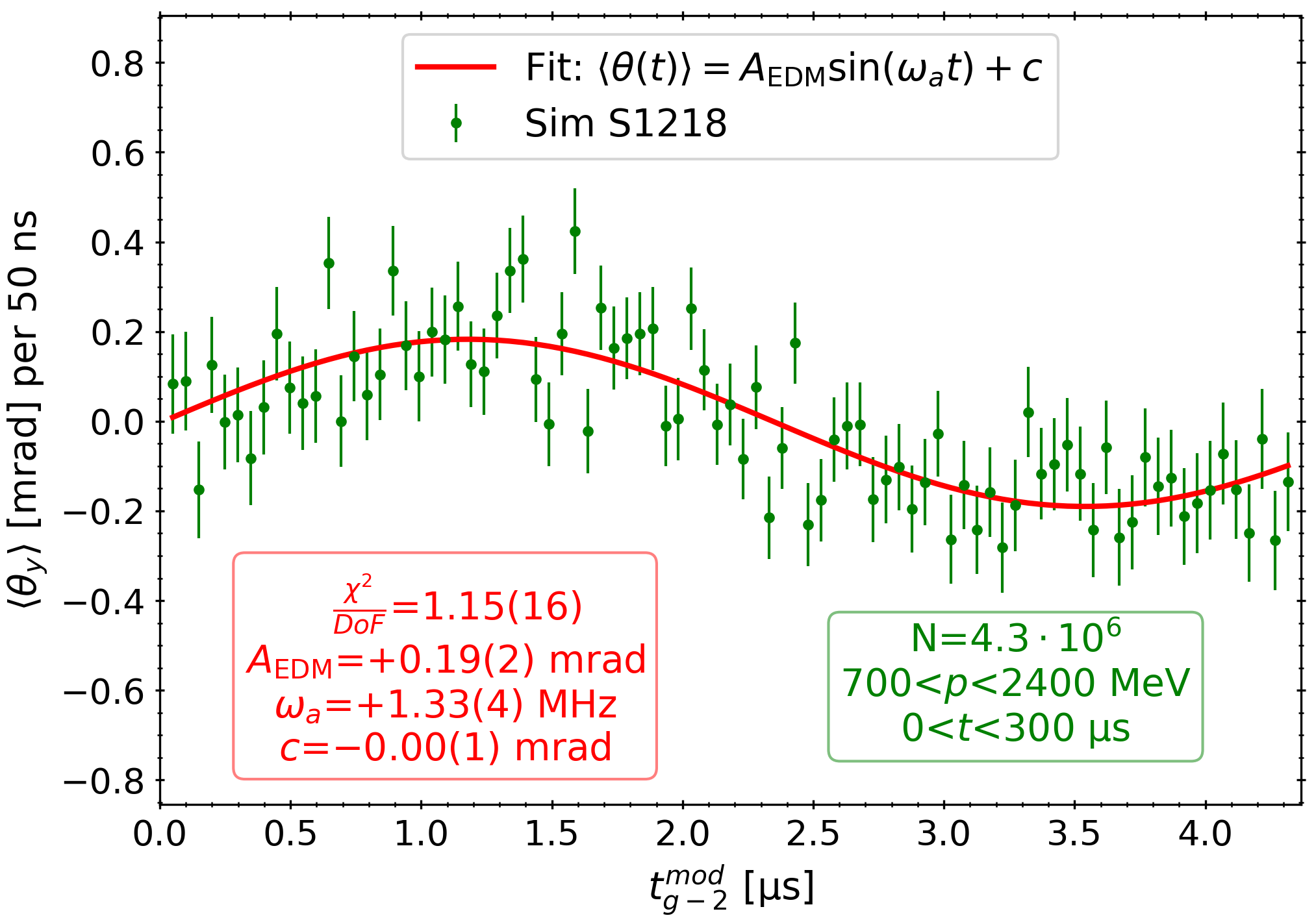}
    \vspace{-0.2cm}
    \caption[Vertical angle oscillation]{Averaged vertical angle oscillation with 4.3M tracks in simulation from both stations (S12 and S18).}
    \label{fig:theta_sim}
\end{figure}

$\theta_y$ was then profiled (i.e. averaged) into $50$~ns bins, as shown in \cref{fig:theta_sim}. The distribution of $\theta_y$ in each time bin is Gaussian, as shown in \cref{fig:theta_y}. The observed oscillation of the averaged vertical angle, $\langle\theta_y\rangle$, is given by
\begin{equation}
  \langle \theta(t) \rangle =  A_{\mathrm{EDM}}\sin(\omega_a t) + c,
  \label{eq:fit_sim}
\end{equation}
where $c$ is the overall offset in $\theta(t)$. \cref{eq:fit_sim} is the fit function used in \cref{fig:theta_sim}. The fitting is done using the \textit{scipy.optimize} \verb!Python! library. The optimal momentum and time cuts for this analysis are discussed in \cref{sub:cuts}. The extracted value of $A_{\mathrm{EDM}}$ is $0.19(2)$~mrad, which agrees well with the prediction from~\cref{eq:pred}. 

\clearpage
\vspace{-0.7cm}
\small
\section{The significance of non-zero magnetic fields}\label{sec:rad_long}
\begin{figure}[htpb]
    \centering
    \subfloat[]{\includegraphics[width=0.45\linewidth]{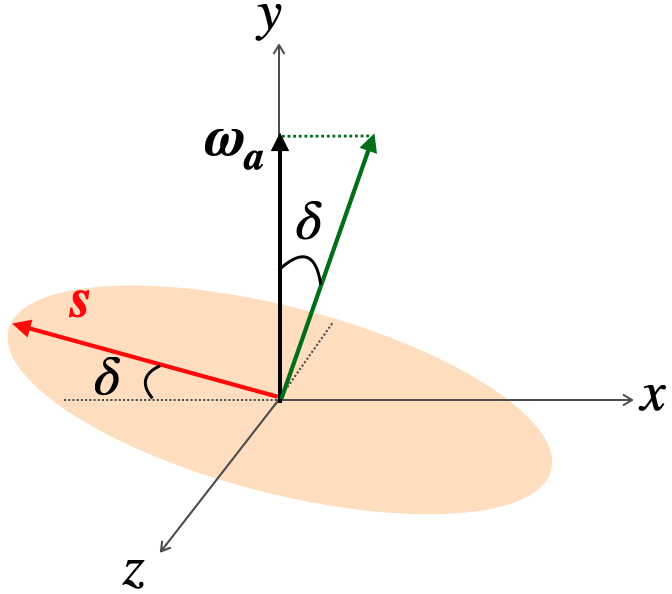}} 
    \subfloat[]{\hspace*{0.08\linewidth}\includegraphics[width=0.45\linewidth]{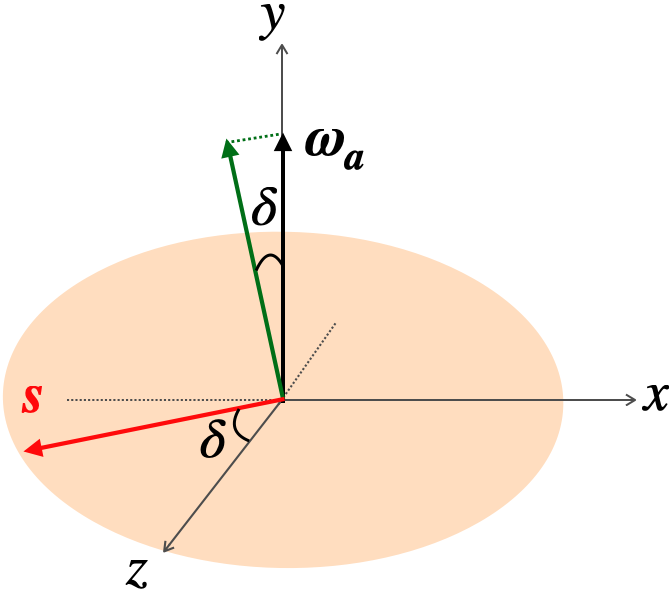}} 
    \vspace{-0.2cm}
    \caption[Precession-plane tilt due to non-zero field components]{The tilt in the precession-plane of the muon due to non-zero field components, in the muon rest-frame.  The spin ($\boldsymbol{s}$) of the muon precesses in the circle that is orthogonal to the vertical magnetic field ($B_y$). The momentum vector ($\beta$) is along the $z$-axis. The centre of the storage ring is along the $x$-axis in the orientation. (a) The tilt due to the radial field ($B_x$), which is along the $x$-axis. (b) The tilt due to a non-zero longitudinal field ($B_z$), which is along the $z$-axis. }
    \label{fig:mf_tilt}
\end{figure}

\normalsize
The \gm2 experiment is designed to have only the vertical component of the magnetic field, with both radial and longitudinal components being zero. In the experiment, the vertical magnetic field, $B_y$, is measured directly by the \ac{NMR} probes to an accuracy of 70 \ac{ppb}, as described in \cref{sc:field}. However, no such direct measurements presently are possible to determine the radial, $B_x$, or longitudinal, $B_z$, fields. Non-zero $B_x$ or $B_z$ can tilt the precession-plane of the muon, as shown in \cref{fig:mf_tilt}. This becomes an important systematic effect for both the \ac{EDM} and $\omega_a$ analyses, as can be seen by adding two aforementioned magnetic field terms to \cref{eq:mub}
\begin{equation}
    \boldsymbol{\omega_a}=a_{\mu}\frac{e}{m_{\mu}}\boldsymbol{B}= a_{\mu}\frac{e}{m_{\mu}} ( B_y\boldsymbol{\hat{y}} + B_z\boldsymbol{\hat{z}} + B_x\boldsymbol{\hat{x}}).
\end{equation}
It is important to note, that $B_x$ would have the same effect on the precession-plane as the \ac{EDM} (c.f. \cref{fig:edm_tilt}). It is, therefore, crucial to estimate $B_x$ in order to not bias the results of a search for the muon \ac{EDM}. Various strategies exist to estimate $B_x$, both direct~\cite{Rachel_mf} and indirect. The approach taken in this thesis is the indirect measurement of $B_x$ using the tracking detectors, as described in \cref{sec:rad}.
\clearpage

\vspace{-0.3cm}
A non-zero $B_z$, on the other hand, would tilt the precession-plane in the direction of the muon momentum vector. This produces an effect that manifests as an \ac{EDM}-like signal that is in-phase with the magnetic dipole moment (i.e. in-phase with $\omega_a$) and out-of-phase with the \ac{EDM} (see \cref{sec:intro_ana}). Therefore, one way to measure $B_z$ is to perform a search for the in-phase signal using data from the tracking detectors, as described in \cref{sec:lon}.

\section{Preliminary estimation of \texorpdfstring{$B_x$}~}\label{sec:rad}

\subsection{Introduction}\label{sub:rad_intro}
\begin{figure}[htpb]
    \centering
    \includegraphics[width=0.65\linewidth]{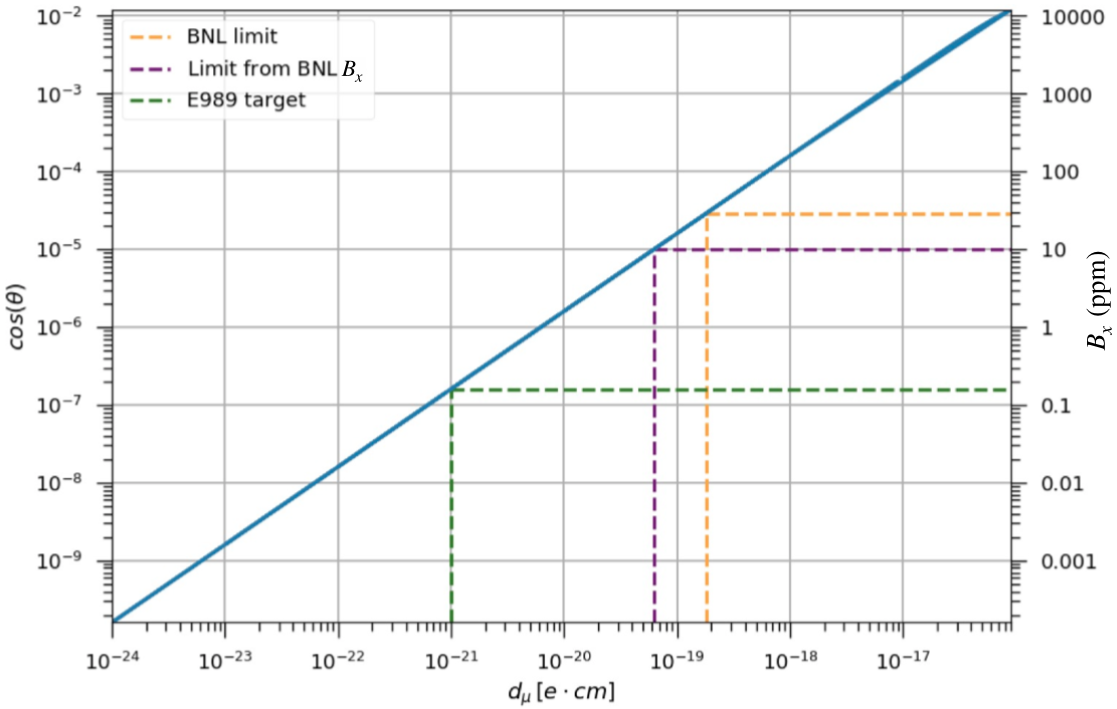}
    \vspace{-0.1cm}
    \caption[Radial field at the BNL]{The value of the EDM, $d_{\mu}$, and the corresponding value of $B_x$. For example, to realise the \gm2 experiment's goal of measuring $d_{\mu}$ to $10^{-21}~e\cdot$cm, $B_x$ must be known to 0.2 \ac{ppm}. Plot courtesy of J.~Price~\cite{Joe_PSI}.}
    \label{fig:br_bnl}
\end{figure}

In the \ac{BNL} experiment, the EDM limit is \say{equivalent} to a 30 \ac{ppm} $B_x$, while the determination of $B_x$ was estimated with a precision of 10 \ac{ppm} \cite{BNL_EDM}. Had the experiment had enough statistics, they would have been limited at $\left|d_{\mu}\right| \sim 4.5 \times 10^{-20} \ e\cdot{\mathrm{cm}}$ due to the uncertainty on $B_x$, as demonstrated in \cref{fig:br_bnl}. Therefore, in the Fermilab \gm2 experiment more precise measurements of $B_x$, both direct and indirect, are motivated to allow for an improved measurement of the EDM. An accuracy of 0.2 ppm is required to probe $d_{\mu}$ at the $10^{-21}~e\cdot$cm level.

\clearpage

The aim of this preliminary study was to estimate $B_x$ by varying the electric field produced by the \ac{ESQ} (see \cref{sc:scraping}) and measuring the resultant mean vertical beam position, $\langle y \rangle$, using the tracking detectors, and using the relation~\cite{Syphers}
\begin{equation}
    B_x = \frac{nB_0\langle y \rangle}{R_0},
    \label{eq:y_ver}
\end{equation}
where $n$ is the field-index (see \cref{sc:scraping}), $R_0$ is the storage ring radius, and $B_0$ is the magnetic field strength. $n$ contains the electric field gradient of the \ac{ESQ}, which is directly proportional to the \ac{HV} applied to the \ac{ESQ} -- QHV. Increasing the value of QHV should result in the observed value of $\langle y \rangle$ decreasing, and vice versa. \cref{eq:y_ver,eq:field_index} allow the estimation~\cite{James_Saskia} of $B_x$ 
\begin{equation}
    B_x \sim (\Delta\langle y \rangle \Delta\mathrm{QHV})
    \label{eq:B_x_ppm}
\end{equation}
where $\mathrm{QHV}$ is the variable, and $\langle y \rangle$ is the observable, and $\Delta\langle y \rangle \Delta\mathrm{QHV}$ can be determined from tracker data. The objective is to verify the viability of such a measurement.

\subsection{Results}
There were two periods in \R2 that had a change in QHV: 20 June 2019 and 24 March 2019. 

\begin{figure}[htpb]
    \centering
    \subfloat[]{\raisebox{1.5mm}{\includegraphics[width=.53\linewidth]{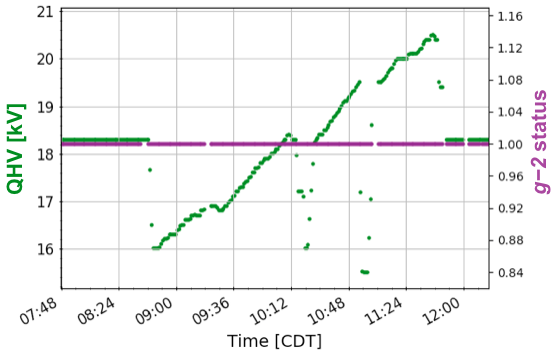}}}
    \subfloat[]{\hspace*{0.02\linewidth}\raisebox{3mm}{\includegraphics[width=.45\linewidth]{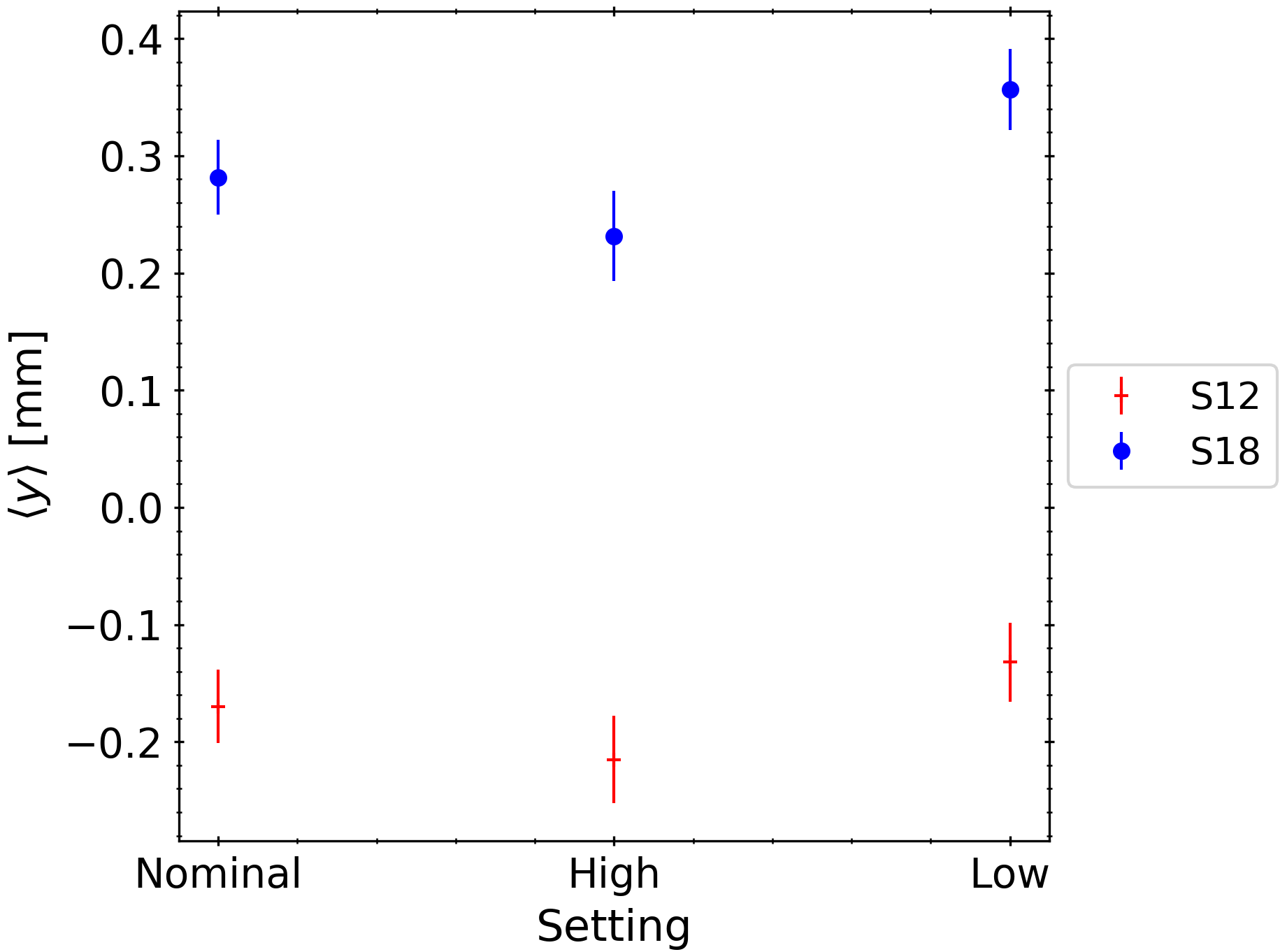}\label{fig:br_1_b}}}
    \vspace{-0.2cm}
    \caption[ESQs voltage versus time]{20 June 2019. (a) QHV versus time. b) The QHV from (a) can be split into three periods of nominal, low, and high QHV values.}
    \label{fig:br_1}
\end{figure}

\clearpage
The QHV settings versus time for 20 June are shown in \cref{fig:br_1}. To ensure the measurement of  $\langle y \rangle$ is only influenced by the QHV, settings of other \gm2 components (e.g. kicker voltages) are required to be constant. The difference in the measured values of $\langle y \rangle$, at a given QHV setting, between the two tracker stations, as seen in~\cref{fig:br_1_b}, are expected and their origin discussed in~\cref{sc:closed_orbit}.
\vspace{-0.2cm}

\begin{figure}[htpb]
    \centering
    \subfloat[]{\includegraphics[width=.49\linewidth]{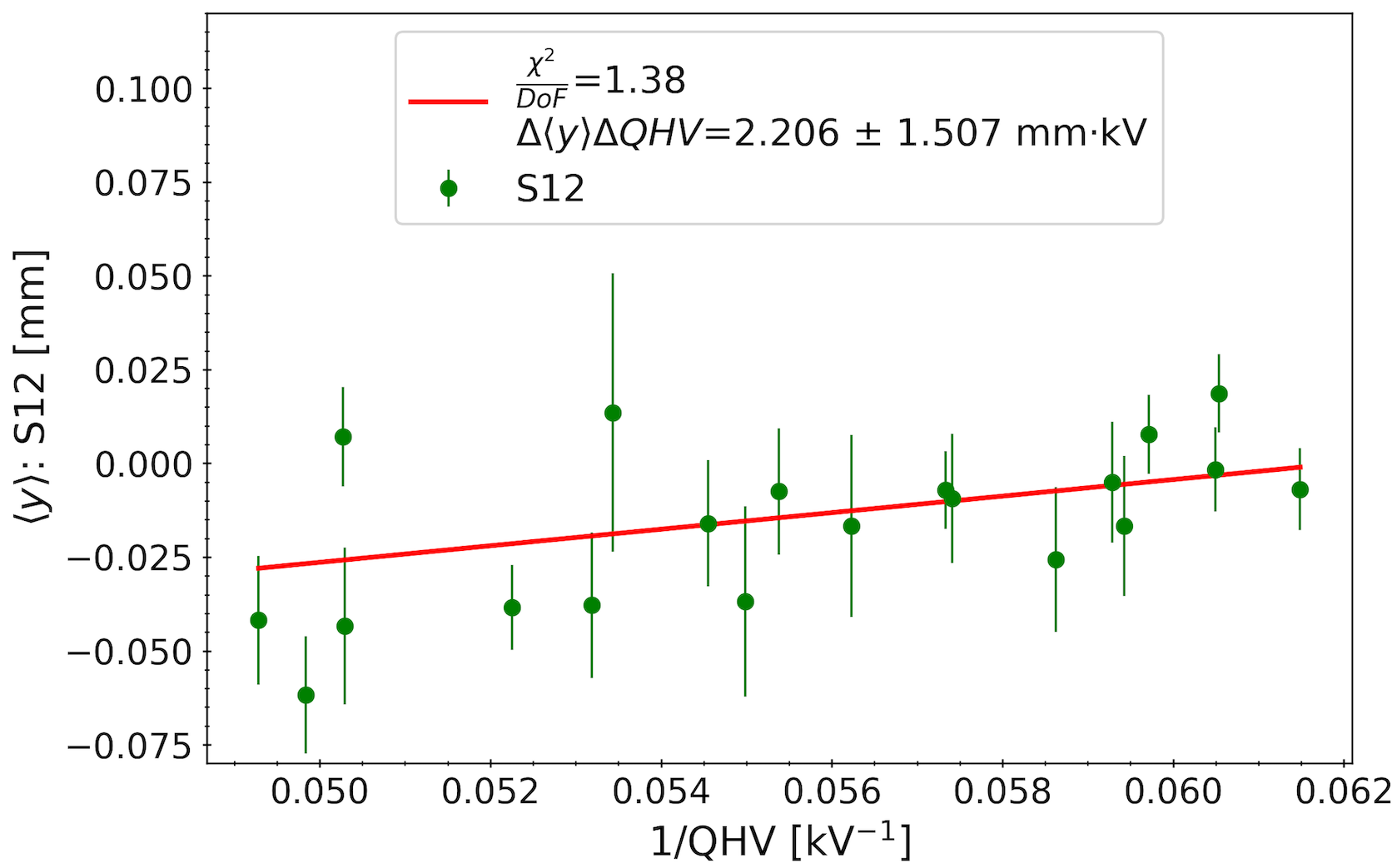} \label{fig:br_2_1}}
    \subfloat[]{\includegraphics[width=.49\linewidth]{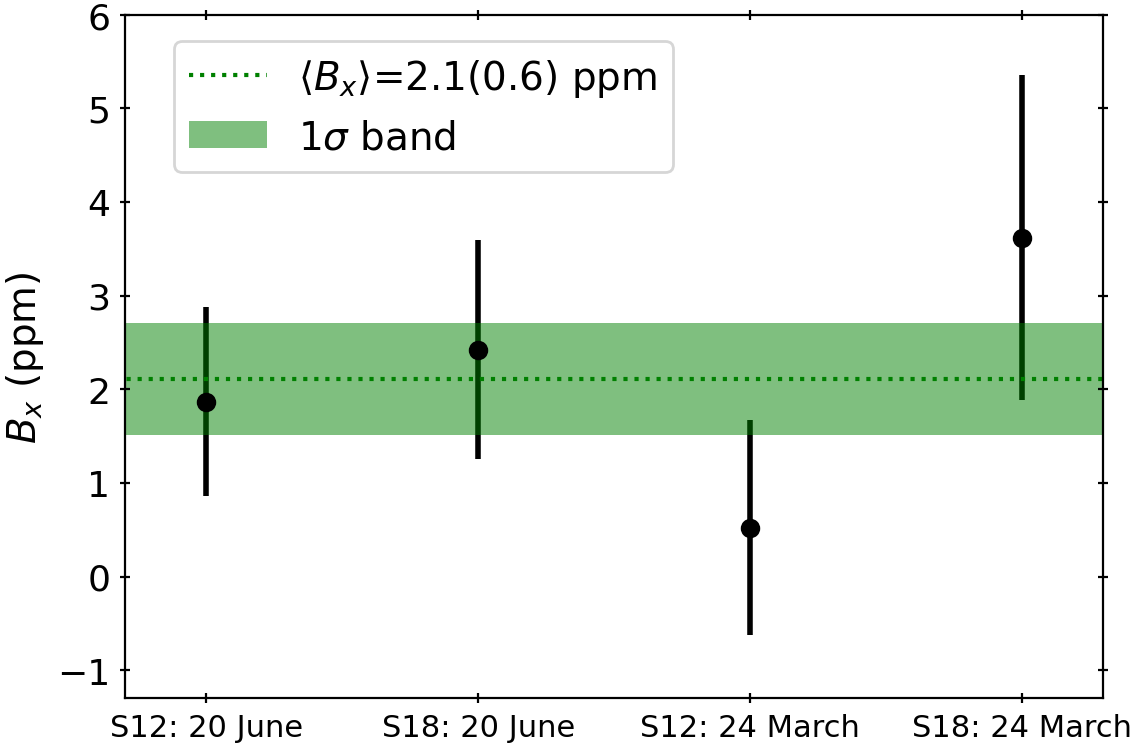}\label{fig:br_2_2}}
    \vspace{-0.2cm}
    \caption[$\langle y \rangle$ versus QHV and $B_x$ results]{(a) $\langle y \rangle$ versus QHV in S12 on 20 June 2019. The slope, $\Delta\langle y \rangle \Delta \mathrm{QHV}$, can be directly extracted from the plot. (b) The estimated values of $B_x$, in ppm, in S12 and S18 on 20 June and 24 March. The weighted mean, $\langle B_x \rangle$, and the $1\sigma$ uncertainty band are shown.}
\end{figure}

\vspace{-0.2cm}
$\langle y \rangle$ in tracker station 12 (20 June) is plotted against (QHV)$^{-1}$ in \cref{fig:br_2_1}. Using the obtained slope from \cref{fig:br_2_1} and \cref{eq:B_x_ppm}, an estimation of $B_x$ is possible. The final result, $\langle B_x \rangle=2.1(6)$~ppm, is summarised in \cref{fig:br_2_2}.
\vspace{-0.2cm}
\clearpage
\subsection{Outlook}
In this study, the preliminary-estimated value of $B_x$ in \R2 of $2.1(6)$~ppm falls short of the required precision of 0.2~ppm, as outlined in \cref{sub:rad_intro}. Therefore, more precise measurements of $B_x$, both direct and indirect, are motivated to allow for a more precise determination of the EDM. An improved method to determine $B_x$ was suggested by B. Kiburg~\cite{Brendan_br}. This method will determine $B_x$ by a variation of QHV together with an applied radial field gradient, $\Delta B_x$, from the correction coils (see \cref{sc:field}), that will induce a large, and known, $\Delta B_x$ 
\small
\begin{equation}
    \langle \Delta y \rangle \sim \frac{1}{\Delta \mathrm{QHV}} \cdot (B_x \pm \Delta B_x),
\end{equation}
\normalsize
where $B_x$ is the unknown variable in this equation. A direct measurement of $B_x$ is also being developed by M. Fertl~\cite{Martin}.

\section{Estimation of \texorpdfstring{$B_z$}~}\label{sec:lon}
The presence of a longitudinal magnetic field, $B_z$, tilts the precession-plane of the muon, as shown in \cref{fig:mf_tilt}. It is therefore imperative to accurately and precisely measure $B_z$. As discussed in \cref{sec:rad_long}, a non-zero $B_z$ introduces an up-down oscillation of the vertical track angle, in-phase with $\omega_a$. This oscillation is used in this analysis to determine $B_z$. This analysis follows a similar methodology as the EDM simulation study described in \cref{sub:edm_measurement_simulation}. Following the method used in the BNL experiment~\cite{Sossong}, two fits are performed. Firstly, the number of quality tracks is time-modulated according to \cref{eq:t_mod}, and $\phi$ is determined using the five-parameter fit of \cref{eq:5_par} but with $\omega_a$ fixed at the value measured by the BNL experiment, $\omega_a=$\SI{1.439311}{\MHz}~\cite{BNL_AMM}. The average vertical angle oscillation, $\langle  \theta_y(t) \rangle$, is then fitted to the function
\small
\begin{equation}
    \langle  \theta_y(t) \rangle =  A_{B_z}\cos(\omega_a t + \phi) + A_{\mathrm{EDM}}\sin(\omega_a t + \phi) + c,
    \label{eq:theta_edm}
\end{equation}
\normalsize
where $A_{B_z}$ is the amplitude due to the longitudinal magnetic field, $A_{\mathrm{EDM}}$ is the EDM amplitude, and $c$ is an overall offset. 
\clearpage

\subsection{EDM blinding}
As seen from~\cref{eq:theta_edm}, the EDM amplitude is directly accessible. This is not desirable when analysing data, as the aim is to measure $A_{B_z}$ without revealing $A_{\mathrm{EDM}}$. A software-level blinding (see \cref{sec:blind}) can be applied to data by injecting an unknown EDM amplitude to $\theta_y$
\small
\begin{equation}
\theta_{y}^{\mathrm{Blinded}} = \theta_y + A_{\mathrm{EDM}}^{\mathrm{Blinded}}(\mathbf{G},S)\cdot\sin( \omega_at^{\mathrm{mod}} + \phi),
\label{eq:blind_edm}
\end{equation}
\normalsize
where $A_{\mathrm{EDM}}^{\mathrm{Blinded}}$ is the unknown injected amplitude, which is a function of a set of parameters, $\mathbf{G}$, and a blinding string, $S$. $\mathbf{G}$ is set to produce a value of $A_{\mathrm{EDM}}$ comparable to the EDM limit set at the \ac{BNL} experiment ($\sim10^{-19} \ e\cdot{\mathrm{cm}}$), while $S$ allows a unique blinding per dataset.

\begin{figure}[htpb]
    \centering
    \subfloat[]{\includegraphics[width=0.49\linewidth]{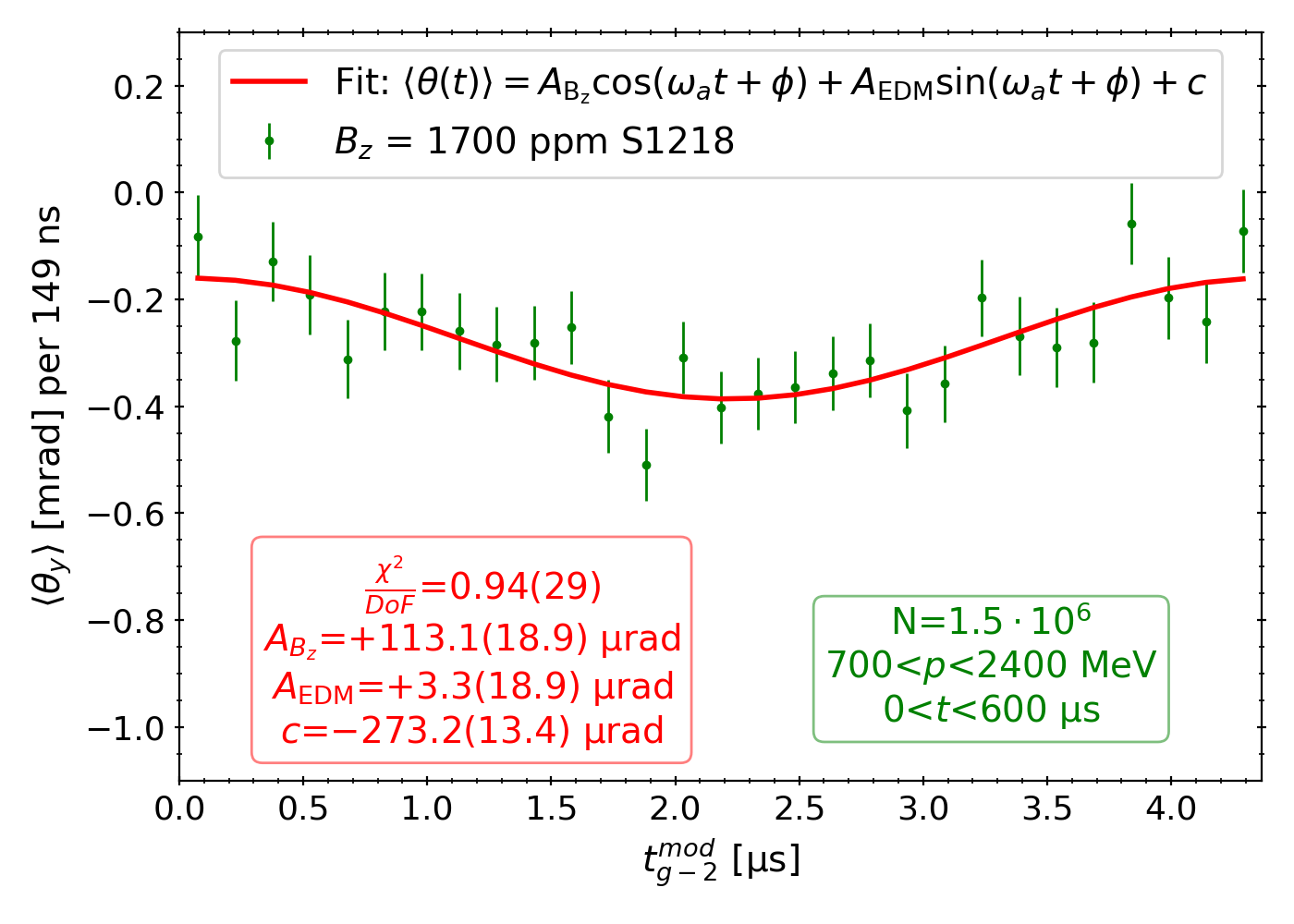}}
    \subfloat[]{\includegraphics[width=0.49\linewidth]{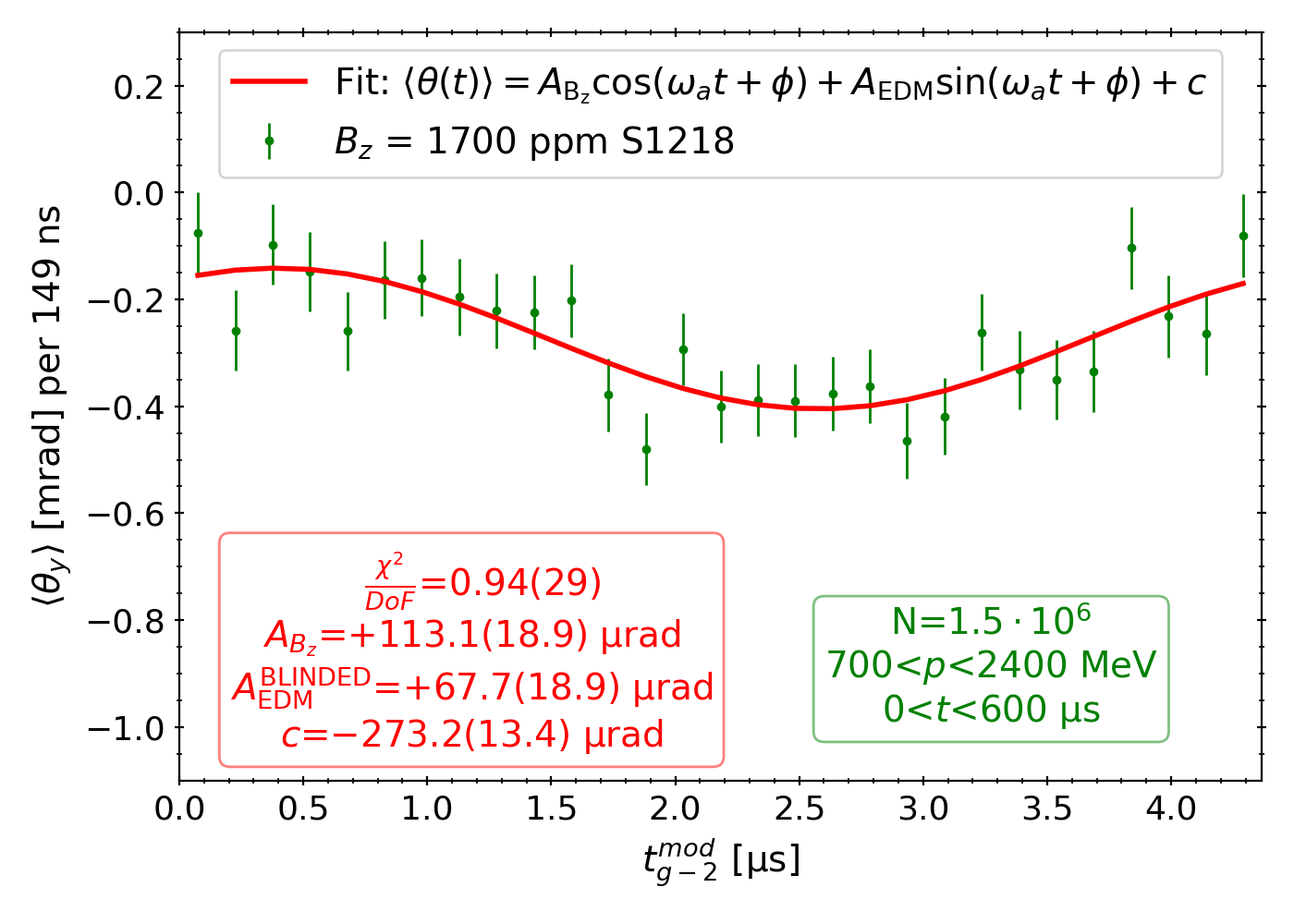}}
    \vspace{-0.2cm}
    \caption[Testing the blinding in the simulation]{Testing the blinding in the simulation with a large truth $B_z$ and no input EDM: a) unblinded, and b) blinded.}
    \label{fig:blind}
\end{figure}

The blinding procedure can be verified directly using the simulation, by comparing the fit-parameters before and after the blinding, as shown in~\cref{fig:blind}. The unblinded fit parameters ($A_{B_z}$ and $c$), as well as $\frac{\chi^2}{\mathrm{DoF}}$, remain unchanged, while $A_{\mathrm{EDM}}$ is blinded.
\vspace{-0.4cm}

\vspace{-0.3cm}
In data, a comparison was made between the blinding from \cref{eq:blind_edm} and a \say{randomised double} blinding: $2 A_{\mathrm{EDM}}^{\mathrm{Blinded}} \times (1 +0.25 \times \mathrm{rand}[0,1])$. This procedure allowed an additional unknown signal, without revealing the original blinding. 
\clearpage

\subsection{Fitting in data}
\begin{figure}[htpb]
    \centering
    \subfloat[]{\includegraphics[width=0.46\linewidth]{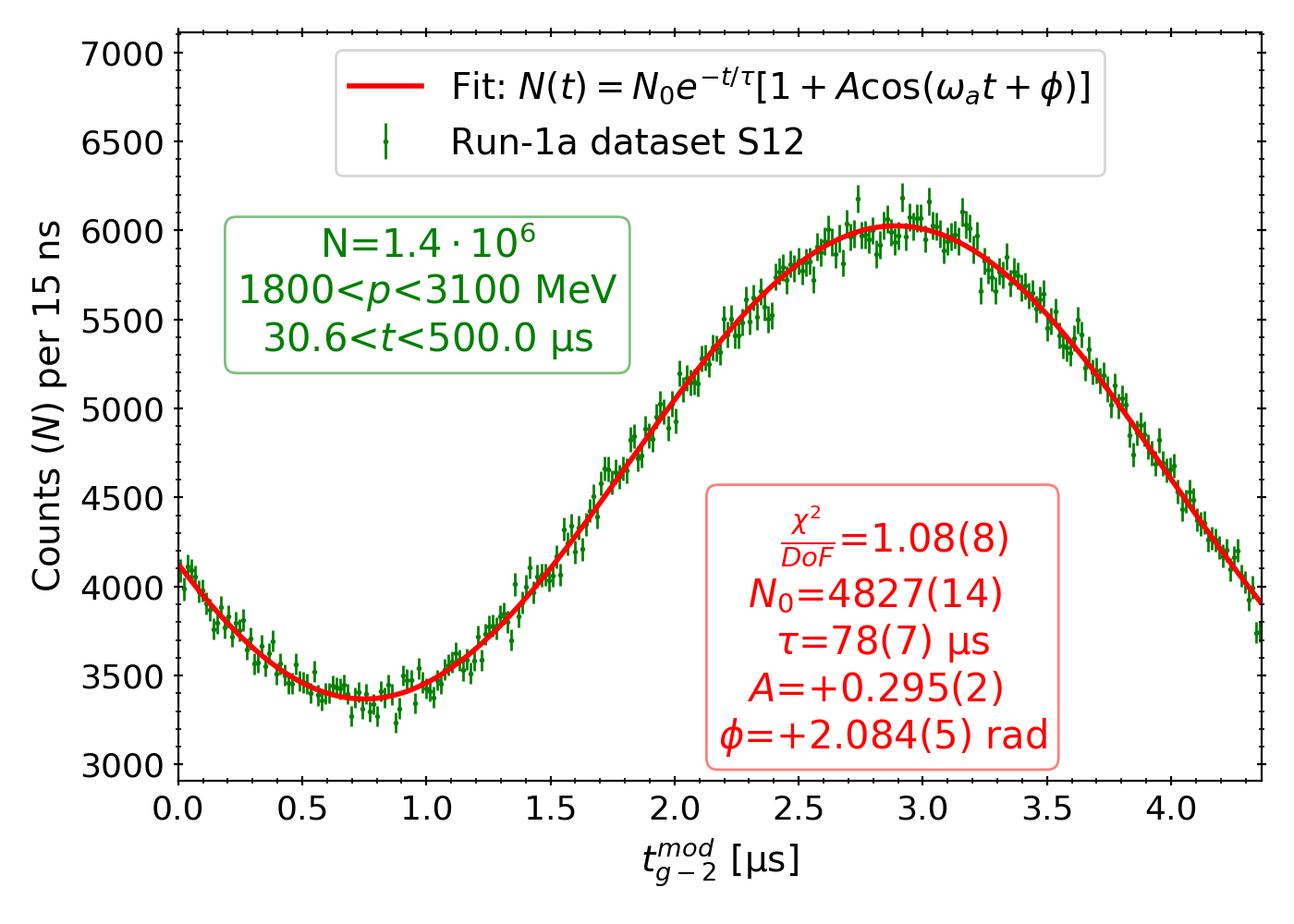}\label{fig:s12_60h_bz_1}}
    \subfloat[]{\includegraphics[width=0.46\linewidth]{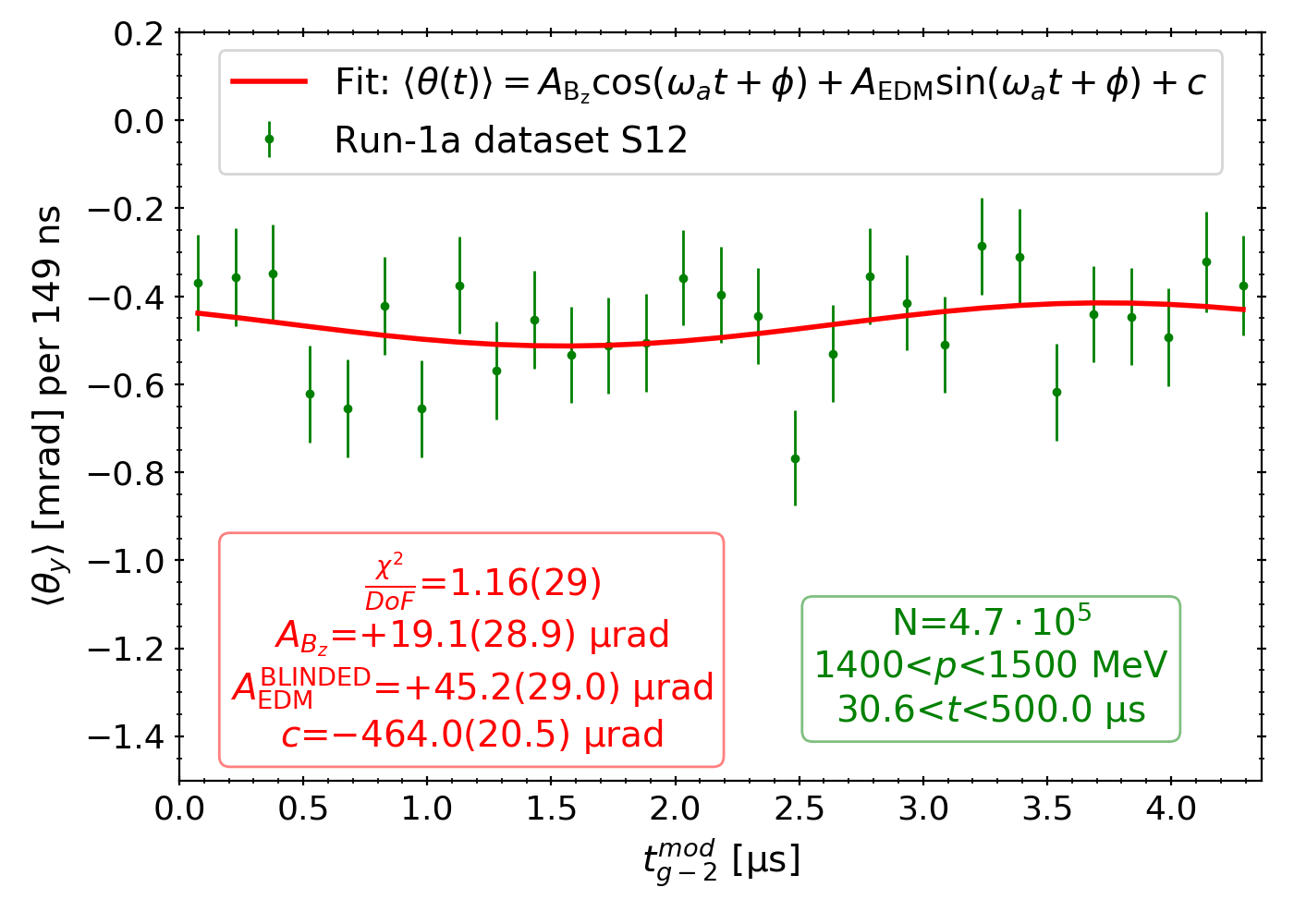}\label{fig:s12_60h_bz_2}} 
    \vspace{-0.2cm}
    \caption[Fit results in the Run-1a dataset (S12)]{Fit results in station 12 in the Run-1a dataset for: a) the number oscillation, and b) $\langle \theta_y \rangle$ oscillation.}
    \label{fig:bz_fits}
\end{figure}

The fit of \cref{eq:5_par} to the number of quality tracks is shown in~\cref{fig:s12_60h_bz_1}. This plot is analogous to the \say{wiggle plot} in \cref{fig:S1218_5par}, but with the time modulation of \cref{eq:t_mod} applied. The subsequent fit of \cref{eq:theta_edm} to the vertical angle oscillation, using the value of $\phi$ from \cref{fig:s12_60h_bz_1}, is shown in~\cref{fig:s12_60h_bz_2}, for tracks in a 100~MeV range.
\vspace{-0.55cm}
\vspace{-0.55cm}
\subsection{Momentum-binned analysis}
\begin{figure}[htpb]
    \centering
    \subfloat[]{\includegraphics[width=0.49\linewidth]{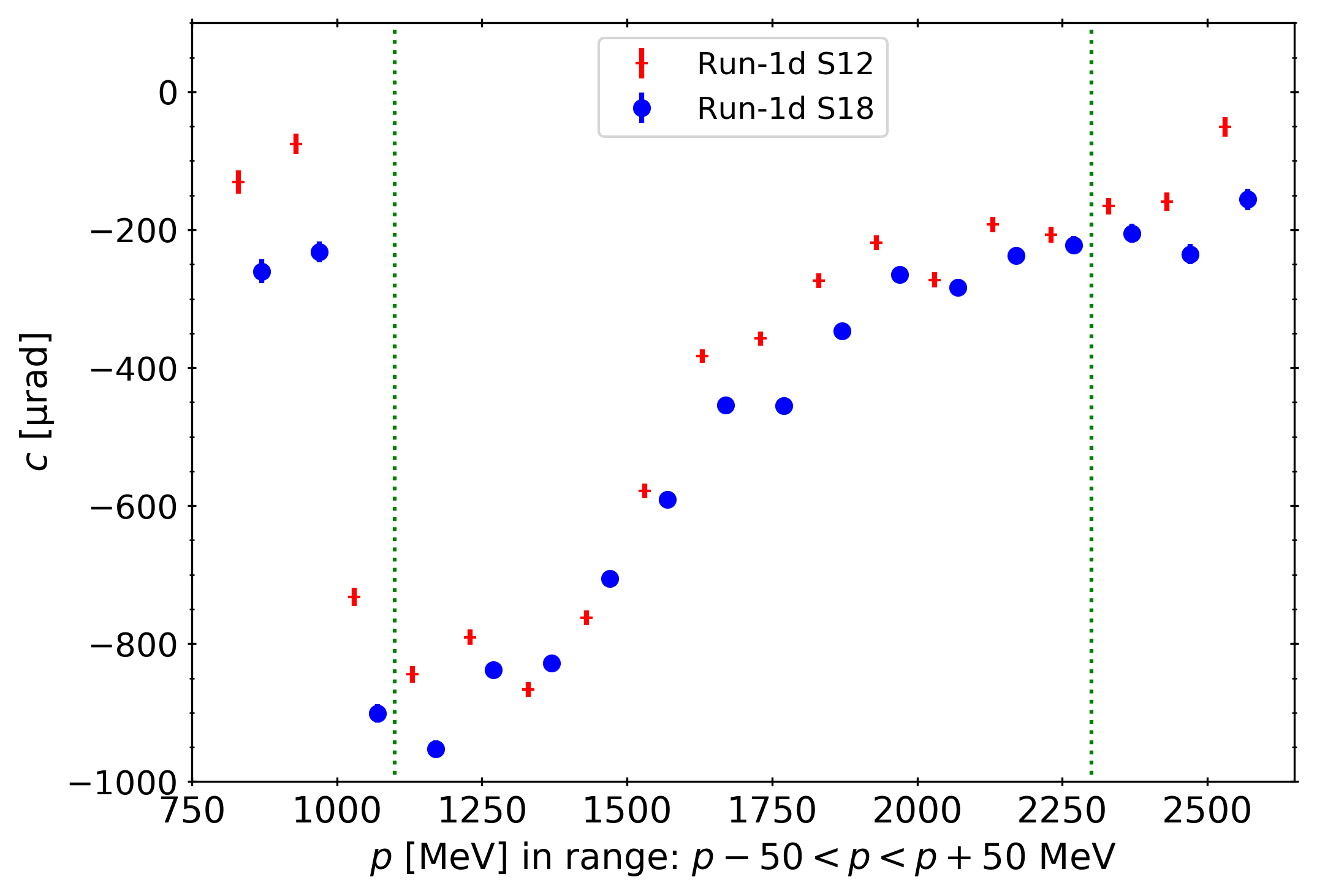} \label{fig:c}}
    \subfloat[]{\includegraphics[width=0.49\linewidth]{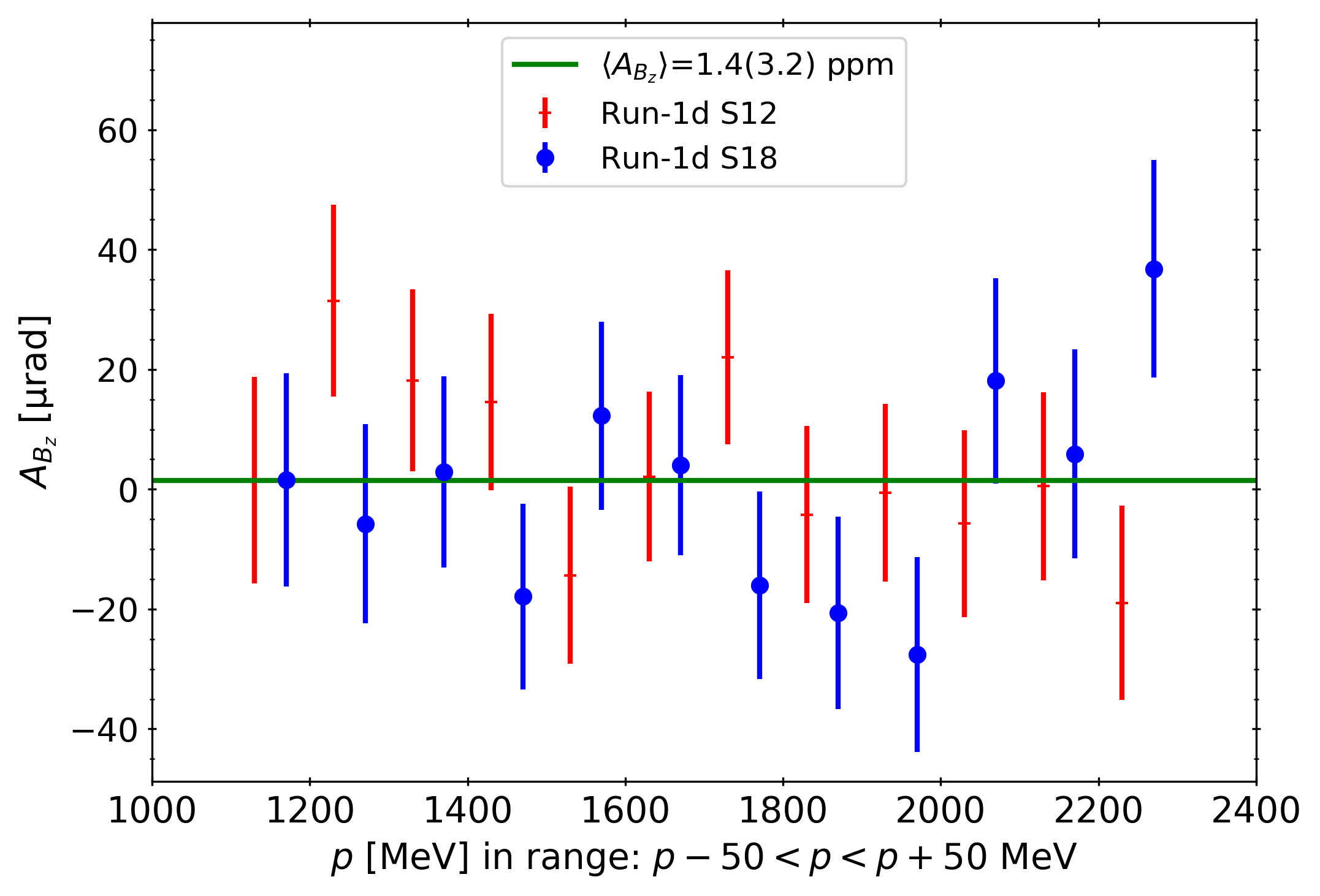} \label{fig:A_B_z}}
    \vspace{-0.2cm}
    \caption[Momentum-binned analysis fit results]{Momentum-binned analysis fit results in Run-1d dataset. The tracks are split by momentum in 100~MeV bins, with the bin centre values displayed on the $x$-axis. (a) The overall offset (parameter $c$) as the function of momentum. (b) $A_{B_z}$ fit parameter in each momentum bin, and the uncertainty-weighted mean ($\langle A_{B_z} \rangle$).}
\end{figure}

The fit in \cref{fig:s12_60h_bz_2} is performed in a narrow momentum range (rather than a simultaneous fit to all data) due to a variation in the overall offset in the vertical angle ($c$) as a function of momentum, as demonstrated in \cref{fig:c}. This is due to the vertical beam motion, discussed in \cref{sc:bd}. Negative $c$ is indicative of the centre of the beam being vertically lower than the centre of the detector. The global alignment effects were excluded as a cause, due to the two stations measuring the same variation in $c$, while having different corrections from the alignment (see \cref{tab:global}).

Tracks of lower momentum exhibit an even more significant change in $c$ and therefore are not considered in this analysis. However, due to the optimal momentum cuts for this analysis (see \cref{sub:momentum_cuts}), and the effect of dilution described in \cref{sub:asymmetry_dilution}, the lowest and highest momentum tracks do not contribute significantly to the measurement of $A_{B_z}$ -- regardless of the observed change in $c$ -- and represent only a small population of all tracks (c.f. \cref{fig:Tracker-Momentum}). 

\clearpage
Therefore, fits in individual momentum bins of 100~MeV are performed to measure $A_{B_z}$, separately for each station and dataset, in the momentum range of 1100~MeV to 2300~MeV, as shown in \cref{fig:A_B_z} for the Run-1d dataset. 
\vspace{-0.2cm}

\begin{figure}[htpb]
    \centering
    \subfloat[]{\includegraphics[width=0.49\linewidth]{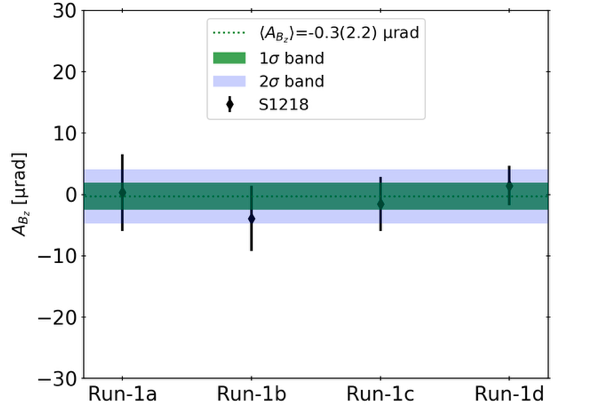}\label{fig:A_bz}}
    \subfloat[]{\includegraphics[width=0.49\linewidth]{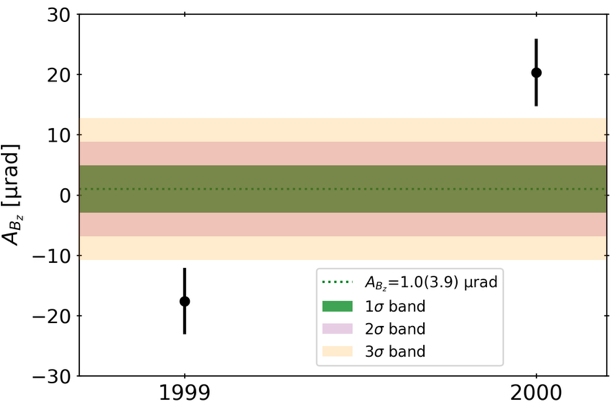}\label{fig:A_bz_BNL}}
    \vspace{-0.2cm}
    \caption[\R1 and the BNL $A_{B_z}$ fitting results]{(a) \R1 $A_{B_z}$ fitting results. The mean result in each dataset is in agreement with the overall \R1 mean within $1\sigma$. (b) The equivalent results from 1999 and 2000 measurements at the BNL \gm2 experiment~\cite{BNL_EDM}.}
\end{figure}

\vspace{-0.2cm}
$\langle A_{B_z} \rangle$ per dataset is determined as the uncertainty-weighted mean. The fitting results in the four \R1 datasets are summarised in \cref{fig:A_bz}, while the equivalent results in 1999 and 2000 at the \ac{BNL}~\cite{BNL_EDM} experiment are summarised in \cref{fig:A_bz_BNL}.
\vspace{-0.2cm}
\clearpage

\subsection{Optimising cuts selection}\label{sub:cuts}
The results in \cref{fig:bz_fits} were obtained with a particular choice of cuts: fit-start and end-times ($30.6~\SI{}{\micro\second} < t < 500.0$~\SI{}{\micro\second}) and momentum ($1800$~MeV $< p < 3100$~MeV, for the number oscillation). Moreover, the analysis relied on a choice of the \gm2 period ($T_{g-2}\approx4.4~\SI{}{\micro\second}$), as well as the phase ($\phi$) extracted from the fit of \cref{eq:5_par}. In this section, these choices will be motivated, and their stability assessed. 

\subsubsection{Dependence of fit on data subset}\label{sub:parameter_scan}
In order to assess the robustness of the fit, fits can be performed to different subsets of the data and the results compared. For example, varying the fit start-time, where data at a later start-time is a subset of data at an earlier time. The difference in a given fit parameter for the two datasets has an allowed difference, $\sigma_{\Delta_{21}}$, where 2 denotes the smaller dataset and 1 the larger, is given by
\begin{equation}
    \langle(x_{1i}-x_{2i})^2\rangle = \sigma^2_{2i} - \sigma^2_{1i}  \rightarrow \sigma_{\Delta_{21}} = \sqrt{\sigma^2_{2i} - \sigma^2_{1i}}.
    \label{eq:Kawall}
\end{equation}
This equation is valid for any fit function, regardless of the number of parameters and possible correlations among them~\cite{BNL_stats}. The results for $A_{B_z}$ with different fit start-times are shown in \cref{fig:start}.  

\subsubsection{Fit start-time}
\begin{figure}[htpb]
    \centering
    \subfloat[]{\includegraphics[width=0.49\linewidth]{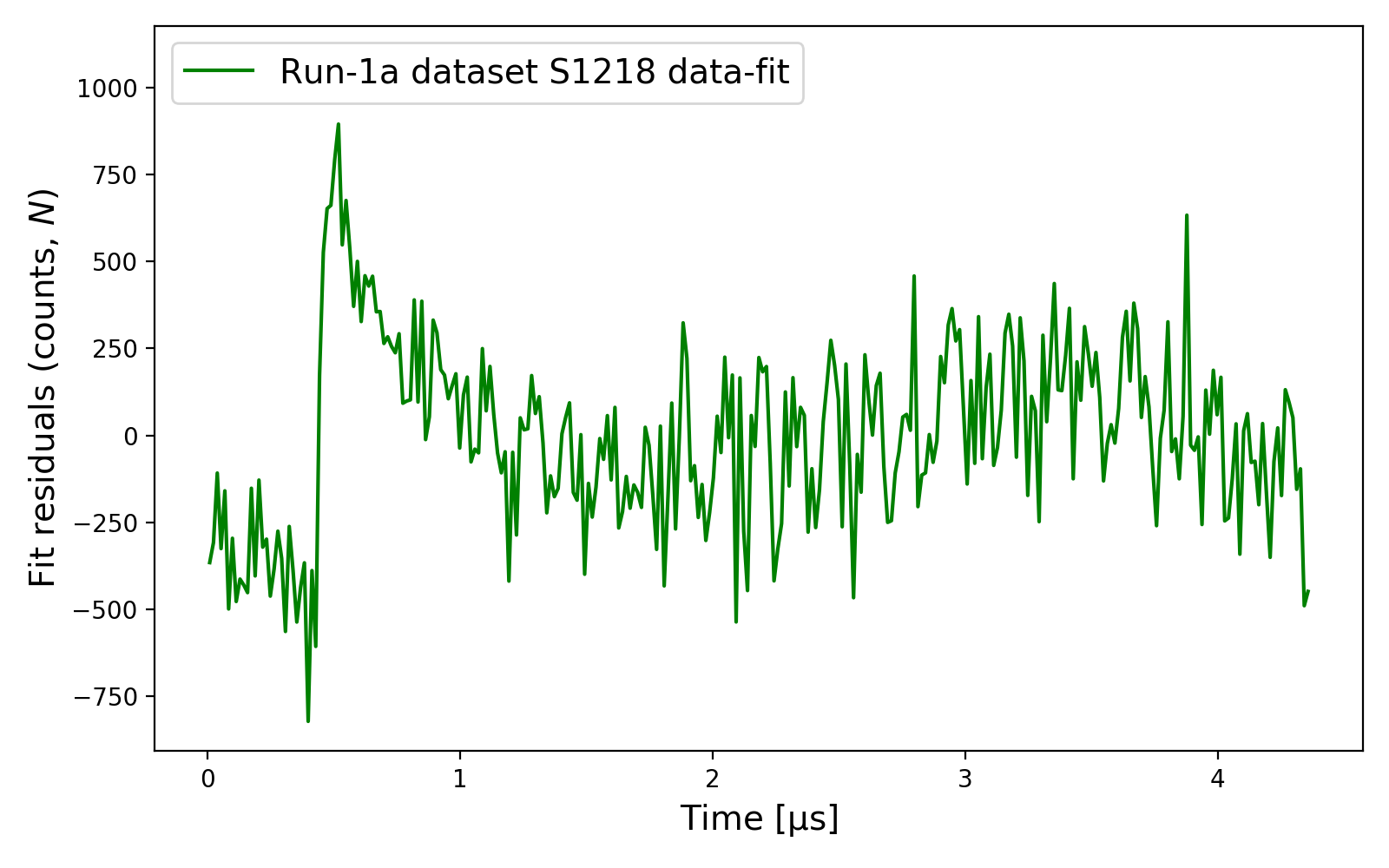}\label{fig:count_res_1}} 
    \subfloat[]{\includegraphics[width=0.49\linewidth]{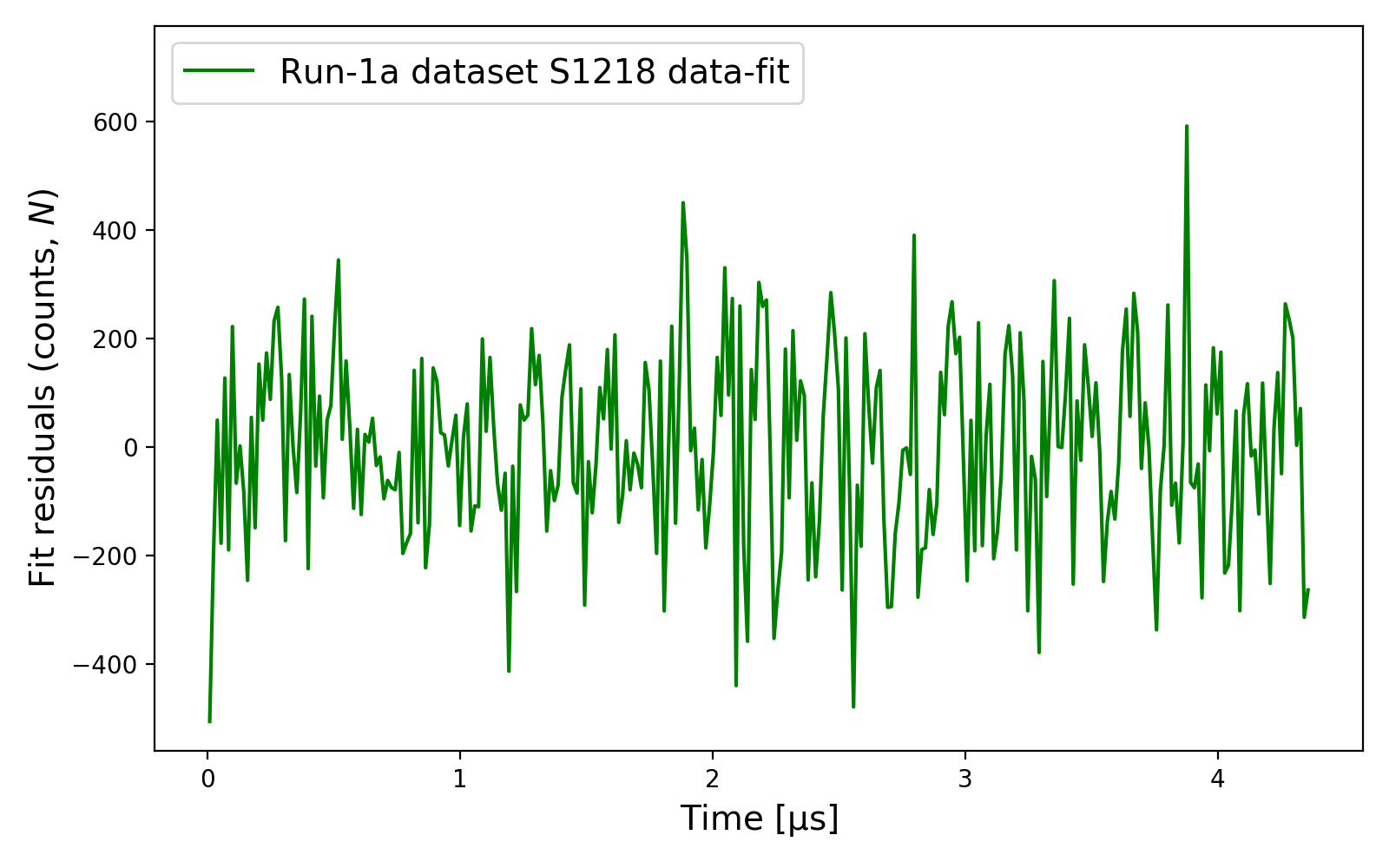}\label{fig:count_res_2}} 
    \caption[Fit residuals]{Fit residuals using different fit start-times. (a) The fit start-time is not a multiple of $T_{g-2}$. This results in a large feature at \SI{0.5}{\micro\second}. (b) The residual spectrum is uniform, with a correct choice of the fit start-time.}
    \label{fig:count_res}
\end{figure}

First of all, a careful choice of the fit start-time is necessary -- it should be a multiple of $T_{g-2}$ because data is modulated using \cref{eq:t_mod}. This is demonstrated in \cref{fig:count_res}.

\begin{figure}[htpb]
    \centering
    \includegraphics[width=0.6\linewidth]{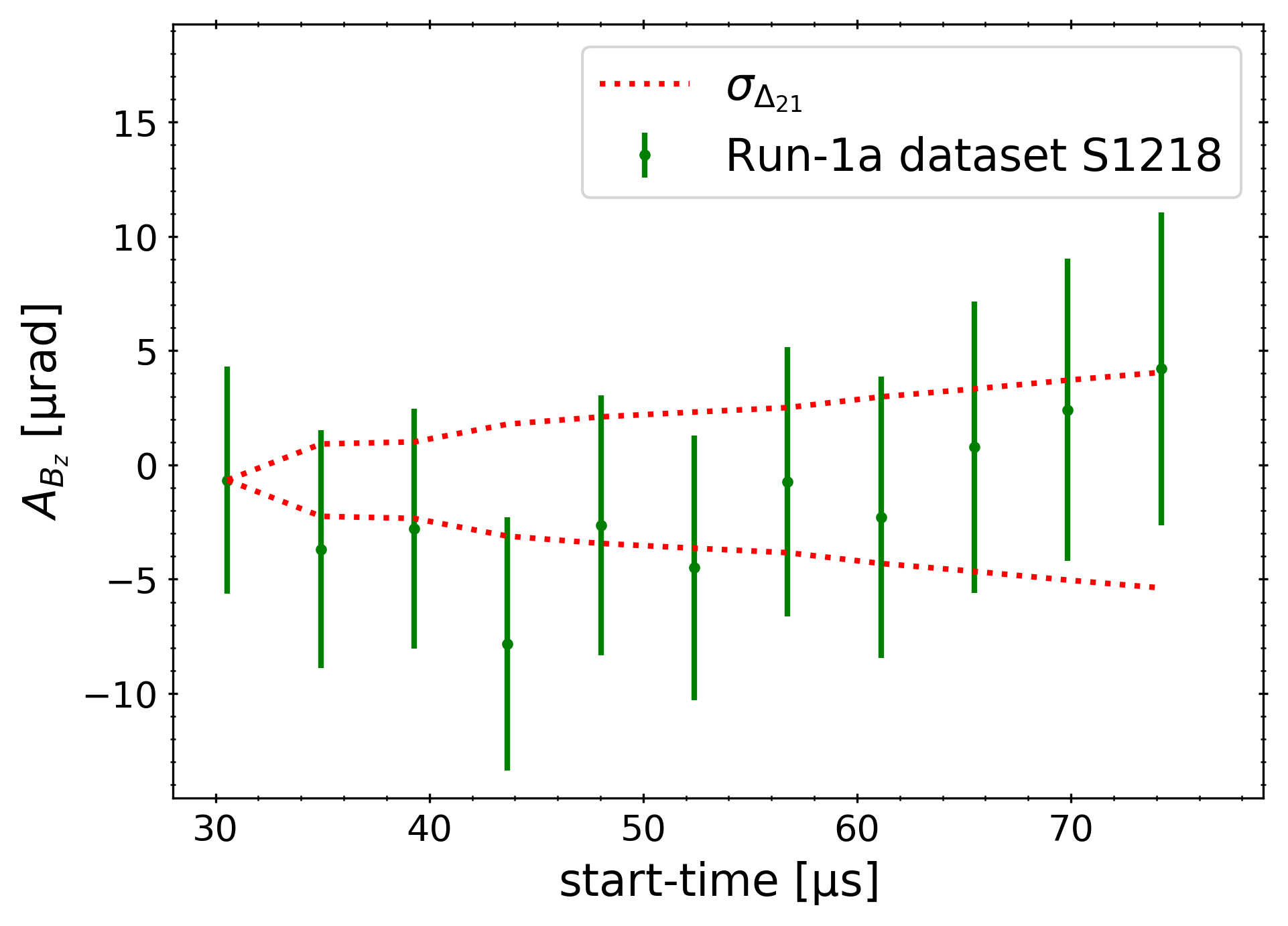}
    \caption[Fit start-time scan]{The variation of $A_{B_z}$ with the the fit start-time. Each fit start-time in this plot is a multiple of $T_{g-2}$. The red band, given by \cref{eq:Kawall}, indicates the allowed set-subset variation.}
    \label{fig:start}
\end{figure}

The second requirement is that the fit start-time must be greater than \SI{30}{\micro\second} (see \cref{sc:scraping}), and should be as low as possible to minimise the statistical uncertainty. A fit start-time of \SI{30.56}{\micro\second} was therefore chosen, with the variation of $A_{B_z}$ at later fit start-times shown in \cref{fig:start}. 

\clearpage

\subsubsection{Momentum cuts}\label{sub:momentum_cuts}
\begin{figure}[htpb]
    \centering
    \subfloat[]{\includegraphics[width=0.49\linewidth]{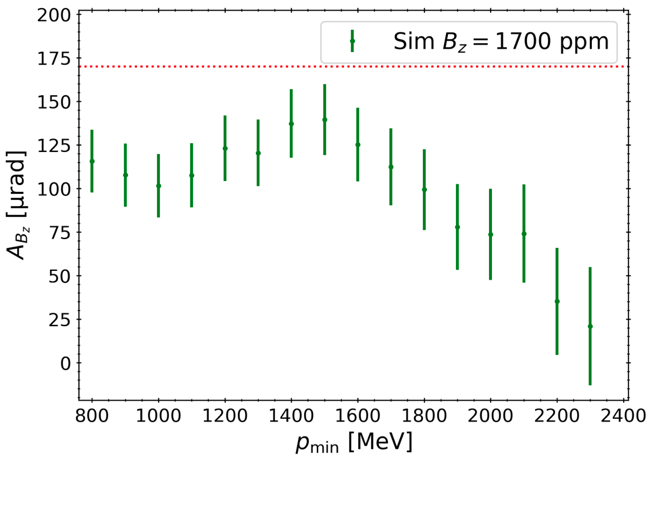}} 
    \subfloat[]{\includegraphics[width=0.49\linewidth]{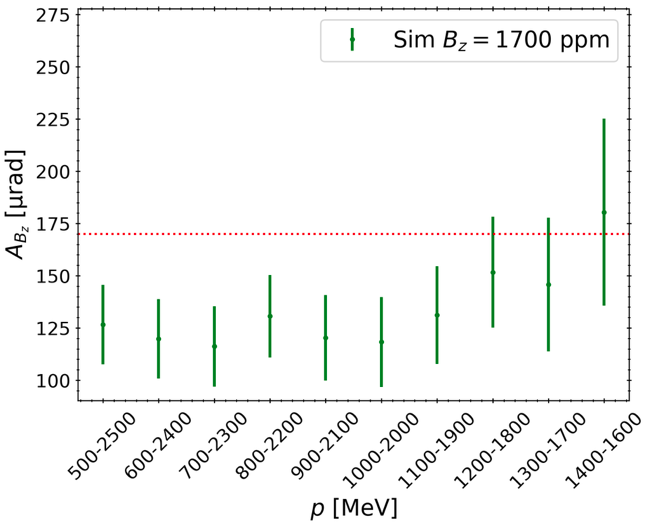}} 
    \caption[Establishing cuts in the simulation]{Simulation results with $B_z=1700$~ppm. $A_{B_z}$ versus momentum cut: a) asymmetric and b) symmetric. Symmetric cuts give a less biased determination of $A_{B_z}$, and particularly when the momentum dependent dilution factor (see \cref{sub:asymmetry_dilution}) is accounted for.}
    \label{fig:cuts_cuts}
\end{figure}

\cref{eq:theta_edm} is fitted in momentum bins of 100~MeV due to the strong momentum dependence of $c$. Using the simulation with a truth value of $B_z=1700$~ppm, it was verified that a symmetric cut on momentum is more appropriate than an asymmetric one (see \cref{sub:momentum_cuts_wiggle}), as shown in \cref{fig:cuts_cuts}.
\clearpage

\begin{figure}[htpb]
    \centering
    \includegraphics[width=0.55\linewidth]{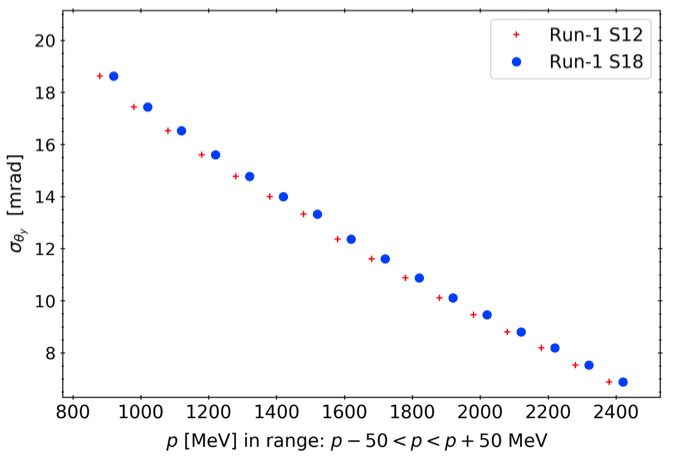}
    \caption[The change in $\sigma_{\theta_y}$ with $p$]{The change in $\sigma_{\theta_y}$ with $p$ in momentum bins of 100~MeV.}
    \label{fig:simga_bz}
\end{figure}

Tracks of higher momentum predominantly originate further away from a station, as seen in \cref{fig:arc}. Therefore, high momentum tracks with large angles ($\theta_y$) miss the tracker station -- this can be thought of as a \say{cone effect}. As a consequence, the \ac{SD} of $\theta_y$ ($\sigma_{\theta_y}$) is narrower for higher momentum tracks, as compared with lower momentum tracks, as measured by the detectors. This is demonstrated in \cref{fig:simga_bz}.

\begin{figure}[htpb]
    \centering
    \subfloat[]{\includegraphics[width=0.53\linewidth]{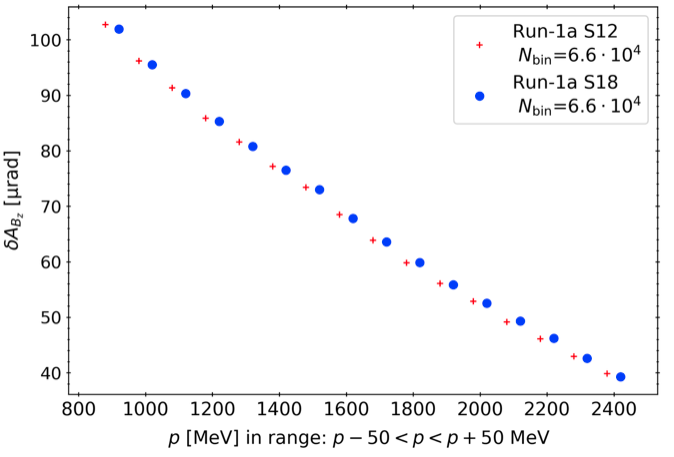}\label{fig:prec_1}} \\
    \subfloat[]{\includegraphics[width=0.53\linewidth]{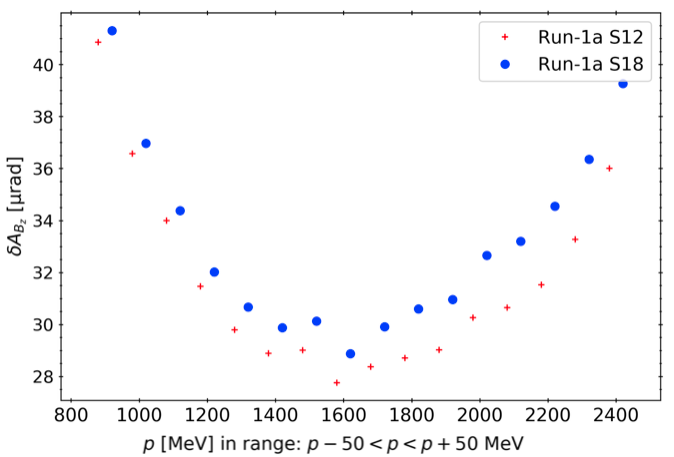}\label{fig:prec_2}} 
    \vspace{-0.2cm}
    \caption[The change in $\delta A_{B_z}$ with $p$]{The change in $\delta A_{B_z}$ with $p$. (a) Fixed number of tracks, $N_{\mathrm{bin}}=6.6\times10^4$, in each momentum bin. (b) The non-fixed number of tracks per bin follows the momentum distribution of \cref{fig:Tracker-Momentum}. }
\end{figure}

This effect directly leads to a reduction in the uncertainty of the $A_{B_z}$ parameter in the fit at high momentum, as shown in \cref{fig:prec_1} where each momentum bin has the same number of tracks.

\clearpage
However, considering a non-fixed number of tracks in each momentum bin results in the distribution shown in \cref{fig:prec_2}, where past 1600~MeV the $\delta A_{B_z}$ is increasing with a deceasing number of tracks per bin.

\vspace{-0.2cm}
The mid-momentum tracks are thus the most sensitive in measuring $A_{B_z}$. A momentum range of 1100~MeV to 2300~MeV is therefore used to fit \cref{eq:theta_edm}, with separate fits performed in 100~MeV momentum bins.

\subsubsection{Assessment of period and phase}
\begin{figure}[htpb]
    \centering
    \subfloat[]{\includegraphics[width=0.49\linewidth]{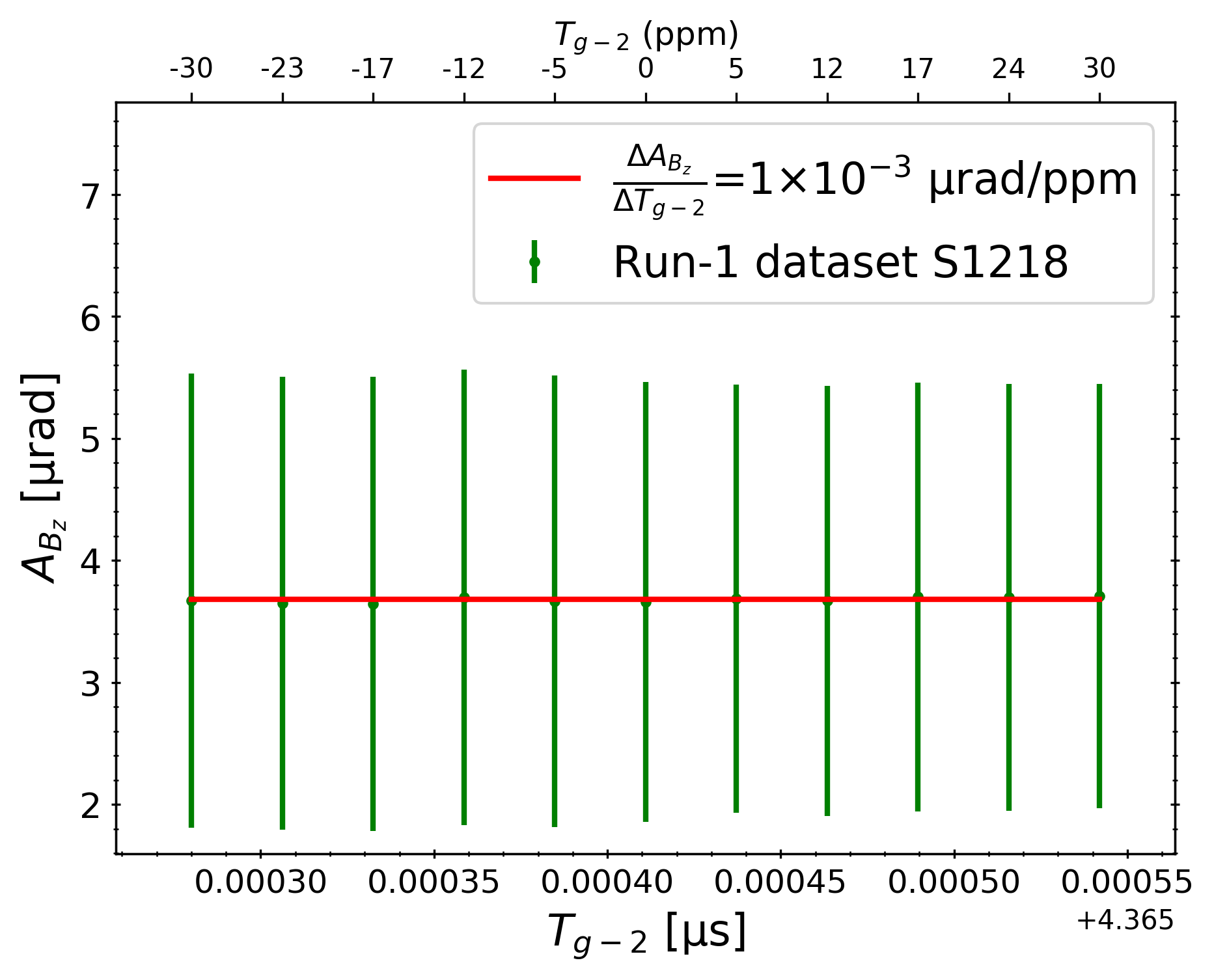}\label{fig:period}}
    \subfloat[]{\includegraphics[width=0.49\linewidth]{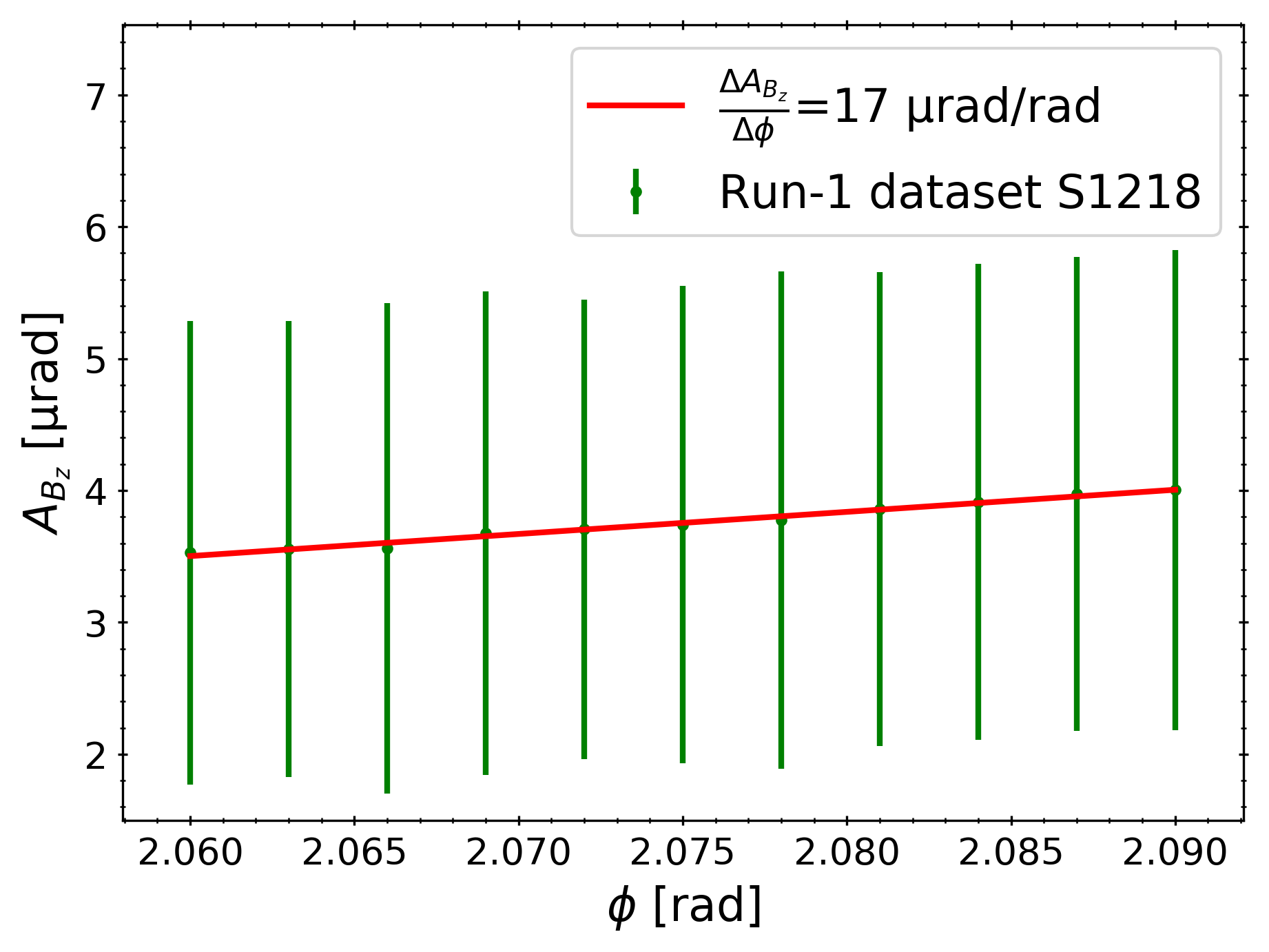}\label{fig:phi}}
    \caption[The change in $A_{B_z}$ due to $T_{g-2}$ and $\phi$]{The change in $A_{B_z}$ due to the variation of: a) $T_{g-2}$ and b) $\phi$.}
\end{figure}

In order to modulate data using \cref{eq:t_mod}, $T_{g-2}$ must be known a priori to a reasonable precision. Using the \gm2 frequency $f_a=0.2290735$~MHz, as measured by the BNL experiment~\cite{BNL_AMM}, yields the value of $T_{g-2}=$~\SI{4.365411}{\micro\second}. This choice can be evaluated by changing $T_{g-2}$ by $-30$~ppm to $+30$~ppm and recording the change in $A_{B_z}$, as shown in \cref{fig:period}. 
\clearpage

The small variations in \cref{fig:period} imply that there is no real systematic effect there. However, extracting the slope of the fit allows an assessment of the magnitude of the systematic uncertainty from the choice of $T_{g-2}$, as described in \cref{sub:syst}, to be made. 

The change in $A_{B_z}$ due to the variation of $\phi$ is shown in \cref{fig:phi}.

\subsubsection{Systematic uncertainty evaluation}\label{sub:syst}
Small variations of $A_{B_z}$ due to the choices of $T_{g-2}$ and $\phi$ were observed. The systematic uncertainty for each, and the total systematic uncertainty, on $A_{B_z}$ are summarised in \cref{tab:a_bz_sys}. Slopes from \cref{fig:period,fig:phi} were used in the estimation, as well as the error from the fit on $\phi$ in \cref{fig:s12_60h_bz_1}. The total systematic uncertainty of $\sim\SI{0.1}{\micro\radian}$ was found to be negligible compared to the statistical uncertainty of $\SI{2.2}{\micro\radian}$ (see \cref{fig:A_bz}). 
\begin{table}[htpb]
\centering  
\begin{tabular}{cc}
\toprule
Systematic source & Value [$\SI{}{\micro\radian}$] \\ \toprule
$T_{g-2}$               & $0.022$  \\ \midrule
$\phi$                  & $0.085$  \\ \midrule \midrule
Total                   & $0.088$  \\ \bottomrule
\end{tabular}
\caption[Systematic uncertainties on $A_{B_z}$]{Systematic uncertainties on $A_{B_z}$.}
\label{tab:a_bz_sys}
\end{table}
\clearpage

\subsection{Conversion of \texorpdfstring{$A_{B_z}$}~into \texorpdfstring{$B_z$}~}\label{sub:asymmetry_dilution}
\begin{figure}[htpb]
    \centering
    \subfloat[]{\includegraphics[width=0.48\linewidth]{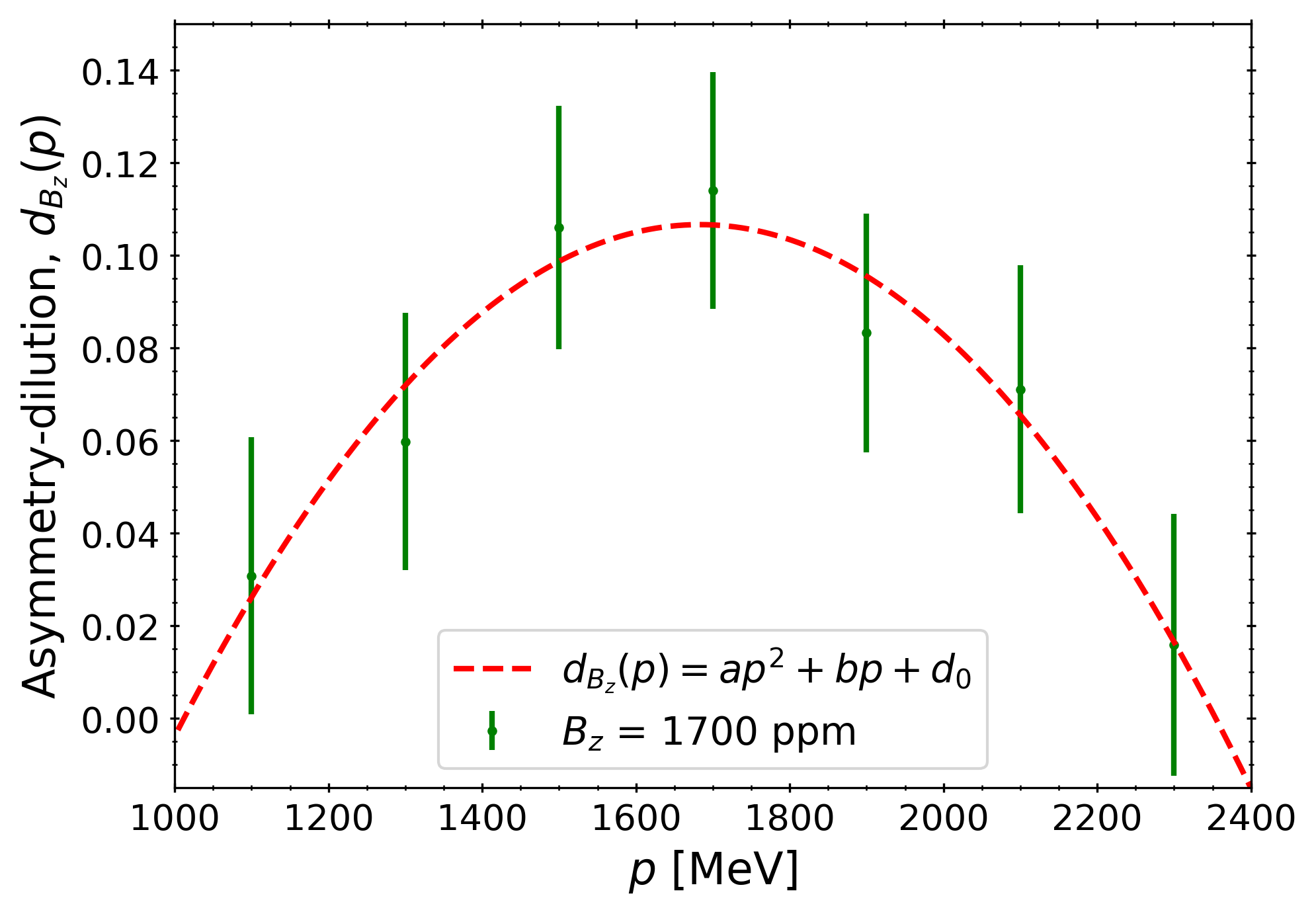}\label{fig:asym}}
    \subfloat[]{\raisebox{3mm}{\includegraphics[width=0.38\linewidth]{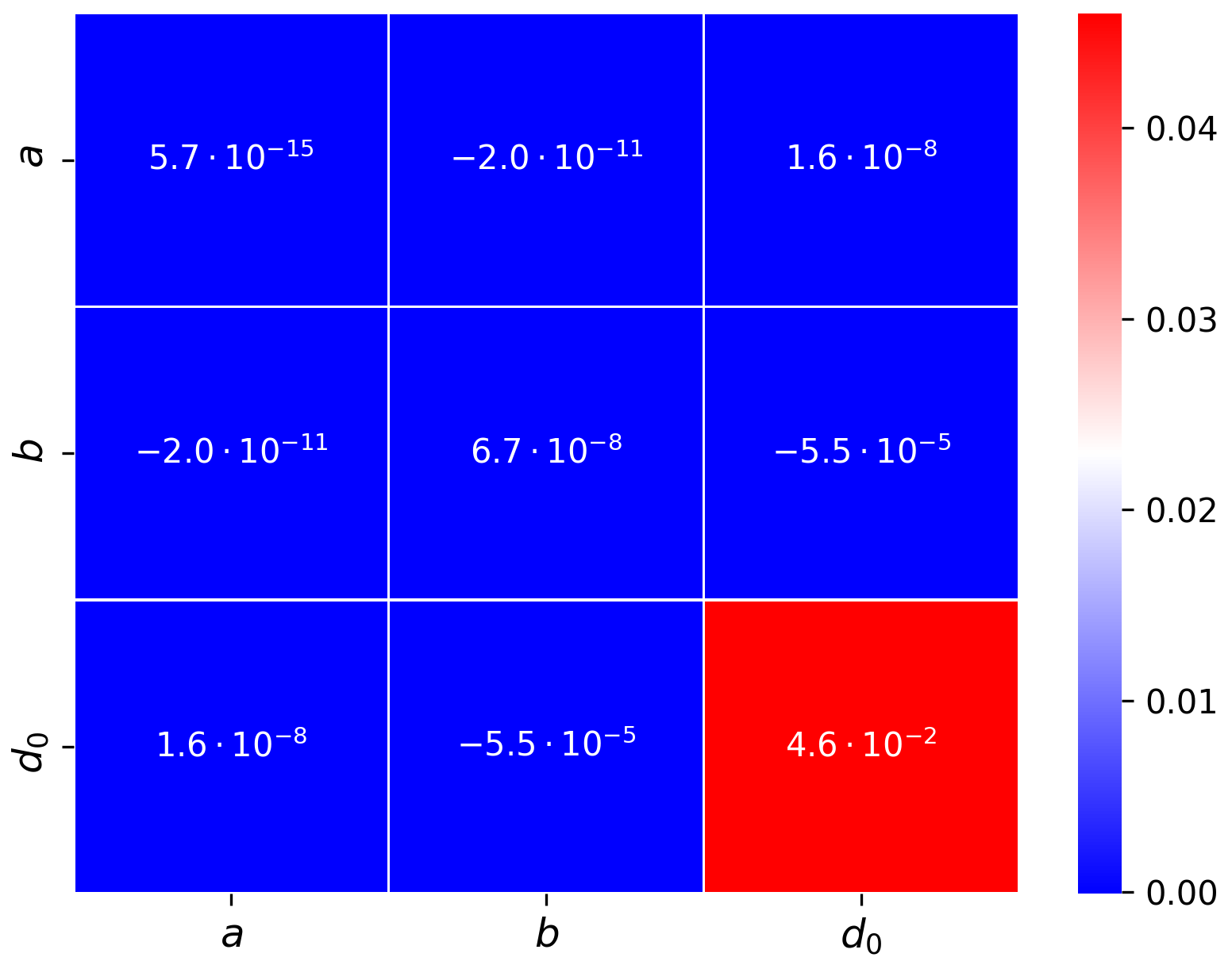}\label{fig:cov}}}
    \vspace{-0.2cm}
    \caption[The estimated asymmetry-dilution]{(a) The estimated asymmetry-dilution ($d_{B_z}(p)$) from the simulation, and a fit of \cref{eq:d}. (b) The covariance matrix of the fit parameters.}
    
\end{figure}

With the optimal cuts justified, and systematic effects on $A_{B_z}$ evaluated, it is now possible to convert the observed amplitude into $B_z$. $A_{B_z}$ is related to the precession-plane tilt in the lab-frame, $\delta '$, in the same way as $A_{\mathrm{EDM}}$ in \cref{eq:a_edm}. However, given that the muon motion is along $z$ (c.f.~\cref{fig:mf_tilt}), there is no Lorentz boost in this direction. Therefore, an estimate of $B_z$ can be derived directly from the measured angle in the lab-frame
\small
\begin{equation}
    \delta ' = \tan\left(\frac{B_z}{B_y}\right)  \approx \frac{B_z}{B_y} = \frac{A_{B_z}}{d_{B_z}(p)},
    \label{eq:bz_ppm}
\end{equation}
\normalsize
for $B_y \gg B_z$ which is the case. In \cref{eq:bz_ppm}, $d_{B_z}(p)$ is the asymmetry-dilution factor, analogous to $a_{\mathrm{EDM}}$ in \cref{eq:a_edm}. This asymmetry factor was estimated from the simulation, as shown in \cref{fig:asym}, where a parabolic function was fitted to the simulation data, given by
\small
\begin{equation}
    d_{B_z}(p)=ap^2+bp+d_0.
    \label{eq:d}
\end{equation}
\normalsize
\vspace{-1.2cm}

\vspace{-0.4cm}
Finally, the uncertainty on $\frac{B_z}{B_y}$ in momentum bin $i$ is given by
\small
\begin{equation}
    \delta \left(\frac{B_z}{B_y}\right)^i = \left|\frac{A^i_{B_z}}{d_{B_z}^i}\right| \sqrt{\left(\frac{\delta A^i_{B_z}}{A^i_{B_z}}\right)^2 + \left(\frac{\delta d^i_{B_z}}{d^i_{B_z}}\right)^2 },
    \label{eq:bz_error}
\end{equation}
\normalsize
where $\delta A^i_{B_z}$ is the fit error (e.g. \cref{fig:s12_60h_bz_2}), and $\delta d^i_{B_z}$ is computed using the variance of the parameters in \cref{fig:cov}.
\clearpage

\begin{figure}[htpb]
    \centering
    \subfloat[]{\includegraphics[width=0.49\linewidth]{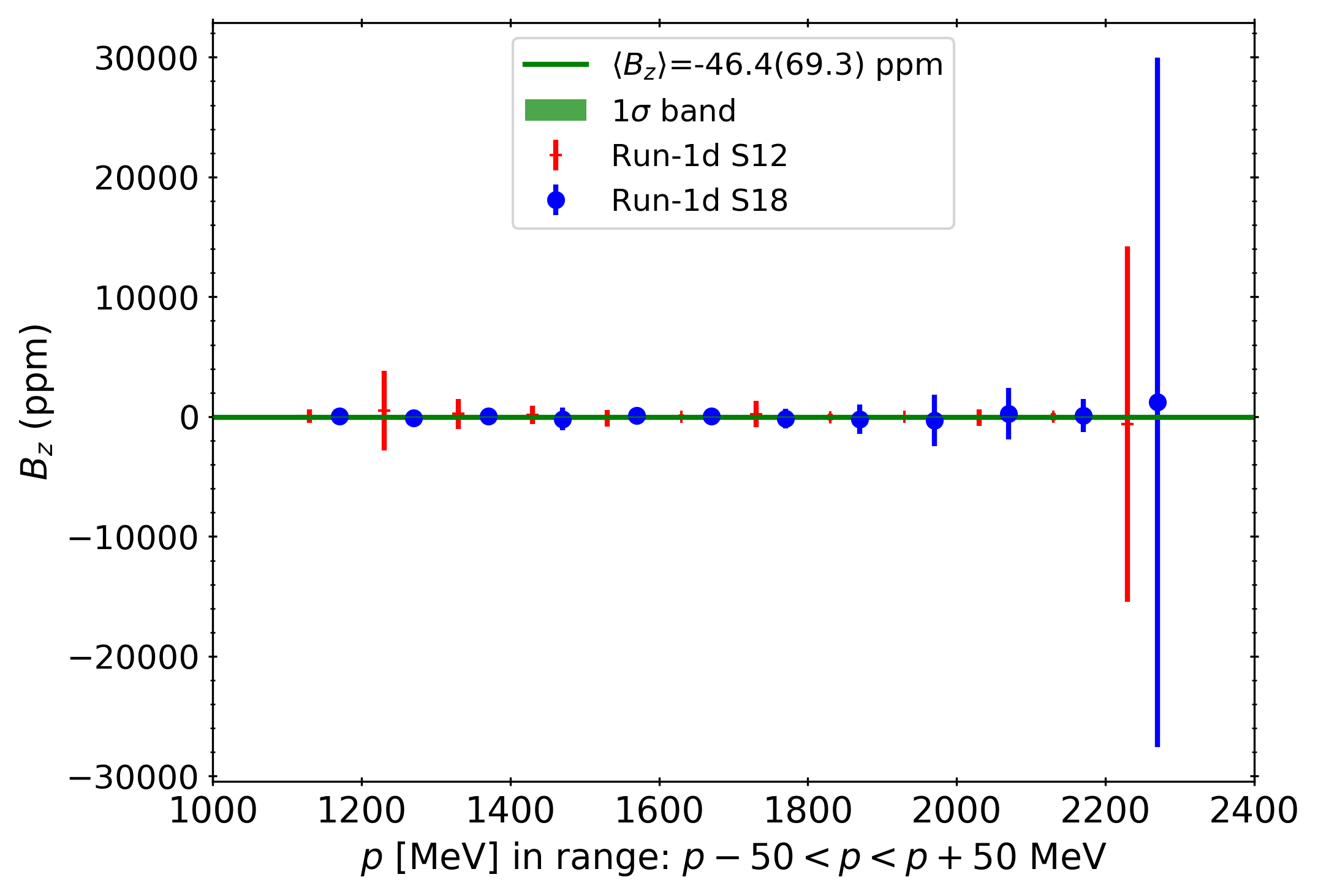}}
    \subfloat[]{\includegraphics[width=0.49\linewidth]{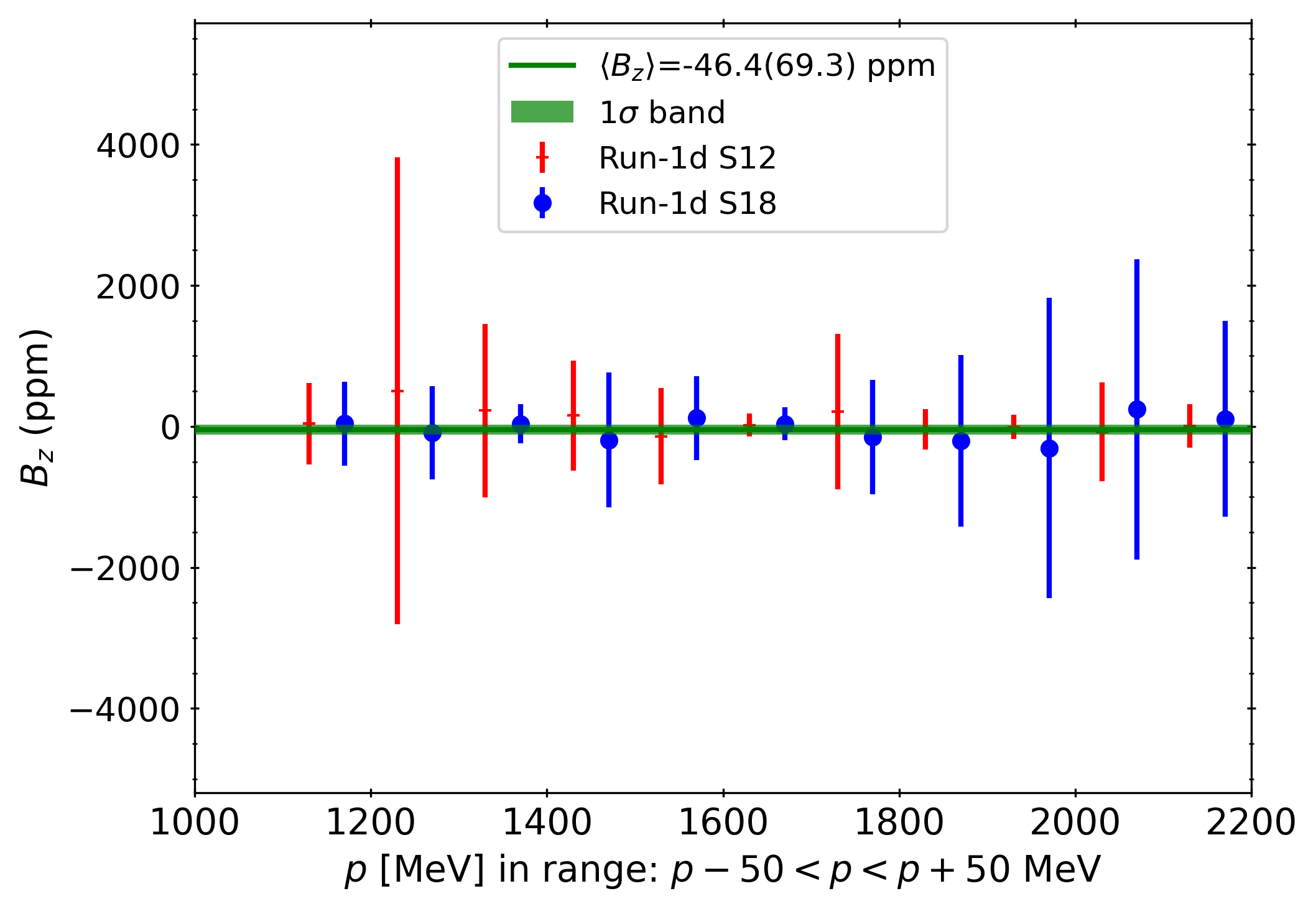}}
    \vspace{-0.2cm}
    \caption[$B_z$ results in Run-1d]{(a) $B_z$ results, in ppm, in each of the momentum bins in the Run-1d dataset. The uncertainty-weighted mean ($\langle {B_z} \rangle$) is indicated. The large uncertainties at high $p$ is the consequence of the small $d^i_{B_z}$ values in \cref{eq:bz_error} -- their contribution to $\langle {B_z} \rangle$ is minimal. (b) The zoomed version of the plot for $p < 2200$~MeV.}
    \label{fig:bz_bins}
\end{figure}

Using \cref{eq:bz_ppm,eq:bz_error}, the measured value of $A^i_{B_z}$ in each momentum bin in \cref{fig:A_B_z}, and the value of $d^i_{B_z}$ from \cref{eq:d}, $\frac{B_z}{B_y}$ was estimated (per momentum bin), as shown in \cref{fig:bz_bins}.
\vspace{-0.2cm}
\vspace{-0.2cm}
\subsection{Analysis results}\label{sub:analysis_results} 
\begin{figure}[htpb]
    \centering
    \includegraphics[width=0.61\linewidth]{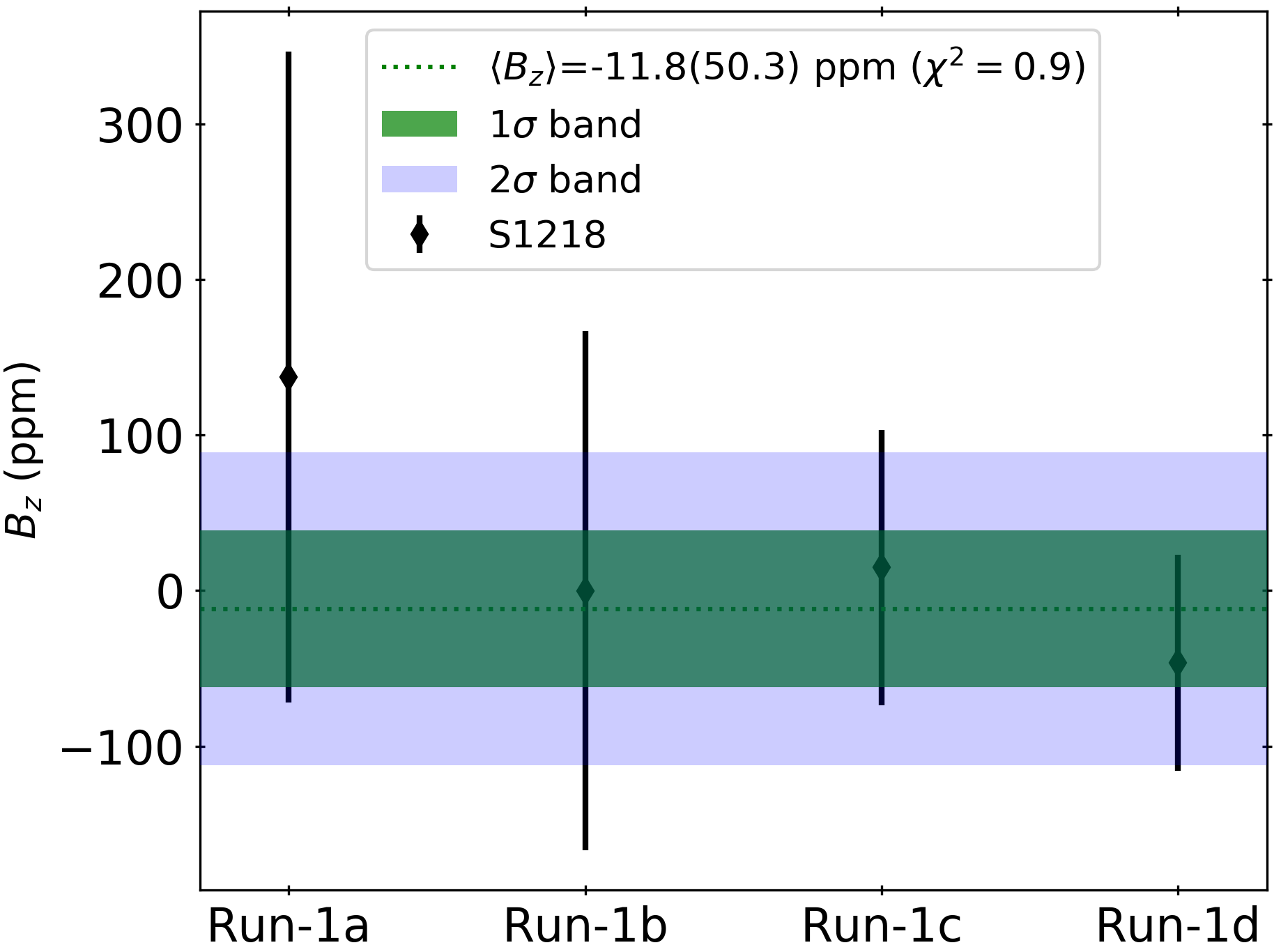}
    \vspace{-0.2cm}
    \caption[$B_z$ results in all \R1 datasets]{$B_z$ results, in ppm, in all \R1 datasets. The mean value of ${B_z}$ was obtained from a weighted mean of the four datasets, its $1\sigma$ uncertainty band is indicated in green and $2\sigma$ band in purple.}
    \label{fig:bz}
\end{figure}

The measured values of $B_z$ from the four \R1 datasets are summarised in \cref{fig:bz}. The results from the individual datasets are compared to the mean \R1 result of $\langle {B_z} \rangle=-11.8(50.3)$~ppm, and are found to be in agreement to better than $1\sigma$.
\vspace{-0.3cm}

\clearpage
\section{Outlook}
The analyses presented in this chapter - measurements of $B_x$ and $B_z$ using the tracking detectors - were not attempted at the BNL experiment. The preliminary measurement of $B_x$ confirmed that such a technique is viable using the tracking detectors, and yielded a value of $2.1(6)$~ppm in \R2.

The measured value of $B_z$ in \R1 is consistent with zero, with a precision of 50~ppm. Additionally, the measurement is in agreement with the direct measurement of $B_z=0.14(36.9)$~ppm~\cite{Rachel_mf} before \R1. 

\begin{figure}[htpb]
    \centering
    \includegraphics[width=\linewidth]{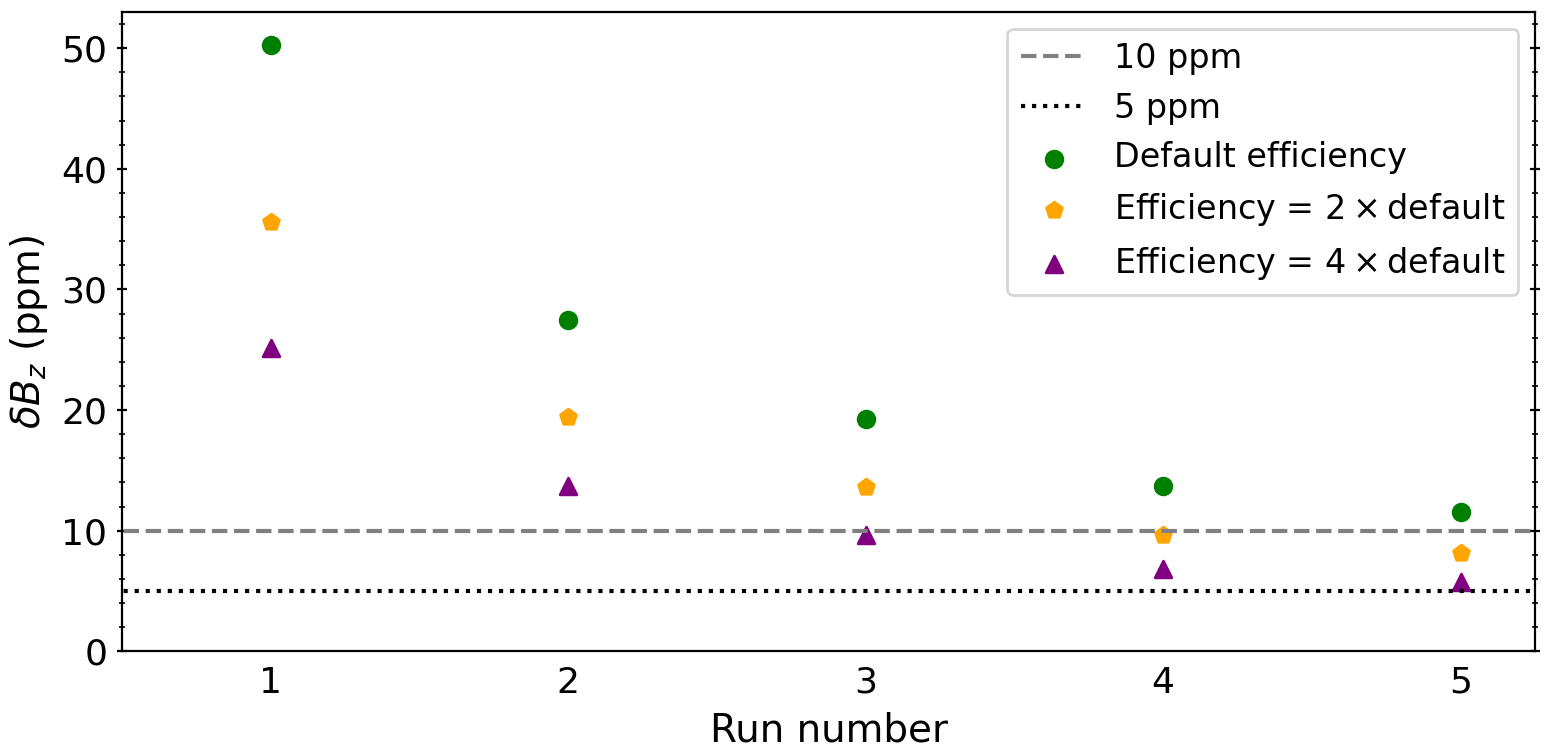} 
    \caption[Projected precision on $B_z$ with tracker data]{Projected precision on $B_z$ with an increase in the data volume.}
    \label{fig:money_bz}
\end{figure}

The precision ($\delta B_z=50$~ppm) will be improved with the increase in the number of reconstructed tracks by adding new data -- \R2 data-reconstruction is currently ongoing. Moreover, the track-reconstruction with an improved tracking efficiency is currently ongoing on the UK grid (see \cref{sub:track_production_in_the_uk}). The final precision is likely to reach 5~ppm by Run-5, as shown in \cref{fig:money_bz}. This is more precise than the design goal of $<7$~ppm, which would correspond to the uncertainty on $\omega_a$ of $<20$~ppb~\cite{Bill}. 

\graphicspath{{fig/}}

\chapter{Conclusions}
\label{ch:end}
The Fermilab Muon \gm2 experiment will measure the muon magnetic anomaly, with a precision of 140~ppb, and will search for the muon \ac{EDM}, with at least a factor of 10 improvement on the world's best measurement. The three data-taking periods, \R1, \R2, and \R3, have been completed, with preparations for \R4 currently ongoing. Essential in reducing the systematic uncertainty on the measurement of $\omega_a$ are the straw tracking detectors, which perform track extrapolation backwards to the muon decay point and forwards to the calorimeters. Moreover, systematic effects, such as the vertical pitch, require a correction that is accessible via a measurement of the vertical width of the beam by the tracking detectors. The beam profile from the tracking detectors is also convoluted with the magnetic field map to find the field experienced by the muons at the point of decay. 

In order to accurately reconstruct the beam profile, the tracking detectors must be correctly aligned. The alignment algorithms were validated using simulation, which converged after three iterations with $\mathcal{O}$($10^5$) tracks to within \SI{3}{\micro\metre} radially and \SI{6}{\micro\metre} vertically. The track-based internal alignment was implemented with data from \R1. The number of reconstructed tracks has increased by $3\%$ due to the position calibration from the alignment. Moreover, an improvement in the mean track \pval of $4\%$ was achieved. After the alignment procedure, the uncertainty contribution from the tracker misalignment to the pitch correction is now negligible. 

\clearpage
An alignment manual~\cite{Gleb_manual} allowing future alignment determinations, has been produced. The derived alignment constants were written into a \verb!PostgreSQL! database, where each set of constants is associated with a given range of runs. Additionally, the stability of the alignment results was verified using \R2 data. 

A potential \ac{EDM} of the muon would increase the observed $\omega_a$ signal and tilt the precession-plane of the muon. The tracking detectors will realise an \ac{EDM} measurement through the direct detection of an oscillation in the average vertical angle of the positron from the muon decay. An observation of a muon \ac{EDM} would be evidence of new physics and would provide a new source of \ac{CP} violation in the charged lepton sector. 

Essential in measuring the \ac{EDM}, as well as $\omega_a$, are accurate and precise estimations of potential non-zero radial and longitudinal magnetic fields, which can tilt the precession-plane of the muon. The radial and longitudinal magnetic fields were estimated using data from the tracking detector. A preliminary estimation of the radial field resulted in a value of $2.1(6)$~ppm. An in-depth analysis of the vertical angle oscillation yielded a value of the longitudinal field consistent with zero: $-11.8(50.3)$~ppm. Moreover, the first analysis of $\omega_a$ with data from the tracking detector was made. The projected precision will allow an independent cross-check of the \ac{BNL} and Fermilab \R1 determinations of $\omega_a$, using the \R1 to \R3 datasets from the tracking detector. 

At the time of writing, the experiment-wide effort on the analysis of \R1 data is nearing completion, with the announcement of the first result expected in the coming months. This could be a fascinating time for the scientific community: only time will tell if the measurement by the experiment will be the harbinger of new physics! 

\graphicspath{{fig/}}

\appendix

\chapter{Global alignment of the tracking detectors} \label{sec:align_global}
\vspace{-1cm}
\textit{This section on the global alignment briefly summarises the work done by Dr~Horst Friedsam~\cite{Horst}, Dr Leah Welty-Rieger~\cite{Leah} and Dr Joe Price~\cite{Joe}.}

The global alignment established an absolute position of the tracker stations relative to the rest of the experiment. This procedure consisted of two parts: physical survey measurements of the modules by the \textit{Alignment and Metrology Department} at Fermilab, and the implementation of these measurements in the software framework.  

\section{Survey alignment}
\begin{figure}[htpb]
    \centering
    \includegraphics[height=4.5 cm]{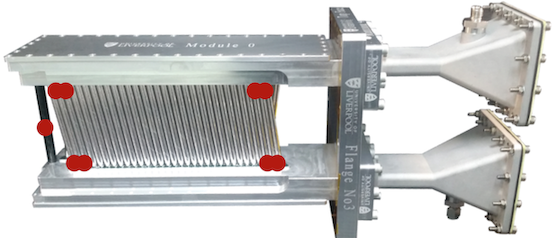}
    \caption[The survey points used in the global alignment]{The survey points used in the global alignment: the centre of the carbon fibre post and top and bottom of first two and last two straws in a layer.}
    \label{fig:SurveyPoints}
\end{figure}

\begin{figure}[htpb]
    \centering
    \subfloat[]{\includegraphics[width=0.48\linewidth]{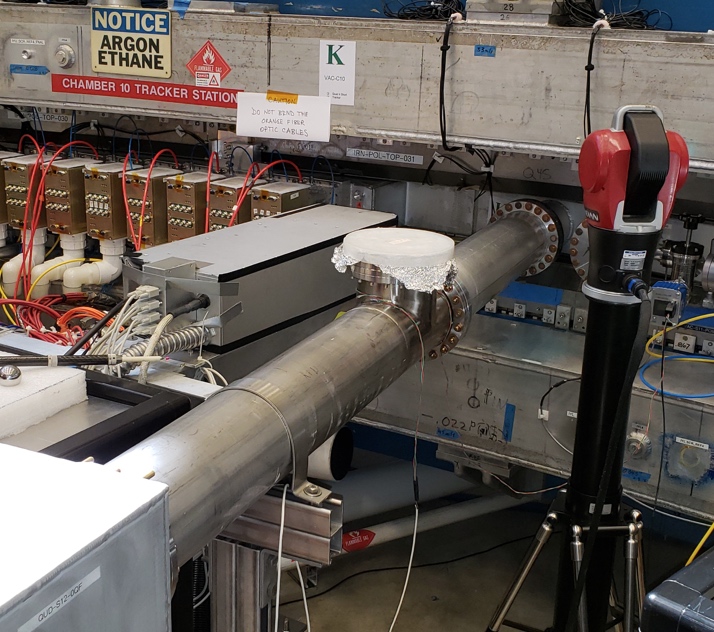}} \hspace{2pt}
    \subfloat[]{\raisebox{7mm}{\includegraphics[width=0.48\linewidth]{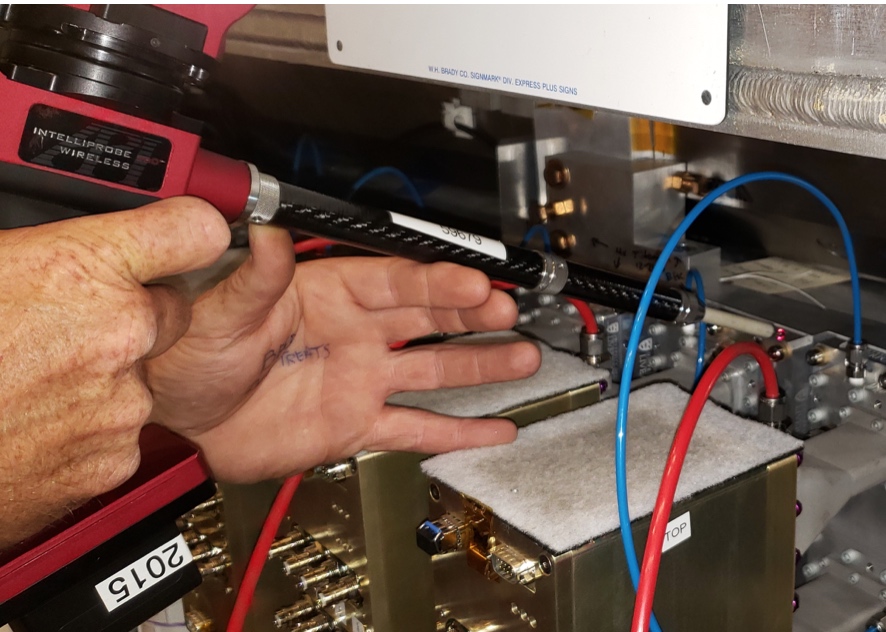}}}
    \caption[The implementation of the survey measurements]{The implementation of the survey measurements: a) the \textit{API Radian Laser Tracker} uses interferometer to accurately measure the vacuum chamber position relative to the rest of the experiment, (b) the \textit{API I-360 Wireless Probe} was used to survey the individual module flanges.}
    \label{fig:laser_align}
\end{figure}

The first step in the survey of the tracker compared measurements of a single tracker module with the design model. Agreement was found within \SI{50}{\micro\metre}, with the survey points used in the comparison indicated in \cref{fig:SurveyPoints}. The design model was then used to transform from a point in a local tracker coordinate system to the global coordinate system.
As the tracker modules are mounted into a vacuum chamber, as shown in \cref{fig:station}, the global alignment requires the radial and vertical positions of the vacuum chambers to be determined. The measurements were performed using the \textit{API Laser Tracker} as shown in \cref{fig:laser_align}. These measurements and the design model was then used to estimate the position of the module flange, the straws themselves, and the carbon fibre post, with the error on the estimated positions determined to be \SI{200}{\micro\metre}. 
\clearpage
\vspace{-0.4cm}
\vspace{-0.5cm}
\section{Global alignment in simulation}
The survey measurements described above resulted in four global corrections per station: a radial translation, a vertical translation, a pitch (radial) rotation, and a roll (vertical) rotation, as summarised in \cref{tab:global}. 
\vspace{-0.1cm}
\begin{table}[htpb] 
  \centering
  \resizebox{\columnwidth}{!}{%
  \begin{tabular}{lcccc}
    \toprule
      &vertical position [\SI{}{\milli\meter}] &vertical angle [rad] &radial position [\SI{}{\milli\meter}]&radial angle [rad]\\ \toprule
    S12 &$-0.55$ &$+4.11 \times 10^{-5}$ &$-1.51$ &$-1.12 \times 10^{-3}$\\ \midrule
    S18 &$-0.61$ &$-4.13 \times 10^{-4}$ &$-2.75$ &$-2.36 \times 10^{-3}$\\ \bottomrule
  \end{tabular}
  } 
  \caption[The corrections from survey alignment]{The corrections from survey alignment applied in the simulation to S12 and S18.}
  \label{tab:global}
\end{table}
\clearpage

\begin{figure}[htpb]
    \centering
    \subfloat[]{\includegraphics[width=0.5\linewidth]{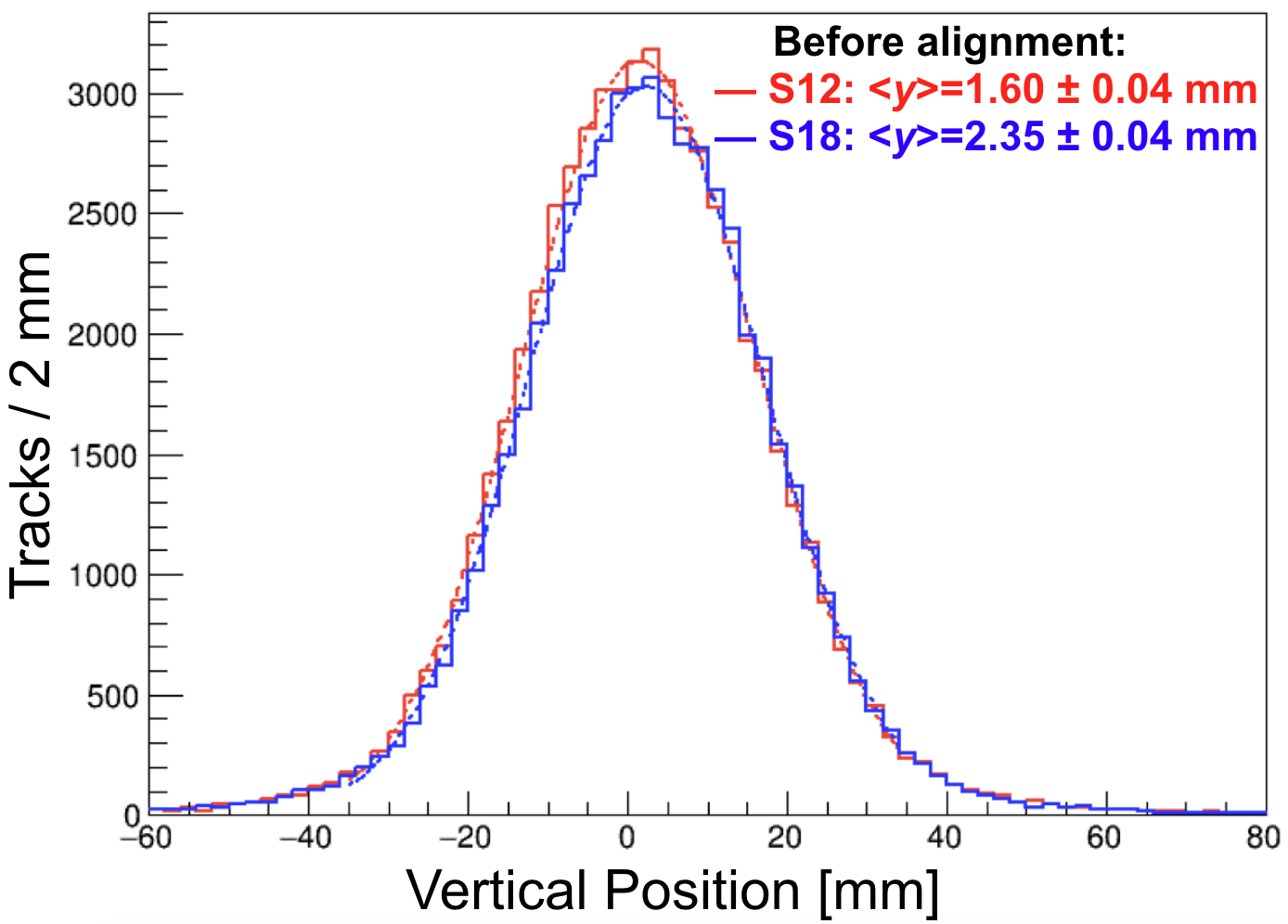}}
    \subfloat[]{\includegraphics[width=0.5\linewidth]{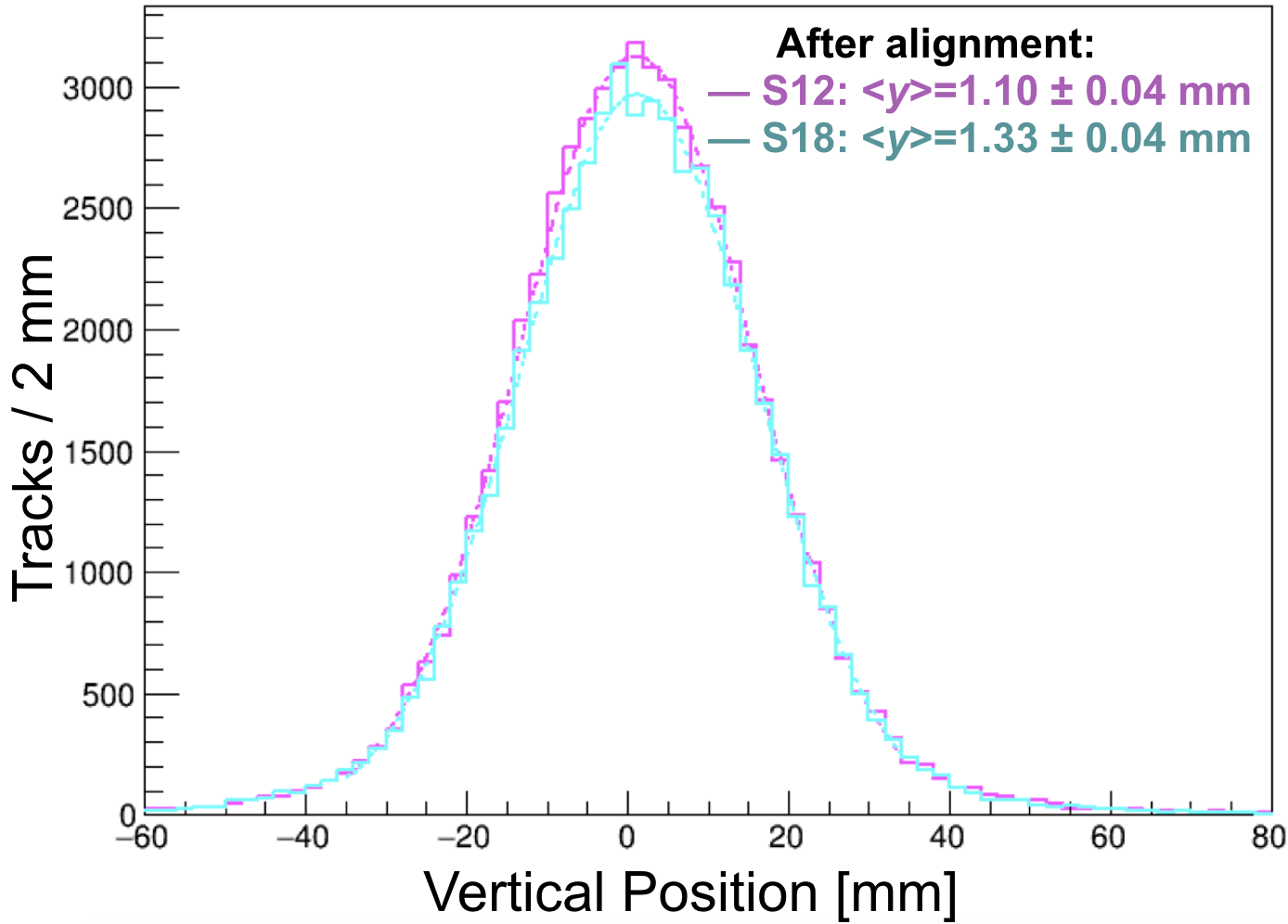}}
    \caption[Mean vertical extrapolated beam positions]{Mean vertical ($\langle y \rangle$) extrapolated beam positions: a) before global alignment, and (b) after global alignment.}
    \label{fig:ver_global}
\end{figure}

A comparison between the extrapolated beam position determined before and after the implementation of the global alignment is shown in \cref{fig:ver_global} for the vertical position. After the alignment, one would not expect the radial or vertical beam distributions from the two stations to match exactly. This is the case due to the closed-orbit effect (see \cref{sc:closed_orbit}), as the two tracker stations see \say{different beams}. The summary of the change in the vertical and radial extrapolated beam position after the global alignment is presented in \cref{tab:global_results}.

\begin{table}[htpb]  
  \centering
  \resizebox{\columnwidth}{!}{%
  \begin{tabular}{cccc}
    \toprule
      $\Delta\langle$vertical$\rangle$ S12 [\SI{}{\milli\meter}] & $\Delta\langle$vertical$\rangle$ S18 [\SI{}{\milli\meter}] &  $\Delta\langle$radial$\rangle$ S12 [\SI{}{\milli\meter}] & $\Delta\langle$radial$\rangle$ S18 [\SI{}{\milli\meter}] \\ \toprule
      $-0.50 \pm 0.06 $ & $-1.02 \pm 0.06$  & $-0.35 \pm 0.06$ & $-0.33 \pm 0.06$ \\ \bottomrule
  \end{tabular}
  } 
  \caption[The change in the extrapolated beam position]{The change in the extrapolated beam position after applying the survey corrections to both stations.}
  \label{tab:global_results}
\end{table}

\section{Global alignment uncertainty contribution}
\begin{figure}[htpb]
    \centering
    \includegraphics[width = 0.6\linewidth]{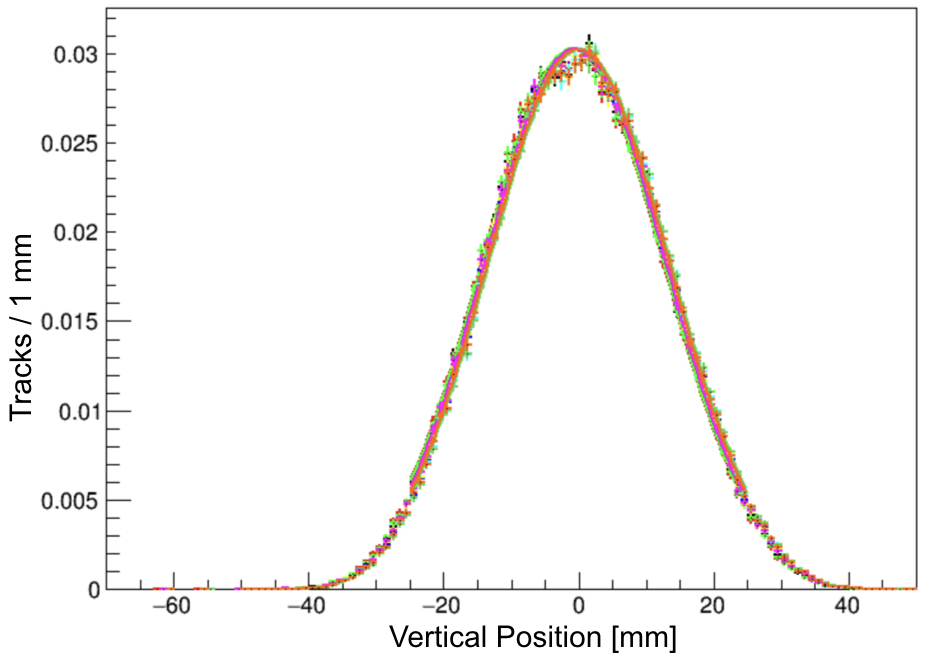}
    \caption[Mahalanobis fits to the extrapolated vertical beam position]{The extrapolated vertical beam position (for run 15922) using parameters from one of the nine Mahalanobis bands. The mean and the width was estimated from a Gaussian fit to each of the nine distributions.}
    \label{fig:LinerMah}
\end{figure}

The Mahalanobis method~\cite{Mah} was used to estimate the uncertainty on the beam extrapolation arising from the uncertainty on the global alignment. Nine Mahalanobis points are derived from the two-parameter straight line-fit. The results from placing the modules in one of these nine positions and determining the extrapolated beam positions are shown in \cref{fig:LinerMah}.

To estimate the uncertainty on the pitch correction from the global alignment, the largest vertical beam width of the eight cases in \cref{fig:LinerMah}, relative to the nominal case,  was used. This change ($\mathrm{\sigma_{dV_{\mathrm{width}}}}$) corresponds to $0.01$~mm. Using \cref{eq:pitch} and the estimated $\mathrm{\sigma_{dV_{\mathrm{width}}}}$ leads to $\Delta C_{\mathrm{pitch}}$ of $0.3$~ppb. This is negligible, compared to the total tracker pitch correction uncertainty of 8.6 ppb (in \R1), as described in \cref{sec:pitch}.

\clearpage
\chapter{Derivations of analytical equations for internal alignment} \label{app:deriv}
\textit{This section provides derivations of the equations used in \cref{ch:align}.}

\section{Straight line-fit in 2D}\label{app:2D}
Derivations of \Crefrange{eq:shear}{eq:res}, describing the effect of a translational misalignment on a line of best fit in 2D is given below, for a line of the form
\begin{equation}
  x=mz+c,
\end{equation}
where $x$ is a hit position (i.e. height) along an infinitesimally thin detector plane, and $z$ is a plane position along the direction of the beam.
A residual, $r_i$, on a detector plane is given by
\begin{equation}
  r = mz_i + c - x_i,
  \label{eq:line_global}
\end{equation}
with the $\chi^2$, as a sum of all the residuals in $P$ planes, defined as
\begin{equation}
\begin{gathered}
  \chi^2 = \sum_i^P\frac{r_i^2}{\sigma^2}=\frac{1}{\sigma^2}\sum_i^P(mz_i + c - x_i)^2 = \\ 
  =\frac{1}{\sigma^2}\sum_i^P(mz_i^2 + c^2 + x_i^2 + 2cx_i + 2mz_ix_i - 2mz_ic),
 \end{gathered}
\end{equation}
for a constant detector resolution, $\sigma$, and a special case of no misalignment and no smearing. The aim is to minimise $\chi^2$ with respect to line parameters $m$ and $c$:
\begin{equation}
  \frac{\partial\chi^2}{\partial m} = \frac{1}{\sigma^2}\sum_i^P(2mz_i^2+2x_iz_i-2z_ic), 
\end{equation}
and
\begin{equation}
  \frac{\partial\chi^2}{\partial c} = \frac{1}{\sigma^2}\sum_i^P(2c+2x_i-2mz_i). 
\end{equation}
Without the loss of generality, one can define the \say{centre-of-mass} of planes to lie at 0 along $z$, so that $\sum_i^Pz_i=0$, which simplifies the above equations to
\begin{equation}
  m=\frac{\sum_i^Pz_ix_i}{\sum_i^Pz_i^2},
  \label{eq:m_mes}
\end{equation}
and 
\begin{equation}
  c=\frac{\sum_i^Px_i}{P}.
\end{equation}
For a realistic case of non-zero \textit{characteristic} misalignment, $M^c_i$, along $x$, and smearing due to the detector resolution, $\delta$, on a plane, one has a hit position defined by
\begin{equation}
   x=m_Tz+c_T+\delta_i + M^c_i,
   \label{eq:x_mis_res+app}
\end{equation}  
where one can define two variables relating the \textit{truth} (e.g. $m_T$) and \textit{measured} line parameters due to the resolution and misalignment effects, substituting for the measured slope, $m$, from \cref{eq:m_mes} into \cref{eq:line_global}
\begin{equation}
   \delta m = m_T-m = \frac{-\sum_i^P(\delta_iz_i+M_i^cz_i)}{\sum_i^Pz_i^2},
   \label{eq:dm}
 \end{equation} 
 and similarly for the intercept
 \begin{equation}
  \delta c = c_T - c = \frac{-\sum_i^P(\delta_i+M_i^c)}{P}.
  \label{eq:dc}
\end{equation} 
Using \cref{eq:x_mis_res+app,eq:dm,eq:dc} the residual in a plane can be defined as
\begin{equation}
  r_i = mz_i + c - x_i = \delta mz_i + \delta c + \delta_i + M^c_i.
\end{equation}
It is now possible to derive an analytical form of the mean residual, $\langle r_p \rangle$ in a detector layer $p$, over $N\rightarrow\infty$ tracks
\begin{equation}
\begin{gathered}
  \langle r_p \rangle = \frac{\sum_n^Nr_{n,p}}{N} = \frac{\sum_n^N(\delta m_nz_p+\delta c_n+\delta_{p,n}+M^c_p)}{N} = \\
  =M^c_p - \frac{\sum_i^PM_i^c}{P} - \frac{z_p\cdot \sum_{i}^{P}{ M^c_{i} \cdot z_i}}{\sum_{i}^{P}{z_i^2}},
  \end{gathered}
  \label{eq:shear_app}
\end{equation}
where $\sum_n^N\delta_{p,n}\rightarrow0$ was used. The second term in \cref{eq:shear_app} gives a hint of an effect of misalignment in other layers on the residual in a particular layer $p$. The analytical form of the standard deviation of the distribution of residuals in a layer $p$, over $N\rightarrow\infty$ tracks is given by
\begin{equation}
\begin{gathered}
  \sigma^2_{r_p} = \frac{\sum_n^N(r_{p,n}- \langle r_p \rangle)^2}{N} = \\ 
  = \frac{1}{N}\left(\delta m_n z_p + \delta c_n + \delta_{p,n} + \frac{\sum_{i}^{P}M^c_i}{P} + z_p\frac{\sum_{i}^{P}M^c_iz_i}{\sum_{i}^{P}z_i}  \right)^2 = \sigma^2  - \frac{z_p^2\sigma^2}{\sum_i^Pz_i^2} - \frac{\sigma^2}{P}.
    \end{gathered}
\end{equation}
Finally, an analytical equation for $\chi^2$ on a fit to a misaligned detector for $N\rightarrow\infty$ tracks can be derived
\begin{equation}
\begin{gathered}
  \chi^2  = \frac{1}{N} \sum_n^N \frac{\sum_i^Pr_i^2}{\sigma^2}=\frac{1}{N\sigma^2} \sum_n^N \sum_i^P (\delta m_nz_i + \delta c_n + \delta_{i,n} + M^c_i)^2 = \\
  = P - 2 +  \frac{
    \sum_{i=1}^{P}(M^{c}_i)^2-\frac{\sum_{i=1}^{P}{(M^{c}_i)^2}}{P} - 
    \frac{2 \cdot \sum_{i=1}^{P}M^{c}_i \cdot z_i}{\sum_{i=1}^{P}(z_i)^2}}
{\sum_{i=1}^{P}(\sigma_i^{\mathrm{det}})^2}.
     \end{gathered}
\end{equation}
The derived equations for $\langle r_p \rangle$, $\sigma^2_{r_p}$, and $\chi^2$ apply for a 2D case of a straight line-fit to a misaligned detector planes along $x$, with a given detector resolution on hits. Nevertheless, the derived equations are quite instructive of the effect of misalignment, and give correct analytical predictions for a 2D toy-model, as shown in \cref{fig:res,,fig:chi2}.

\section{Residual derivatives with circle-fit in 2D}\label{app:2D_rot}
In the case of a circle-fit using straight tracks the residual, $r$, is given by
\begin{equation}    
r = \mathrm{DCA}(x,m,c) - h = \frac{ |c+mz-x| }  { \sqrt{m^2+1} } - h,
\end{equation}
with two local derivatives,
\begin{equation}
\frac{\partial r}{\partial c} = \frac{ c+mz-x }  { \sqrt{m^2+1} |c+mz-x| },
\label{eq:2D_DLC1}
\end{equation}
and
\begin{equation}
\frac{ \partial r}{\partial m} = \frac{ (m^2+1)z(c+mz-x) - m|c+mz-x|^2 }{ (m^2+1)^{3/2}|c+mz-x|  }.
\label{eq:2D_DLC2}
\end{equation}
The action of an anticlockwise rotation is given by
\begin{equation}
\begin{pmatrix}z_m'\\x_m'\\\end{pmatrix}=\begin{pmatrix}\cos \phi &-\sin \phi \\\sin \phi &\cos \phi \\\end{pmatrix} \begin{pmatrix}z_m\\x_m\\\end{pmatrix} = \begin{pmatrix}z_m\cos \phi -x_m\sin \phi \\z_m\sin \phi +x_m\cos \phi \\\end{pmatrix}.
\end{equation}
The transformation back to the global coordinates is given by
\begin{equation}
\begin{pmatrix}z\\x\\\end{pmatrix}=\begin{pmatrix}z_m'+z^{\mathrm{centre}}\\x_m'+x^{\mathrm{centre}}\\\end{pmatrix}=\begin{pmatrix}z_m\cos \phi -x_m\sin \phi + z^{\mathrm{centre}} \\z_m\sin \phi +x_m\cos \phi + x^{\mathrm{centre}} \\\end{pmatrix}. 
\label{eq:global}
\end{equation}
The equation for a residual, under a 2D rotation, can be written using \cref{eq:global}, and is given by
\begin{equation}
\begin{gathered}
r = DCA(z(\phi),x(\phi),m,c) - r = \frac{ |c+mz(\phi)-x(\phi)| }  { \sqrt{m^2+1} } -r = \\
= \frac{ |c+mz_m\cos \phi -mx_m\sin \phi + mz^{\mathrm{centre}} -z_m\sin \phi - x_m\cos \phi - x^{\mathrm{centre}}| }  { \sqrt{m^2+1} } -r.
\label{eq:rot2D_res}
\end{gathered}
\end{equation}

\clearpage
Ultimately, an expression for $\frac{\partial r}{\partial\phi}$ in terms of global and local parameters is used 
\begin{equation}
\frac{\partial r}{\partial\phi} = \frac{\partial \mathrm{DCA}(z(\phi),x(\phi),m,c)}{\partial\phi}.
\end{equation}
This can be obtained using the chain rule
\begin{equation}
\frac{\partial r}{\partial\phi} = \frac{\partial \mathrm{DCA}(\phi)}{\partial\phi} = \frac{\partial \mathrm{DCA}(d)}{\partial d}\frac{\partial d(\phi)}{\partial \phi},  \label{eq:partialRM}
\end{equation}
where $d$ represents a measurement (e.g. straw displacement in $z$ or $x$). Using \cref{eq:partialRM} one can write the equation for the residual derivative expressed without the use of the truth parameters (i.e. $\phi$)
\small
\begin{equation}
\begin{gathered}
\frac{\partial r}{\partial\phi} = \frac{\partial r}{\partial z}\frac{\partial z}{\partial \phi} + \frac{\partial r}{\partial x}\frac{\partial x}{\partial \phi} = \frac{ m(c+mz-x) }  { \sqrt{m^2+1} \cdot |c+mz-x| } \times (-z_m\sin \phi - x_m\cos \phi) \ + \\ \frac{ c+mz-x }  { \sqrt{m^2+1} \cdot |c+mz-x| } \times (z_m\cos \phi - x_m\sin \phi),
\end{gathered}
\label{eq:2D_drdphi}
\end{equation}
\normalsize
at this point, a substitution from \cref{eq:global} can be used to simplify the expression to
\small
\begin{equation}
\begin{gathered}
\frac{\partial r}{\partial\phi} = \frac{ m(c+mz-x) }  { \sqrt{m^2+1} \cdot |c+mz-x| } \times (-x'_m) + \frac{ c+mz-x }  { \sqrt{m^2+1} \cdot |c+mz-x| } \times (z'_m) = \\
\frac{\partial r}{\partial z} (-x + x^{\mathrm{centre}}) + \frac{\partial r}{\partial x} (z - z^{\mathrm{centre}}),
\end{gathered}
\label{eq:2D_drdphi_2}
\end{equation}
\normalsize
where all inputs come either from measurements or assumption of the ideal geometry. 

\clearpage
\section{Residual derivatives in 3D}\label{app:3D}
The straw wire can be parametrised as 
\begin{equation}
\boldsymbol{r}_W(s) = \begin{pmatrix}x\\y\\z\\\end{pmatrix} = \begin{pmatrix}x_W + m_W s\\ s\\  z_W\\\end{pmatrix} = \boldsymbol{u}s+\boldsymbol{r}_W^0=\begin{pmatrix} m_W \\ 1\\  0\\\end{pmatrix}s + \begin{pmatrix}x_W  \\0 \\  z_W\\\end{pmatrix},
\label{eq:wire}
\end{equation}
as the position on the straw is constant with $z$, and the slope in $x$ is due to the \textit{stereo angle} on the straw wire, $\theta_W$, of $\sim 7.5^{\circ}$ with $m_W = \tan(\theta_W)$. A geometrical approach can be taken to express the requirements on the \ac{DCA} from \cref{eq:DCA_geom_req} as
\begin{equation}
\mathrm{DCA} = \frac{|\boldsymbol{u} \times \boldsymbol{v} \cdot (\boldsymbol{r}^0_W - \boldsymbol{r}^0_T)|}{|\boldsymbol{u} \times \boldsymbol{v}|} = |(\boldsymbol{r}^0_W - \boldsymbol{r}^0_T)\cdot \boldsymbol{n}| ,
\label{eq:DCA_3D}
\end{equation}
which projects the distance between the intercept of the lines, to the normal, $\boldsymbol{n}$, between the two lines.  

Using the track parametrisation from \cref{eq:track} the DCA in 3D is given by
\small
\begin{equation}
\openup 1\jot
\begin{gathered}
DCA(x_T, y_T, m_x, m_y, x_W, z_W, m_W) = \frac{|\begin{pmatrix} 1 \\ -m_W \\ m_Wm_y - m_x  \end{pmatrix} \cdot
\begin{pmatrix} x_W-x_T \\ -y_T \\ z_W  \end{pmatrix} |}{|\begin{pmatrix} 1 \\ -m_W \\ m_Wm_y - m_x  \end{pmatrix}|}  = \\
= \frac{| m_ym_Wz_W-m_xz_W + m_Wy_T + x_W - x_T |}{\sqrt{(m_W m_y-m_x)^2+(m_W)^2+1}},
\label{eq:DCAGeom_app}
\end{gathered}
\end{equation}
and the residual is simply 
\begin{equation}
r=\mathrm{DCA(x_T, y_T, m_x, m_y, x_W, z_W, m_W)} - h.
\end{equation}
\normalsize
The four local derivatives $\frac{\partial r}{\partial x_T}$, $\frac{ \partial r}{\partial y_T}$, $\frac{ \partial r}{\partial m_x}$, and $\frac{ \partial r}{\partial m_y}$ are then given by:\\
For the \say{$x$-intercept}
\small
\begin{equation}
\frac{\partial r}{\partial x_T} = \frac{m_W m_y z_W+m_W y_T-m_x z_W-x_T+x_W}{-|m_W m_y z_W+m_W y_T-m_x z_W-x_T+x_W|\sqrt{\left(m_W m_y-m_x\right){}^2+m_W^2+1}}
, \label{eq:dlc1}
\end{equation}
\normalsize
for the \say{$y$-intercept}
\small
\begin{equation}
\begin{gathered}
\frac{ \partial r}{\partial y_T} = \frac{m_W (m_W m_y z_W+m_W y_T-m_x z_W-x_T+x_W)}{|m_W m_y z_W+m_W y_T-m_x z_W-x_T+x_W|\sqrt{\left(m_W m_y-m_x\right){}^2+m_W^2+1}}. \label{eq:dlc2}
\end{gathered}
\end{equation}
\normalsize
The \say{slope} (defined in \cref{eq:m_i})
\small
\begin{equation}
\begin{gathered}
\frac{ \partial r}{\partial m_x} = \frac{\left(m_W m_y-m_x\right) |m_W m_y z_W+m_W y_T-m_x z_W-x_T+x_W|}{\left(\left(m_W m_y-m_x\right){}^2+m_W^2+1\right){}^{3/2}}-\\-
\frac{z_W(m_W m_y z_W+m_W y_T-m_x z_W-x_T+x_W)}{|m_W m_y z_W+m_W y_T-m_x z_W-x_T+x_W|\sqrt{\left(m_W m_y-m_x\right){}^2+m_W^2+1}}, \label{eq:dlc3}
\end{gathered}
\end{equation}
\normalsize
and
\small
\begin{equation}
\begin{gathered}
\frac{ \partial r}{\partial m_y} = \frac{m_W z_W(m_W m_y z_W+m_W y_T-m_x z_W-x_T+x_W)}{|m_W m_y z_W+m_W y_T-m_x z_W-x_T+x_W|\sqrt{\left(m_W m_y-m_x\right){}^2+m_W^2+1}}-\\-
\frac{m_W \left(m_W m_y-m_x\right) |m_W m_y z_W+m_W y_T-m_x z_W-x_T+x_W|}{\left(\left(m_W m_y-m_x\right){}^2+m_W^2+1\right){}^{3/2}}. \label{eq:dlc4}
\end{gathered}
\end{equation}
\normalsize
And the two global derivatives for radial and vertical translations are 
\small
\begin{equation}    
\frac{\partial r}{\partial x_W} = - \frac{\partial r}{\partial x_T},
\end{equation}
\normalsize
and
\small
\begin{equation}    
\frac{\partial r}{\partial y_W} = \frac{\partial r}{\partial y_T}.
\end{equation}
\normalsize
The 3D rotation matrix is given by
\begin{equation}
\scriptstyle
\boldsymbol{R}=\begin{pmatrix} \cos\phi\cos\psi & \cos\theta\sin\psi + \sin\theta\sin\phi\cos\psi & \sin\theta\sin\psi - \cos\theta\sin\phi\cos\psi \\
-\cos\phi\sin\psi & \cos\theta\cos\psi - \sin\theta\sin\phi\sin\psi & \sin\theta\cos\psi + \cos\theta\sin\phi\cos\psi\\
\sin\phi & - \sin\theta\cos\phi & \cos\theta\cos\phi\\\end{pmatrix},
\label{eq:rot_matrix}
\end{equation}
\clearpage
where $\theta$, $\phi$ and $\psi$ are the Euler angles. There are three possible rotations (see \cref{fig:Rotation}) around the centre of the tracker module. A rotation around the $y$-axis is the same as considered previously in \cref{eq:2D_drdphi}, but needs to be extended to a 3D geometry. With the constraint that a point along the straw will have the same vertical height ($y$) before and after the above rotation with the derivatives for the anticlockwise rotation $\phi$ along the $y$-axis given by
\small
\begin{equation}
\begin{pmatrix}x\\y\\z\end{pmatrix}=\begin{pmatrix}x_m'+x^{\mathrm{centre}}\\ y_m + y^{\mathrm{centre}} \\z_m'+z^{\mathrm{centre}}\\\end{pmatrix}=\begin{pmatrix}z_m\sin \phi +x_m\cos \phi + x^{\mathrm{centre}} \\ y_m + y^{\mathrm{centre}} \\ z_m\cos \phi -x_m\sin \phi + z^{\mathrm{centre}} \\\end{pmatrix}. \label{eq:global_3D}
\end{equation}
\normalsize
The global derivatives for the anticlockwise rotation $\phi$ along the $y$-axis is
\small
\begin{equation}
   \frac{\partial r}{\partial \phi} =  \frac{\partial r}{\partial z_W} (-x_W + x^{\mathrm{centre}}) + \frac{\partial r}{\partial x_W} (z_W - z^{\mathrm{centre}}).    
\end{equation}
\normalsize
and similarly for the rotation along the $z$-axis
\small
\begin{equation}
\frac{ \partial r}{\partial \psi} = \frac{ \partial r}{\partial x_W} (-y_W + y_{\mathrm{centre}}) + \frac{ \partial r}{\partial y_W} (x_W - x_{\mathrm{centre}}), 
\end{equation}
\normalsize
and $x$-axis
\small
\begin{equation}
\frac{ \partial r}{\partial \theta} = \frac{ \partial r}{\partial y_W} (-z_W + z_{\mathrm{centre}}) + \frac{ \partial r}{\partial z_W} (y_W - y_{\mathrm{centre}}).
\end{equation}
\normalsize
The fifth local derivative in the presence of a magnetic field is given by
\small
\begin{equation}
\frac{ \partial r}{\partial \kappa} = \frac{ \partial r}{\partial \left(\frac{1}{p_z}\right)} = \frac{ \partial r}{\partial m_x}\frac{ \partial m_x}{\partial \kappa} + \frac{ \partial r}{\partial m_y}\frac{ \partial m_y}{\partial \kappa} = \frac{ \partial r}{\partial m_x}p_x + \frac{ \partial r}{\partial m_y}p_y. 
\label{eq:dlc5}
\end{equation}
\normalsize
The vector containing derivatives of the residuals with respect to the fitted track parameters is given by
\small
\begin{equation}
\frac{\partial r}{\partial \boldsymbol{b}} = \begin{pmatrix} \frac{\partial r}{\partial x_T} \\ \frac{\partial r}{\partial y_T} \\ \frac{\partial r}{\partial m_x} \\ \frac{\partial r}{\partial m_y} \\ \frac{ \partial r}{\partial \kappa}  \end{pmatrix}. 
\end{equation}
\normalsize

\clearpage
\chapter{Control and steering of \pede} \label{app:steer}
\textit{This section summarises the steering and constraint options used in \pede.}

The \pede steering file is given by
\begin{lstlisting}[basicstyle=\scriptsize,language=Bash]
* g-2 Tracker Alignment: PEDE Steering File
ConstraintFile.txt ! Constraints text file
Cfiles ! Following bin files are Cfiles
Data.bin ! Binary data file
method inversion 10 0.01 
\end{lstlisting}
The constraint equation is given by
\begin{equation}
        c = \sum_i^N l_i\cdot f_l,
\end{equation}
where $c$ is is the constraint value, $l$ is the global parameter label, $f$ is the parameter factor, and the summation is taken over the parameters contributing to the constraint. The constraints given in \cref{sc:align_constr} can be written as the inputs to \pede as follows:
\begin{enumerate} \itemsep -2pt 
    \item Constraint on the overall radial translation 
    \begin{equation}
        0 = \sum_{i=1}^N l_i\cdot 1 = 1211 \cdot 1 + 1221 \cdot 1 + ... 1281 \cdot 1,
        \label{eq:pede_rad}
    \end{equation}
    where, for example, label $1281$ corresponds to station number 12, module number 8 and parameter number 1 (i.e.~radial shift). The factor of $1$ simply means equal weighting for all parameters.
    \clearpage 

    \cref{eq:pede_rad} is written in the \pede format as follows
\begin{lstlisting}[basicstyle=\scriptsize,language=Bash]
! Radial translational fixing (no overall movement):
Constraint 0
1211 1
1221 1
1231 1
1241 1
1251 1
1261 1
1271 1
1281 1
\end{lstlisting}
    \item Constraint on the overall vertical rotation
    \begin{equation}
        0 = \sum_{i=1}^N l_i\cdot (N+1-2\cdot i)  = 1212 \cdot 7 + 1222 \cdot 5 + ... 1282 \cdot -7,
    \end{equation}
    \vspace{-1cm}
\begin{lstlisting}[basicstyle=\scriptsize,language=Bash]
! Vertical rotational constraint 
! around the centre of the station:
Constraint 0
1212 7
1222 5
1232 3
1242 1
1252 -1
1262 -3
1272 -5
1282 -7
\end{lstlisting}
    \item Constraint on the radial bowing effect
     \begin{equation}
        0 = \sum_i^N l_i\cdot (N+1-2\cdot i)^2 = 1211 \cdot 49 + 1221 \cdot 25 + ... 1281 \cdot 49, \label{eq:bowing}
    \end{equation}
    \vspace{-1cm}
\begin{lstlisting}[basicstyle=\scriptsize,language=Bash]
! Bowing effect: parabloic radial term 
! describing the track curvature 
! around the centre of the station:
Constraint 0
1211 49
1221 25
1231 9
1241 1
1251 1
1261 9
1271 25
1281 49
\end{lstlisting}
\end{enumerate}

\clearpage
\chapter{Ancillary analysis plots and derivations} \label{app:plots}
\textit{This section contains accompanying plots and derivations to \cref{ch:wiggle,ch:edm}.}

\section{Ancillary wiggle plots}\label{sec:ancillary_plots_wa}
\begin{figure}[htpb]
    \centering
    \includegraphics[width=0.69\linewidth]{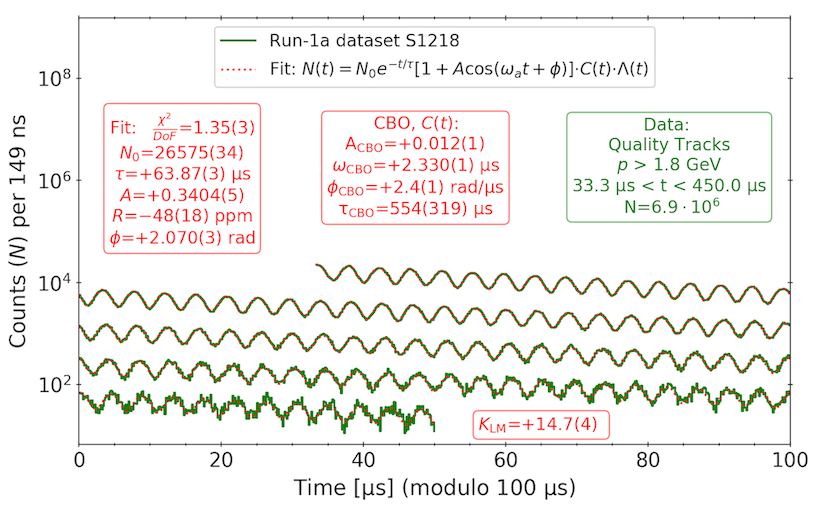}
    \caption[Ten-parameter fit in the Run-1a dataset]{Ten-parameter fit in the Run-1a dataset.}
    \label{fig:1a_10par}
\end{figure}

\begin{figure}[htpb]
    \centering
    \includegraphics[width=0.69\linewidth]{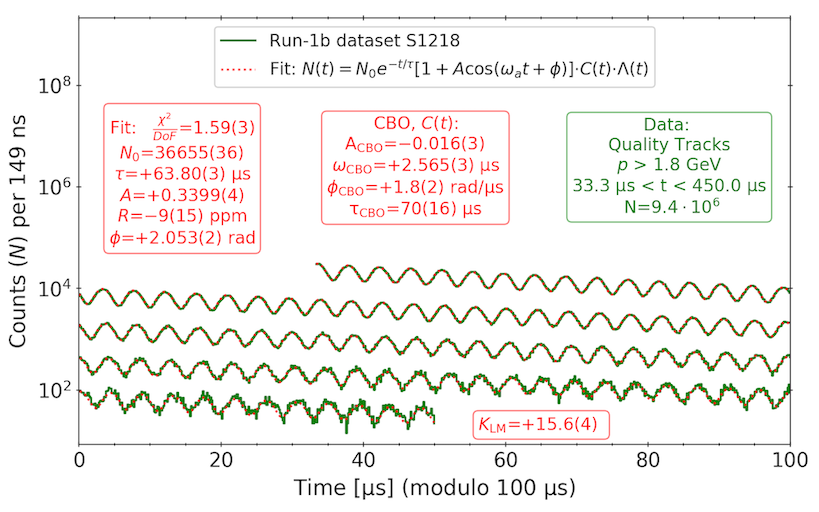}
    \vspace{-0.5cm}
    \caption[Ten-parameter fit in the Run-1b dataset]{Ten-parameter fit in the Run-1b dataset.}
    \label{fig:HK_10par}
\end{figure}

\begin{figure}[htpb]
    \centering
    \includegraphics[width=0.69\linewidth]{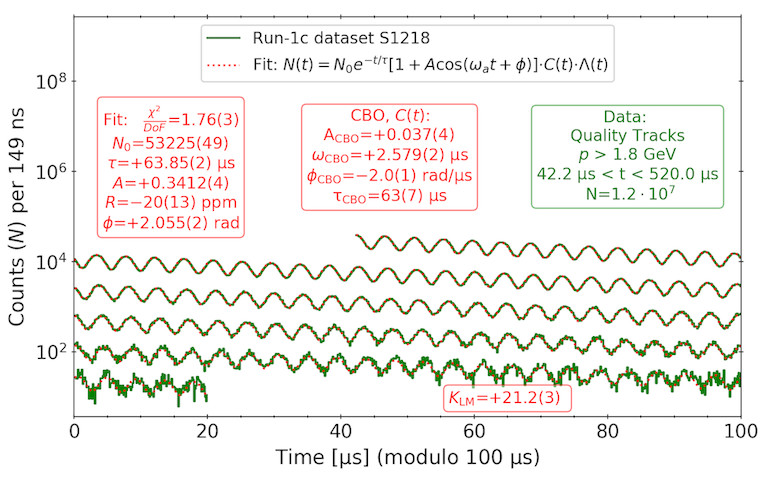}
    \vspace{-0.5cm}
    \caption[Ten-parameter fit in the Run-1c dataset]{Ten-parameter fit in the Run-1c dataset.}
    \label{fig:1c_10par}
\end{figure}

\begin{figure}[htpb]
    \centering
    \includegraphics[width=0.69\linewidth]{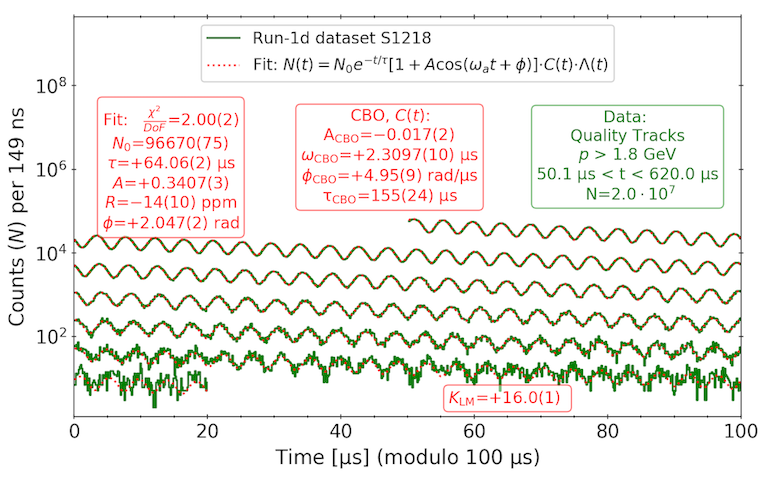}
    \vspace{-0.5cm}
    \caption[Ten-parameter fit in the Run-1d dataset]{Ten-parameter fit in the Run-1d dataset.}
    \label{fig:EG_10par}
\end{figure}

Wiggle plots and ten-parameter fits across the four datasets: Run-1a (\cref{fig:1a_10par}), Run-1b (\cref{fig:HK_10par}), Run-1c (\cref{fig:1c_10par}), and Run-1d (\cref{fig:EG_10par}).

\clearpage

\vspace{-0.6cm}
\vspace{-0.6cm}

\clearpage
\section{Momentum cuts used in the wiggle plots}\label{sub:momentum_cuts_wiggle}
Imposing a momentum cut, $p_{\mathrm{min}}$, on data in \cref{fig:S1218_5par} is motivated by the muon decay asymmetry ($A$) and the number of tracks ($N$) used in the fit, such that the optimal value of $p_{\mathrm{min}}$ corresponds to the quantity $NA^2$ reaching the maximum~\cite{BNL_stats}. That is, the statistical uncertainty on $\omega_a$ (or $R$) is minimised when $NA^2$ is maximised.

\begin{figure}[htpb]
    \centering
    \subfloat[]{\includegraphics[width=0.42\linewidth]{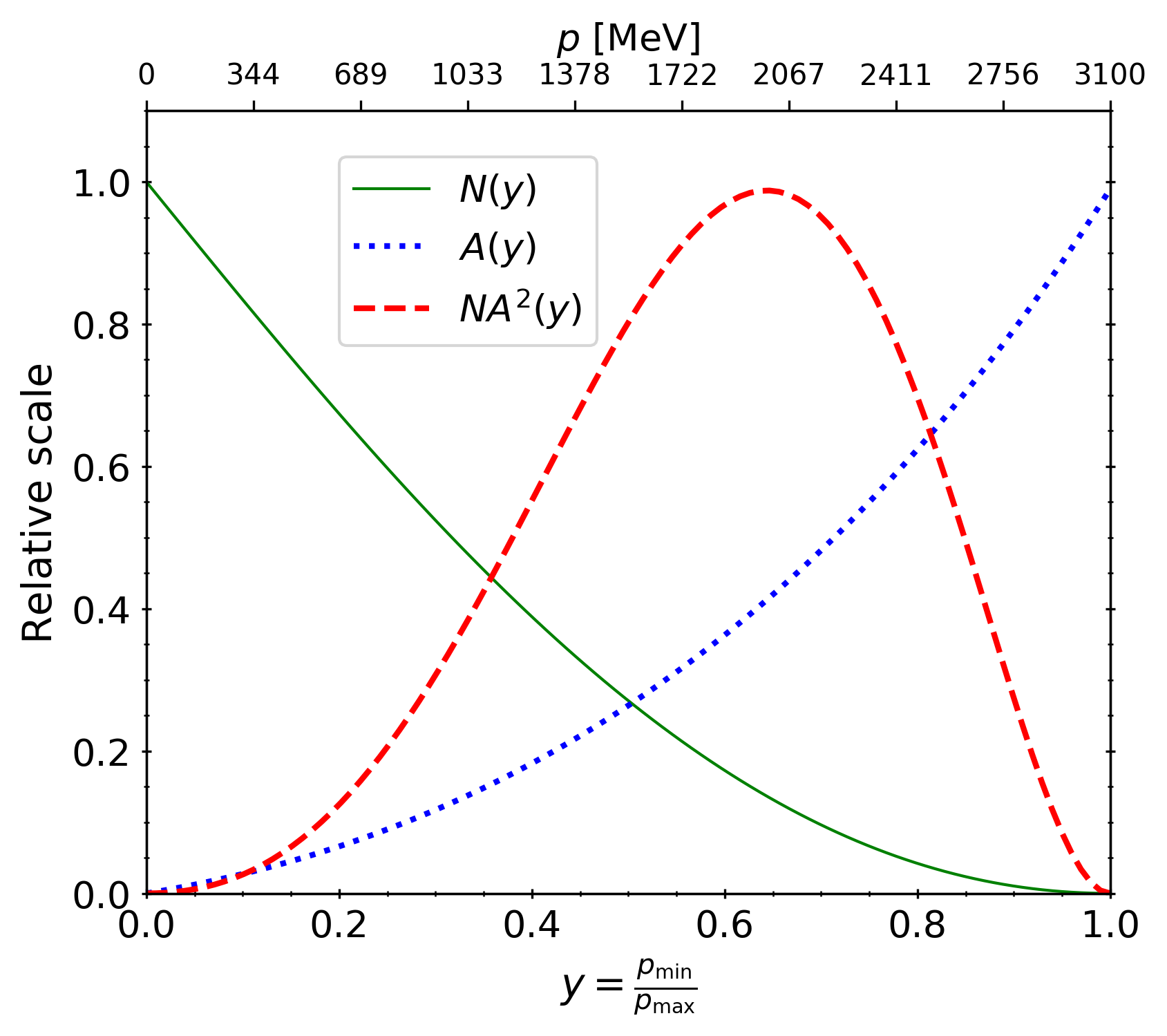}\label{fig:NA2_1}}
    \subfloat[]{\raisebox{3mm}{\includegraphics[width=0.44\linewidth]{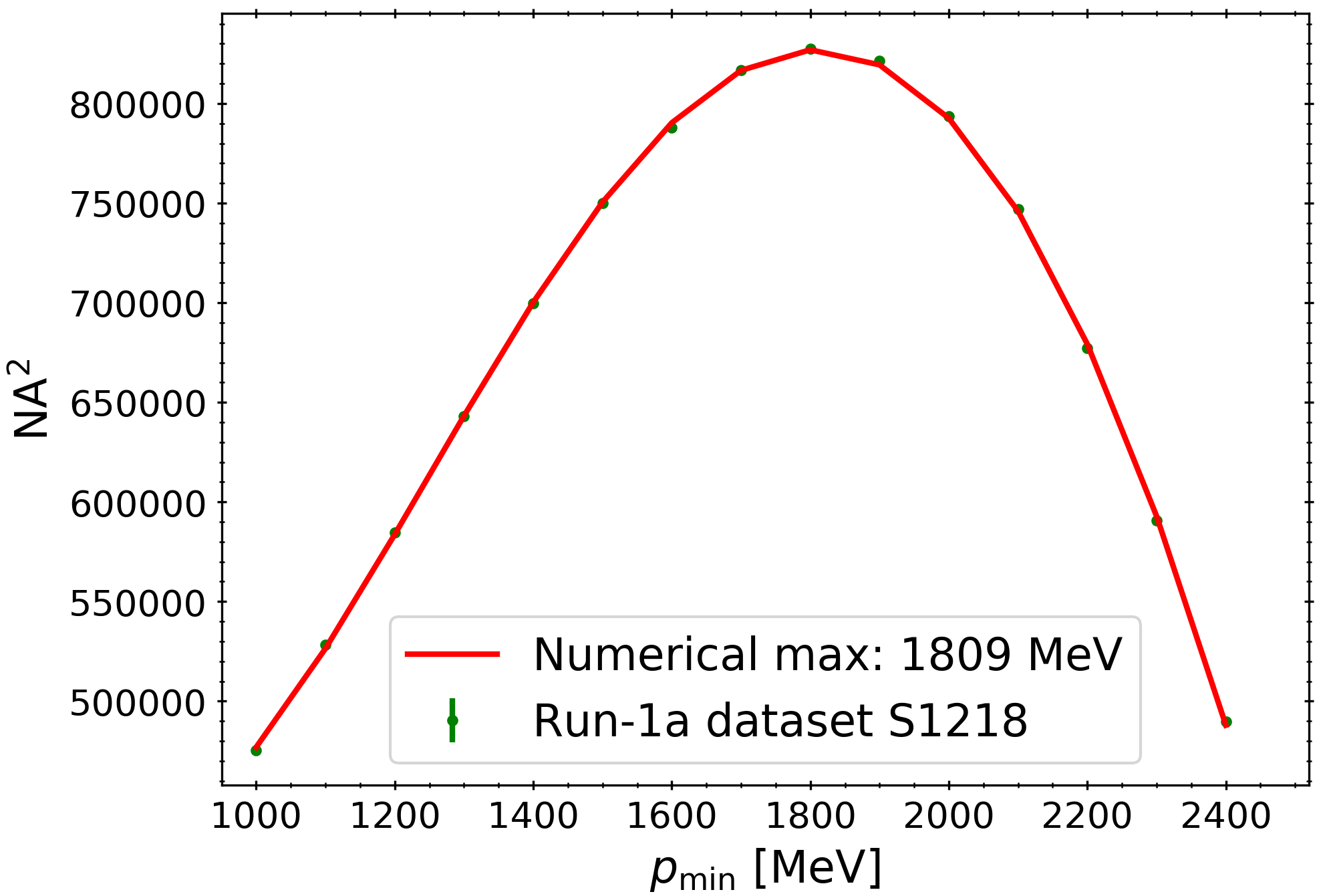}\label{fig:mom_1}}}
    \vspace{-0.2cm}
    \caption[The $NA^2$ term change with the track momentum]{(a) The analytical change in $NA^2$ terms as the function of $y$ in the lab-frame, not including the effect of the detector acceptance. (b) The measured change in $NA^2$ with $p_{\mathrm{min}}$, as determined from the Run1-a dataset.}
\end{figure}

If the momentum fraction, $y$, is defined to be
\begin{equation}
  y = \frac{p_{\mathrm{min}}}{p_{\mathrm{max}}},
\end{equation}
then, in the lab-frame, both $N$ and $A$ can be expressed~\cite{FNAL_TDR} as
\begin{equation}
  N(y) \propto (y-1)^2(-y^2+y+3),
\end{equation}
and
\begin{equation}
  A(y) = \frac{y(2y+1)}{-y^2+y+3}.
\end{equation}
The above expressions are plotted in \cref{fig:NA2_1}. Analytically, the optimal asymmetric momentum cut ($p_{\mathrm{min}}$) occurs at $y\sim0.6$ or $p_{\mathrm{min}} \sim 1800$~MeV. This was empirically verified using the Run1-a dataset, with the optimal momentum cut determined in \cref{fig:mom_1}.
\vspace{-0.2cm}  
\clearpage

\section{Lorentz boost of the tilt angle}\label{sec:lorentz_boost}
\begin{figure}[htpb]
    \centering
    \includegraphics[width = 0.8\linewidth]{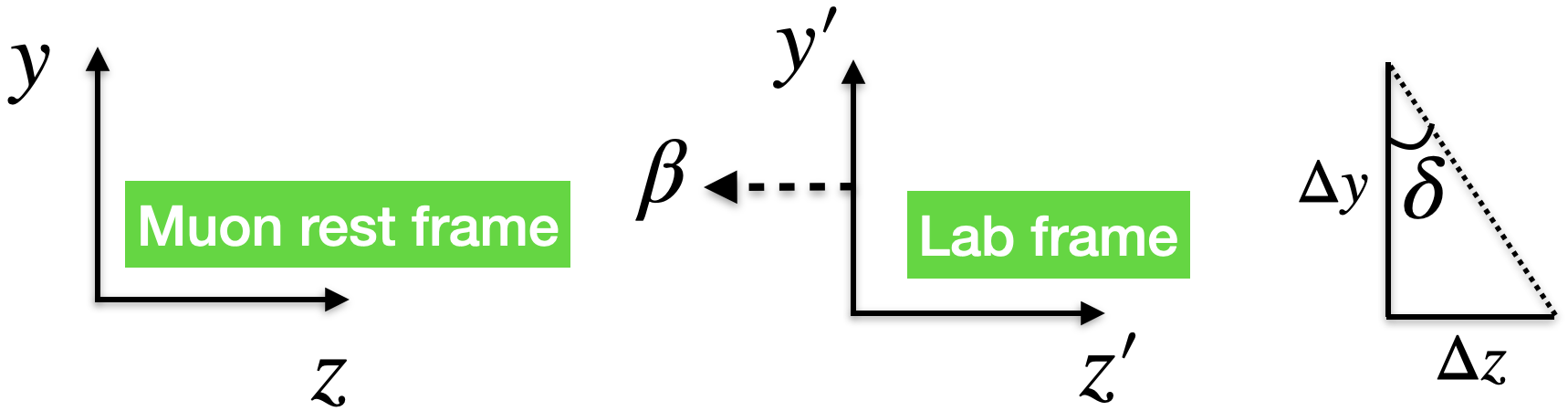}
    \caption[Lorentz boost of the tilt angle]{The muon rest-frame and the lab-frame (i.e. the detector-frame) are shown. The coordinate components of the lab-frame are \say{primed} ($'$) for clarity.} 
    \label{fig:boost}
\end{figure}

The tilt angle of the muon spin precession-plane is reduced by the relativistic effect of length contraction~\cite{Joe_Saskia}. This is demonstrated in~\cref{fig:boost}.

The tilt angle, in the muon rest-frame, is given by
\begin{equation}
  \tan(\delta) = \frac{\Delta z}{\Delta y},
\end{equation}
which, in the lab-frame, is equivalent to
\begin{equation}
  \Delta z ' = \frac{\Delta z}{\gamma},
\end{equation}
and
\begin{equation}
  \Delta y ' = \Delta y.
\end{equation}
These allow for the tilt angle in the lab-frame to be expressed as
\begin{equation}
  \tan(\delta ') = \frac{\Delta z '}{\Delta y '} = \frac{\Delta z}{\gamma\Delta y}.
\end{equation}
Therefore, 
\begin{equation}
\delta ' = \tan^{-1}\left(\frac{\tan(\delta)}{\gamma}\right).
\end{equation}


\addcontentsline{toc}{chapter}{Bibliography}
\begin{spacing}{1.0}
\small
\bibliography{ref} 
\normalsize
\end{spacing}

\end{document}